\documentclass[english,a4paper,12pt,twoside]{book}
\usepackage[english]{babel}
\usepackage[utf8]{inputenc}
\usepackage{amsmath}
\usepackage{amsfonts}
\usepackage{amssymb}
\usepackage{amsthm}
\usepackage{cancel}
\usepackage{slashed}
\usepackage{graphics}
\usepackage{graphicx}
\usepackage{fancyhdr}
\usepackage{multirow}
\usepackage{hyperref}
\usepackage{fancyhdr}
\usepackage{bookmark}
\usepackage{bm}
\usepackage[left=3cm,top=2.25cm,bottom=2.75cm,right=2.5cm,includehead]{geometry}
\usepackage[titletoc]{appendix}
\usepackage{youngtab}
\usepackage{dsfont}
\usepackage{tikz}
\usetikzlibrary{shapes,arrows}
\usepackage{bbm}
\usepackage{mcite}
\usepackage{cite}
\usepackage{verbatim}
\usepackage{indentfirst}

\usetikzlibrary{shapes,snakes,shadows,arrows}
\def\ellip{(0,0) ellipse (12 and 6)}
\def\ellipp{(2,0) ellipse (6 and 3)}
\tikzstyle{P_3} = [draw,very thick]
\tikzstyle{P_4} = [draw,fill=blue!20,very thick]

\def\re#1{\text{Re}\left( #1 \right)}
\def\im#1{\text{Im}\left( #1 \right)}
\def\lie#1{\text{Lie } #1}
\def\diag#1{\text{diag}\left( #1 \right)}

\def\0{\mathbf{0}}
\def\1{\mathbf{1}}

\let\origdoublepage\cleardoublepage
\newcommand{\clearemptydoublepage}{%
  \clearpage
  {\pagestyle{empty}\origdoublepage}%
}
\let\cleardoublepage\clearemptydoublepage

\fancypagestyle{plain}{%
\fancyhf{} 
\fancyfoot{} 

}

\renewcommand{\chaptermark}[1]%
{\markboth{{\thechapter.\ #1}}{}}
\renewcommand{\sectionmark}[1]%
{\markright{{\thesection.\ #1}}}

\begin{document}

\title{Gaugings and other aspects in supergravity}
\author{Jose Juan Fern\'andez Melgarejo}
\date{\today}

 \begin{titlepage}
 \begin{flushright}
 \small ~~
 \end{flushright}

 \bigskip

 \begin{center}

 \vskip 0cm

 {\LARGE \bf Gaugings and other aspects\\[2mm] in supergravity}\footnote{Based on the author's Ph.D. thesis, defended on September 20, 2013.}\\[6mm]
 \vskip 0.5cm

 {\bf J. J. Fern\'andez-Melgarejo}\\

\vskip 25pt

 {\em FISPAC, Departamento de F\'isica, \\
 Universidad de Murcia,\\
 Campus de Espinardo, E-30100-Murcia, Spain\\
 {\small {\tt jj.fernandezmelgarejo@um.es}}}

\vskip 0.8cm

 \end{center}

 \vskip 1cm

 \begin{center}

 {\bf ABSTRACT}\\[3ex]

 \begin{minipage}{13cm}
 \small

\qquad We discuss various topics in supergravity: gaugings, double field theory and $N=2$ $D=4$ BPS multicenter black holes.

\qquad We introduce the main features of supergravity, focusing on the aspects of gauged supergravities. We study the embedding-tensor formalism as a tool that facilitates the construction of gauged supergravities due to its covariant formulation as well as its relevance in the description of the magnetic higher-rank field potentials.

\qquad In particular, we present a full study of the general gaugings of maximal $d=9$ as an example in which this formalism is applied. We obtain all the possible gaugings of the theory and its extended field content.

\qquad We also classify the orbits of gaugings of maximal and half-maximal $d=9,8,7$ supergravity and study their (non-)geometric origins by means of double field theory. By performing a generalized Scherk-Schwarz dimensional reduction of this T-duality-invariant formalism, we reproduce the orbits found by means of the embedding-tensor formalism.

\qquad Finally, we study a formalism to describe BPS multicenter solutions for $N=2$, $D=4$ theories with quadratic prepotentials. Based on the charge vector space, this approach allows for the treatment of these solutions in a more general way.

 \end{minipage}

 \end{center}

 \vfill

 \end{titlepage}

\newpage
\thispagestyle{empty}
\mbox{}
\newpage

\thispagestyle{empty}
\setcounter{page}{1}

\begin{center}

\vspace*{0.1cm}

{\textbf{\Huge Gaugings and other aspects \vspace{3mm} \\  in supergravity}}

\vspace{2cm}

\includegraphics[scale=.65]{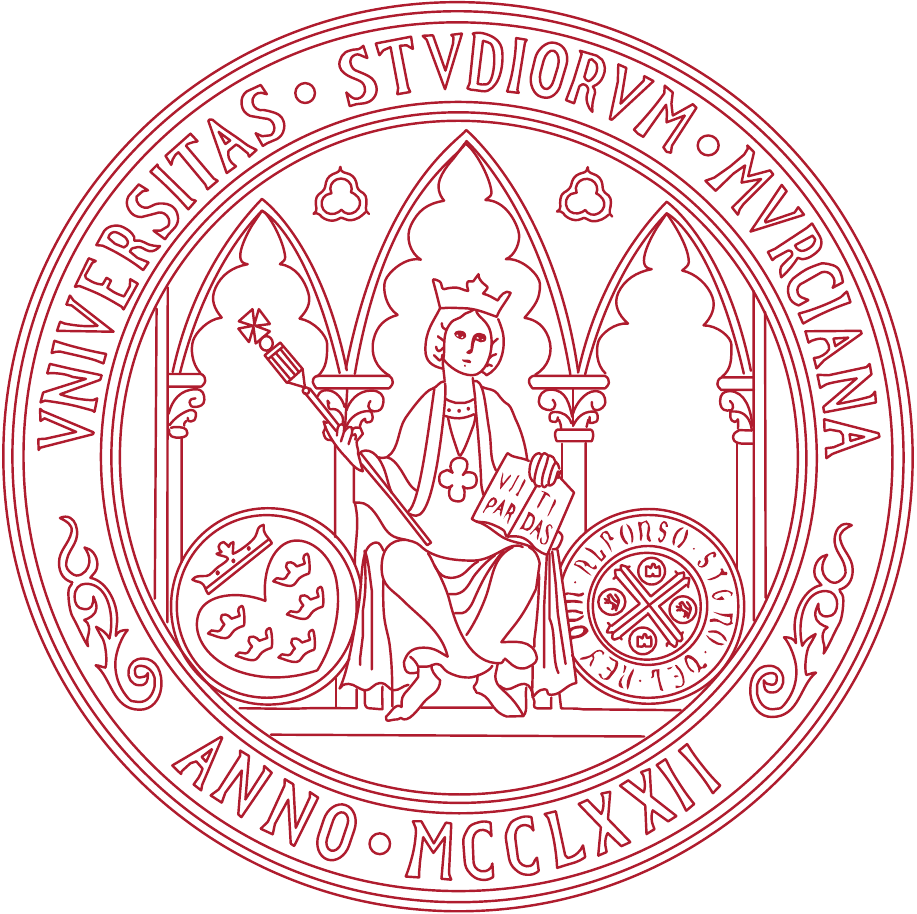}

\vspace{2cm}

\emph{A dissertation submitted to\\ the University of Murcia for the\\ degree of Doctor of Physics}

\vspace{1.5cm}

{\LARGE Jose Juan Fern\'andez Melgarejo}

\vspace{1cm}

{\Large Supervised by Emilio Torrente Luj\'an}

\vspace{1cm}

{\large September 2013}

\vspace{1cm}

{\large Department of Physics
\vspace{2mm}
\\
University of Murcia
}

\end{center}


\newpage
\thispagestyle{empty}
\mbox{}
\newpage
\thispagestyle{empty}
\mbox{}

\tableofcontents

	 \chapter*{Introduction\markboth{Introduction}{}}
  \addcontentsline{toc}{chapter}{Introduction}

In this thesis we treat three well defined  aspects of
supergravity (SUGRA), being the possible gaugings of
supergravity theories the central subject of study.
Other relevant topics in
the framework of supergravity, as specific topics of
Double Field Theory formalism or multicenter black hole solutions in different scenarios, are also addressed throughout this thesis.

In order to contextualize the significance of these topics,
we present a broad introduction
with the main purpose of  explaining
 the r\^ole that supergravity plays from the theoretical
 viewpoint of the current research in Physics.
More specific and in-depth introductions of each of the
topics treated in this work
are presented at the beginning of the corresponding
chapters.

Born in the late 1970s \cite{Freedman:1976xh,Deser:1976eh,Freedman:1976py}, supergravity is a quantum field theory that accounts gravity as a dynamical field. Namely, it is a theory that describes the gravitational interaction by means of a massless spin-2 particle. In addition, it is able to host additional fundamental interactions as internal symmetries.

However, the historical development and the reasons why  supergravity is studied nowadays are completely different. In its origin, this theory was studied as a UV completion of the General Relativity theory that could host the internal symmetries of all the known interactions in the Universe. It was supposed to be divergenceless at high energies and that motivated its interest during the late 1970s and the 1980s decade. For $N=1$ $d=4$, it was proven that its divergences could not be avoided\cite{Deser:1977nt}. However, a recent result  \cite{Bern:2007hh} strongly points  towards the ultraviolet finiteness of $N=8$ $D=4$ supergravity.

\subsubsection*{Supergravity as an effective theory}

The appearance of string theory supposed an important leap in the resurrection of supergravity. Supergravity strongly reappeared when it was discovered that the behavior of the superstring theories at a certain regime were equivalent to some specific supergravities (which, moreover, preserved the same amount of supersymmetries). In particular, supergravity describes the massless sector of the superstring theories.  Let us see this explicitly.

The most general non-linear $\sigma$-model that describes a string coupled to different non-trivial backgrounds is
\begin{align}
\label{eq:gen_ST_action}
S
&=
-\frac{T}{2}\int_\Sigma
d^2\sigma\sqrt{|\gamma|}\left[
	\left(
		\gamma^{ij} g_{\mu\nu}(X)
		+\epsilon^{ij}B_{\mu\nu}(X)
		\right) \partial_i X^\mu\partial_j X^\nu
	-\alpha'\phi(X)R(\gamma)
	\right]
\, .
\end{align}
Here, $X^\mu$ are the spacetime coordinates of the string. The background fields are the spacetime metric $g_{\mu\nu}$, a 2-form gauge potential $B_{\mu\nu}$ and a scalar field $\phi$. The worldsheet is parametrized by coordinates $\sigma^i=(\sigma,\tau)$, $\gamma_{ij}$ is the induced metric on the worldsheet and $R(\gamma)$ is the scalar curvature of the worldsheet.

The action \eqref{eq:gen_ST_action} is not conformally invariant.
This is a dilemma, since scale invariance is a necessary requirement for the consistency of the theory when this $\sigma$-model is quantized. Thus, we can wonder what constraints have to be imposed on the fields such that Weyl invariance remains unbroken. Namely, we want to know  what field configurations are the ones that guarantee this scale invariance. For this purpose, inspired by the problem of the dimensionful running coupling constants in other theories, a renormalization procedure can be performed,\footnote{In this case, the fields play the r\^ole of running coupling constants.} in which the $\beta$ functions calculated for each one of the fields have to vanish \cite{Callan:1985ia},
\begin{align}
\label{eq:beta_g}
\beta_{\mu\nu}(g)
&=
\alpha'\left[
	R_{\mu\nu}
	-2\nabla_\mu\nabla_\nu\phi
	+\frac{1}{4}H_\mu{}^{\alpha\beta}H_{\nu\alpha\beta}
	\right]
	+\mathcal{O}(\alpha'^2)
\, ,
\\
\label{eq:beta_B}
\beta_{\mu\nu}(B)
&=
\frac{\alpha'}{2}e^{2\phi}\nabla^\rho(e^{-2\phi}H_{\rho\mu\nu})
+\mathcal{O}(\alpha'^2)
\, ,
\\
\label{eq:beta_phi}
\beta(\phi)
&=
\frac{d-26}{6}-\frac{\alpha'}{2}\left[
	\nabla^2\phi
	-(\partial\phi)^2
	-\frac{1}{4}R
	-\frac{1}{48}H^2
	\right]
+\mathcal{O}(\alpha'^2)
\, .
\end{align}
For the critical dimenison $d=26$, $\beta(\phi)$ vanishes.
Recasting properly these equations on the fields, we can construct an action that contains the fields $\{g_{\mu\nu},B_{\mu\nu},\phi\}$ in such a way that the equations of motion arising from it are equivalent to these constraints. Namely, the minimization of the action
\begin{align}
S=\frac{g^2}{16\pi G_N^{(d)}}\int d^d x \sqrt{|g|}e^{-2\phi}\left[
	R
	-4(\partial\phi)^2
	+\frac{1}{2\cdot 3!} H_{\mu\nu\rho}H^{\mu\nu\rho}
	-\frac{4}{3\alpha'}(d-26)
	\right]
\, ,
\label{eq:effective_ST}
\end{align}
with respect to the fields $\{g_{\mu\nu},B_{\mu\nu},\phi\}$ is equivalent to the vanishing of the three $\beta$ functions \eqref{eq:beta_g}-\eqref{eq:beta_phi}, $\beta_{\mu\nu}(g)=\beta_{\mu\nu}(B)=\beta(\phi)=0$. In particular, we obtain
\begin{align}
\frac{16\pi G_N^{(d)}e^{2(\phi-\phi_0)}}{\sqrt{|g|}}\frac{\delta S}{\delta g^{\mu\nu}}
&\sim
\frac{1}{\alpha'}[\beta_{\mu\nu}(g)-4g_{\mu\nu}\beta(\phi)]
+\mathcal{O}(\alpha'^2)
\, ,
\\
\frac{16\pi G_N^{(d)}e^{2(\phi-\phi_0)}}{\sqrt{|g|}}\frac{\delta S}{\delta \phi}
&\sim
-\frac{16}{\alpha'}\beta(\phi)
+\mathcal{O}(\alpha'^2)
\, ,
\\
\frac{16\pi G_N^{(d)}e^{2(\phi-\phi_0)}}{\sqrt{|g|}}\frac{\delta S}{\delta B^{\mu\nu}}
&\sim
-\frac{1}{\alpha'}\beta_{\mu\nu}(B)
+\mathcal{O}(\alpha'^2)
\, .
\end{align}

By means of a conformal scaling on the metric, we define the so-called \emph{Einstein-frame metric} $g_{E\mu\nu}$ as
\begin{align}
g_{\mu\nu}
&=
e^{\frac{4}{d-2}\phi}g_{E\mu\nu}
\, .
\end{align}
This new metric allows us to get rid of the scalar field factor $e^{-2\phi}$. Hence, in this Einstein frame, the action is rewritten as
\begin{align}
S
&=
\frac{1}{16\pi G_N^{(d)}}
\int
d^d x\sqrt{|g_E|}\left[
	R_E
	+\frac{4}{d-2}(\partial\phi)^2
	+\frac{1}{2\cdot 3!} e^{-\frac{8}{d-2}\phi}H_{\mu\nu\rho}H^{\mu\nu\rho}
	-\frac{2(d-26)}{3\alpha'}e^{\frac{4}{d-2}\phi}
	\right]
\, .
\end{align}
Then, we obtain that the low-energy limit effective action of the string common sector \eqref{eq:effective_ST}, coincides with the action of the NSNS sector of supergravities. 

The possibility of projecting the low-energy behavior of string theory on a quantum field theory provides a useful scenario to investigate and understand different issues of string theory. For instance, the search of non-perturbative solutions or vacua of string theories is one of the main aspects that are exploited in the context of supergravity and that can be extrapolated to the string sector.

In the cases of interest for us,  the low-energy effective action of the string theory is a supergravity theory. 
For example, the effective actions of type IIA and IIB string theories coincide with the $N=2A$ and $N=2B$ supergravity theories. In addition, type I and the two versions of the heterotic string (the ones with $SO(32)$ and $E_8\times E_8$ gauge symmetries) coincide with different versions of $N=1$ $D=10$ supergravity.

Another intriguing aspect that supergravity possesses is its connection with the so-called \emph{M theory}. $N=2A/2B$ SUGRAs have a UV completion, the type IIA/IIB string theories, respectively. These full theories not only include the massless modes of SUGRA, but also extended objects as strings (or branes).

This UV completion is a basis to conjecture the existence of a theory which, analogous to the string theories, entails $D=11$ SUGRA as its low-energy effective theory. This is the hypothetical M theory. In addition, a relationship between M theory and string theory (and hence, SUGRA) strongly supports its existence. It is proven that the action of $N=2A$ SUGRA theory is obtained by performing a dimensional reduction of $D=11$ SUGRA on a circle \cite{Giani:1984wc,Huq:1983im,Campbell:1984zc}. We can compare the factors of $N=2A$ SUGRA and $D=11$ SUGRA compactified on a circle. Hence, due to their different origins, we wee obtain the following relation:
\begin{align}
R_{11}
&=
\ell_s g_A
\, ,
\end{align}
where $R_{11}$ is the radius of compactification of the 11th dimension, $\ell_s$ is the characteristic string length and $g_A$ is the coupling constant of type IIA string theory. We see that at small radius, that is, when taking $N=2A$ SUGRA description, the coupling is weak. However, at the strong-coupling regime, the radius grows and a new dimension becomes macroscopic. Therefore, since the UV completion of $N=2A$ SUGRA is type IIA theory, this suggests that the UV completion of $D=11$ SUGRA corresponds to the strong coupling limit of type IIA theory, the conjectured M theory.

In addition, since $D=11$ SUGRA has no scalar fields nor dimensionful coupling constants, M theory must also exhibit these features. This means that, unlike string theory, M theory does not have a perturbative expansion and therefore, its treatment is more difficult.

Once we have justified the importance and the r\^ole of supergravity  from a stringy viewpoint, we are going to dissect the main features of supergravity theories.

Supergravity is a quantum field theory in which local supersymmetry and General Relativity coexist. The transformation parameter of supersymmetry is a spacetime dependent spinor $\epsilon(x)$. The local character of this symmetry necessarily requires the introduction of a corresponding gauge field, which in this case must be a spinor. Then, it is the gravitino, $\psi_\mu(x)$, a spin-$\tfrac{3}{2}$ particle, the fermionic field that carries out this action. However, this is not all what we need. The supergravity algebra, the so-called superPoincar\'e algebra, implies the following anticommutating relation,
\begin{align}
\{Q^i{}_\alpha, Q^j{}_\beta \}
&=
(\gamma^\mu C^{-1})_{\alpha\beta} P_\mu \delta^{ij}
\, .
\end{align}
Here, $Q^i{}_\alpha$ are the supercharges, where $i$ is an index that denotes the number of supercharges and $\alpha$ is a spinorial index, $C$ is a charge conjugation matrix and $P_\mu$ are the generators of the translations. We expect that gauging of supersymmetry leads to gauging of translations. Then, since local translations are part of the general coordinate transformations, we also expect that the gravitational field $g_{\mu\nu}(x)$ (or alternatively, the vielbein $e_\mu{}^a$) behaves as a gauge field.

Hence, we see how superPoincar\'e algebra is the tailor that sews local supersymmetry and general coordinate transformations together so that supergravity results properly outfitted.

From the phenomenological viewpoint, one wonders whether it is worth considering such a scenario like supergravity when one wants to obtain results that could be tested by particle experiments in laboratories. Despite of the relative recent result on the renormalizability of $N=8$ $D=4$ supergravity and  the absence of results for $N<8$\footnote{For a very pedagogical explanation, we recommend \cite{Deser:1999uy}.}, the answer to this question is positive provided that  supergravity is considered an effective phenomenological theory arising from a UV completing theory.

This scenario is similar to the one of the old Fermi theory \cite{Fermi:1934hr}, in which the weak interaction is described by means of a dimensionful coupling constant $[G_F]=-2$. It is know that the 4-fermion interaction is valid for carrying out a description of the weak interaction at the scale energy $E\approx M_W$, where $M_W$ is the mass of the $W^-$ boson, one of the three force carriers of the weak force. However, for $E\gg M_W$, this theory breaks down and the genuine Glashow-Weinberg-Salam theory \cite{Glashow:1961tr,Weinberg:1967tq,Salam:1968rm} is required for a suitable description of Nature. That is, although the Fermi theory is non-renormalizable, its results are correct at a certain regime.

In a similar fashion, for $E\approx M_P$, being $M_P$ the Planck mass, one must use the UV completion of supergravity: the superstring or M theories. However, for the regime $E\ll M_P$ is a good approximation to work with supergravity. Consequently, from these arguments we conclude that the study of supergravity becomes crucial. It is the link between the possible final theory of elementary particles, strings or any other extended object and the low-energy effective theory which has to reproduce, at least, the $SU(3)\times SU(2)\times U(1)$ \emph{Standard Model} that describes the electromagnetic, weak and strong interactions that exist in our Universe \cite{Beringer:1900zz}.

\subsubsection*{Dimensional reduction and hidden symmetries}

Aside from the existing parallelism between string/M theories and supergravity, we can formulate SUGRA theories for dimensions $d\le11$. Starting out from higher-dimensional supergravities, one obtains new supergravity theories in lower dimensions by means of a dimensional reduction mechanism.\footnote{Very complete lectures on this topic are \cite{Duff:1986hr,Roest:2004pk}.} The symmetry structure of the resulting theory  depends very much on the geometric properties of the internal manifold on which we compactify.

\emph{Maximal supergravities} (namely, the supergravity theories that host the maximum number of supercharges) in several dimensions are related by dimensional reduction. When compactifying a $D$-dimensional supergravity theory on a $T^n$ $n$-torus, we obtain a $d$-dimensional supergravity, with $d=D-n$. In order to consistently construct  the lower-dimensional theory, one has to decompose the higher-dimensional fields into fields that transform covariantly under gauge symmetries and diffeomorphisms of the lower-dimensional theory. This rearrangement of the degrees of freedom is necessary for building and classifying the supermultiplets in a covariant way.

Let us briefly discuss the dimensional reduction of a $D$-dimensional toy-model based on gravity coupled to an antisymmetric 2-form gauge field on a $T^n$ $n$-torus. The model is given by
\begin{align}
\mathcal{L}
&\propto
e\left(
 	\frac{1}{2}\hat R
 	+\frac{1}{4}\hat H_{\hat\mu\hat\nu\hat\rho} H^{\hat\mu\hat\nu\hat\rho}
 	\right)
\, ,
\end{align}
where $\hat e$ is the determinant of the vielbein, $\hat R (\hat g)$ is the scalar curvature $\hat H_{\hat\mu\hat\nu\hat\rho}= 3\partial_{[\hat\mu} \hat B_{\hat\nu\hat\rho]}$ and $\hat\mu,\hat\nu,\hat\rho=1,\ldots,D$. The dimensional reduction scheme consists of the following redefinitions: the $\frac{(d+n)(d+n+1)}{2}$ degrees of freedom of $\hat g_{\hat\mu\hat\nu}$ are decomposed into
\begin{align}
\hat g_{\hat\mu\hat\nu}
\quad
&\longrightarrow
\quad
g_{\mu\nu}
\, ,
\quad
A^m{}_\mu
\, ,
\quad
g_{mn}
\, ,
\nonumber
\\
\tfrac{D(D+1)}{2}
\quad
&
\qquad
\quad
\tfrac{d(d+1)}{2}
\quad
d\times n
\quad
\tfrac{n(n+1)}{2}
\end{align}
and similarly, the $\frac{(d+n)(d+n-1)}{2}$ degrees of freedom of $\hat B_{\hat\mu\hat\nu}$ are rearranged into
\begin{align}
\hat B_{\hat\mu\hat\nu}
\quad
&\longrightarrow
\quad
B_{\mu\nu}
\, ,
\quad
B^m{}_\mu
\, ,
\quad
B_{mn}
\, .
\nonumber
\\
\tfrac{D(D-1)}{2}
\quad
&
\qquad
\quad
\tfrac{d(d-1)}{2}
\quad
d\times n
\quad
\tfrac{n(n-1)}{2}
\end{align}
Here, $g_{\mu\nu}$ and $B_{\mu\nu}$ are the $d$-dimensional metric and gauge potential, respectively. $A^m{}_\mu$ and $B^m{}_\mu$ are $d\times n$ vector fields whereas $g_{mn}$ and $b_{mn}$ are two symmetric and antisymmetric $n\times n$ scalar matrices, respectively.

The diffeomorphisms acting on the torus coordinates $\{x^m\}_{m=1,n}$,
\begin{align}
x^m\rightarrow U^m{}_n x^n
\, ,
\end{align}
act on the scalar matrices as follows:
\begin{align}
g
&\rightarrow
U^TgU
\, ,
&
B
&\rightarrow
U^TBU
\, .
\label{eq:diffeo}
\end{align}
The $U$ matrices generate the $GL(n)$ group, which contains the rotation group $SO(n)$ as a subgroup. In addition, special gauge transformations whose transformation parameter is proportional to $\Lambda_{mn} x^n$ induce a shift on $B_{mn}$,
\begin{align}
B_{mn}
&\rightarrow
B_{mn}
+\Lambda_{[mn]}
\, .
\label{eq:gauge_B}
\end{align}
Hence, we are able to identify $n^2+\frac{1}{2}n(n-1)=\frac{1}{2}n(3n-1)$ transformations. However, it turns out that there exist additional transformations which do not have a higher-dimensional origin \cite{deWit:2002vz}. These $\frac{1}{2}n(n-1)$ extra transformations,  combined with the previous ones, imply an enhancement on the global symmetry of the lower-dimensional theory, so that the action becomes $SO(n,n)$ invariant \cite{Maharana:1992my}. In addition, the coset space parametrized by the scalar fields also results improved to $SO(n,n)/SO(n)\times SO(n)$.

These so-called \emph{hidden symmetries} occur in many other scenarios and some attempts have been done to try to justify them. The most successful ones are methods that have to do with decompositions of Ka\v{c}-Moody algebras. These approaches are the $E_{10}$ and $E_{11}$ formalisms. By means of the decomposition of $E_{10}$ or $E_{11}$ (at the level of the algebras) into the global symmetry group $G_D$ of the $D$-dimensional SUGRA times a residual factor $A_{D-1}$,\footnote{$A_{D-1}$ represents the diffeomorphisms of the $D$-dimensional spacetime. Further work on this aspect can be found in \cite{Bergshoeff:2007qi,Bergshoeff:2008qd,Bergshoeff:2008xv}.}
\begin{align}
E_{11}
\quad
&=
\quad
G_D
\quad
\times
\quad A_{D-1}
\, ,
\end{align}
they provide the full field content of a given supergravity theory.

In particular, the case of the $SO(n,n)$ symmetry is understood as a realization of the T-duality symmetry that takes place at the level of string theory, which interchanges string momenta and winding modes. 


\subsubsection*{Gauged supergravities}

As it was seen, we can formulate diverse supergravity theories for dimensions $d\le11$. By means of a dimensional reduction mechanism, we can construct new supergravities with different features and these features depend on the geometry of the compactified manifold.

For instance, if we compactify a $D$-dimensional supergravity (namely, 11- or 10-dimensional theories) over a $T^n$ torus, we obtain a $d$-dimensional supergravity with a larger global symmetry group and an Abelian gauge symmetry. The Abelian character of the local symmetry has to do with the commutativity of the generators of the group manifold $T^n$. In this way, we obtain the so-called \emph{ungauged theories} for every dimension. The local symmetry of the ungauged supergravities corresponds to the $U(1)^n$ Abelian gauge symmetry of the vector fields
\begin{align}
\delta_\Lambda A_\mu{}^M
&=
d\Lambda^{(0)M}
\, ,
\label{eq:abelian_deformation}
\end{align}
where $\Lambda^{(0)M}$ is a 0-form gauge parameter. The number of generators of the gauge group corresponds to the number of vector fields. A  \emph{gauging} or \emph{gauged deformation} turns this Abelian local symmetry into a non-Abelian local one.

If we consider more geometrically complicated compactification manifolds, the new lower-dimensional theories that emerge will enjoy a non-Abelian gauge symmetry. These are the so-called \emph{gauged supergravities}, which entail the main part of this manuscript. Gauged supergravities are the only supersymmetric deformations of maximal supergravity that preserve supersymmetry\footnote{Up to now, the only known exceptions are the massive IIA Romans' supergravity \cite{Romans:1985tz} and a massive deformation for the $N=4$ $D=6$ supergravity \cite{Romans:1985tw}.}. Whatever the dimensional reduction scheme it is, the gauge parameters of the theory must depend on the compactification parameters. Namely, if the gauging arises from a compactification with non-trivial fluxes (i.e., background values for the higher-dimensional gauge fields), a certain brane configuration or any kind of torsion of the compactification manifold, the gauge parameters must exhibit a dependence on the variables that govern these phenomena. This scenario is schematically illustrated in Figure \ref{fig:flow_chart}.

Up to now, the general statement is that given a certain compactification scheme, a certain gauged supergravity arises. However, despite of flux compactification is a confident and straightforward mechanism to generate gauged supergravities, deformations can also be done without following this path. Indeed, the first deformed  theories \cite{deWit:1981eq,deWit:1982ig,Gunaydin:1984qu,Pernici:1984xx,Hull:1984yy} were constructed by adding the ingredients that the theory required step by step and assembling them properly.

From the decade of the 1980s to late 2000s, a wide variety of gaugings for different supergravities were found. However, since their search was mainly inspired in group theoretical arguments, the quest became harder as long as the global symmetry group grew. Taking into account that the only restriction is that te gauged symmetry group has to be a subgroup of the global symmetry group, the set of all the possible gaugings (including non-semisimple algebras) still result too broad to perform a complete analysis.
Nevertheless, there appeared a new tool that, applied on a certain ungauged theory, systematically scanned all the possible gaugings of that supergravity. This is the so-called \emph{embedding tensor mechanism} \cite{Cordaro:1998tx,deWit:2002vt,deWit:2003hq,deWit:2005hv,deWit:2005ub}, which basically promotes a certain subgroup $G$ of the global symmetry $G_0$ to be gauged in a covariant way. If we denote by $t_\alpha$ the generators of $\mathfrak{g}=\text{Lie }G$ and let $M=1,\ldots,n_V$ label the $n_V$ vector fields of the ungauged theory, then the embedding tensor, $\vartheta$, describes the embedding of $G$ into $G_0$ by means of the gauge generators
\begin{align}
X_M
&=
\vartheta_M{}^\alpha t_\alpha
\, .
\end{align}
Then, the deformation parameters are identified with the non-zero components of the embedding tensor. $X_M$, as the gauge generators of the theory, appear in the covariant derivative,
\begin{align}
D_\mu
&=
\partial_\mu
+g A_\mu{}^M X_M
\, .
\label{eq:cov_derivative}
\end{align}
That is, the embedding tensor formalism, acts as a caretaker of the covariance, deciding (by means of the constraints that act on it) what linear combinations of the global symmetry generators are the ones that preserve the covariance and the supersymmetry of the theory.

On the other hand, the fact of deforming a supergravity theory implies certain collateral adjustments that are essential for keeping covariance and supersymmetry unbroken. The main ones are the following:
\begin{itemize}
\item The standard derivatives have to be replaced by the covariant derivatives \eqref{eq:cov_derivative} to provide the local character of the promoted subgroup. Therefore, some quantities, as the field strengths, result modified.
\item The modification of the field strengths by means of the covariant derivatives and the condition that they have to transform covariantly imply the introduction of new couplings in their definition and in the gauge transformation of the fields, the so-called \emph{St\"uckelberg couplings}.
\item The SUSY transformations of the fermion fields are modified by the addition of the so-called \emph{fermion shifts}, which are linear in the deformation parameters.
\item A scalar potential is generated and can be expressed as a sum of the squares of the fermion shifts. Thus, it is quadratic in the deformation parameters.
\end{itemize}

Let us focus on the modified field strengths. As we said in the second item, it is necessary the introduction of St\"uckelberg couplings to guarantee the covariance of the field strengths. Schematically, the `deformed' field strength and gauge transformation of an arbitrary $p$-form gauge field $C^{(p)\alpha}$, where $\alpha$ is an index of the representation of $G$ under which $C^{(p)}$ transforms, result
\begin{align}
\mathcal{F}^{(p+1)\alpha}
&=
\mathcal{D}C^{(p)\alpha}
+\cdots
+ Z^{\alpha}{}_I C^{(p+1)I}
\, ,
\nonumber
\\
\delta_\Lambda C^{(p)\alpha}
&=
\mathcal{D}\Lambda^{(p-1)\alpha}
+\cdots
+Z^{\alpha}{}_I \Lambda^{(p)I}
\, .
\end{align}
The St\"uckelberg couplings are the tensors $Z^\alpha{}_I$, which are supposed to be linear in the embedding tensor, $C^{(p+1)I}$ is the $(p+1)$-form gauge field realized on a certain representation of $G$, denoted by the index $I$. The tensors $\Lambda^{(p-1)\alpha}$ and $\Lambda^{(p)I}$ are $(p-1)$- and $p$-form gauge parameters living in their respective representations and can be understood as the generalizations of the 0-form gauge parameter that appears in \eqref{eq:abelian_deformation}. Hence, the St\"uckelberg couplings connect the leading-order $(p+1)$-form with the field strength of a $p$-form. The completion of this structure from the vector fields to the top-forms is the so-called \emph{tensor hierarchy} and, by means of it, we have access to the full field content of the theory, including the magnetic dual gauge fields. In particular, we obtain information  about the $(d-1)$- and $d$-form gauge fields, which are related to the parameter deformations and the constraints that filter the valid gaugings, respectively.

In summary, following the many subtleties of the embedding tensor formalism, we are able to obtain all the possible gaugings (as well as all their possible combinations) of a given ungauged supergravity.

\tikzstyle{block} = [rectangle, draw, fill=blue!20,
    text width=7em, text centered, rounded corners, minimum height=4em]
\tikzstyle{line} = [draw, -latex']
\tikzstyle{cloud} = [draw, ellipse,fill=red!20, text centered,minimum height=1.8em,text width=6em]
\begin{figure}[t]
\centering
\begin{tikzpicture}[node distance = 2.75cm, auto]
    \node [cloud] (et) {Embedding tensor};
    \node [block, above of=et, left of=et,yshift=-2em] (low_sugras) {maximal lower-dimensional SUGRAs};
    \node [block, above of=et, right of=et,yshift=-2em] (gauged_sugras) {Gauged SUGRAs};
    \node [block, right of=gauged_sugras,xshift=5em] (new_gauged_sugras) {New gauged SUGRAs};
    \node [block, above of=low_sugras,node distance = 2.75cm] (high_sugras) {$D=11$,\\ $N=2A/2B$ SUGRAs};
    \node [block, right of=high_sugras,node distance = 2.75cm,xshift=12em] (DFT) {T-duality constructions};

    \path [line] (high_sugras) -- node  {$T^n$} (low_sugras);
    \path [line,ultra thick] (high_sugras) -- node[align=center]  {SS and  others} (gauged_sugras);
    \path [line,double] (high_sugras) -- (DFT);
    \path [line,densely dashed] (low_sugras) |- (et);
    \path [line,densely dashed] (et) -| (gauged_sugras);
    \path [line,densely dashed] (et) -| (new_gauged_sugras);
    \path [line,double] (DFT) -- (new_gauged_sugras);
    \path [line,double] (DFT) -- (gauged_sugras);
\end{tikzpicture}
{\it \caption{Scheme about how gaugings are obtained from higher-dimensional theories. The solid line represents the dimensional reduction  of $D$-dimensional supergravities on a $T^n$ $n$-torus that gives rise to the ungauged SUGRAs in $d=D-n$ dimensions. The thick line points towards the gaugings obtained by SS dimensional reductions of the higher-dimensional SUGRAs. The dashed lines show how the embedding tensor formalism provides a mechanism to obtain two different kinds of gaugings: the ones that are already obtained by SS compactification and new deformations associated to non-geometric fluxes. The double arrows suggest the T-duality invariant constructions as a mechanism to obtain both types of gaugings by means of SS reductions.}
\label{fig:flow_chart}}
\end{figure}
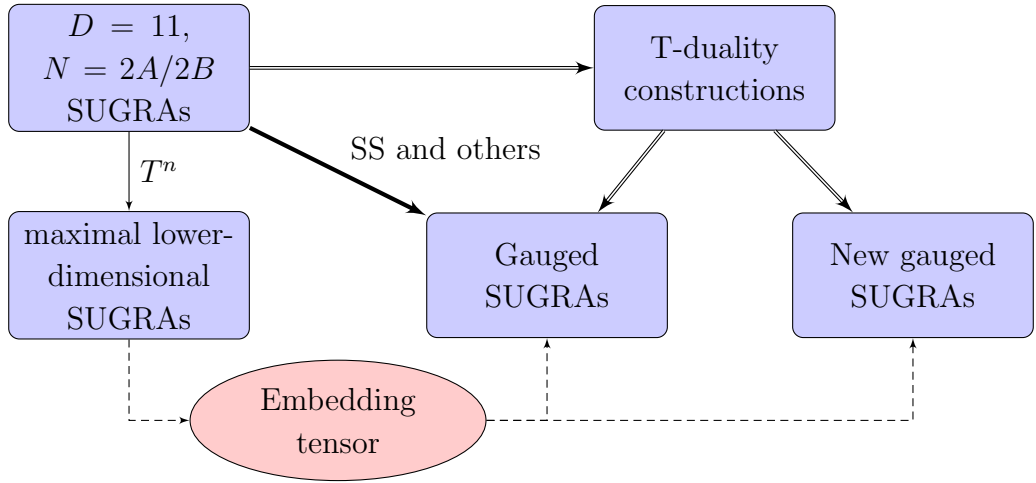

\subsubsection*{Non-geometric fluxes and T-duality constructions}

Thus, once we possess a mechanism that provides all the possible non-Abelian deformations of an ungauged supergravity, it seems natural to compare these results with the ones obtained by dimensional reduction of the higher-dimensional supergravities. When this analysis is done, the situation is the following: not all the gaugings arising from the embedding tensor formalism can be obtained by means of a dimensional reduction procedure. Namely, despite of using a wide variety of compactification schemes \cite{Kaloper:1999yr,Hull:2005hk,Hull:2006tp}, there is no higher-dimensional geometric explanation for some of the non-Abelian deformations.

However, inspired by duality covariance arguments,  it was proven \cite{Shelton:2005cf} that by performing T-duality transformations on the gauge and geometric fluxes (the ones associated to non-trivial $H_{\mu\nu\rho}$ and $g_{\mu\nu}$ backgrounds, respectively), there appeared new fluxes that could not be reached by dimensional reduction but which, nevertheless, were found by means of the embedding tensor formalism. These are the so-called \emph{non-geometric fluxes}. The transformations that were applied to the gauge and geometric fluxes are given by  the so-called \emph{Buscher's rules}, which relate two different string backgrounds. That is, by applying T-duality transformations on fluxes that have a higher-dimensional origin, we obtain new fluxes that cannot be justified by compactification.

Therefore, the situation is the following: roused by T-duality arguments, there are some new fluxes that seem to be associated to the gaugings that are not reached by the standard dimensional reduction procedure. Are we missing any ingredient in the compactification procedure that has to do with T-duality? The appearance of non-geometric fluxes in \cite{Shelton:2005cf} suggests the realization of T-duality as a fundamental symmetry of the theory, rather than just being a symmetry of the compactified versions of string theory. Thus, in order to provide a scenario in which non-geometric fluxes naturally emerge (and, hence the missed supergravity deformations associated to them), new constructions have been investigated in the literature whose main feature is the inclusion of the T-duality symmetry group as a true symmetry of the theory.

The origin of T-duality could give us a hint on the importance of the winding modes in the compactification procedure. Let us assume that we have two different backgrounds in a $(d+1)$-dimensional theory, $\{g,B,\phi\}$ and $\{g',B',\phi'\}$. Then, if we perform a dimensional reduction on a circle of radius $R$ of the theory turning on the unprimed background and a dimensional reduction on a circle of radius $R'$ of the theory with the primed background turned on, we obtain an equivalent theory. Hence, what T-duality does, by means of the Buscher's rules, is to establish a relation between $\{g,B,\phi\}$ and $\{g',B',\phi'\}$. At the string theory level, this can be understood as an interchange of winding modes and momentum modes.

At this point, we wonder whether the fact of not considering the winding modes in a compactification scheme could lead to the loss of any flux generated by the wrapping of the closed string around a certain geometry or around nontrivial 1-cycles in spacetime. This could qualitatively justify the necessity of taking the winding modes of the string in our compactification scheme into account.


There exist two main approaches to promote T-duality from a hidden symmetry to a truly global symmetry. One direction  is the so-called \emph{generalized complex geometry} \cite{Grana:2006hr,Grana:2008yw,Coimbra:2011nw}, in which the internal manifold enjoys a particular bundle structure such that the corresponding gauge fields span the whole T-duality symmetry group. Another construction consists of doubling the internal coordinates by adding winding modes as the dual spacetime  coordinates. Na\"{\i}vely, this winding would contribute to the generation of the non-geometric fluxes in the same way as a compactification on a twisted double torus does \cite{Dall'Agata:2007sr}.

The last procedure has been recently improved into the so-called \emph{Double Field Theory} (DFT). This construction is a T-duality invariant reformulation of supergravity in $10+10$ dimensions, where the new set of coordinates are related to winding modes. The first formulation of DFT involved the $(10+10)$-dimensional metric $g_{ij}$, a 2-form field $B_{ij}$ and a scalar dilaton field $\phi$, which correspond to the field content of the bosonic common sector of strings. Later on, motivated by the search of an $O(D,D)$ invariant theory, these fields were encoded into the so-called \emph{generalized metric} $\mathcal{H}_{MN}$,
\begin{align}
\mathcal{H}
&=
\left(\begin{array}{cc}
g^{ij} & -g^{ik}b_{kj}
\\
b_{ik}g^{kj} & g_{ij} - b_{ik}g^{kl}b_{lj}
\end{array}\right)
\, ,
\end{align}
which is $O(D,D)$ invariant by construction. The scalar dilaton, multiplied by the determinant of the metric, becomes T-duality invariant,
\begin{align}
e^{-2d}
&=
\sqrt{|g|}e^{-2\phi}
\, .
\end{align}
Once the metric and the 2-form are unified into this generalized metric, one wonders whether it is possible to define certain `generalized diffeomorphisms' such that the diffeomorphisms and the gauge symmetry that act on the metric and the 2-form, respectively, result unified in a similar way. This leads to the definition of a generalized gauge parameter $\xi^M$,
\begin{align}
\xi^M
&=
(\tilde \Lambda_i,\Lambda^i)
\, ,
\end{align}
made out of the parameters of both symmetries. Hence, we can think of a generalized Lie derivative,
\begin{align}
\mathcal{L}_\xi V^M
&=
\xi^P\partial_P V^M
+(\partial^M\xi_P-\partial_P\xi^M)V^P
\, ,
\end{align}
which will bring up the definition of a suitable generalized bracket such that the generalized Lie derivatives, which are associated to the gauge transformations of the fields, close properly.

Once that DFT is properly defined, one can perform Scherk-Schwarz dimensional reductions on it, so that one expects to catch the gaugings associated to the non-geometric fluxes. First attempts resulted successful to formally reproduce $N=4=D$ supergravities \cite{Aldazabal:2011nj,Geissbuhler:2011mx}. However, it was shown \cite{Grana:2012rr} that a relaxation of some of the constraints of DFT formalism enabled a full description of the gaugings of a theory. The predictability power of DFT for describing  gaugings of maximal and half-maximal supergravities in dimensions $d=9,8,7$ was confirmed in \cite{Dibitetto:2012rk}

\subsubsection*{Supergravity extremal black holes}

Once we have presented some key features of supergravity
theories and their understanding from a string theory
perspective, we can study the existence of
solutions, how solutions in supergravity are modified to those
in pure gravity and how symmetric they are
with respect to the underlying theory: what is the influence
of preserving a fraction of the supersymmetric charges in our
solution.

In particular, due to its simplicity and its conceptual richness,
extremal black hole solutions
 in $D=4$ will be our guinea pigs to carry out this
brief primer. They are also the subject of the third part
of this work.
Extremal black holes are particular solutions of (super)gravity and possess the minimal amount of mass allowed by their charge,  $M=\sqrt{Q^2+P^2}$ (see Chapter \ref{ch:bh} for details).
Their temperature vanishes, making them stable solutions under Hawking radiation.
Solutions as the
 Bertotti-Robinson metric or, for multicenter
configurations, the Majumdar-Papapetrou metric,
\begin{align}
ds^2
&=
-H^{-2}(x) dt^2
+H^2(x) dx^i dx^i
\, ,
\end{align}
represent  extremal black holes.
In the last expression, $H(x)$ is a harmonic function, $\Delta_3 H=0$.

Let us see 
some consequences of having local supersymmetry in our theory.
For every extended supersymmetry, the algebra contains central charges, \emph{i.e.} operators that commute with all the generators of the algebra. For the case of $N=2$ $D=4$, which will be the case we will treat, there is one complex central charge $Z$. From the supersymmetry algebra, we can infer that all massive representations satisfy a mass bound, which is given by this central charge,
\begin{align}
M &\ge|Z|\, .
\end{align}
States that saturate this bound, $M=|Z|$, are the \emph{BPS states}.

Minimal $N=2$ $D=4$ SUGRA is a suitable scenario in which Einstein-Maxwell theory is naturally embedded.
The pure supergravity multiplet is spanned by the graviton
(vielbein) $e_\mu{}^a$, the gravitini $\psi_\mu{}^i$, $i=1,2$
and gauge field $A_\mu$, which in this context is usually
called \emph{graviphoton}.

The central charge transformations are $U(1)$ symmetries and
the graviphoton plays the r\^ole of the gauged field to guarantee its
local nature. This central charge is related to the electric
and magnetic charges by the relation  \cite{Teitelboim:1977hc}
\begin{align}
Z&=Q+iP
\, .
\end{align}
Hence, the classical bound $M=\sqrt{Q^2+P^2}$ translates into
the supersymmetric one, $M=|Z|$.

Typically, a non-trivial field configuration
has less symmetry than the vacuum.
However, there exist certain solutions that preserve
different portions of the symmetries.
For instance, axisymmetric black holes are solutions of
Einstein-Maxwell theory that keep rotations as isometries
of the metric. These isometries are generated by the
Killing vectors.
The situation in the context of supersymmetry transformations
is comparable to this one. If we are able to find
supersymmetric transformation parameters $\epsilon(x)$
(in the previous case, the transformation parameters are the
Killing vectors) such that a particular field configuration is
invariant under these transformations, we have the fermionic
analogue of an isometry. Due to their fermionic nature, these
parameters are called \emph{Killing spinors}.

For the above case, proving the invariance of the fields means
to put to zero the supersymmetric transformations of all the
fields of the theory,
\begin{align}
\delta_\epsilon\left\{
	e_\mu{}^a\, ,\,\,  A_\mu\, ,\,\, \psi_\mu^i
	\right\}
&=0\, .
\end{align}
Since we are interested in purely bosonic solutions, we truncate the fermion fields to zero,
\begin{align}
\psi_\mu{}^i&=0\, .
\end{align}
On the other hand, the fact of having a spinorial transformation parameter implies that all the supersymmetric transformations of the bosonic fields are odd in fermions. Hence, we automatically have
\begin{align}
\delta_\epsilon(\text{boson})
&=
0
\, ,
\end{align}
so that the remaining non-trivial condition is
\begin{align}
\label{eq:kse}
\delta_\epsilon\psi_\mu{}^i
&=
0
\, ,
\end{align}
which is an equation for $\epsilon(x)$ (the so-called \emph{Killing spinor equation}).
In order to make this problem more tractable, some assumptions
 on $e_\mu{}^a$ and $A_\mu$ can be done.
For instance, we can assume  static ansatze for the
gauge field and the metric.
One can check that the aforementioned Bertotti-Robinson and Majumdar-Papapetrou metrics allows the existence of some spinor $\epsilon(x)$  that satisfies \eqref{eq:kse}. In the former case, the full set of supercharges remains unbroken, whereas the latter only preserves half of it.

From a general point of view, we can consider black holes as
solitonic solutions. Solitons are broadly defined as
time-independent, non-perturbative, non-singular\footnote{In
gravity contexts, solutions whose singularities are covered
by event horizons are also admitted.}, localized solutions of
classical equations of motion with finite energy in a field
theory \cite{Callan:1991at}.
The method presented above
can be used to construct different supersymmetric
solitonic solutions of supergravity theories in various
dimensions.
Such solutions in $d$ spacetime dimensions are alternatively
called $p$-branes \cite{Duff:1994an} if they are localized
in $d - 1 - p$ spatial coordinates and independent of the
other $p$ spatial coordinates, where $p < d - 1$.
The $p = 0 $ case (0-brane) corresponds to a point particle; $p = 1$ case is
called a string; $p \ge 2$ cases are known as membranes.

The discovery of string dualities in the
\emph{second string revolution} led to a new picture in the
knowledge of solutions of the theories.
The knowledge of non-perturbative solitons in
these scenarios is essential for the understanding of
different regimes of string/M theories by means of dualities applied on them.
An example of this is the work done in \cite{Bergshoeff:2011se,Bergshoeff:2012ex,Bergshoeff:2012xc,Bergshoeff:2013sxa}.
In particular, S-duality provided a bridge between the strong coupling limit
of a given string theory and its dual theory that turned out
to be weakly coupled.
The result of applying dualities to black hole is a dual
description of string excitations.
The so-called \emph{string-black hole correspondence}
\cite{Susskind:1993ws,Horowitz:1996nw} predicts the black hole
entropy in terms of string states and gives an explanation
for the final state of a Schwarzschild BH.

In general, Supergravity reproduces the Einstein-Hilbert
action coupled to a certain number of matter fields, whose
specific
content depends on the theory under analysis.
Generically, these fields are a bunch of scalar fields
(moduli), spin-1/2 fermions, spin-1 gauge fields and $N$
gravitini, where the latter behave as the gauge fields of
the local supersymmetry.
At this point, we can wonder what are the simplest BH
solutions in this scenario where additional fields are
present and their relation to the pure
gravity ones.

The so-called \emph{attractor mechanism}
may be  described as follows in simple terms:
the scalar fields approach fixed values at the BH horizon,
that are only determined by the charge configuration.
The asymptotic values of the moduli are forgotten even though
the dynamics is completely valid and the fixed point
represents the equilibrium of the system.
In fact, the flow of the scalars towards the horizon behaves
as a gradient flow towards a fixed point, which is the
minimum of a function called \emph{black hole potential}.
The entropy is shown to be
proportional to the black hole potential evaluated in the
horizon \cite{Ferrara:1997tw}.
On the other hand, non-supersymmetric extremal solutions
(non-BPS states) also exhibit this attractor behavior.
However, in this case not all the scalar fields of the vector
multiplets become stabilized in terms of the BH conserved
charges at the event horizon.
Some of them generate flat directions at the minimum of the potential
\cite{Ferrara:2007tu}.
Even though, the entropy of non-BPS BHs also depends on
the dyonic charges, as in the supersymmetric case \cite{Nampuri:2007gv,Ferrara:2007tu}.

Once we have reviewed from a general
perspective
some topics of supergravity
that are going to be treated in this manuscript,
we are going to describe how this dissertation is organized as well as the content of each of its sections.

\subsection*{Outline of this work}

The work of this thesis is arranged by following a comprehensible progression. In the following paragraphs, we will enumerate each chapter, with a brief description of its content.

Chapter \ref{ch:sugra_primer} is an introductory presentation of the main features of ungauged supergravities, emphasizing in supersymmetry as one of its pillars. We give a digest of the maximal higher-dimensional supergravities due to their importance in the following chapters.

In Chapter \ref{ch:gauged_sugra}, gauged supergravities and the embedding tensor formalism are studied. We will dissect the structure of these deformed theories as well as the consequences of inserting the embedding tensor inside an ungauged supergravity.

After that, we give a full example of how the embedding tensor scans all the possible gaugings of $D=9$ maximal supergravity. This is done in Chapter \ref{ch:gaugings_d9}. We construct the tensor hierarchy of the gauged theory and compare our results with the ones based in the $E_{11}$ formalism.

Chapter \ref{ch:duality_orbits} deals about flux compactifications and how do they motivate one of the clashes between string theory and supergravity. As we said, the gauged supergravities that the embedding tensor allows us to construct and the ones generated by flux compactification of higher-dimensional theories do not coincide. In this chapter, we use Double Field Theory to solve this problem, at least, for all maximal and half-maximal $D=9,8,7$ supergravities.

In Chapter \ref{ch:bh}, we present a study of
some aspects of extremal multicenter
black hole solutions
 in certain $N=2$ $D=4$ supergravity models.
In the context of special geometry,
we provide a formalism to obtain explicit composite black
hole solutions with an arbitrary number of centers for any arbitrary quadratic prepotential.

Finally, Chapter \ref{ch:conclusions} summarizes and synthesizes the main results and conclusions of the work done in this dissertation. Various prospects and further projects are shown as possible candidate ideas to address in a near future.

Several appendices are included.
%
Appendix \ref{sec:Tduality} treats general aspects of T duality.
Appendix \ref{app:gaugings_d9} includes general notation and
 definitions and more results obtained in Chapter \ref{ch:gaugings_d9}.
Appendix \ref{app:duality_orbits} shows some technical material used in the development of the calculus of Chapter \ref{ch:duality_orbits}.

In page \pageref{publications}, we provide the list of publications on which the thesis is based, as well as other works that have been done during the PhD period.
\chapter{Supergravity: a primer}
\label{ch:sugra_primer}

In this chapter, we will introduce some basic aspects of 
supersymmetry and supergravity theories. 
We will show how supersymmetry restricts and cast the field 
content of the theory depending on the dimension in which we formulate our theory. 
Furthermore, we will show a catalog of the maximal higher-dimensional 
theories in  $D=11,10,9$.

\section{Supersymmetry essentials}

It is generally assumed that the exact or approximate 
symmetry groups of the known fundamental laws of nature
 are (at least locally) isomorphic to direct products of 
the spacetime Poincar\'e group and compact Lie 
groups\cite{Beringer:1900zz} representing internal symmetries.
The internal symmetry concept (as \emph{isospin}) 
was initially introduced in Physics by Heisenberg  in 
1932 \cite{Heisenberg:1932dw}
and quickly expanded by the
$SU(4)\supset SU(2)_{s}\times SU(2)_{iso-s}$ 
Wigner  model \cite{Wigner:1936dx}.
%
However, prior to the establishment of QCD and the current 
Standard Model (SM) of particles and interactions, 
symmetries that extended Poincar\'e symmetry 
(or its non-relativistic limit including spin) in a 
non-trivial way were suggested in the early 1960s as a way 
of formulating a viable theory of hadronic physics 
(see for example 
\cite{Sakita:1964qq,Gursey:1992dc,Dashen:1965zz,MacFarlane:1965wq,Michel:1965qs,Sakita:1964qr,Sakita:1965qt},
also \cite{Wigner:1936dx}).
%
%
%
%
%

In one of these extensions the older, non-relativistic, 
$SU(4)$ Wigner `supermultiplet' model  was extended 
to  $SU(6)$ (see \cite{Sakita:1964qq} 
and references therein).\footnote{The term `supermultiplet', 
as in `the $SU(4)$ Wigner supermultiplet', 
apparently appears for the first time in 1964 \protect\cite{Sakita:1964qq}, having nothing to do with the concept of `supersymmetry'.}
This group  has a subgroup $SU(2)\times SU(3)$ identified 
with the direct product of the  non-relativistic 
spin group $SU(2)$ and a $SU(3)$ internal-symmetry group. 
The full $SU(6)$ theory proposes to treat the ordinary spin
on the same footing as the isotopic spin and hypercharge.
It mixes the spin and $SU(3)$  coordinates so that particles
with different spin as well as with different isospin
and strangeness can lie in the same supermultiplet: 
 quarks $u$, $d$, $s$, with spin up and down belong to the 
fundamental representation {\bf 6}. Mesons and baryons 
belong to the {\bf 35} and {\bf 56} representations 
obtained from the product $q\overline{q}$ and 
$qqq$, respectively.

Later on, the (partial) success of the $ SU(6)$ theories
in explaining some aspects of the classification and  
properties of hadrons raised the possibility of a relativistic 
symmetry group which was not simply a direct product of 
Poincar\'e and internal symmetry. The extension of the 
theories to include special relativity was however very 
problematic.\footnote{In a relativistic wave equation, the spin 
indices are so tightly related to the coordinates (as we may 
see in the Dirac equation) that it is impossible to decouple 
the spin variables from the coordinates even in the 
free Hamiltonian.}
A  way of extension was by searching a larger group which 
included the $SU(6)$ and the Lorentz groups as subgroups. 
One fitting candidate of this kind was found to be
 the $SL(6,\mathbb{C})$ group \cite{Sakita:1964qq}, 
which contains $SL(2)\times SL(3)$ as a subgroup. 
However, this extension  seemed to be impossible without 
considering a higher 36-dimensional spacetime. Moreover, it 
would have to admit either an infinite number of 
one-particle states  or a continuous mass distribution 
for a given particle state \cite{Sakita:1965qt}.

Other group structures were explored, as for example 
the $\tilde{U}(8)$ and $\tilde{U}(12)$ theories 
\cite{Sakita:1965qt,MacFarlane:1965wq}. 
They were based on a covariant merging of isospin and 
spacetime symmetries
including higher-dimensional gamma matrices  generators.\footnote{These $\tilde{U}$ groups are generated by 
Lie algebras including gamma matrices 
 and $su(2)$ or $su(3)$ algebras.}
However, these models became very problematic; since the free Lagrangian 
was not invariant with respect to the symmetry group 
(only the interaction part resulted invariant), the physical states did not form a unitary representation of it.

All attempts to find such a group were clearly unsuccessful.
At the same time, there appeared a set of no-go theorems
 \cite{Coleman:1967ad,PhysRev.138.B1262,PhysRev.139.B597},  
the \emph{Coleman-Mandula theorem}  \cite{Coleman:1967ad} 
the strongest among them, which showed that 
the symmetry 
group of a consistent 4-dimensional relativistic 
quantum field theory with a finite number of massive 
particles is necessarily the direct 
product of the internal symmetry group and 
the Poincar\'e group. 

Typically, these theorems  showed that a physical field 
theory with a finite number of definite mass particles and 
with an analytical $S$ matrix without any of these groups 
as symmetries, did not allow anything but trivial 
scattering, in the forward and backward directions.

Superalgebras, as a way of avoiding the no-go theorems
 and extending the concept of symmetry,
were introduced in particle physics  for the first time 
in 1966 by Miyazawa \cite{Miyazawa:1966mfa,Miyazawa:1968zz}. They were used to introduce spinor currents, in addition to the algebra 
of vector  currents. These currents are  bilinear 
combinations of both bosonic and fermionic fields.
As a result  mesons  and baryons of different spins appear
in an unified way in the same multiplet.\footnote{Lie superalgebras appeared, though not in a central role, in some mathematical contexts in the 1960s \cite{Frolicher:1956nb,Milnor65onthe}.} 


The four-dimensional Poincar\'e 
superalgebra was developed in  1971 by
 Gol'fand \& Likthman developed 
\cite{Golfand:1971iw}.
Ramond \cite{Ramond:1971gb} and Neveu \& Schwarz \cite{Neveu:1971rx} 
developed superstrings and the supersymmetric extensions of a non Lie 
algebra, the Virasoro algebra. 
Volkov \& Akulov \cite{Volkov:1973ix} and Wess \& Zumino 
\cite{Wess:1973kz} wrote different realizations of supersymmetric 
field theories, even without being aware of the earlier work done by 
Gol'fand \& Likthman. 
In particular,\footnote{At this time,  QCD and the full SM 
with their symmetry product of the Poincar\'e group and 
local Lie groups were well established 
and apparently there was not need for further developments.} the Wess-Zumino model
\cite{Wess:1974tw,Ferrara:1974ac,Wess:1974jb,Wess:1973kz}
 was the first widely 
known example of an interacting 4-dimensional quantum 
field theory with supersymmetry.\footnote{The model consists of
 a single chiral superfield (composed of a complex scalar and 
a spinor fermion) whose cubic superpotential leads to a 
renormalizable theory. 
The action of the free massless Wess-Zumino model is invariant under 
the transformations generated by a superalgebra allowing both commuting 
and anticommuting symmetry generators.} 
Superspace formalism was introduced in 1974 \cite{Salam:1974gu}.

In 1975,  Haag, {\L}opusza\'nski, and Sohnius 
published \cite{Haag:1974qh}
a general proof that weakened the assumptions of the Coleman-Mandula 
theorem by allowing both commuting and anticommuting symmetry generators. 
There is a nontrivial extension of the Poincar\'e algebra, \emph{the 
supersymmetry algebra}, which is the most general symmetry of the 
$S$ matrix of a quantum field theory.
More in detail, the theorem may be summarized as follows: the most 
general  Lie algebra of generators of supersymmetries and ordinary 
symmetries of the $S$ matrix in a massive theory involves the following 
Bose type operators:
the energy-momentum operators $P_\mu$; 
the generators of the homogeneous Lorentz group $M_{\mu\nu}$; 
and a finite number of scalar charges. 
It will involve, in addition, Fermi-type operators, all of which commute 
with the translations and transform like spinors  under the homogeneous 
Lorentz group. 

\subsection{Clifford algebras and spinors}

The transformation properties of Bose and Fermi generators under the 
Lorentz group imply restrictions on the number of each of these 
types of generators and, indirectly, on the number and signature 
of spacetime dimensions. 
These restrictions are trivial for the case of Bose generators: 
a vector representation in any $D$-dimensional spacetime has 
always $D$ components.
The situation is less trivial for the Fermi generators.
They carry a spinorial representation of the Lorentz group which makes
convenient the detailed study of the representation theory of 
Clifford algebras.

Clifford algebras are relevant in Physics due to the fact 
that their 
representations can be used to construct  specific representations 
of symmetry groups, the spinorial representations.
In particular, a representation of the $D$-dimensional Clifford algebra 
can be used to construct a representation of the $D$-dimensional Lorentz 
algebra $\mathfrak{so}(1,D-1)$.
More in detail, if we define a set of gamma matrices 
$\left\{\gamma_\mu\right\}_{\mu=0,\ldots,D-1}$ which satisfy a 
Clifford algebra with associated metric $\eta_{\mu\nu}=\diag{-,+,\ldots,+}$,
\begin{align}
\left\{ \gamma_\mu,\gamma_\nu\right\}&=2\eta_{\mu\nu}\mathds{1}
\, ,
\label{eq:clifford}
\end{align}
then the matrices
\begin{align}
\Sigma_{\mu\nu}^S&=\frac{1}{2}[\gamma_\mu,\gamma_\nu]\, 
\end{align}
are   generators for a spinorial representation $S$ of the Lorentz group.
The exponentiation of these generators gives a  Lorentz transformation
\begin{align}
\Gamma^S(\Lambda)&=
\exp\left(\tfrac{1}{2}\omega^{\mu\nu}\Sigma_{\mu\nu}^S\right)
\, .
\end{align}
It can be shown that for a given dimension $D$, there is only one inequivalent irreducible 
representation of the Clifford algebra whose dimension is $2^{[D/2]}$. The elements of this
  $2^{[D/2]}$-dimensional vector representation space, 
where the algebra acts are the \emph{Dirac spinors}.

Irreducible representations of Clifford algebras may lead to 
 reducible Lorentz representations depending on the dimension of 
the spacetime. For instance, even dimensions allow the existence 
of $2^{[D/2]-1}$-dimensional irreducible representations. 
This can be easily seen by defining a  matrix $\gamma^{D+1}$,
\begin{align}
\gamma^{D+1}&=i(-1)^{\frac{D-2}{4}-1}\gamma^0\cdots\gamma^{D-1}
\, .
\end{align}
This \emph{chirality matrix} is traceless, squares to unity, 
half of its eigenvalues are +1s and the other half are -1s. 
It is natural then to split  Dirac spinors into the direct sum of 
the subspaces of spinors with different eigenvalues. 
The elements of each of these subspaces are called \emph{Weyl spinors} and 
satisfy, by definition, the Weyl or chirality condition,
\begin{align}
\frac{1}{2}\left(
	1
	\pm\gamma^{D+1}
	\right)\chi
&=
\chi
\, ,
\end{align}
where $\chi$ is an arbitrary spinor. The so-called \emph{left-} and \emph{right-handed} spinors correspond to the eigenvectors with eigenvalues +1 and -1, respectively.

We can also reduce Dirac spinors using the fact that, since $\gamma_\mu$ 
satisfies \eqref{eq:clifford}, then $\gamma_\mu^*$ and $\gamma_\mu^T$ do as well. 
This implies the existence of some isomorphisms relating these 
representations. One isomorphism, represented by $C$, relates
\begin{align}
C\gamma_{\mu\nu} C^{-1}
&=
-\gamma_{\mu\nu}^T
\, .
\end{align}
The matrix $C$ is called a \emph{charge conjugate matrix} and allows to define a charge-conjugate spinor,
\begin{align}
\hat \lambda&=\lambda^T C \, .
\end{align}
We can look for spinors whose charge-conjugate spinors are proportional 
to their Dirac conjugate $\bar\lambda$ defined by
\begin{align}
\bar\lambda &\equiv i\lambda^\dagger\gamma^0 \, .
\end{align}
That is, spinors satisfying
\begin{align}
\hat\lambda=\alpha\bar\lambda=\lambda^T C &
=\alpha i\lambda^\dagger\gamma^0 \, .
\end{align}
This is a `reality' condition for the spinors. The ones that fulfill 
it are called \emph{Majorana spinors}.
Sometimes chirality and Majorana conditions may be
 simultaneously satisfied. 

We schematically show in Table \ref{tab:spinors} some characteristics of 
the spinorial  irreducible representations for any dimension.

\begin{table}[t]
\begin{center}
\begin{tabular}{cccc}
\hline
$D\mod 8$ & spinor irreps & real components & $R$-symmetry
\\
\hline\hline
1, 3 & M & $2^{(D-1)/2}$ & $SO(N)$
\\
2    & MW & $2^{D/2-1}$ & $SO(N_L)\times SO(N_R)$
\\
4, 8 & M & $2^{D/2}$ & $U(N)$
\\
5, 7 & D & $2^{(D+1)/2}$ & $USp(2N)$
\\
6    & W & $2^{D/2}$  & $USp(2N_L)\times USp(2N_R)$\\
\hline
\end{tabular}

{\it
\caption{We show the different irreducible spinorial representations for every dimension $D$ and the number of real components. Depending on the dimension $D\text{ (mod }8)$, we can have Dirac (D), Weyl (W), Majorana (M) or Majorana-Weyl (MW) representations. In addition, we show the $R$-symmetry group for every dimension, where $N$ and $(N_L,N_F)$ denote the number of supersymmetric charges preserved.}
\label{tab:spinors}
}
\end{center}
\end{table}

\subsection{SUSY algebras and their representations}

From a mathematical point of view, a Lie superalgebra is 
an algebra based on 
a $\mathbb{Z}_2$ graded vector space
\cite{Frolicher:1956nb,Milnor65onthe,Miyazawa:1968zz}. 
The physical Bose and Fermi elements will be, respectively,
the grade 0 and grade 1 algebra vectors.

A Lie superalgebra $\mathfrak{s}$ satisfies the following properties:
\begin{itemize}
\item \emph{$\mathfrak{s}$ is a mod 2 graded vector space over $\mathbb{C}$}. \emph{I.e.}, it admits a map
\begin{align}
\text{gr}:\mathfrak{s}\longrightarrow \mathbb{Z}_2
\, ,
\end{align}
which decomposes $\mathfrak{s}$ into $\mathfrak{s}^{(0)}$ and $\mathfrak{s}^{(1)}$ in such a way that
\begin{align}
\text{gr}(B)
&=0\mod 2\, ,&\forall B\in \mathfrak{s}^{(0)}\, ,\nonumber\\
\text{gr}(F)&=1\mod 2\, ,&\forall F\in \mathfrak{s}^{(1)}\, .
\end{align}
\item \emph{$\mathfrak{s}$ is endowed with a binary operation, the bracket $\{,]$, which is bilinear, superanticommutative and mod 2 grade additive}, 

This means that, given $A,B\in\mathfrak{s}$, we have $[A,B]=-[B,A]$ in all cases but one, where both $A$ and $B$ are Fermi in which case $[A,B]=+[B,A]$. The mod 2 grade additivity means that denoting the grades $a,b,c$ of $A,B,C\in\mathfrak{s}$ respectively, if we have $[A,B]=C$ then $a+b=c \text{ (mod }2)$ has to be satisfied.
\begin{align}
\left\{ A,B\right]
&=
(-1)^{1+\text{gr}(A)\text{gr}(B)}\left\{B,A\right]
\, ,
\nonumber
\\
\text{gr}(\{A,B])
&=
\text{gr}(A)
+\text{gr}(B)
\, .
\end{align}

\item \emph{The bracket operation obeys the superJacobi identity}
\begin{align*}
(-1)^{1+\text{gr}(C)\text{gr}(A)}\{\{A,B],C]&\\
+(-1)^{1+\text{gr}(A)\text{gr}(B)}\{\{B,C],A]&\\
+(-1)^{1+\text{gr}(B)\text{gr}(C)}\{\{C,A],B]
&=
0
\, .
\end{align*}
This reduces to the ordinary Jacobi identity in all cases but one: when any two of the elements $A,B,C$ are Fermi and the third one is Bose, in which case one of the three usual Jacobi terms has its sign flipped.
\end{itemize}


The simple finite-dimensional Lie superalgebras over 
$\mathbb{C}$ are fully classified \cite{Kac:1975qb,Kac:1977qb}.
 There are eight infinite families, a 
continuum $D(2| 1;\alpha)$ of 17-dimensional exceptional 
superalgebras, and one exceptional superalgebra each 
in dimensions 31 and 40. 
The special linear $\mathfrak{sl}(m| n)$ and the 
orthosymplectic $\mathfrak{osp}(m| n)$ superalgebras are 
the most relevant ones from the physical point of view. 
The superalgebra $\mathfrak{osp}(4| N)$, which has 
as bosonic Lie algebra 
$\mathfrak{so}(3,2)\times\mathfrak{so}(N)$, 
corresponds to the AdS superalgebra. 
The superconformal one is $\mathfrak{su}(2,2| N)$, 
which has as Lie algebra $\mathfrak{so}(4,2) \times \mathfrak{su}(N)\times \mathfrak{u}(1)$.


It is of interest to us  superalgebras which include the Poincar\'e group. 
The Poincar\'e superalgebra (the superalgebra whose bosonic sector is 
strictly the Poincar\'e algebra) 
is spanned by the generators $\{P_\mu,M_{\mu\nu},Q^i{}_\alpha\}$. These generators satisfy the following relations:
\begin{align}
\label{eq:superPoincare}
[M_{\mu\nu},M^{\rho\sigma}]&=-2\delta_{[\mu}^{\left[\rho\right.}M_{\nu]}{}^{\sigma]}\, ,
&[P_\mu,M_{\nu\rho}]&=\eta_{\mu[\nu} P_{\rho]}\, ,&
[P_\mu,P_\nu]&=0\, ,\nonumber\\
[M_{\mu\nu},Q^i{}_\alpha]&=-\frac{1}{4}(\gamma_{\mu\nu})_\alpha{}^\beta Q^i{}_\beta\, ,&[P_\mu,Q^i{}_\alpha]&=0\, , \nonumber \\
\{Q^i{}_\alpha, Q^j{}_\beta \}&=(\gamma^\mu C^{-1})_{\alpha\beta} P_\mu \delta^{ij}\, .
\end{align}
The last term implies that two internal fermionic transformations lead 
to a spacetime translation. Here it is realized the basic feature of SUSY, 
the interplay between spacetime and some other internal symmetry.\footnote{Incidentally we observe here the spin-statistics 
connection at work: Fermi half-integer spin generators have 
to be anticommuting.}  

For superalgebras including the Poincar\'e group, the number of 
supercharges (or grade 1 generators) turns out  to be a 
multiple of the number of real components of an irreducible 
spinor. This is required by Lorentz invariance itself, 
since in this case the components of an irreducible 
spinor transform into each other. 
Thus, the supercharges $Q^i{}_\alpha$ carry two 
indices: $i=1,\ldots,N$, where $N$ is, in principle, 
an arbitrary integer, and $\alpha$ is an irreducible 
spinor index.

\subsubsection*{Some elementary properties}

Some well-known important properties can be straightforwardly inferred from the Fermi sector of the 
Poincar\'e superalgebra.
Let us take as an example the simplest ($N=1$) supersymmetric 
extension of the Poincar\'e algebra, which
 can be written in terms of  two complex Weyl spinors 
and their conjugates with the following anti-commutation 
relations:
\begin{align}
\{ Q_\alpha,  Q_{\beta}\}&=\{ Q_{\dot\alpha}^\dagger,  Q_{\dot\beta}^\dagger\}=0
\, ,
\nonumber\\
\{ Q_\alpha, \bar Q_{\dot\beta}\}&=2(\sigma^\mu)_{\alpha\dot\beta} P_\mu
\, .
\end{align}
Contracting the first relation 
with $(\bar\sigma^\nu)^{\dot\beta\alpha}$, we have
\begin{align}
4P_\nu&=(\bar\sigma^\nu)^{\dot\beta\alpha}\{ Q_\alpha, \bar Q_{\dot\beta}\}
\, .
\end{align}
In a quantum theory, the superalgebra generators are operators
in the Hilbert space of the system which includes bosonic and
fermionic states.
Single particle states fall into irreducible representations
of the algebra, the supermultiplets. 
Since the fermionic generators commute with 
$P^\mu P_\mu$, all particles in a supermultiplet have the 
same mass.\footnote{They have, in addition, the same charge corresponding
to any possible gauge symmetry.}

The time component $P_0$  corresponds to  
the Hamiltonian operator, which can be written as
\begin{align}
4P_0=4H
&=
\sum_\alpha \{ Q_\alpha, \bar Q_{\dot\alpha}\}=
\sum_\alpha \{ Q_\alpha,  Q_{\alpha}^\dagger\}
\nonumber\\
&=
\sum_\alpha (Q_\alpha Q_{\alpha}^\dagger+ Q_{\alpha}^\dagger Q_\alpha)
\, .
\end{align}

The expected value of the Hamiltonian in an arbitrary 
state $|s\rangle$ is given by
\begin{align}
\langle s |H|s\rangle
&=
\frac{1}{4}\sum_\alpha \langle s|(Q_\alpha Q_{\alpha}^\dagger+ Q_{\alpha}^\dagger Q_\alpha)|s\rangle
\nonumber\\
&=
\frac{1}{4}\sum_\alpha \sum_{s'} \langle s|Q_\alpha|s'\rangle\langle s'| Q_{\alpha}^\dagger|s \rangle+ \langle s|Q_{\alpha}^\dagger |s'\rangle\langle s'|Q_\alpha|s\rangle
\nonumber\\
&=
\frac{1}{2}\sum_\alpha\sum_{s'}
\left|\langle s' |Q_\alpha|s\rangle \right|^2
\geq 0 \, ,
\end{align} 
where we have introduced 
the closure relation 
$\mathds{1}=\sum_{s'}|s'\rangle\langle s'|$. 
Thus, we conclude that in a supersymmetric quantum
 theory, any physical state $|s\rangle$ must have 
non-negative energy. The inequality saturates if 
the ground (or vacuum) state 
denoted by $|0\rangle$
 is annihilated by a SUSY generator $ Q_\alpha|0\rangle =0$.
In this case, one talks of absence of spontaneous 
SUSY symmetry breaking.

Since  $Q_\alpha$ has spinorial indices, when it acts on a 
bosonic state of the Hilbert space it produces a spinor, 
fermionic state. Hence, any supermultiplet has 
both bosonic and fermionic states. 
Moreover, one can show that the number of bosonic states is equal to the 
number of fermionic ones for each supermultiplet
with non-zero energy.


Using the SUSY algebra properties, one can construct 
the corresponding algebra representations, that is, the detailed 
particle supermultiplet content \cite{Terning:2006bq}.
Since all the particles in the supermultiplet have the same 
mass, one can independently study the massive and massless 
cases. In both cases, the SUSY algebra reduces to a Clifford
algebra of raising and lowering anticommuting operators.
By combining the Clifford algebra representation theory and
maximal weight techniques, one can construct the entire 
massive or massless multiplets repeatedly applying 
``raising'' $Q^\dagger$ operators to  a given maximal spin state.
As an example,  the so-called massive (massless) `chiral' multiplet is formed by starting with 
a spin-0 state: it contains a Majorana (Weyl) 
fermion and a complex scalar. 
The massive vector multiplet is formed from a spin-$\tfrac{1}{2}$ initial
state and contains two Majorana fermions, a massive spin-1 vector
and a real scalar. The massless vector multiplet turns out to
be composed of a Weyl fermion and a massless spin-1 boson.

There exists a physical upper bound for $N$, the number of
spinorial charges. 
If $N \geq 9$, massless representations necessarily contain 
some undesirable particles of higher spin $s\geq 5/2$. 

If we restrict ourselves to theories with particles of spin
$s\le2$ (and not more than one time-like coordinate), the maximum number of supercharges that 
we can have is 32 and the theory may live in dimensions no higher than 11. 
We will refer to these SUSY theories with the maximal number of
supercharges as \emph{maximal}. 
%
In $D=1, 3, 4, 5, 7, 8, 9, 11$, the supersymmetric algebra is 
classified by a positive integer $N$, whereas for $D=2, 6, 10$, the SUSY algebra is classified by 
two integers $(N_L, N_R)$ which, at least, one of them has to be non-zero.
$N_L$ and $N_R$ represent the number of left-handed and right-handed supersymmetries, respectively.


For $N=1$, the super-Poincar\'e algebra is invariant under a multiplication of the fermionic charges $Q_\alpha$ by a 
phase. The corresponding symmetry group, called 
$U(1)_R$,  is the simplest example  of an additional symmetry at the 
level of the supercharges. The so-called \emph{$R$-symmetry} is a 
automorphism of the fermionic sector, which transforms different supercharges 
into each other. 
 For extended SUSY ($N\geq 2$), it becomes a non-Abelian group. 
Formally, it is defined as the largest subgroup of the 
automorphism group of the SUSY algebra that commutes with 
Lorentz transformations. In Table \ref{tab:spinors}, it is shown the $R$-symmetry group for any dimension.


\subsubsection*{Central charges}

SUSY algebras with $N\geq 2$ can be   extended 
by adding  `central charge' operators.
These can be Lorentz scalar 
 $Z^{ij}$ \cite{Haag:1974qh} 
or 
`tensorial' $Z^{ij}_{\mu_1\mu_2\dots}$ 
{\em central charges} 
\cite{deAzcarraga:1989gm,deAzcarraga:1989gu}.
They appear in the anticommutator of two 
SUSY generators as
\begin{align}
\{Q^i{}_\alpha, Q^j{}_\beta \}&=
(\gamma_\mu C^{-1})_{\alpha\beta} P_\mu \delta^{ij}
+\sum_k(\Gamma^{\mu_1\cdots\mu_k} C)_{\alpha\beta} Z^{ij}_{\mu_1\cdots\mu_p}
\, .
\end{align}
The possible combinations of central extensions will depend 
on the dimension and characteristics of the theory. 
For example, for $D=11$, we have \cite{vanHolten:1982mx}
\begin{align}
\{Q_\alpha, Q_\beta \}
&=
(\gamma_\mu C^{-1})_{\alpha\beta} P_\mu \delta^{ij}
+\sum_n(\Gamma^{\mu_1\mu_2} C)_{\alpha\beta} Z_{\mu_1\mu_2}
+\sum_n(\Gamma^{\mu_1\cdots\mu_5} C)_{\alpha\beta} Z_{\mu_1\cdots\mu_5}
\, .
\end{align}
For a pure scalar central charge we have \cite{Haag:1974qh}
($\epsilon= i \sigma^2$):
 \begin{align}
\{ Q_\alpha^i,  Q_{\beta}^j\}& = 2\epsilon_{\alpha\beta} Z^{ij} 
   \, , 
   \nonumber\\
\{ Q_{\dot\alpha}^i {}^\dagger,  Q_{\dot\beta}^j{}^\dagger\}&= 
2\epsilon_{\dot\alpha\dot\beta} Z^{ij}{}^\dagger 
\, ,\\
\{ Q_\alpha, \bar Q_{\dot\beta}\}&
=2(\sigma^\mu)_{\alpha\dot\beta} P_\mu\, .
\nonumber
\end{align}
It is possible to choose a set of states 
 $\left\{| M, Z^{ij}\rangle\right\}_{i<j}$, which are simultaneously eigenstates of $P^\mu P_\mu$ 
and $Z^{ij}$. The corresponding $N\times N$ matrix $(Z^{ij})$ is 
antisymmetric in its indices. This  matrix can be
 skew-diagonalized to $N/2$ real eigenvalues. 
Thus, for example, for $N=2$ one can write 
\begin{eqnarray}
(Z^{ij}) &=& Z (\epsilon^{ij}),
\end{eqnarray}
where $Z$ is a real parameter which can be assigned, 
in addition to the mass, to any corresponding multiplet. 
By a redefinition of the supercharges, and  ensuring that
all the states of the supermultiplet have 
 non-negative norm, one arrives to the
inequality \cite{Wess:1992cp} (see \cite{Terning:2006bq} 
for a simple example) 
\begin{eqnarray}
M &\ge& |Z| .
\end{eqnarray}
This is an example of a
\emph{Bogmol'nyi-Prasad-Sommerfeld (BPS)} 
bound \cite{Bogomolny:1975de,Prasad:1975kr}. 
In particular, for massless states, we have $Z=0$. The states 
that saturate the inequality, $M=|Z|$, have zero norm. Namely, they are annihilated by a fraction (a half, a quarter or an eighth) of the supercharges. 

The structure of the unitary representations of the algebra is 
different for the cases $M>|Z|$ and $M=|Z|$: the supermultiplets 
with $M=|Z|$ are smaller (called \emph{short multiplets}) 
than those corresponding to $M>|Z|$ (\emph{long multiplets}).
The short multiplets are also called
\emph{BPS multiplets} because they are related to BPS monopoles
 \cite{Bogomolny:1975de,Prasad:1975kr,deAzcarraga:1989gm,deAzcarraga:1989gu}.

\section{Supergravity}

Supergravity theories are  field theories that remain invariant 
under local supersymmetry \cite{Gliozzi:1976qd,Grisaru:1976nn,Ferrara:1976fu,vanNieuwenhuizen:1976bg,Chamseddine:1976bf,Nath:1976ci,Deser:1976eh,Freedman:1976xh}, \emph{i.e.} under super-Poincar\'e transformations with spacetime dependent commuting and anticommuting parameters.
Because of the underlying supersymmetry algebra, the invariance
under local supersymmetry implies the invariance under spacetime
diffeomorphisms. 
Therefore these theories are necessarily theories of gravity.
Supergravity
\footnote{It can be considered that supergravity was initially proposed 
in 1973 by  Volkov in 
\protect\cite{Volkov:1973ix} where there appears an action invariant
under local spacetime transformations with 10 commuting
and 4 anticommuting parameters.}
 was quickly generalized for several  dimensions and for
additional $N$ supersymmetric charges. 
The number of supercharges in a spinor depends on the 
dimension and the signature of spacetime. 
Supergravity theories  do not contain any fields 
that transform as symmetric tensors of rank higher than 
two under Lorentz transformations. 
Thus, the limit on the number of supercharges cannot be 
satisfied in a spacetime of arbitrary dimension. 
Supergravity can be formulated, in spacetimes with 
Lorentz signatures, in any number of dimensions up 
to 11 \cite{Nahm:1977tg}.

About supermultiplets, the most common ones that appear in supergravity are the following:
\begin{description}
\item[Gravity supermultiplet.] The field content satisfies $s_{max}=2$. It hosts the graviton plus $N$ gravitini at least. 

\item[Vector/gauge supermultiplet.] Here $s_{max}=1$. They exist for $N\leq4$ theories. The gauge fields of those multiplets can gauge an extra Yang-Mills-like group that commutes with supercharges and it is not part of the superalgebra.

\item[Chiral supermultiplet.] $s_{max}=1/2$. In $D=4$ theories, they only exist for $N=1$. Supersymmetry requires the scalars 
to span a K\"ahler-Hodge manifold. 
They must transform under the gauge group defined by the vector multiplet.

\item[Hypermultiplets.] They are the equivalent chiral multiplets for $N=2$. They also must transform under the gauge group. In this case, the scalars must parametrize a quaternionic K\"ahler manifold.

\item[Tensor multiplets.] They include  antisymmetric tensors 
$T_{\mu\nu...}$. In some cases, they can be dualized to scalar or vector fields and thus, be included in the other multiplets. 

\end{description}


\section{Maximal higher-dimensional supergravities}

In the next sections we are going to inspect the $D=9,D=10$ 
and $D=11$ maximal supergravities.

\subsection{$D=11$ supergravity}

In 1978, Cremmer, Julia and Scherk (CJS) \cite{Cremmer:1978km}
found the classical action for an 11-dimensional supergravity 
theory. Up to now, this is the only known 
classical 11-dimensional theory with local supersymmetry 
and no fields of spin higher than two. 

Other 11-dimensional theories are known that are 
quantum-mechanically inequivalent to the CJS theory, but 
classically equivalent. That is, they reduce to the CJS theory when one imposes the classical equations of motion. 
For example, in \cite{deWit:1986mz}, it is found a  $D=11$ supergravity 
with local $SU(8)$ invariance. 

The field content of CJS $D=11$ supergravity is 
\begin{align}
\left\{
	e_\mu{}^a
	\, ,
	C_{\mu\nu\rho}
	\, ,
	\psi_\mu 
	\right\}
\, .
\end{align}
That is, there is an `elfbein' $e_\mu{}^a$, a Majorana gravitino field $\psi_\mu$ and a 3-rank antisymmetric gauge field $C_{\mu\nu\rho}$.  Together with chiral $(2,0)$ supergravity  in $D=6$, it is the only $Q\geq16$ theory  without a scalar field. Its full action reads
\begin{align}
S&=
\frac{1}{2\kappa^2}\int d^{11}x e [
	e^{a\mu}e^{b\nu}R_{\mu\nu a b}
	-\bar\psi_\mu \gamma^{\mu\nu\rho}
D_\nu
\psi_\rho
	-\frac{1}{24}F^{\mu\nu\rho\sigma}F_{\mu\nu\rho\sigma}
\nonumber
\\
& 
-\frac{\sqrt2}{192}\bar\psi_\nu\left(
		\gamma^{\alpha\beta\gamma\delta\nu\rho}
		+12\gamma^{\alpha\beta}g^{\gamma\nu}g^{\delta\rho}
		\right)\psi_\rho(F_{\alpha\beta\gamma\delta}+\tilde F_{alpha\beta\gamma\delta})
\nonumber
\\
&		
-\frac{2\sqrt2}{(144)^2}e^{-1}\epsilon^{\alpha'\beta'\gamma'\delta'\alpha\beta\gamma\delta\mu\nu\rho} F_{\alpha'\beta'\gamma'\delta'} F_{\alpha\beta\gamma\delta} C_{\mu\nu\rho}
]
\, ,
\label{eq:D11_SUGRA}
\end{align}
where the Ricci scalar and the covariant 
derivative, respectively $R=R(\omega)$ 
and $D_\mu=D_\mu\left(\frac{1}{2}(\omega+\tilde\omega)\right)$,
depend on the spinorial connection 
$\omega$ and its supercovariant version $\tilde\omega$.
In components, we have for these and other quantities,
\begin{align}
\omega_{\mu a b}&=\omega_{\mu ab}(e)+K_{\mu a b}\, ,
\nonumber\\
\tilde\omega_{\mu a b}&=\omega_{\mu a b}(e)
+K_{\mu a b}
-\frac{1}{8}\bar\psi_\nu\gamma^{\nu\rho}{}_{\mu a b}\psi_\rho
\, ,
\nonumber
\\
K_{\mu a b}&=
-\frac{1}{4}\left(
	\bar\psi_\mu \gamma_a\psi_b
	-\bar\psi_a\gamma_\mu\psi_b
	+\bar\psi_b\gamma_a\psi_\mu
	\right)
+\frac{1}{8}\bar\psi_\nu\gamma^{\nu\rho}{}_{\mu a b}\psi_\rho
\, ,
\nonumber\\
\tilde F_{\mu\nu\rho\sigma}&=4\partial_{[\mu}C_{\nu\rho\sigma]}
+\frac{3\sqrt2}{2}\bar\psi_{[\mu}\gamma_{\nu\rho}\psi_{\sigma]}
\, ,
\end{align}
where
$\psi_b= e_b{}^\mu \psi_\mu$, 
and the covariant derivative $D_\mu$ acts on the spinors as follows:
\begin{align}
D_\mu\psi_\nu
&=
\partial_\mu\psi_\nu
+
\frac{1}{8}(\omega+\tilde\omega)_{\mu ab}
\gamma^{ab}\psi_\nu
\, .
\end{align}

Taking into account that $F=dC$ is the field strength of the 3-form $C_{\mu\nu\rho}$, the bosonic field equations and the Bianchi identity are
\begin{align}
\label{eq:eomsD11}
R_{\mu\nu}
&=
\frac{1}{72}g_{\mu\nu} F_{\rho\sigma\lambda\tau} F^{\rho\sigma\lambda\tau}
-\frac{1}{6}F_{\mu\rho\sigma\lambda}F_\nu{}^{\rho\sigma\lambda}
\, ,
\nonumber
\\
\partial_{\mu}(e F^{\mu\nu\rho\sigma})
&= 
\frac{1}{1152} \sqrt{2}\varepsilon^{\nu\rho\sigma\lambda\tau\alpha\beta\gamma\delta\kappa\pi } F_{\lambda\tau\alpha\beta } F_{\gamma\delta\kappa\pi }
\, ,
\\
\partial_{[\mu}F_{\nu\rho\sigma\lambda]}
&=
0
\, .
\nonumber
\end{align}
An alternative form for the second equation is
\begin{align}
\label{eq:D11-field-eq}
\partial_{[\mu}H_{\nu\rho\sigma\lambda\tau\alpha\beta]}
&=
0
\, ,
\end{align}
where $H_{\mu\nu\rho\sigma\lambda\tau\alpha}$ is the dual field strength,
\begin{align}
H_{\mu\nu\rho\sigma\lambda\tau\alpha }
&=
\frac{1}{7!}e \varepsilon_{\mu\nu\rho\sigma\lambda\tau\alpha\beta\gamma\delta\kappa } F^{\beta\gamma\delta\kappa} 
-\frac{1}{ \sqrt{2}} F_{[\mu\nu\rho\sigma } C_{\lambda\tau\alpha ]}
\, .
\end{align}

Let us analyze 
the constant $\kappa_{11}^{-2}$ that multiplies 
the Lagrangian and carries dimension $[\text{mass}]^9$. 
We can see that, in principle, it is undetermined and 
depends on fixing some length scale. If we apply the 
following shift on the fields (an $\mathbb{R}^+$ symmetry),
\begin{align}
e_\mu{}^a &\rightarrow e^{-\alpha}e_\mu{}^a
\, ,
&
\psi_\mu
&\rightarrow
e^{-\alpha/2}\psi_\mu
\, ,
&
C_{\mu\nu\rho}
&\rightarrow
e^{-3\alpha}C_{\mu\nu\rho}
\, ,
\end{align}
the Lagrangian rescales as
\begin{align}
\mathcal{L}_{11}
&\rightarrow
e^{-9\alpha}\mathcal{L}_{11}
\, .
\end{align}
This is the so-called \emph{trombone symmetry} \cite{Cremmer:1997xj} and 
it is manifest only at the level of the equations of motion. 
This scaling could be reabsorbed into a redefinition of $\kappa_{11}^{-2}$,
\begin{align}
\kappa_{11}^2
&\rightarrow
e^{-9\alpha} \kappa_{11}^2
\, .
\end{align}
For other supergravities in arbitrary $D$ dimensions, we observe a 
similar behavior.\footnote{Lower-dimensional theories inherit the trombone symmetry. This can be justified by dimensional reduction arguments \cite{Roest:2004pk}.} In general, we could make the following 
redefinitions:
\begin{align}
g_{\mu\nu}    &\rightarrow e^{-2\alpha}g_{\mu\nu}\, ,&
\mathcal{L}_D&\rightarrow e^{(2-D)\alpha}\mathcal{L}_D \, , &
\kappa_D^2   &\rightarrow e^{(2-D)\alpha}\kappa_D^2 \, .
\end{align}


\subsection{$D=10$ supergravities}
 In $D=10$  we have Majorana-Weyl (MW) irreducible spinors. 
The maximal supersymmetry is $N=2$, which gives rise to 
two discrete and inequivalent possibilities, $N=(1,1)$ with 
opposite chiralities and $N=(2,0)$, with same chirality. 
They correspond to the $N=2A$ and $N=2B$ theories, 
respectively.


\subsubsection{$N=2A$ supergravity}

The $N=2A$ 10-dimensional theory can be obtained by 
dimensional reduction of $D=11$ supergravity on a circle. 
Its field content is given by
\begin{align}
\left\{
	g_{\mu \nu}
	\, ,
\phi \, ,
B_{\mu \nu} \, ,
C^{(3)}_{ \mu \nu \rho} \, ,
C^{(1)}_{ \mu }    \, ,
\psi _{\mu} ^\pm \, ,
\chi^\pm
\right\}
\, .
\end{align}
The bosonic fields are split into the NSNS sector (the graviton $g_{\mu\nu}$, the dilaton $\phi$ and the 2-form $B$) and the RR sector (the 3-form $C^{(3)}$ and the graviphoton $C^{(1)}$), whereas the fermionic content consists of 2 MW gravitini $\psi_\mu{}^\pm$ and 2 MW dilatinos $\chi^\pm$. The gravitini and the two dilatinos
have opposite chiralities.

The bosonic part of the Lagrangian  is
\begin{align}
\mathcal{L}_{\text{2A}}
&=e
\left\{
	R
	-\tfrac{1}{2}(\partial\phi)^2
	-\tfrac{1}{2}e^{-\phi}|H|^2
	-\tfrac{1}{2}\sum_{d=1,3} e^{(4-d)\phi/2} |G^{(d+1)}|^2
	-\tfrac{1}{2}\star\left(
		dC^{(3)}\wedge dC^{(3)}\wedge B
		\right)
	\right\}
\, ,
\label{eq:2Aaction}
\end{align}
where $H=dB$ is the field strength associated to the NSNS
 2-form $B$ and $G^{(d+1)}$ is the modified field strength of $C^{(d)}$,
\begin{align}
G^{(d+1)}&=dC^{(d)}-dB \wedge C^{(d-2)}\, ,
\end{align}
for $d=1,3$. In this case we have two different $\mathbb{R}^+$ symmetries: 
one is  the trombone symmetry, a symmetry of the 
field equations
analogous to the existing one in $D=11$ supergravity and 
the other one 
is a symmetry of the Lagrangian, which acts on the field 
as follows:
\begin{align}
e^\phi
&\rightarrow
\lambda e^\phi
\, ,
&
B
&\rightarrow
\lambda^{1/2} B
\, ,
&
C^{(1)}
&\rightarrow
\lambda^{1/2} C^{(1)}
\, ,
&
C^{(3)}
&\rightarrow
\lambda^{-1/4} C^{(3)}
\, .
\end{align}

\subsubsection{$N=2B$ supergravity}

The field content of $N=2B$ $D=10$  supergravity is given by
\begin{align}
\left\{
g_{\mu \nu}   \, ,
B_{\mu \nu},   \, ,
\phi        \, ,
C^{(0)}      \, ,
C^{(2)}      \, ,
C_{\mu\nu\rho\sigma}^{(4)}{}_{SD} \, ,
\psi _{\mu \alpha} ^I \, ,
\lambda _\alpha ^I 
\right\}
\, ,
\end{align}
where $I=1,2$. The bosonic fields are contained in the NSNS common sector (the graviton $g_{\mu\nu}$, the dilaton $\phi$ and the 2-form $B$) and the RR sector (the axion $C^{(0)}$, the 2-form $C^{(2)}$ and the 4-form $C^{(4)}$), whereas the fermionic sector consists of 2 MW gravitini $\psi_\mu{}^\pm$ and 2 MW dilatinos $\chi^\pm$. The rank-4 antisymmetric tensor is supposed to have
 a self-dual field strength. 
Since this is a $N=(2,0)$ theory, both gravitini have the same
chirality. Both
 dilatinos also have the same chirality but opposite to that 
of the gravitini.

%

The Lagrangian of the bosonic sector is given by
\begin{align}
\mathcal{L}_{\text{2B}}
&=
e\left\{
	R
	-\tfrac{1}{2}(\partial\phi)^2
	-\tfrac{1}{2}e^{-\phi}|H|^2
	-\tfrac{1}{2}\sum_{d=0,2,4} |G^{(d+1)}|^2
	-\tfrac{1}{2}\star\left(
		C^{(4)}\wedge dC^{(2)}\wedge B
		\right)
	\right\}
\, ,
\label{eq:2Baction}
\end{align}
where $H=dB$ is, again, the field strength of $B$ and $G^{(d+1)}$ is given by
\begin{align}
G^{(d+1)}&=dC^{(d)}-dB \wedge C^{(d-2)}\, ,
\end{align}
for $d=0,2,4$.

The 5-form field strength $G^{(5)}$ satisfies a self-duality condition,
\begin{align}
G^{(5)}
&=
\star G^{(5)}
\, .
\end{align}
This condition does not follow from the equations of motion the $N=2B$ action, but has to be imposed as an extra constraint \cite{Bergshoeff:1995as}.

The $N=2B$ theory enjoys two symmetries: 
a trombone scaling symmetry and 
a $SL(2,\mathbb{R})$ symmetry. The former, as 
in the 11-dimensional case, is only realized on-shell 
whereas the latter is realized at the level of the Lagrangian and
acts on the fields as follows. 
Considering an $SL(2,\mathbb{R})$ element
\begin{align}
\Lambda =
\left(\begin{array}{cc}
a & b
\\
c & d
\end{array}
\right)
\in SL(2,\mathbb{R})
\, ,
\end{align}
the fields transform as
\begin{align}
\tau
&\rightarrow
\frac{a\tau+b}{c\tau+d}
\, ,
&
B^\alpha
&\rightarrow
(\Lambda^{-1})_\beta{}^\alpha B^\beta
\, ,
&
C^{(4)}
&\rightarrow
C^{(4)}
\, ,
\nonumber\\
\psi_\mu
&\rightarrow
\left(
	\frac{c\tau^*+d}{c\tau+d}
	\right)^{1/4}\psi_\mu 
\, ,
&
\chi
&\rightarrow
\left(
	\frac{c\tau^*+d}{c\tau+d}
	\right)^{3/4}\chi 
\, ,
\end{align}
where the complex scalar $\tau$ encodes the two real ones\footnote{Type IIB string theory breaks $SL(2,\mathbb{R})$ into its discrete 
subgroup $SL(2,\mathbb{Z})$. This group contains the 
so-called $S$-duality transformation that flips the sign of 
the dilaton $\phi$ in a background with vanishing axion 
$C^{(0)}$. Explicitly, this is done by 
choosing $a=d=0$ and $b=-c=1$ in the $SL(2,\mathbb{R})$ 
transformation. Because of its very definition, S-duality 
turns out to be a non-perturbative duality relating the strong- and weak-coupling regimes. 
}
\begin{align}
\tau
&\equiv
C^{(0)}
+ie^{-\phi}
\, ,
\end{align} 
and the pair of 2-form fields are arranged into a doublet
\begin{align}
B^\alpha
&\equiv (-B,C^{(2)})
\, .
\end{align}

\subsection{$N=2$ $D=9$ supergravity}
\label{sec-d9-primer}

Next, we are going to study  the maximal supergravity in 
$D=9$.
There is only one undeformed (\textit{i.e.}~ungauged, massless) maximal
(\textit{i.e.}~$N=2$, containing no dimensionful parameters in their action,
apart from the overall Newton constant) 9-dimensional supergravity
\cite{Gates:1984kr}.

The  theory has as (classical) global symmetry group
$SL(2,\mathbb{R})\times (\mathbb{R}^{+})^{2}$. The $(\mathbb{R}^{+})^{2}$
symmetries correspond to scalings of the fields, the first of which, that we
will denote by $\alpha$\footnote{This discussion follows closely that of
  Ref.~\cite{Bergshoeff:2002nv} in which the higher-dimensional origin of each
  symmetry is also studied. In particular, we use the same names and
  definitions for the scaling symmetries and we reproduce the table of scaling
  weights for the electric fields.}, acts on the metric and only leaves the
equations of motion invariant while the second of them, which we will denote
by $\beta$, leaves invariant both the metric and the action. The $\alpha$
rescaling corresponds to a
 \textit{trombone symmetry}.

Both the dimensional reduction of the massless
$N=2A,d=10$ theory and that of the $N=2B,d=10$ theory on a circle give the
same  $N=2,d=9$ theory
\footnote{
This is a property related to the T~duality between
type~IIA and~IIB string theories compactified on circles
\protect\cite{Dai:1989ua,Dine:1989vu} and from which 
the type~II Buscher's rules can be derived \protect\cite{Bergshoeff:1995as}.}.

The fundamental (\textit{electric}) fields of this theory are, 
\begin{equation}
  \left\{ e_{\mu}{}^{a},  \varphi, \tau\equiv  \chi+ie^{-\phi},
    A^{I}{}_{\mu}, B^{i}{}_{\mu\nu}, C_{\mu\nu\rho}, 
\psi_{\mu}, \tilde{\lambda}, \lambda,  \right\}\, .
\end{equation}

\noindent
where $I=0,\mathbf{i}$, with $\mathbf{i,j,k}=1,2$ and
$i,j,k=1,2$\footnote{Sometimes we need to distinguish the indices $1,2$ of the
  1-forms (and their dual 6-forms) from those of the 2-forms (and their dual
  5-forms). We will use boldface indices for the former and their associated
  gauge parameters.}.  The complex scalar $\tau$ parametrizes an
$SL(2,\mathbb{R})/U(1)$ coset that can also be described through the symmetric
$SL(2,\mathbb{R})$ matrix

\begin{equation}
\mathcal{M} 
\equiv 
e^{\phi}  
\left(
\begin{array}{cc}
|\tau|^{2}  & \chi \\
& \\
\chi & 1 \\
\end{array}
\right)\, ,
\hspace{1cm}
\mathcal{M}^{-1} 
\equiv 
e^{\phi}  
\left(
\begin{array}{cc}
1 & - \chi  \\
& \\
- \chi & |\tau|^{2}   \\
\end{array}
\right)\, .
\end{equation}

The field strengths of the electric $p$-forms are, in our
conventions
\footnote{We use the shorthand notation $A^{IJ}\equiv A^{I}\wedge
  A^{J}$, $B^{ijk}\equiv B^{i}\wedge B^{j}\wedge B^{k}$
  etc.}${}^{,}$\footnote{The relation between these fields and those of
  Refs.~\protect\cite{Meessen:1998qm} and \protect\cite{Bergshoeff:2002nv} are given in
  Appendix~\protect\ref{d9:sec-relationwithotherconventions}.}
\begin{eqnarray}
F^{I}& = & dA^{I}\, ,\\
\label{eq:Hiundeformed}
H^{i}& = & dB^{i}
+\tfrac{1}{2}\delta^{i}{}_{\mathbf{i} }(A^{0}\wedge F^{\mathbf{i}}
+A^{\mathbf{i}}\wedge F^{0})\, ,\\
G& = & 
d[C -\tfrac{1}{6} \varepsilon_{\mathbf{ij}}A^{0\mathbf{ij}}] 
-\varepsilon_{\mathbf{i}j}
F^{\mathbf{i}}  \wedge\left( B^{j} 
  +\tfrac{1}{2}\delta^{j}{}_{\mathbf{j}}A^{0\mathbf{j}} \right)\, ,
\end{eqnarray}
and are invariant under the  gauge transformations
\begin{eqnarray}
\delta_{\Lambda} A^{I} & = & -d\Lambda^{I}\, ,\\
\delta_{\Lambda} B^{i} & = & 
-d\Lambda^{i}
+\delta^{i}{}_{\mathbf{i}} \left[
\Lambda^{\mathbf{i}} F^{0}
+\Lambda^{0} F^{\mathbf{i}}
+\tfrac{1}{2} 
\left(A^{0} \wedge \delta_{\Lambda}A^{\mathbf{i}} 
+A^{\mathbf{i}} \wedge \delta_{\Lambda}A^{0}\right)
\right]
\, , \qquad\\
\delta_{\Lambda} [C -\tfrac{1}{6} \varepsilon_{\mathbf{ij}}A^{0\mathbf{ij}}]
& = & -d\Lambda
-\varepsilon_{\mathbf{i}j} 
\left( 
F^{\mathbf{i}}\wedge \Lambda^{j}
+\Lambda^{\mathbf{i}}\wedge H^{j}
- \delta_{\Lambda}A^{\mathbf{i}}\wedge B^{j}
\right.
\nonumber \\
& & \left.
+\tfrac{1}{2} \delta^{j}{}_{\mathbf{j}} A^{0\mathbf{i}}\wedge \delta_{\Lambda}A^{\mathbf{j}}
\right)\, .
\end{eqnarray}

The bosonic action is, in these conventions, given by
\begin{equation}
\label{eq:undeformedaction}
\begin{array}{rcl}
S & = & 
{\displaystyle \int}
\biggl \{
-\star R +\tfrac{1}{2}d\varphi \wedge \star d\varphi   
+\tfrac{1}{2}\left[ d\phi \wedge \star d\phi +e^{2\phi} d\chi \wedge \star d\chi\right]
+\tfrac{1}{2} e^{\frac{4}{\sqrt{7}}\varphi} F^{0}\wedge \star F^{0}
\\
& & \\
& & 
+\tfrac{1}{2} e^{\frac{3}{\sqrt{7}}\varphi} (\mathcal{M}^{-1})_{\mathbf{ij}}
F^{\mathbf{i}}\wedge \star F^{\mathbf{j}}
+\tfrac{1}{2} e^{-\frac{1}{\sqrt{7}}\varphi} (\mathcal{M}^{-1})_{ij}
H^{i}\wedge \star H^{j}
+\tfrac{1}{2} e^{\frac{2}{\sqrt{7}}\varphi} G \wedge \star G
\\
& & \\
& & 
-\tfrac{1}{2}
\left[
G+\varepsilon_{\mathbf{i}j}A^{\mathbf{i}}\wedge 
\left(H^{j} -\tfrac{1}{2}\delta^{j}{}_{\mathbf{j}}A^{\mathbf{j}}\wedge F^{0}\right)
\right] \wedge
\left\{
\left[
G+\varepsilon_{\mathbf{i}j}A^{\mathbf{i}}\wedge 
\left(H^{j} -\tfrac{1}{2}\delta^{j}{}_{\mathbf{j}}A^{\mathbf{j}}\wedge F^{0}\right)
\right]\wedge A^{0}
\right.
\\
& & \\
& & 
\left.
-\varepsilon_{ij}
\left(H^{i} -\delta^{i}{}_{\mathbf{i}}A^{\mathbf{i}}\wedge F^{0}\right)
\wedge
\left(B^{j} -\tfrac{1}{2}\delta^{j}{}_{\mathbf{j}}A^{0\mathbf{j}}\right)
\right\} 
\biggr \}\, .
\end{array}
\end{equation}

The kinetic term for the $SL(2,\mathbb{R})$ scalars $\phi$ and $\chi$ can be
written in the alternative forms

\begin{equation}
\tfrac{1}{2}\left[ d\phi \wedge \star d\phi +e^{2\phi} d\chi \wedge \star d\chi\right]
=
\frac{d\tau \wedge \star d\bar{\tau}}{2(\Im {\rm m}\tau)^{2}}
= 
\tfrac{1}{4} \mathrm{Tr}\left[ d\mathcal{M}\mathcal{M}^{-1} \wedge \star 
d\mathcal{M}\mathcal{M}^{-1}\right]\, ,
\end{equation}

\noindent
the last of which is manifestly $SL(2,\mathbb{R})$-invariant. The Chern-Simons
term of the action 
(the last two lines of Eq.~(\ref{eq:undeformedaction})) can
also be written in the alternative form

\begin{equation}
\begin{array}{l}
-\tfrac{1}{2}
d\left[ C -\tfrac{1}{6}\varepsilon_{\mathbf{ij}}A^{0\mathbf{ij}}
-\varepsilon_{\mathbf{i}j}A^{\mathbf{i}}\wedge B^{j}
\right] \wedge
\left\{
d\left[ C -\tfrac{1}{6}\varepsilon_{\mathbf{ij}}A^{0\mathbf{ij}}
-\varepsilon_{\mathbf{i}j}A^{\mathbf{i}}\wedge B^{j}
\right]\wedge A^{0}
\right.
\\
\\
\left.
-\varepsilon_{ij}
d\left(B^{i} -\tfrac{1}{2}\delta^{i}{}_{\mathbf{i}}A^{0\mathbf{i}}\right)
\wedge
\left(B^{j} -\tfrac{1}{2}\delta^{j}{}_{\mathbf{j}}A^{0\mathbf{j}}\right)
\right\} 
\, ,
\end{array}  
\end{equation}

\noindent
that has an evident 11-dimensional origin.

The equations of motion of the scalars, derived from the action above, are

\begin{eqnarray}
d\star d \varphi
-\tfrac{2}{\sqrt{7}} e^{\frac{4}{\sqrt{7}}\varphi} F^{0}\wedge \star F^{0}
-\tfrac{3}{2\sqrt{7}} e^{\frac{3}{\sqrt{7}}\varphi} 
(\mathcal{M}^{-1})_{\mathbf{ij}}F^{\mathbf{i}}\wedge \star F^{\mathbf{j}}
& & \nonumber \\
& & \nonumber \\
+\tfrac{1}{2\sqrt{7}} e^{-\frac{1}{\sqrt{7}}\varphi} (\mathcal{M}^{-1})_{ij}
H^{i}\wedge \star H^{j}
-\tfrac{1}{\sqrt{7}} e^{\frac{2}{\sqrt{7}}\varphi} G \wedge \star G
& = & 0\, ,\\
& & \nonumber \\
d \left[\star\frac{d\bar{\tau}}{(\Im {\rm m} \tau)^{2}} \right]
-i \frac{d\tau \wedge \star d\bar{\tau}}{(\Im {\rm m}\tau)^{3}}
-\partial_{\tau}(\mathcal{M}^{-1})_{\mathbf{ij}}
\left[F^{\mathbf{i}}\wedge \star F^{\mathbf{j}}+H^{i}\wedge \star H^{j} \right]
& = & 0\, ,
\end{eqnarray}

\noindent
and those of the fundamental $p$-forms ($p\geq 1$), after some algebraic
manipulations, take the form

\begin{align}
\label{eq:eomA0}
d\left(e^{\frac{4}{\sqrt{7}}\varphi}\star F^{0} \right)
& = 
-e^{-\frac{1}{\sqrt{7}}\varphi} \mathcal{M}^{-1}_{\mathbf{i}j} 
F^{\mathbf{i}} \wedge \star H^{j} +\tfrac{1}{2}G \wedge G\, ,
\\  
& \nonumber \\
\label{eq:eomAi}
d\left(e^{\frac{3}{\sqrt{7}}\varphi}\mathcal{M}^{-1}_{\mathbf{ij}}
\star F^{\mathbf{j}} \right)
& = 
-e^{\frac{3}{\sqrt{7}}\varphi} \mathcal{M}^{-1}_{\mathbf{i}j} 
F^{0} \wedge \star H^{j}
+\varepsilon_{\mathbf{i}j} e^{\frac{2}{\sqrt{7}}\varphi}H^{j}\wedge \star G\, ,
\\  
& \nonumber \\
\label{eq:eomBi}
d\left(e^{-\frac{1}{\sqrt{7}}\varphi}\mathcal{M}^{-1}_{ij}\star H^{j} \right)
& = 
\varepsilon_{\mathbf{i}j} e^{\frac{2}{\sqrt{7}}\varphi}F^{\mathbf{j}} \wedge \star G
-\varepsilon_{ij}H^{j}\wedge G\, ,
\\
& \nonumber \\ 
\label{eq:eomC} 
d\left(e^{\frac{2}{\sqrt{7}}\varphi}\star G \right)
& = 
F^{0} \wedge G+ \tfrac{1}{2}\varepsilon_{ij}H^{i}\wedge H^{j}\, .
\end{align}

The study of this theory and its possible deformations will be extensively addressed in Chapter~\ref{ch:gaugings_d9}.

\chapter{Gauged supergravities and the embedding tensor}
\label{ch:gauged_sugra}

At any dimension, with the remarkable exception of $D=11$,
there are some \emph{deformations} of the known basic supergravity
theories.
These deformations may consist of the introduction of
a superpotential and modifications of kinetic terms, a 
gauging of the R-symmetry group in extended supergravities,
or of the global symmetries of the theory, etc.

We restrict ourselves to gauged supergravities, \emph{i.e.} supergravities
in which vector fields gauge a Yang-Mills group.
In this case the number of generators of the gauge group
(including Abelian components) equals the number of
\cite{Freedman:2012zz}.


Important examples of known basic supergravities are those
constructed by dimensional reduction. For example,
the dimensional reduction of the
common sector of the $10$-dimensional  supergravities  on a $T^n$ $n$-torus,
provides a theory with  Abelian gauge symmetry and a
manifest global symmetry, $O(n,n)$ in this case.
The Kaluza-Klein vector fields  and the genuine vector fields combine
into $2n$ vector fields  which transform as the fundamental representation of
$G\equiv O(n,n)$. The scalar fields take values in the coset
$O(n, n)/O(n)\times O(n)$.

One is  typically  interested in gauged supergravities that arise
as deformations of this theory. One way of deforming the theory
is to perform the dimensional reduction on twisted tori with fluxes.
Another possibility consists of promoting
a subgroup $G_0\subset G$ to a local symmetry gauged by the
existing vector fields in the theory. 


A possible systematic approach to the problem of gauging a theory consistentlyis provided
by the embedding tensor formalism\footnote{For recent reviews see  refs.~\cite{Trigiante:2007ki,Weidner:2006rp,Samtleben:2008pe}.}.
This formalism, introduced in
refs.~\cite{Cordaro:1998tx,deWit:2002vt,deWit:2003hq,deWit:2005hv,deWit:2005ub}
allows the study of the most general deformations of field
theories and, in particular, of supergravity theories
\cite{deWit:2004nw,Samtleben:2005bp,Schon:2006kz,deWit:2007mt,Bergshoeff:2007vb,deWit:2008ta,Bergshoeff:2008bh,Hartong:2009az,Huebscher:2010ib}.
In this formalism, if the generators of $G$ are denoted by
$t_\alpha$,
then the generators $X_M$ of the subgroup $G_0\subset G$ to be gauged
are conveniently specified
by the \emph{embedding tensor}
$\vartheta_M{}^\alpha$,
so that
\begin{align}
X_M
&=
\vartheta_M{}^\alpha t_\alpha
\, .
\end{align}


All the terms in the (purely bosonic) deformed action, except a
possible scalar potential,  are completely determined by gauge
invariance (that is, by $G_0$ throughout the embedding tensor)
and by the requirement of recovering the undeformed action
in the ungauged limit ($\vartheta_M{}^\alpha\to 0$). The scalar potential
could be in principle any $G$-invariant function of the scalar fields.
In the supersymmetric case, this is the bosonic sector of
a gauged supergravity action. In addition, the fermionic sector results modified by the addition of fermionic mass terms to keep supersymetry preserved.

Supersymmetry leads to
further restrictions. In general, it rules out some of the
possible gauge groups (by imposing certain linear constraints on
the components of the embedding tensor) and determines the
form of the scalar potential.
The ungauged action is manifestly invariant under $G$
 global transformations, while
the couplings of the gauged action would break this to a subgroup.
However, it becomes invariant if the embedding tensor behaves as a spurionic object and is allowed to transform under the gauge group according to its index structure.

\subsubsection{Extended objects and gauged SUGRAS}

The construction of gauged supergravities is not only important from
the viewpoint of enlarging the catalogue of known
supergravities, but also as a way of completing our knowledge
of extended objects in supergravity and string theory. Gauged
supergravities become relevant in the study of string
configurations and flux compactifications. They are a reliable
scenario under which string theory results can be tested.

The discovery of the relation between RR $(p+1)$-form potentials
in 10-dimensional type~II supergravity theories and D-branes
\cite{Polchinski:1995mt} made it possible to associate most of
the fields of the string low-energy effective field theories
(supergravity theories in
general) to extended objects (\textit{branes}) of diverse kinds: fundamental,
Dirichlet, solitonic, Kaluza-Klein, etc. This association has been fruitfully
used in two directions: to infer the existence of new supergravity fields from
the known existence in the String Theory of a given brane
or string state and
\textit{vice versa}. Thus, the knowledge of the existence of D$p$-branes with
large values of $p$ made it necessary to learn how to deal consistently with
the magnetic duals of the RR fields that were present in the standard
formulations of the supergravity theories constructed decades before, because
in general it is impossible to dualize and rewrite the theory in terms of the
dual magnetic fields. The existence of NSNS $(p+1)$-forms in the supergravity
theories that could also be dualized made it necessary to include solitonic
branes dual to the fundamental ones (strings, basically).


The search for all the extended states of string theory has
motivated the search for all the fields that can be
consistently introduced in the corresponding supergravity
theories, a problem that has no simple answer for the $d$-, $(d-1)$ and $(d-2)$-form fields, which are not the duals of electric
fields already present in the standard formulation,
at least in any obvious way. The branes that would couple
to them can play important r\^oles in String Theory models,
which makes this search more interesting.

 U-duality arguments, systematic studies of the possible
consistent supersymmetry transformation rules for $p$-forms
 in the 10-dimensional maximal supergravities ~\cite{Bergshoeff:1999bx,Bergshoeff:2001pv,Bergshoeff:2005ac,Bergshoeff:2006qw,Bergshoeff:2010mv,Greitz:2011da}
or the use of the conjectured infinite dimensional
$E_{11}$ symmetry \cite{Julia:1997cy,West:2001as,Riccioni:2009xr}
have been used to determine the bosonic extended field
content of maximal supergravity in different dimensions.

Another possible systematic approach to this problem  is provided
by the embedding-tensor gauging formalism presented before.
One of the main features of this formalism is that it requires
the systematic introduction of new higher-rank potentials which
are related by St\"uckelberg gauge transformations.
This structure is known as the \textit{tensor  hierarchy} of
the theory
\cite{deWit:2005hv,deWit:2005ub,deWit:2008ta,Bergshoeff:2009ph,deWit:2009zv,Hartong:2009vc}
and can be taken as the (bosonic) extended field content of
the theory. In Supergravity Theories one may need to take into
account additional constraints on the possible gaugings, but,
if the gauging is allowed by supersymmetry,  gauge invariance
will require the introduction of all the fields in the associated tensor hierarchy

This formalism  cannot be used in the most interesting
cases, $N=1,d=11$ and $N=2A,B,d=10$ Supergravity, because these
theories cannot be gauged because they do not have
1-forms ($N=1,d=11$ and $N=2B,d=10$) or the 1-form transforms
under the only (Abelian) global symmetry ($N=2A,d=10$).
Only $N=2A,d=10$ can be deformed through the introduction of
Romans' mass parameter, but the consistency of this deformation
does not seem to require the introduction of any higher-rank
potentials. The dimensional reduction to $d=9$ of these theories,
 the unique $d=9$ maximal supergravity, though, has 3 vector
fields, and their embedding tensor formalism can be used to
study all its possible gaugings and find its extended field
content.


This chapter is dedicated to the introduction
of the basic aspects of the embedding tensor formalism.
We will study  how to gauge a given supergravity theory,
\emph{i.e.} we will choose a subgroup $G_0\subset G$ and
promote it to a local symmetry. This is  a covariant formalism
that preserves and guarantees the covariance of the final theory.
In the next chapter this formalism will be employed to find
all the possible gaugings of the $d=9$ maximal supergravity.

\section{The embedding tensor}

The r\^ole of the vector fields already existing in a theory
are crucial in any gauging procedure.
The vectors $A_\mu{}^M$ of a typical ungauged theory transform
under a group  $G$ of global transformations and under an Abelian
gauge symmetry $U(1)^{n_V}$, where $n_V$ is the number of vector
fields in the theory.
If $\xi^\alpha$ is the transformation parameter of the global
symmetry $G$ and $\Lambda^M$ is the transformation parameter
of $U(1)^{n_V}$ symmetry, the fields transform as
\begin{align}
\delta_\xi A_\mu{}^M&=
-\xi^\alpha(t_\alpha)_N{}^M A_\mu{}^N
\, ,
&
\delta_\Lambda A_\mu{}^M
&=
\partial_\mu \Lambda^M
\, ,
\end{align}
where $t_\alpha$ are the generators of $G$, $M=1,\ldots,n_V$ is
an index of the fundamental representation and
$\alpha=1,\ldots,\dim G$ is an index of the adjoint
representation.

In general, any other generic field $V^{(r)}$ of the theory
transforming under a certain representation (symbolically denoted by the superindex $(r)$) of the global symmetry group $G$, will transform as
\begin{align}
\delta {V}^{(r)}
&=
\Lambda^\alpha t_\alpha{}^{(r)} {V}
\, ,
\end{align}
where $t_\alpha{}^{(r)}$ are the group generators in the
corresponding representation.


The aim of the gauging procedure is to promote an undetermined
subgroup (or subgroups) $G_0\subset G$ to a local symmetry.
Let us assume a subset of generators
$X_M\subset\mathfrak{g}=\lie{G}$
to be the candidates to be gauged.
The explicit embedding of $G_0$ into $G$ is given by a  $(n_V\times\dim G)$
matrix $\vartheta_M{}^\alpha$. This is the so-called
\emph{embedding tensor}, which describes the relation between
the global and the gauge candidate generators
\begin{align}
X_M&=\vartheta_M{}^\alpha t_\alpha\in\mathfrak{g}\, .
\end{align}
This is a relation at the level of the
abstract 
Lie algebras.
For any specific representation $(r)$ of the algebra,
of the algebra this is
translated to a relation of the form
\begin{align}
X_M^{(r)}&=\vartheta_M{}^\alpha t_\alpha{}^{(r)}\, .
\end{align}
In particular, for the fundamental representation, we have
\footnote{The representation label will be
 generally suppressed if there is no ambiguity.}
\begin{align}
X_{MN}{}^P = \vartheta_M{}^\alpha (t_\alpha)_N{}^P
\, .
\end{align}

In general, the closure of the algebra generated by the
gauge generators $X_M$ is not guaranteed. In principle, we have
\begin{align}
\label{eq:notclosureX}
[X_M,X_N]&=
-Z_{MN}{}^PX_P
-X_{MN}{}^PX_P
\, ,
\end{align}
where the gauge generators are split into their symmetric and antisymmetric parts, respectively,
\begin{align}
X_{MN}{}^P
&=
Z_{(MN)}{}^P
+
X_{[MN]}{}^P
\, .
\end{align}
In the following paragraphs, we will see how this situation is solved.

On the other hand, the local symmetry
 that we want to establish is implemented by means of the following covariant derivative
\begin{align}
\partial_\mu
&\rightarrow
D_\mu = \partial_\mu -  A_\mu{}^M X_M
\, ,
\end{align}
where $X_M$ is realized in the corresponding representation of the object
on which the derivative is applied upon.
The new covariant derivatives will guarantee the covariance
of the theory under the local symmetry group.
The global covariance of the theory is also preserved along
the procedure. Only when we choose a particular gauge
group $G_0$, \emph{i.e.} a particular $\vartheta_M{}^\alpha$ configuration,
$G$ gets broken. Indeed, the dimension of the final gauge group is the rank
of $\vartheta_M{}^\alpha$, which fulfills $\text{rank}(\vartheta_M{}^\alpha)\le \min(n_V,\text{dim } G)$.
In order to respect the global $G$ covariance of the theory, the
embedding tensor is considered as a spurionic field
with global and local transformation properties.
Thus, gauge transformations are supposed to
act on it in the corresponding way:
\begin{align}
\label{eq:QC}
\delta_\Lambda \vartheta_M{}^\alpha
&=
\Lambda^N X_{NM}{}^P\vartheta_P{}^\alpha
-\Lambda^N X_{N\beta}{}^\alpha \vartheta_M{}^\beta
\nonumber\\
&=
\Lambda^N \vartheta_N{}^\beta\left(
	t_{\beta M}{}^P\vartheta_P{}^\alpha
	-f_{\beta\gamma}{}^\alpha\vartheta_M{}^\gamma
	\right)
\, ,
\end{align}
where we have used that, in the adjoint representation,
$X_{N\gamma}{}^\alpha =\vartheta_M{}^\beta f_{\beta\gamma}{}^\alpha$
and $f_{\alpha\beta}{}^\gamma$ are the structure constants
of $G$. The constraints \eqref{eq:QC}, obtained
by demanding the gauge invariance of the embedding tensor, is a
set of second degree constraints in $\vartheta_M{}^\alpha$,
the so-called \emph{quadratic constraints} (QC).
The QC guarantee the closure of the algebra of the gauge generators for any representation. After its imposition, we have
\begin{align}
\label{eq:closureX}
[X_M,X_N]&=-X_{MN}{}^PX_P\, .
\end{align}
so we can check that
\begin{align}
\label{eq:ZXcancel}
Z^P{}_{MN} X_P
&=
0
\, .
\end{align}
However, if we define $X_{[MN]}{}^P$ to be the structure constants, we realize that
\begin{align}
X_{[MN]}{}^P X_{[QP]}{}^R
+X_{[QM]}{}^P X_{[NP]}{}^R
+X_{[NQ]}{}^P X_{[MP]}{}^R
&=
-Z^R{}_{P[Q} X_{MN]}{}^P
\, .
\end{align}
That is, Jacobi identity is satisfied upon  contracting with $X_R$, due to the condition \eqref{eq:ZXcancel}. This is enough for  the QC \eqref{eq:QC} to be satisfied.


The embedding tensor components can be decomposed into
irreducible representations. In general, we have
\begin{align}
\vartheta_M{}^\alpha:
&=
V'\otimes \mathfrak{g}_0
=
\theta_1\oplus\theta_2\oplus\cdots\oplus\vartheta_k
\, ,
\label{eq:linear_constraint}
\end{align}
where $V'$ is the conjugate representation of the
fundamental $V$, $\mathfrak{g}_0$ is the adjoint representation
and $\theta_i$ are several irreps.
In a theory with a given number of bosonic $p$-form fields transforming in
different representations, gauge consistency of the tensor
hierarchy usually implies the existence of additional linear
constraints on the embedding tensor. In addition, there may also
exists a \emph{linear constraint} (LC) arising from supersymmetry
(this constraint is not necessarily independent of the linear
constraints arising from the bosonic sector).
That is, SUSY kills some of the representations of the embedding
tensor. In Chapter \ref{ch:gaugings_d9}, we will see how this
restriction explicitly appears when we study the closure of
the supersymmetric transformations of the fields in $D=9$ maximal
supergravity.

Then, the linear constraints restrict the r.h.s. of \eqref{eq:linear_constraint}. In Table \ref{tab:ETrepresentations} we have the resulting
representations of the embedding tensor in maximal theories.
For half-maximal supergravities, the structure is similar.
Thus, the classification of all the possible gaugings of a
given theory reduces to the search and analysis of solutions
of the quadratic and linear constraints.
 Moreover, the counting of inequivalent gaugings
or identification of the different orbits
is also a non-trivial problem to be solved.

\begin{table}[t!]
\begin{center}
\begin{tabular}{cccccc}
\hline
$D$ & $G$ & $H$ & $\sharp$ scalars & vectors & $\vartheta$\\
\hline\hline
9 & $\mathbb{R}^+\times SL(2)$ & $SO(2)$ & 3 & $\mathbf{1}_{+4}+\mathbf{2}_{-3}$ & $\mathbf{2}_{+3}+\mathbf{3}_{-4}$\\
8 & $SL(2)\times SL(3)$ & $SO(2)\times SO(3)$ & 7 & $(\mathbf{2},\mathbf{3}')$ & $(\mathbf{2},\mathbf{3})+(\mathbf{2},\mathbf{6}')$\\
7 & $SL(5)$ & $SO(5)$ & 14 & $\mathbf{10}'$ & $\mathbf{15}+\mathbf{40}'$\\
6 & $S0(5,5)$ & $SO(5)\times SO(5)$ & 25 & $\mathbf{16}$ & $\mathbf{144}$\\
5 & $E_{6(6)}$ & $USp(8)$ & 42 & $\mathbf{27}'$ & $\mathbf{351}$\\
4 & $E_{7(7)}$ & $SU(8)$ & 70 & $\mathbf{56}$ & $\mathbf{912}$\\
\hline
\end{tabular}
{\it
\caption{In this table we show some aspects of maximal supergravities in various dimensions. $G$ is the global symmetry group and $H$ is its maximal compact subgroup. We show the representation of the vector fields and the embedding tensor, where subindices refer to the weights of the corresponding representation with respect to the $\mathbb{R}^+$ scaling.}
\label{tab:ETrepresentations}
}
\end{center}
\end{table}


\section{Deformed tensor gauge algebra}

\qquad Once we have introduced a covariant derivative,
it might be a  natural  ansatz to define a generalized
field strength by the expression
\begin{align}
\label{eq:wrongF}
F_{\mu\nu}{}^M&=2\partial_{[\mu}A_{\nu]}{}^M+X_{[NP]}{}^M A_\mu{}^N A_\nu{}^P\, .
\end{align}
However this is a too na\"{\i}ve hypothesis, since it  does not transform covariantly,
\begin{align}
\label{eq:wrongFtransf}
\delta_\Lambda F_{\mu\nu}{}^M
&=
-\Lambda^P X_{PN}{}^M F_{\mu\nu}{}^N
+2Z^M{}_{PQ}\left(
	\Lambda^P F_{\mu\nu}{}^Q
	-A_{[\mu}{}^P\delta A_{\nu]}{}^Q
	\right)
\, .
\end{align}
Only when $Z^M{}_{PQ}$ vanishes, the field strength transforms covariantly.

The condition of keeping $G$ covariance is the responsible of
this situation. We are performing a redundant description of
the gauging in terms of the $n_V$ generators $X_M$.
In general, since the dimension of the gauge group is smaller
than that of the global symmetry group, $n_V$, not all of
the $X_M$ generators are linearly independent.
For some cases, we can split the vector fields into two groups
\begin{itemize}
\item $A_\mu{}^m$, which transform in the adjoint of $G_0$,
\item $A_\mu{}^i$, which transform in some representation of $G_0$,
\end{itemize}
so that $Z^m{}_{PQ}=0$, and $Z^i{}_{PQ}\neq0$.\footnote{For some explicit examples, see \cite{Samtleben:2008pe}.}
For some particular examples, this can be done and the
problem can be circumvented. However, a general procedure
is required.

Let us now define the generalized field strength by the
expression
\begin{align}symbolically fu
\mathcal{F}^{(1)}_{\mu\nu}{}^M&=F_{\mu\nu}{}^M+Z^M{}_{PQ} B_{\mu\nu}{}^{PQ}\, ,
\label{eq:goodF}
\end{align}
where $B_{\mu\nu}{}^{PQ}$ are  2-forms which maybe belong to
 the field content of the corresponding theory.
Then, we can balance the contribution of the non-covariant
terms of \eqref{eq:wrongFtransf} if the gauge
transformations of the 1- and 2-form fields are
\begin{align}
\delta A_\mu{}^M&=D_\mu\Lambda^M-Z^M{}_{PQ} \Xi_\mu{}^{PQ}\, ,
\nonumber\\
\delta B_{\mu\nu}{}^{MN}&=
2 D_{[\mu} \Xi_{\nu]}{}^{MN}-2 \Lambda^{(M} \mathcal{F}^{(1)}_{\mu\nu}{}^{N)}
+2 A_{[\mu}{}^{(M}\delta A_{\nu]}{}^{N)}
\, ,
\label{eq:gauge_transf_fields}
\end{align}
where $\Xi_\mu{}^{MN}$ is a 1-form gauge parameter.
Thus, we have a St\"uckelberg-type coupling between the vector fields and the antisymmetric 2-forms (see
\cite{Stueckelberg:1938zz} for the original introduction
of the \emph{St\"uckelberg mechanism}).
This is a typical situation  in massive deformations of supergravities \cite{Romans:1985tz}.

Moreover, the quantity $Z^M{}_{PQ}$ is restricted to live in
the representation in which the $B_{\mu\nu}{}^{MN}$ do.
Then, since $Z^M{}_{PQ}$ depends on the embedding tensor by construction, this condition entails a linear  restriction on the embedding tensor and its allowed representations.


%

This procedure can be extended to the existing higher-order
rank $p$-forms of the theory. As a consequence, a new set of
3-forms have to be properly added to the field strength of
the 2-form $B_{\mu\nu}$ and its gauge transformation.
This mechanism necessarily brings to light all the
$p$-form fields of a given theory.
 Schematically, we have a tower of relations as
\begin{align}
\mathcal{F}^{(2)M}&=D A^M+\cdots+Z^{MI}C^{(2)}_{ I}\, ,\nonumber\\
\mathcal{F}^{(3) I}&=D C^{(2) I}+\cdots+Z^{IA}C^{(3)}_{ A}\, ,\nonumber\\
&\dots \nonumber\\
\mathcal{F}^{(n) P}&=D C^{(n-1) P}+\cdots+Z^{PW}C^{(D)}_{W}\, ,
&
\end{align}
where the indices $\{M,I,A,P,W\}$ denote the different representations of $G$ under which the 1-, the 2-, the 3- , the $(n-1)$- and the $n$-form fields transform, respectively. The gauge variations of the previous field
strengths would be
\begin{align}
\delta \mathcal{F}^{(2)M}
&=
D (\delta A^M)+\cdots+Z^{MI}
\delta C^{(2)}_{ I}\, ,
\nonumber\\
\delta\mathcal{F}^{(3) I}
&=
D(\delta C^{(2) I})+\cdots+Z^{IA}\delta C^{(3)}_{A}\, ,
\nonumber\\
&
\dots
\nonumber\\
\delta \mathcal{F}^{(n) P}
&=
D (\delta C^{(n-1) P})+\cdots+Z^{PW}\delta C^{(n)}_{W}
\, ,
\end{align}
where the gauge transformation of the fields are given by
\begin{align}
\delta A^M&=D \Lambda^M+\cdots-Z^{MI}\Xi^{(1)}{}_{ I }\, ,\nonumber\\
\delta C^{(2)I}&=D \Xi^{(1)I}+\cdots-Z^{IA}\Sigma^{(2)}_A\, ,\nonumber\\
&\dots \nonumber\\
\delta C^{(n-1)P}&=D \Delta^{(n-2)P}+\cdots-Z^{PW}\Delta^{(n-1)}_W
\, .
\end{align}
The elements $\Lambda^M,\Xi^I,\Sigma^A,\Delta^P,\Delta^W$ are 0-, 1-, 2-, $(n-2)$- and $(n-1)$-form
 gauge parameters, respectively.

We thus realize  that not only covariant derivatives are
necessary as new ingredients to gauge a theory, but also
St\"uckelberg-like couplings between $p$-forms and $(p+1)$-forms
become crucial. In particular, they are essential to construct suitable field
transformations and guarantee the covariance of the field
strengths.


Another consequence of the new gaugings is that the new field
strengths do not satisfy the standard Bianchi identities.
It can be seen that they  satisfy a hierarchy of coupled
\emph{deformed Bianchi identities}, which schematically has the following structure:
\begin{align}
D \mathcal{F}^{(2)M}&=Z^{MI} \mathcal{F}^{(3)}{}_{I}\, ,\nonumber\\
D \mathcal{F}^{(3)I}&=\cdots+ Z^{IA} \mathcal{F}^{(4)}{}_{A}\, ,\nonumber\\
&\dots
\nonumber
\\
D \mathcal{F}^{(n-1)P}&=\cdots+ Z^{PQ} \mathcal{F}^{(n)}_{Q}
\, .
\end{align}


A detailed analysis of the higher rank tensor gauge
transformations allows us to determine the full field
content of the theory, including the $D$- and $(D-1)$-forms,
which are non-propagating fields.


\subsection{The deformed Lagrangian}

Once we have studied the impact of the gaugings in the group
structure of supergravities, let us focus on the Lagrangian
of the deformed gauged theory. This study is valid for theories that admit a Lagrangian description, otherwise this treatment is performed in a similar fashion at the level of the equations of motion.
The first, straightforward, modifications to be introduced are
the covariantization of the derivatives and the replacement
of  the Abelian field strengths by the fully covariant ones.
Next, it is  necessary the modification of the topological
Chern-Simons terms of the ungauged version and the addition of a potential.

Concerning the fermionic sector, we require the addition of new mass terms for the spinorial fields in order to keep SUSY invariance. On the other hand, it is indispensable the modification of the supersymmetric variations of the fermions by means of the so-called \emph{fermion shifts}. These two subtle enhancements ensure the supersymmetric invariance of the action (or, alternatively, the equations of motion).

We have seen how the St\"uckelberg couplings connect
the $p$- and $(p+1)$-forms throughout the field strengths
of the former.
This could be, in principle, problematic, since
 they could imply new equations of motion.
However, these contributions combine into first order
equations of motions, which show nothing but the fact that
they are the on-shell dual fields of the ungauged theory.
They enter as Lagrange multipliers-like equations in the Lagrangian.

It is important to point out the conceptual difference between
this situation and the so-called
\emph{democratic formulations} of supergravities
\cite{Bergshoeff:2001pv}, in which all the dual fields are
introduced in the action in an egalitarian way and the duality
relations must be added by hand.

Once the gaugings are properly implemented in the theory,
local supersymmetry invariance of the Lagrangian has to be
imposed. The SUSY variations of the new St\"uckelberg couplings of
the field strengths have to be canceled by new terms of
the lagrangian. Let us consider a truly covariant field strength of the form $\mathcal{F}^{(p)}\sim F^{(p)}+ZC^{(p)}$ (\emph{c.f.} \eqref{eq:goodF}). Let us focus on a generic kinetic term  $\mathcal{F}^{(p)}\wedge \star \mathcal{F}^{(p)}$, which is schematically given by
\begin{align}
\mathcal{F}^{(p)}\wedge \star \mathcal{F}_{(p)}
&\sim
F^{(p)}\wedge\star F_{(p)}+2ZC_{(p)}\wedge\star F^{(p)}+ZZ C^{(p)}\wedge\star  C_{(p)}
\, .
\end{align}
In general, the SUSY variations of the field strengths and $p$-forms, at second-order in fermions, have the following structures:
\begin{align}
\delta_\epsilon F^{(p)} &\sim
A d(\bar\epsilon\gamma\overset{(p-1)}{\cdots\cdots}\gamma\lambda)
+B d(\bar\epsilon\gamma\overset{(p-2)}{\cdots\cdots}\gamma\psi_\mu)
\, ,\nonumber\\
\delta_\epsilon C^{(p)} &\sim D \bar\epsilon\gamma\overset{(p)}{\cdots\cdots}\gamma\lambda
+E \bar\epsilon\gamma\overset{(p-1)}{\cdots\cdots}\gamma\psi_\mu
\, ,
\end{align}
where $\lambda$ is an arbitrary spin-1/2 field, $\psi_\mu$ is
a gravitino and $A,B,C,D$ are  functions that may depend
on the scalar fields.
Then, if we focus on the  SUSY variation of the Lagrangian kinetic
term for $\mathcal{F}$, we obtain that
\begin{multline}
\delta_\epsilon (\mathcal{F}^{(p)}\wedge \star\mathcal{F}_{(p)})
=
2\left[
	F^{(p)}\wedge \star\delta_\epsilon F_{(p)}
	+F^{(p)}\wedge \star Z\delta_\epsilon C_{(p)}
\right.
\\
\left.
	+\delta_\epsilon F^{(p)}\wedge \star Z C_{(p)}
	+ZZ C^{(p)}\wedge \star \delta_\epsilon C_{(p)}
	\right]
\, .
\end{multline}
The first term also appears in the ungauged theory and does
not imply any problem (as we can check by demanding $\vartheta\to 0 \Rightarrow Z\to 0$). The rest of the terms depend on the embedding tensor, so
one possibility to cancel them is the addition of some \emph{fermionic mass terms} that explicitly depend linearly on the embedding tensor. Generically, these terms have the following structure
\begin{align}
\label{eq:Lfm}
\mathcal{L}_{\text{fm}}&=
	\bar\psi_\mu{}^a A^{\mu\nu}{}_{ab}\psi_\nu{}^b
	+\bar\chi^m B^{\mu}{}_{ma} \psi_\mu^a
	+\bar\chi^m C_{ab}\chi^n
+h.c.
\, ,
\end{align}
where  $\psi_\mu{}^a$ and $\chi^m$ are generic gravitini
and spin-$\tfrac{1}{2}$ fermions, respectively.
The indices $a,b$ and $m,n$ belong to some representations of the maximal compact subgroup $H$ of $G$. Hence, the tensors $A_{\mu\nu}{}^{ab}$,  $B_\mu{}^{ma}$ and $ C_{mn}$ which, by construction, depend on the embedding tensor and may depend on the scalar fields, transform under $H\subset G$. In addition, the presence of these new terms requires
the modification of the supersymmetric transformation rules of the fermion fields.\footnote{For instance, if we study the supersymmetric transformation of the gravitini mass term $\bar\psi_\mu{}^a A^{\mu\nu}{}_{ ab}\psi_\nu{}^b$, we obtain the following pattern:
\begin{align}
\delta_\epsilon (\bar\psi_\mu{}^a A^{\mu\nu}{}_{ ab}\psi_\nu{}^b)
\sim
2 (\bar\psi_\mu{}^a A^{\mu\nu}{}_{ ab}\delta_\epsilon\psi_\nu{}^b)
\sim
2(\bar\psi_\mu{}^a A^{\mu\nu}{}_{ ab}D_\mu\epsilon^b)+\cdots
\, .
\end{align}
Then, considering the structure of the fermionic transformation rules, the only cancellation of this term arises from $\delta_\epsilon(\bar\psi_\mu \gamma^{\mu\nu\rho}D_\nu\psi_\rho)$. However, there is not any contribution in $\delta_\epsilon\psi_\mu$ proportional to $\vartheta_M{}^\alpha$. Thus, we need to modify the supersymmetric rules of the fermion fields to include these terms.
}
The new terms are required to depend linearly on the embedding tensor. The appropriate modifications result to depend on  $A_{\mu\nu}{}^{ab}$ and $B_\mu{}^{ma}$:
\begin{align}
\delta_\epsilon\psi_\mu{}^a
&=
\delta_0\psi_\mu{}^a
+A_\mu^{ab}\epsilon_b
\, ,
\nonumber
\\
\delta_\epsilon \chi^m
&=
\delta_0 \chi^m
+B^{ma}\epsilon_a
\, ,
\end{align}
where $\delta_0$ denotes the supersymmetric transformation of the ungauged theory. These extra terms are known as \emph{fermion shifts}.
As a consequence, new terms proportional to $\vartheta^2$
are generated by the action of the fermion shifts
on \eqref{eq:Lfm}. This requires the inclusion of one more term, a
scalar potential, which schematically has the form
\begin{align}
\mathcal{L}_{pot}&=-eV=-e\left(
	B_\mu{}^{ma}B^\mu{}_{ma}
	-A_{\mu}{}^{ab}A^{\mu}{}_{ab}
	\right)
\, .
\end{align}
$V$ can be rewritten in terms of the embedding tensor.
In general, it can be expressed as
\begin{align}
V&=V^{MN}{}_{\alpha\beta}\vartheta_M{}^\alpha\vartheta_N{}^\beta\, ,
\end{align}
where $V^{MN}{}_{\alpha\beta}$ is a scalar dependent matrix.

Summarizing, in this chapter we have introduced some basic  aspects of gauged
supergravities and the embedding tensor formalism.
In the next chapter, we will show an exhaustive study of all of the gauged supergravities that the maximal $D=9$ supergravity can host, by using the embedding tensor to scan all the valid gaugings.

\chapter{Gaugings in $N=2$ $d=9$ supergravity}
\label{ch:gaugings_d9}


After having studied the embedding tenso formalism, it is illustrative to apply it to a non-trivial theory. In this case, we have chosen $N=2$ $d=9$ supergravity, since the size of its global symmetry group, $\mathbb{R}^+\times SL(2,\mathbb{R})$, allows to carry out the full implementation of the formalism.

\section{Introduction}

We use the embedding tensor method to construct the most general maximal gauged (massive) supergravity in $d=9$ dimensions and to determine its    extended field content.
%
Some gaugings of the maximal $d=9$ supergravity have been obtained in the past
by generalized dimensional reduction \cite{Scherk:1979zr} of the
10-dimensional theories with respect to the $SL(2,\mathbb{R})$ global symmetry
of the $N=2B$ theory \cite{Lavrinenko:1997qa,Meessen:1998qm,Gheerardyn:2001jj}
or other rescaling symmetries \cite{Howe:1997qt}\footnote{An $SO(2)$-gauged
  version of the theory was directly constructed in
  Ref.~\cite{Nishino:2002zi}.}. All these possibilities were systematically
and separately studied in Ref.~\cite{Bergshoeff:2002nv}, taking into account
the dualities that relate the possible deformation parameters introduced with
the generalized dimensional reductions. However, the possible combinations of
deformations were not studied, and, as we will explain, some of the
higher-rank fields are associated to the constraints on the combinations of
deformations. Furthermore, we do not know if other deformations, with no
higher-dimensional origin (such as Romans' massive deformation of the
$N=2A,d=10$ supergravity) are possible.

Our goal in this chapter will be to make a systematic study of all these
possibilities using the embedding-tensor formalism plus supersymmetry to
identify the extended-field content of the theory, finding the r\^ole played
by the possible 7-, 8- and 9-form potentials, and compare the results with the
prediction of the $E_{11}$ approach. We expect to get at least compatible
results, as in the $N=2,d=4,5,6$ cases studied in \cite{Huebscher:2010ib} and
\cite{Kleinschmidt:2008jj}.

This chapter is organized as follows: in Section~\ref{sec-undeformed} we review
the undeformed maximal 9-dimensional supergravity and its global
symmetries. In Section~\ref{sec-deformation} we study the possible
deformations of the theory using the embedding-tensor formalism and checking
the closure of the local supersymmetry algebra for each electric $p$-form of
the theory. In Section~\ref{sec-summary} we summarize the results of the
previous section describing the possible deformations and the constraints they
must satisfy. We discuss the relations between those results and the possible
7- 8- and 9-form potentials of the theory and how these results compare with
those obtained in the literature using the $E_{11}$ approach.
Section~\ref{sec-conclusions} contains our conclusions.  Our conventions are
briefly discussed in Appendix~\ref{d9:app-conventions}. The Noether currents of
the undeformed theory are given in Appendix~\ref{d9:sec-noether}. A summary of
our results for the deformed theory (deformed field strengths, gauge
transformations and covariant derivatives, supersymmetry transformations etc.)
is contained in Appendix\ref{d9:sec-final}.


\section{More on the maximal ungauged $d=9$ supergravity}
\label{sec-undeformed}

We have seen some aspects of the only undeformed maximal $N=2$ 9-dimensional supergravity in Section \ref{sec-d9-primer}. Now, we are going to perform a detailed analysis of its symmetries and its magnetic field content.

\subsection{Global symmetries}
\label{sec-global}

The undeformed theory has as (classical) global symmetry group
$SL(2,\mathbb{R})\times (\mathbb{R}^{+})^{2}$. The $(\mathbb{R}^{+})^{2}$
symmetries correspond to scalings of the fields, the first of which, that we
will denote by $\alpha$\footnote{This discussion follows closely that of
  Ref.~\cite{Bergshoeff:2002nv} in which the higher-dimensional origin of each
  symmetry is also studied. In particular, we use the same names and
  definitions for the scaling symmetries and we reproduce the table of scaling
  weights for the electric fields.}, acts on the metric and only leaves the
equations of motion invariant while the second of them, which we will denote
by $\beta$, leaves invariant both the metric and the action. The $\beta$
rescaling corresponds to the so-called \textit{trombone symmetry} which may
not survive to higher-derivative string corrections. 

One can also discuss two more scaling symmetries $\gamma$ and $\delta$, but
$\gamma$ is just a subgroup of $SL(2,\mathbb{R})$ and $\delta$ is related to
the other scaling symmetries by

\begin{equation}
\tfrac{4}{9} \alpha - \tfrac{8}{3} \beta -\gamma -\tfrac{1}{2} \delta=0\, .
\label{eq:rel}
\end{equation}

We will take $\alpha$ and $\beta$ as the independent symmetries.  The weights
of the electric fields under all the scaling symmetries are given in
Table~\ref{fundamental9d_weights}. We can see that each of the three gauge
fields $A^{I}{}_{\mu}$ has zero weight under {\it two} (linear combinations)
of these three symmetries: one is a symmetry of the action, the other is a
symmetry of the equations of motion only. The 1-form that has zero weight
under a given rescaling is precisely the one that can be used to gauge that
rescaling, but this kind of conditions are automatically taken into account by
the embedding-tensor formalism and we will not have to discuss them in
detail.

\begin{table}[t!]
\centering
{\scriptsize
\begin{tabular}{cccccccccccccccc}
\hline 
  $\mathbb{R}^+$ & $e_\mu{}^a$  & $e^\varphi$ & $e^\phi$ & $\chi$ & 
$A^0$ &  $A^{1}$ & $A^{2}$ & $B^{1}$ & $B^{2}$ & $C$ & $\psi_\mu$ & 
$\lambda$ & $\tilde \lambda$ & $\epsilon$ & $\mathcal{L}$  \\
\hline 
\hline 
$\alpha$ & $9/7$ & $6/\sqrt{7}$ & $0$ & $0$ & $3$ & $0$ & $0$ &
 $3$ & $3$ & $3$ & $9/14$ & $-9/14$ & $-9/14$ & $9/14$ & $9$ \\

$\beta$ & $0$ & $\sqrt{7}/4$ & $3/4$ & $-3/4$ & $1/2$ & $-3/4$ & $0$ & $-1/4$ & 
$1/2$ & $-1/4$ & $0$ & $0$ & $0$ & $0$ &  $0$  \\

$\gamma$ & $0$ & $0$ & $-2$ & $2$ & $0$ & $1$ & $-1$ & $1$ & $-1$ & $0$ & 
$0$ & $0$ & $0$ & $0$ &  $0$   \\

$\delta$ & $8/7$ & $-4/\sqrt{7}$ & $0$ & $0$ & $0$ & $2$ & 
$2$ & $2$ & $2$ & $4$ & $4/7$ & $-4/7$ & $-4/7$ & $4/7$ &  $8$   \\

\hline 
\end{tabular}
}
{\it \caption{The scaling weights of the electric fields of
  maximal $d=9$ supergravity.}
\label{fundamental9d_weights}
}
\end{table}

The action of the element of $SL(2,\mathbb{R})$ given by the matrix

\begin{equation}
\left(\Omega_{j}^{i}\right)  
=
\left( \begin{array}{cc} a&b\\ c&d \end{array} \right)\, ,
\hspace{1cm}
ad-bc=1\, ,
\end{equation}

\noindent
on the fields of the theory is

\begin{equation}
\label{SL2R9D}
\begin{array}{rclrcl}
\tau^{\prime} 
& = & 
{\displaystyle\frac{a {\tau} +b}{c{\tau} +d}}\, , 
\hspace{1.5cm} &
\mathcal{M}^{\prime}_{ij}
& = & 
\Omega_{i}{}^{k} \mathcal{M}_{kl}\Omega_{j}{}^{l}\, ,\\
& & & & &  \\
A^{\mathbf{i}\, \prime} 
& =  &  
\Omega_{\mathbf{j}}{}^{\mathbf{i}} A^{\mathbf{j}}\, , &
B^{i\, \prime} 
& =  &  
\Omega_{j}{}^{i} B^{j}\, ,\\
& & & & &  \\
\psi^{\prime}_{\mu} 
& = &
e^{\frac{i}{2}l}\psi_{{\mu}}\, ,&
{\lambda} 
& = & 
e^{\frac{3i}{2}l} \lambda\, , \\
& & & & &  \\
\tilde{\lambda}^{\prime} 
& = & 
e^{-\frac{i}{2}l} \tilde{\lambda}\, ,&
\epsilon^{\prime} 
& = & 
e^{\frac{i}{2}l} \epsilon\, .\\
\end{array}
\end{equation}

\noindent
where 

\begin{equation}
e^{2il} \equiv \frac{c \, {\tau}^*+d}{c\, {\tau}+d}\, .
\end{equation}

\noindent
The rest of the fields ($e^{a}{}_{\mu},\varphi,A^{0}{}_{\mu},C_{\mu\nu\rho}$),
are invariant under $SL(2,\mathbb{R})$.

We are going to label the 5 generators of these global symmetries by
$T_{A}$, $A=1,\cdots,5$. $\{T_{1},T_{2},T_{3}\}$ will be the 3
generators of $SL(2,\mathbb{R})$ (collectively denoted by $\{T_{m}\}$,
$m=1,2,3$), and $T_{4}$ and $T_{5}$ will be, respectively, the
generators of the rescalings $\alpha$ and $\beta$.  Our choice for the
generators of $SL(2,\mathbb{R})$ acting on the doublets of 1-forms
$A^{\mathbf{i}}$ and 2-forms $B^{i}$ is

\begin{equation}
T_{1} = \tfrac{1}{2}\sigma^{3}\, ,
\hspace{1cm}
T_{2} = \tfrac{1}{2}\sigma^{1}\, ,
\hspace{1cm}
T_{3} = \tfrac{i}{2}\sigma^{2}\, ,
\end{equation}

\noindent
where the $\sigma^{m}$ are the standard Pauli matrices, so 

\begin{equation}
[T_{1},T_{2}] = T_{3}\, ,
\hspace{1cm}
[T_{2},T_{3}] = -T_{1}\, ,
\hspace{1cm}
[T_{3},T_{1}] = -T_{2}\, .
\end{equation}

\noindent
Then, the $3\times 3$ matrices corresponding to generators acting
(contravariantly) on the 3 1-forms $A^{I}$ (and covariantly on their dual
6-forms $\tilde{A}_{I}$ to be introduced later) are

\begin{equation}
  \begin{array}{ccc}
\left((T_{1})_{J}{}^{I}\right)
=
\tfrac{1}{2}
\left(
  \begin{array}{c|c}
 0  & 0 \\  \hline 0  & \sigma^{3}  \\ 
  \end{array}
\right)\, ,
\hspace{.5cm}
&
\left((T_{2})_{J}{}^{I}\right)
=
\tfrac{1}{2}
\left(
  \begin{array}{c|c}
 0  & 0 \\  \hline 0  & \sigma^{1}  \\ 
  \end{array}
\right)\, ,
\hspace{.5cm}
&
\left((T_{3})_{J}{}^{I}\right)
=
\tfrac{1}{2}
\left(
  \begin{array}{c|c}
 0  & 0 \\  \hline 0  & i\sigma^{2}  \\ 
  \end{array}
\right)\, ,
\\
& & \\
\left((T_{4})_{J}{}^{I}\right)
=
\mathrm{diag}(3,0,0)\, ,
\hspace{.5cm}
&
\left((T_{5})_{J}{}^{I}\right)
=
\mathrm{diag}(1/2,-3/4,0)\, .
&
\\
\end{array}
\end{equation}

\noindent
We will sometimes denote this representation by $T^{(3)}_{A}$.  The $2\times
2$ matrices corresponding to generators acting (contravariantly) on the
doublet of 2-forms $B^{i}$ (and covariantly on their dual 5-forms
$\tilde{B}_{i}$ to be introduced later) are

\begin{equation}
  \begin{array}{ccc}
\left((T_{1})_{j}{}^{i}\right)
=
\tfrac{1}{2}\sigma^{3}\, ,
\hspace{.5cm}
&
\left((T_{2})_{j}{}^{i}\right)
=
\tfrac{1}{2}\sigma^{1}\, ,
\hspace{.5cm}
&
\left((T_{3})_{j}{}^{i}\right)
=
\tfrac{i}{2}\sigma^{2}\, ,
\\
& & \\
\left((T_{4})_{j}{}^{i}\right)
=
\mathrm{diag}(3,3)\, ,
\hspace{.5cm}
&
\left((T_{5})_{j}{}^{i}\right)
=
\mathrm{diag}(-1/4,1/2)\, .
&
\\
\end{array}
\end{equation}

\noindent
We will denote this representation by $T^{(2)}_{A}$.  The generators
that act on the 3-form $C$ (sometimes denoted by $T^{(1)}_{A}$) are

\begin{equation}
T_{1}= T_{2}= T_{3} = 0\, ,  
\hspace{.5cm}
T_{4}= 3\, ,  
\hspace{.5cm}
T_{5}= -1/4\, .  
\end{equation}

\noindent
We will also need the generators that act on the magnetic 4-form $\tilde{C}$, also denoted by $T^{(\tilde{1})}_{A}$

\begin{equation}
\tilde{T}_{1}= \tilde{T}_{2}= \tilde{T}_{3} = 0\, ,  
\hspace{.5cm}
\tilde{T}_{4}= 6\, ,  
\hspace{.5cm}
\tilde{T}_{5}= 1/4\, .  
\end{equation}

We define the structure constants $f_{AB}{}^{C}$ by

\begin{equation}
[T_{A},T_{B}] = f_{AB}{}^{C} T_{C}\, .  
\end{equation}

The symmetries of the theory are isometries of the scalar manifold
($\mathbb{R}\times SL(2,\mathbb{R}/U(1)$). The Killing vector associated to
the generator $T_{A}$ will be denoted by $k_{A}$ and will be normalized so
that their Lie brackets are given by

\begin{equation}
[k_{A},k_{B}] = -f_{AB}{}^{C} k_{C}\, .  
\end{equation}

The $SL(2,\mathbb{R})/U(1)$ factor of the scalar manifold is a K\"ahler space
with K\"ahler potential, K\"ahler metric and K\"ahler 1-form, respectively
given by

\begin{equation}
\mathcal{K}= -\log \Im{\rm m}\tau =\phi\, ,  
\hspace{.5cm}
\mathcal{G}_{\tau\tau^{*}}
=
\partial_{\tau}  
\partial_{\tau^{*}}\mathcal{K} = \tfrac{1}{4}e^{2\phi}\, ,  
\hspace{.5cm}
\mathcal{Q} 
= \tfrac{1}{2i}\left(\partial_{\tau}\mathcal{K}d\tau - \mathrm{c.c.}\right)
= \tfrac{1}{2}e^{\phi}d\chi\, .  
\end{equation}

In general, the isometries of the K\"ahler metric only leave invariant the
K\"ahler potential up to K\"ahler transformations :

\begin{equation}
\pounds_{k_{m}}\mathcal{K} 
= 
k_{m}{}^{\tau}\partial_{\tau}\mathcal{K}  +\mathrm{c.c.}
=
\lambda_{m}(\tau) +\mathrm{c.c.}\, ,
\hspace{1cm}
\pounds_{k_{m}}\mathcal{Q} = -\tfrac{i}{2}d\lambda_{m}\, , 
\end{equation}

\noindent
where the $\lambda_{m}$ are holomorphic functions of the coordinates that
satisfy the equivariance property

\begin{equation}
\pounds_{k_{m}}\lambda_{n} 
-  
\pounds_{k_{n}}\lambda_{m}
=
-f_{mn}{}^{p}\lambda_{p}\, . 
\end{equation}

Then, for each of the $SL(2,\mathbb{R})$ Killing vectors $k_{m}$, $m=1,2,3$,
it is possible to find a real \textit{Killing prepotential} or \textit{momentum
  map} $\mathcal{P}_{m}$ such that 

\begin{equation}
\label{eq:momentummapsdef}
  \begin{array}{rcl}
k_{m\, \tau^{*}} 
& = & 
\mathcal{G}_{\tau^{*}\tau}k_{m}{}^{\tau} 
=i\partial_{\tau^{*}}\mathcal{P}_{m}\, ,
\\
& & \\
k_{m}{}^{\tau} \partial_{\tau}\mathcal{K} 
& = & 
i\mathcal{P}_{m}+\lambda_{m}\, , 
\\
& & \\
\pounds_{k_{m}}\mathcal{P}_{n}
& = & 
-f_{mn}{}^{p}\mathcal{P}_{p}\, .
\end{array}
\end{equation}

The non-vanishing components of all the Killing vectors are\footnote{The
  holomorphic and anti-holomorphic components are defined by
  $k=k^{\tau}\partial_{\tau} +\mathrm{c.c.}= k^{\chi}\partial_{\chi}+
  k^{\phi}\partial_{\phi}$.}

\begin{equation}
\label{eq:taukillingvectors}
k_{1}{}^{\tau}=\tau\, ,
\hspace{.7cm}  
k_{2}{}^{\tau}=\tfrac{1}{2}(1-\tau^{2})\, ,
\hspace{.7cm}  
k_{3}{}^{\tau}=\tfrac{1}{2}(1+\tau^{2})\, ,
\hspace{.7cm}  
k_{4}{}^{\tau}=0\, ,
\hspace{.7cm}  
k_{5}{}^{\tau}=-\tfrac{3}{4}\tau\, .
\end{equation}

\noindent
and

\begin{equation}
\label{eq:varphikillingvectors}
k_{4}{}^{\varphi}=6/\sqrt{7}\, ,
\hspace{1.5cm}  
k_{5}{}^{\varphi}=\sqrt{7}/4\, .
\end{equation}

\noindent
The holomorphic functions $\lambda_{m}(\tau)$ take the values

\begin{equation}
\label{eq:lambdam}
\lambda_{1}=-\tfrac{1}{2}\, ,
\hspace{1cm}
\lambda_{2}=\tfrac{1}{2}\tau\, ,
\hspace{1cm}
\lambda_{3}=-\tfrac{1}{2}\tau\, ,
\end{equation}

\noindent
and the momentum maps are given by:

\begin{equation}
\label{eq:momentummaps}
\mathcal{P}_{1}= \tfrac{1}{2}e^{\phi}\chi\, ,
\hspace{1cm}
\mathcal{P}_{2}= \tfrac{1}{4}e^{\phi}(1-|\tau|^{2})\, ,
\hspace{1cm}
\mathcal{P}_{3}= \tfrac{1}{4}e^{\phi}(1+|\tau|^{2})\, .
\end{equation}

These objects will be used in the construction of $SL(2,\mathbb{R})$-covariant
derivatives for the fermions.


\subsection{Magnetic fields}
\label{sec-magnetic}

As it is well known, for each $p$-form potential with $p>0$ one can define a
\textit{magnetic} dual which in $d-9$ dimensions will be a $(7-p)$-form
potential. Then, we will have magnetic 4-, 5- and 6-form potentials in the
theory. 

A possible way to define those potentials and identify their $(8-p)$-form
field strengths consists in writing the equations of motion of the $p$-forms
as total derivatives. Let us take, for instance,  the equation of motion of
the 3-form $C$ Eq.~(\ref{eq:eomC}). It can be written as 

\begin{equation}
\label{eq:eomC2}
  \begin{array}{rcl}
d{\displaystyle\frac{\partial \mathcal{L}}{\partial  G}}   
= 
d \biggl \{
e^{\frac{2}{\sqrt{7}}\varphi}  \star G
-
\left[
G+\varepsilon_{\mathbf{i}j}A^{\mathbf{i}}\wedge 
\left(H^{j} -\tfrac{1}{2}\delta^{j}{}_{\mathbf{j}}A^{\mathbf{j}}\wedge F^{0}\right)
\right]\wedge A^{0}
& & \\
& & \\
+\tfrac{1}{2}\varepsilon_{ij}
\left(H^{i} -\delta^{i}{}_{\mathbf{i}}A^{\mathbf{i}}\wedge F^{0}\right)
\wedge
\left(B^{j} -\tfrac{1}{2}\delta^{j}{}_{\mathbf{j}}A^{0\mathbf{j}}\right)
\biggr \}
& = & 0\, .  
\end{array}
\end{equation}

\noindent 
We can transform this equation of motion into a Bianchi identity by replacing
the combination of fields on which the total derivative acts by the total
derivative of a 4-form which we choose for the sake of
convenience\footnote{With this definition $\tilde{G}$ will have exactly the
  same form that we will obtain from the embedding tensor formalism.}

\begin{equation}
  \begin{array}{rcl}
d\left[\tilde{C} -C\wedge A^{0} -\tfrac{3}{4}
\varepsilon_{\mathbf{i}j}A^{0\mathbf{i}}\wedge B^{j} \right]
& \equiv &
e^{\frac{2}{\sqrt{7}}\varphi}  \star G
-
\left[
G+\varepsilon_{\mathbf{i}j}A^{\mathbf{i}}\wedge 
\left(H^{j} -\tfrac{1}{2}\delta^{j}{}_{\mathbf{j}}A^{\mathbf{j}}\wedge F^{0}\right)
\right]\wedge A^{0}
\\
& & \\
& & 
+\tfrac{1}{2}\varepsilon_{ij}
\left(H^{i} -\delta^{i}{}_{\mathbf{i}}A^{\mathbf{i}}\wedge F^{0}\right)
\wedge
\left(B^{j} -\tfrac{1}{2}\delta^{j}{}_{\mathbf{j}}A^{0\mathbf{j}}\right)\, ,
\end{array}
\end{equation}

\noindent
where $\tilde{C}$ will be the magnetic 4-form. This relation can be put in the
form of a duality relation

\begin{equation}
e^{\frac{2}{\sqrt{7}}\varphi}  \star G
=
\tilde{G}\, ,
\end{equation}

\noindent
where we have defined the magnetic 5-form field strength

\begin{equation}
\label{eq:tildeGundeformed}
\tilde{G}
\equiv
d\tilde{C}
+C \wedge F^{0}   
-\tfrac{1}{24}\varepsilon_{\mathbf{ij}} A^{0\mathbf{ij}} \wedge F^{0} 
- \varepsilon_{ij} \left(H^{i} -\tfrac{1}{2} dB^{i}\right) \wedge B^{j}\, .
\end{equation}

\noindent
The equation of motion for $\tilde{C}$ is just the Bianchi identity of $G$
rewritten in terms of $\tilde{G}$.







In a similar fashion we can define a doublet of 5-forms $\tilde{B}_{i}$ with
field strengths denoted by $\tilde{H}_{i}$, and a singlet and a doublet of
6-forms $\tilde{A}_{0},\tilde{A}_{\mathbf{i}}$ with field strengths denoted,
respectively, by $\tilde{F}_{0}$ and $\tilde{F}_{\mathbf{i}}$. The field
strengths can be chosen to have the form

\begin{align}
\tilde{H}_{i} 
& = 
d\tilde{B}_{i} -\delta_{ij} B^{j}\wedge G
+\delta_{i\mathbf{j}}\tilde{C} \wedge F^{\mathbf{j}}
+\tfrac{1}{2}\delta_{i\mathbf{j}} \left(A^{0} \wedge F^{\mathbf{j}}
+A^{\mathbf{j}} \wedge F^{0}\right)\wedge C
\nonumber \\
& \nonumber \\
& 
+\tfrac{1}{2}\delta_{ij}\varepsilon_{k\mathbf{l}}B^{jk} \wedge F^{\mathbf{l}}
\, ,
\\
& \nonumber \\
\tilde{F}_{0} 
& =  
d\tilde{A}_{0} +\tfrac{1}{2} C\wedge G
-\varepsilon_{\mathbf{i}j}F^{\mathbf{i}} \wedge 
\left(\delta^{jk}\tilde{B}_{k} -\tfrac{2}{3}B^{j}\wedge C \right)
\nonumber \\
& \nonumber \\
& 
-\tfrac{1}{18} \varepsilon_{\mathbf{ij}} A^{\mathbf{ij}} \wedge 
\left(
\tilde{G} -F^{0} \wedge C -\tfrac{1}{2} \varepsilon_{kl}B^{k} \wedge H^{l}
\right)
\nonumber \\
& \nonumber \\
& 
-\tfrac{1}{6}\varepsilon_{\mathbf{i}j}
A^{\mathbf{i}}\wedge \left(
B^{j} \wedge G - C \wedge H^{j} -\tfrac{2}{3} \delta^{j}{}_{\mathbf{j}}
\tilde{C} \wedge F^{\mathbf{j}}
-\varepsilon_{k\mathbf{l}}B^{jk} \wedge F^{\mathbf{l}}
\right)\, ,
\\
& \nonumber \\
\tilde{F}_{\mathbf{i}}  
& = 
d\tilde{A}_{\mathbf{i}} 
+\delta_{\mathbf{i}j} \left(
B^{j} +\tfrac{7}{18} \delta^{j}{}_{\mathbf{k}}A^{0\mathbf{k}} 
\right) 
\wedge \tilde{G}
-\delta_{\mathbf{i}}{}^{j}F^{0} \wedge \tilde{B}_{j}
-\tfrac{1}{9}
\delta_{\mathbf{ij}}\left(
8A^{0} \wedge F^{\mathbf{j}} +  A^{\mathbf{j}}\wedge F^{0}
\right) 
\wedge \tilde{C}
\nonumber \\
& \nonumber \\
& 
-\tfrac{1}{3}\delta_{\mathbf{i}j} \varepsilon_{lm}
 \left(
B^{j} +\tfrac{1}{3} \delta^{j}{}_{\mathbf{k}}A^{0\mathbf{k}} 
\right) 
\wedge B^{l} \wedge H^{m}
-\tfrac{1}{6}\delta_{\mathbf{i}j} \varepsilon_{\mathbf{k}l}
\left(
A^{0} \wedge H^{j} -B^{j} \wedge F^{0}
\right)
\wedge A^{\mathbf{k}} \wedge B^{l}
\nonumber \\
& \nonumber \\
& 
-\tfrac{1}{9} A^{0} \wedge F^{0} \wedge
\delta_{\mathbf{ij}}\left(
\tfrac{7}{2} A^{\mathbf{j}} \wedge C
+\delta^{\mathbf{j}}{}_{k}\varepsilon_{\mathbf{lm}}
A^{\mathbf{lm}} \wedge B^{k}
\right)\, ,
\end{align}

\noindent
and the duality relations are

\begin{align}
 \tilde{H}_{i}
& =
e^{-\frac{1}{\sqrt{7}}\varphi}\mathcal{M}^{-1}_{ij}\star H^{j}\, ,
\\
& \nonumber \\
\tilde{F}_{0}
& = 
e^{\frac{4}{\sqrt{7}}\varphi}\star F^{0}\, ,
\\
& \nonumber \\
\tilde{F}_{\mathbf{i}}
& = 
e^{\frac{3}{\sqrt{7}}\varphi}\mathcal{M}^{-1}_{\mathbf{ij}}
\star F^{\mathbf{j}}\, .
\end{align}

The situation is summarized in Table~\ref{xxx1}. The scaling weights of the
magnetic fields are given in Table~\ref{dual9d_weights}.

\begin{table}[t!]
\begin{center}
\begin{tabular}{cccccccccc}
\hline
0 & 1 & 2 & 3 & 4 & 5 & 6 & 7 & 8 & 9\\ \hline \hline
$j_{A}$ &  $ A^{I}$ & $B^{i}$ & $C$ & $\tilde{C}$ & $\tilde{B}_{i}$ & $\tilde{A}_{I}$ 
& $\tilde{A}_{(7)}^{A}$ & $\tilde{A}_{(8)}$ & $\tilde{A}_{(9)}$ \\ 

 & $ F^{I}$ & $H^{i}$ & $G$ & $\tilde{G}$ & $\tilde{H}_{i}$ & $\tilde{F}_{I}$ 
& $\tilde{F}_{(8)}^{A}$ & $\tilde{F}_{(9)}$ & \\ 
\hline 
\end{tabular}
{\it \caption{\it Electric and magnetic forms and their field strengths.}
\label{xxx1}
}
\end{center}
\end{table}

\begin{table}[t!]
\centering
\begin{tabular}{ccccccc}
\hline 
  $\mathbb{R}^{+}$ &  $\tilde{C}$ &  
$\tilde{B}_{2}$ & $\tilde{B}_{1}$ & $\tilde{A}_{2}$ & $\tilde{A}_{1}$ & 
$\tilde{A}_{0}$  \\
 \hline
\hline 
$\alpha$ & 6 &6 &6 &9 &9 &6  \\ 

$\beta$ & $1/4$ & $-1/2$&$+1/4$&0&$+3/4$ &$-1/2$  \\ 

$\gamma$ & 0&1 &-1 &1 &-1 &0      \\  

$\delta$ & 4&6 &6 &6 &6 &8        \\ 
\hline
\end{tabular}
{\it \caption{\it The scaling weights of the  magnetic fields of
  maximal $d=9$ supergravity can be determined by requiring that
  the sum of the weights of the electric and magnetic potentials 
  equals that of the Lagrangian. The scaling weights of the 7-, 8-
  and 9-forms can be determined in the same way after we find the entities
  they are dual to (Noether currents, embedding-tensor components and
  constraints, see Section~\ref{sec-summary}).}
\label{dual9d_weights}
}
\end{table}

This dualization procedure is made possible by the gauge symmetries associated
to all the $p$-form potentials for $p>0$ (actually, by the existence of gauge
transformations with constant parameters) and, therefore, it always works for
massless $p$-forms with $p>0$ and generically fails for 0-form
fields. However, in maximal supergravity theories at least, there is a global
symmetry group that acts on the scalar manifold and whose dimension is larger
than that of the scalar manifold. Therefore, there is one Noether 1-form
current $j_{A}$ associated to each of the generators of the global symmetries
of the theory $T_{A}$. These currents are conserved on-shell,
\textit{i.~e.}~they satisfy

\begin{displaymath}
d\star j_{A}=0\, ,  
\end{displaymath}

\noindent
on-shell, and we can define a $(d-2)$-form potential $\tilde{A}_{(d-2)}^{A}$ by

\begin{displaymath}
d\tilde{A}_{(d-2)}^{A}= G^{AB}\star j_{B}\, ,  
\end{displaymath}

\noindent
where $G^{AB}$ is the inverse Killing metric of the global symmetry group, so
that the conservation law (dynamical) becomes a Bianchi identity.

Thus, while the dualization procedure indicates that for each electric
$p$-form with $p>0$ there is a dual magnetic $(7-p)$-form transforming in the
conjugate representation, it tells us that there are as many magnetic
$(d-2)$-form duals of the scalars as the dimension of the global group (and
not of as the dimension of the scalar manifold) and that they transform in the
co-adjoint representation. Actually, since there is no need to have scalar
fields in order to have global symmetries, it is possible to define magnetic
$(d-2)$-form potentials even in the total absence of scalars\footnote{See
  Refs.~\cite{Hartong:2009az,Huebscher:2010ib} for examples.}.

According to these general arguments, which are in agreement with the general
results of the embedding-tensor formalism
\cite{Bergshoeff:2009ph,Hartong:2009vc,Hartong:2009az,Huebscher:2010ib}, we
expect a triplet of 7-form potentials $\tilde{A}_{(7)}^{m}$ associated to the
$SL(2,\mathbb{R})$ factor of the global symmetry group \cite{Meessen:1998qm}
and two singlets $\tilde{A}_{(7)}^{4},\tilde{A}_{(7)}^{5}$ associated to the
rescalings $\alpha,\beta$ (see Table~\ref{xxx1}).

Finding or just determining the possible magnetic $(d-1)$- and $d$-form
potentials in a given theory is more complicated. In the embedding-tensor
formalism it is natural to expect as many $(d-1)$-form potentials as
deformation parameters (embedding-tensor components, mass parameters etc.) can
be introduced in the theory since the r\^ole of the $(d-1)$-forms in the
action is that of being Lagrange multipliers enforcing their
constancy\footnote{The embedding-tensor formalism gives us a reason to
  introduce the $(d-1)$-form potentials based on the deformation parameters
  but the $(d-1)$-form potentials do not disappear when the deformation
  parameters are set equal to zero.}. The number of deformation parameters
that can be introduced in this theory is, as we are going to see, very large,
but there are many constraints that they have to satisfy to preserve gauge and
supersymmetry invariance. Furthermore, there are many St\"uckelberg shift
symmetries acting on the possible $(d-1)$-form potentials.  Solving the
constraints leaves us with the independent deformation parameters that we can
denote by $m_{\sharp}$ and, correspondingly, with a reduced number of
$(d-1)$-form potentials $\tilde{A}_{(d-1)}^{\sharp}$ on which only a few St\"uckelberg
symmetries (or none at all) act\footnote{The $(d-1)$-form potentials that
  ``disappear'' when we solve the constraints are evidently associated to the
  gauge-fixing of the missing St\"uckelberg symmetries.}.

The $d$-form field strengths $\tilde{F}_{(d)}^{\sharp}$ are related to the
scalar potential of the theory through the expression
\cite{Bergshoeff:2009ph,Hartong:2009vc,Hartong:2009az,Huebscher:2010ib}

\begin{equation}
  \tilde{F}_{(d)}^{\sharp}
  =
  \tfrac{1}{2}\star\frac{\partial V}{\partial m_{\sharp}}\, . 
\end{equation}

Thus, in order to find the possible 8-form potentials of this theory we need
to study its independent consistent deformations $m_{\sharp}$. We will
consider this problem in the following section.

In the embedding-tensor formalism, the $d$-form potentials are associated to
constraints of the deformation parameters since they would be the Lagrange
multipliers enforcing them in the action \cite{Bergshoeff:2007vb}. If we do
not solve any of the constraints there will be many $d$-form potentials but
there will be many St\"uckelberg symmetries acting on them as well. Thus, only
a small number of \textit{irreducible} constraints that cannot be
solved\footnote{In general, the quadratic constraints cannot be used to solve
  some deformation parameters in terms of the rest. For instance, in this
  sense, if $a$ and $b$ are two of them, a constraint of the form $ab=0$
  cannot be solved and we can call it \textit{irreducible}.} and of associated
$d$-forms may be expected in the end, but we have to go through the whole
procedure to identify them. This identification will be one of the main
results of the following section.

However, this is not the end of the story for the possible 9-forms. As it was
shown in Ref.~\cite{Huebscher:2010ib} in 4- 5- and 6-dimensional cases, in the
ungauged case one can find more $d$-forms with consistent supersymmetric
transformation rules than predicted by the embedding-tensor formalism. Those
additional fields are predicted by the Ka\v{c}-Moody approach
\cite{Kleinschmidt:2008jj}. However, after gauging, the new fields do not have
consistent, independent, supersymmetry transformation rules to all orders in
fermions\footnote{The insufficiency of first-order in fermions checks was
  first noticed in Ref.~\cite{Bergshoeff:2010mv}.}, and have to be combined
with other $d$-forms, so that, in the end, only the number of $d$-forms
predicted by the embedding-tensor formalism survive. 

This means that the results obtained via the embedding-tensor formalism for
the 9-forms have to be interpreted with special care and have to be compared
with the results obtained with other approaches.

The closure of the local supersymmetry algebra needs to be checked on all the
fields in the tensor hierarchy predicted by the embedding-tensor formalism
and, in particular, on the 9-forms to all orders in fermions.  However, given
that gauge invariance is requirement for local supersymmetry invariance, we
expect consistency in essentially all cases with the possible exception of the
9-forms, according to the above discussion. In the following section we will do
this for the electric fields of the theory.


\section{Deforming the maximal $d=9$ supergravity}
\label{sec-deformation}

In this section we are going to study the possible deformations of $d=9$
supergravity, starting from its possible gaugings using the embedding-tensor
formalism and constructing the corresponding tensor hierarchy
\cite{Cordaro:1998tx,deWit:2002vt,deWit:2003hq,deWit:2005hv,deWit:2005ub,Bergshoeff:2009ph,Hartong:2009vc}
up to the 4-form potentials.

If we denote by $\Lambda^{I}(x)$ the scalar parameters of the gauge
transformations of the 1-forms $A^{I}$ and by $\alpha^{A}$ the constant
parameters of the global symmetries, we want to promote 

\begin{equation}
\label{eq:gauging}
\alpha^{A} \longrightarrow \Lambda^{I}(x)\vartheta_{I}{}^{A}\, ,
\end{equation}

\noindent
where $\vartheta_{I}{}^{A}$ is the \textit{embedding tensor}, in the
transformation rules of all the fields, and we are going to require the theory
to be covariant under the new local transformations using the 1-forms as gauge
fields. 

To achieve this goal, starting with the transformations of the
scalars, the successive introduction of higher-rank $p$-form potentials is
required, which results in the construction of a tensor hierarchy. Most of
these fields are already present in the supergravity theory or can be
identified with their magnetic duals but this procedure allows us to introduce
consistently the highest-rank fields (the $d$-, $(d-1)$- and $(d-2)$-form
potentials), which are not dual to any of the original electric
fields. Actually, as explained in Section~\ref{sec-magnetic}, the highest-rank
potentials are related to the symmetries (Noether currents), the independent
deformation parameters and the constraints that they satisfy, but we need to
determine these, which requires going through this procedure checking the
consistency with gauge and supersymmetry invariance at each step.

Thus, we are going to require invariance under the new gauge transformations
for the scalar fields and we are going to find that we need new couplings to
the gauge 1-form fields (as usual). Then we will study the modifications of
the supersymmetry transformation rules of the scalars and fermion fields which
are needed to ensure the closure of the local supersymmetry algebra on the
scalars. Usually we do not expect modifications in the bosons' supersymmetry
transformations, but the fermions' transformations need to be modified by
replacing derivatives and field strengths by covariant derivatives and
covariant field strengths and, furthermore, by adding \textit{fermion
  shifts}. The local supersymmetry algebra will close provided that we impose
certain constraints on the embedding tensor components and on the fermion
shifts. 

Repeating this procedure on the 1-forms (which requires the coupling to the
2-forms) etc.~we will find a set of constraints that we can solve, determining
the independent components of the deformation tensors\footnote{As we are going
  to see, besides the embedding tensor, one can introduce many other
  deformation tensors.} and the fermions shifts. Some constraints (typically
quadratic in deformation parameters) have to be left unsolved and we will have
to take them into account towards the end of this procedure.

As a result we will identify the independent deformations of the theory and
the constraints that they satisfy. From this we will be able to extract
information about the highest-rank potentials in the tensor hierarchy.


\subsection{The 0-forms $\varphi, \tau$}

Under the global symmetry group, the scalars transform according to

\begin{equation}
\delta_{\alpha}\varphi =\alpha^{A}k_{A}{}^{\varphi}\, ,
\hspace{1cm}
\delta_{\alpha}\tau =\alpha^{A}k_{A}{}^{\tau}\, , 
\end{equation}

\noindent
where the $\alpha^{A}$ are the constant parameters of the transformations,
labeled by $A=1,\cdots,5$, and where $k_{A}{}^{\varphi}$ and $k_{A}{}^{\tau}$
are the corresponding components of the Killing vectors of the scalar
manifold, given in Eq.~(\ref{eq:varphikillingvectors})
(Eq.~(\ref{eq:taukillingvectors})).

According to the general prescription Eq.~(\ref{eq:gauging}), we want to gauge
these symmetries making the theory invariant under the local transformations

\begin{equation}
\label{eq:scalargaugetransformations}
\delta_{\Lambda}\varphi =\Lambda^{I}\vartheta_{I}{}^{A}k_{A}{}^{\varphi}\, ,
\hspace{1cm}
\delta_{\Lambda}\tau =\Lambda^{I}\vartheta_{I}{}^{A}k_{A}{}^{\tau}\, , 
\end{equation}

\noindent
where $\Lambda^{I}(x)$, $I=0,\mathbf{1},\mathbf{2}$, are the 0-form gauge
parameters of the 1-form gauge fields $A^{I}$ and $\vartheta_{I}{}^{A}$ is the
embedding tensor. 

To construct gauge-covariant field strengths for the scalars it is enough to
replace their derivatives by covariant derivatives.


\subsubsection{Covariant derivatives}

The covariant derivatives of the scalars have the standard form

\begin{equation}
\mathfrak{D}\varphi = d\varphi +A^{I}\vartheta_{I}^{A}k_{A}{}^{\varphi}\, ,  
\hspace{1cm}
\mathfrak{D}\tau = d\tau +A^{I}\vartheta_{I}^{A}k_{A}{}^{\tau}\, ,  
\end{equation}

\noindent
and they transform covariantly provided that the 1-form gauge fields transform
as

\begin{equation}
\label{eq:deltaAI}
\delta_{\Lambda}A^{I}
= 
-\mathfrak{D}\Lambda^{I} +Z^{I}{}_{i}\Lambda^{i}\, ,  
\end{equation}

\noindent
where the $\Lambda^{i}$, $i=1,2$, are two possible 1-form gauge parameters and
$Z^{I}{}_{i}$ is a possible new deformation parameter that must satisfy
the orthogonality constraint

\begin{equation}
\label{eq:constraint2}
\vartheta_{I}{}^{A}Z^{I}{}_{i} =0\, .  
\end{equation}

\noindent
Furthermore, it is necessary that the embedding tensor satisfies the standard
quadratic constraint

\begin{equation}
\label{eq:standardquadraticconstraint}
\vartheta_{I}{}^{A}T_{A\,  J}{}^{K}\vartheta_{K}{}^{C}
-\vartheta_{I}{}^{A}\vartheta_{J}{}^{B}f_{AB}{}^{C}=0\, ,  
\end{equation}

\noindent
that expresses the gauge-invariance of the embedding tensor. 

As a general rule, all the deformation tensors have to be gauge-invariant and
we can anticipate that we will have to impose the constraint that expresses
the gauge-invariance of $Z^{I}{}_{i}$, namely 

\begin{equation}
\label{eq:constraint3}
X_{J\, K}{}^{I}Z^{K}{}_{i}
-X_{J\, i}{}^{j}Z^{I}{}_{j} =  0\, ,
\end{equation}

\noindent
where 

\begin{equation}
\label{eq:XIJKXIjk}
X_{I\, J}{}^{K} \equiv \vartheta_{I}{}^{A}T_{A\, J}{}^{K}\, ,
\hspace{1cm}
X_{J\, i}{}^{j} \equiv \vartheta_{J}{}^{A}T_{A\, i}{}^{j}\, .
\end{equation}


\subsubsection{Supersymmetry transformations of the fermion fields}

We will assume for simplicity that the supersymmetry transformations of the
fermion fields in the deformed theory have essentially the same form as in the
undeformed theory but covariantized (derivatives and field strengths) and,
possibly, with the addition of fermion shifts which we add in the most general
form:

\begin{eqnarray}
\delta_{\epsilon}\psi_{\mu}
& = & 
\mathfrak{D}_{\mu}\epsilon
+f\gamma_{\mu} \epsilon
+k\gamma_{\mu} \epsilon^{*}
+\tfrac{i}{8\cdot 2!}e^{-\frac{2}{\sqrt{7}}\varphi}
\left(\tfrac{5}{7}\gamma_{\mu}\gamma^{(2)} 
-\gamma^{(2)}\gamma_{\mu} \right)F^{0}\epsilon  
\nonumber \\
& & \nonumber \\
& & 
-\tfrac{1}{8\cdot 2!}e^{\frac{3}{2\sqrt{7}}\varphi+\frac{1}{2}\phi}
\left(\tfrac{5}{7}\gamma_{\mu}\gamma^{(2)} 
-\gamma^{(2)}\gamma_{\mu} \right)(F^{1}-\tau F^{2})\epsilon^{*}  
\nonumber \\
& & \nonumber \\
& & 
-\tfrac{i}{8\cdot 3!}e^{-\frac{1}{2\sqrt{7}}\varphi}
\left(\tfrac{3}{7}\gamma_{\mu}\gamma^{(3)} 
+\gamma^{(3)}\gamma_{\mu} \right)(H^{1}-\tau H^{2})\epsilon^{*}  
\nonumber \\
& & \nonumber \\
& & 
-\tfrac{1}{8\cdot 4!}e^{\frac{1}{\sqrt{7}}\varphi}
\left(\tfrac{1}{7}\gamma_{\mu}\gamma^{(4)} 
-\gamma^{(4)}\gamma_{\mu} \right)G \epsilon\, ,
\end{eqnarray}

\begin{eqnarray}
\delta_{\epsilon}\tilde{\lambda}
& = & 
i\not\!\!\mathfrak{D} \varphi\epsilon^{*} +\tilde{g}\epsilon  +\tilde{h}\epsilon^{*}
-\tfrac{1}{\sqrt{7}}e^{-\frac{2}{\sqrt{7}}\varphi}\not\! F^{0}\epsilon^{*}
-\tfrac{3i}{2\cdot 2!\sqrt{7}}e^{\frac{3}{2\sqrt{7}}\varphi +\frac{1}{2}\phi}
(\not\! F^{1}-\tau^{*}\not\! F^{2})\epsilon
\nonumber \\
& & \nonumber \\
& & 
-\tfrac{1}{2\cdot 3!\sqrt{7}}e^{-\frac{1}{2\sqrt{7}}\varphi +\frac{1}{2}\phi}
(\not\!\! H^{1}-\tau^{*}\not\!\! H^{2})\epsilon
-\tfrac{i}{4!\sqrt{7}}e^{\frac{1}{\sqrt{7}}\varphi}\not\! G\epsilon^{*}\, ,
\end{eqnarray}

\begin{eqnarray}
\delta_{\epsilon}\lambda
& = & 
-e^{\phi}\not\!\!\mathfrak{D} \tau\epsilon^{*} +g\epsilon  +h\epsilon^{*}
-\tfrac{i}{2\cdot 2!}e^{\frac{3}{2\sqrt{7}}\varphi +\frac{1}{2}\phi}
(\not\! F^{1}-\tau\not\! F^{2})\epsilon
\nonumber \\
& & \nonumber \\
& & 
+\tfrac{1}{2\cdot 3!}e^{-\frac{1}{2\sqrt{7}}\varphi +\frac{1}{2}\phi}
(\not\!\! H^{1}-\tau\not\!\! H^{2})\epsilon\, .
\end{eqnarray}

\noindent
In these expressions, $f,k,g,h,\tilde{g},\tilde{h}$ are six functions of the
scalars and deformation parameters to be determined, the covariant field
strengths have the general form predicted by the tensor hierarchy (to be
determined) and the covariant derivatives of the scalars have the forms given
above. Furthermore, in $\delta_{\epsilon}\psi_{\mu}$, 
$\mathfrak{D}_{\mu}\epsilon$ stands for the Lorentz- and gauge-covariant
derivative of the supersymmetry parameter, which turns out to be given by

\begin{equation}
\mathfrak{D}_{\mu}\epsilon  
\equiv
\left\{ 
\nabla_{\mu}
+\tfrac{i}{2}
\left[
\tfrac{1}{2}e^{\phi}
\mathfrak{D}^{5}_{\mu}\chi
+A^{I}{}_{\mu}\vartheta_{I}{}^{m}\mathcal{P}_{m}
\right]
+\tfrac{9}{14}\gamma_{\mu}\not\!\!A^{I}\vartheta_{I}{}^{4}
\right\}\epsilon
\end{equation}

\noindent
where $\mathcal{P}_{m}$ $1,2,3$ are the momentum maps of the holomorphic
Killing vectors of $SL(2,\mathbb{R})$, defined in
Eq.~(\ref{eq:momentummapsdef}) and given in Eq.~(\ref{eq:momentummaps}),
$\nabla_{\mu}$ is the Lorentz-covariant derivative and

\begin{equation}
\mathfrak{D}^{5}_{\mu}\chi
\equiv
\partial_{\mu}\chi
-\tfrac{3}{4}A^{I}{}_{\mu}\vartheta_{I}{}^{5} \chi
\end{equation}

\noindent
is the derivative of $\chi$ covariant only with respect to the $\beta$
rescalings. it can be checked that $\mathfrak{D}_{\mu}\epsilon$ transforms
covariantly under gauge transformations if and only if the embedding tensor
satisfies the standard quadratic constraint
Eq.~(\ref{eq:standardquadraticconstraint}).

An equivalent expression for it is

\begin{equation}
\mathfrak{D}_{\mu}\epsilon  
=
\left\{ 
\nabla_{\mu}
+\tfrac{i}{2}
\left[
\tfrac{1}{2}e^{\phi}
\mathfrak{D}_{\mu}\chi
-A^{I}{}_{\mu}\vartheta_{I}{}^{m}\Im{\rm m}\lambda_{m}
\right]
+\tfrac{9}{14}\gamma_{\mu}\not\!\!A^{I}\vartheta_{I}{}^{4}
\right\}\epsilon\, ,
\end{equation}

\noindent
where the $\lambda_{m}$, $m=1,2,3$, of $SL(2,\mathbb{R})$ and defined in
Eq.~(\ref{eq:momentummapsdef}) and given in Eq.~(\ref{eq:lambdam}) and where
now

\begin{equation}
\mathfrak{D}_{\mu}\chi
\equiv
\partial_{\mu}\chi +A^{I}{}_{\mu}\vartheta_{I}{}^{A} k_{A}{}^{\chi}\, ,
\end{equation}

\noindent
is the total covariant derivative of $\chi$ (which is invariant under both the
$\alpha$ and $\beta$ scaling symmetries as well as under $SL(2,\mathbb{R})$).

The actual form of the $(p+1)$-form field strengths will not be needed until
the moment in which study the closure of the supersymmetry algebra on the
corresponding $p$-form potential. 


\subsubsection{Closure of the supersymmetry algebra on the
 0-forms $\varphi,\tau$}

We assume that the supersymmetry transformations of the scalars are the same as
in the undeformed theory

\begin{eqnarray}
\delta_{\epsilon}\varphi 
& = & 
-\tfrac{i}{4}\bar{\epsilon}\tilde{\lambda}^{*}+\mathrm{h.c.}\, ,
\\
& & \nonumber \\  
\delta_{\epsilon}\tau 
& = & 
-\tfrac{1}{2}e^{-\phi}\bar{\epsilon}^{*}\lambda\, .
\end{eqnarray}

To lowest order in fermions, the commutator of two supersymmetry
transformations gives


\begin{eqnarray}
\label{eq:commutatoronvarphi1}
\left[\delta_{\epsilon_{1}},\delta_{\epsilon_{2}}\right]\varphi
& = &   
\xi^{\mu}\mathfrak{D}_{\mu}\varphi 
+\Re{\rm e}(\tilde{h}) b
-\Im{\rm m}(\tilde{g})c +\Re{\rm e}(\tilde{g}) d\, ,
\\
& & \nonumber \\
\label{eq:commutatorontau1}
\left[\delta_{\epsilon_{1}},\delta_{\epsilon_{2}}\right]\tau
& = &   
\xi^{\mu}\mathfrak{D}_{\mu}\tau +e^{-\phi}\left[g(c-id) -ihb\right]\, ,
\end{eqnarray}

\noindent
where $\xi^{\mu}$ is one of the spinor bilinears defined in
Appendix~\ref{d9:sec-bilinears} that clearly plays the r\^ole of parameter of the
general coordinate transformations and $a,b,c,d$ are the scalar bilinears
defined in the same appendix.

In the right hand side of these commutators, to lowest order in fermions, we
expect a general coordinate transformation (the Lie derivative $\pounds_{\xi}$
of the scalars with respect to $\xi^{\mu}$) and a gauge transformation which
has the form of Eq.~(\ref{eq:scalargaugetransformations}) for the
scalars. Therefore, the above expressions should be compared with

\begin{eqnarray}
\label{eq:commutatoronvarphi2}
\left[\delta_{\epsilon_{1}},\delta_{\epsilon_{2}}\right]\varphi
& = &   
\pounds_{\xi}\varphi +\Lambda^{I}\vartheta_{I}{}^{A}k_{A}{}^{\varphi}\, ,
\\
& & \nonumber \\
\label{eq:commutatorontau2}
\left[\delta_{\epsilon_{1}},\delta_{\epsilon_{2}}\right]\tau
& = &   
\pounds_{\xi}\tau +\Lambda^{I}\vartheta_{I}{}^{A}k_{A}{}^{\tau}\, ,
\end{eqnarray}

\noindent
from which we get the relations

\begin{eqnarray}
\label{eq:relation1}
\Re{\rm e}(\tilde{h}) b
-\Im{\rm m}(\tilde{g})c +\Re{\rm e}(\tilde{g}) d  
& = & 
(\Lambda^{I}-a^{I})\vartheta_{I}{}^{A}k_{A}{}^{\varphi}
\, ,
\\
& & \nonumber \\
\label{eq:relation2}
g(c-id) -ihb  
& = & 
e^{\phi}(\Lambda^{I}-a^{I})\vartheta_{I}{}^{A}k_{A}{}^{\tau}
\, ,
\end{eqnarray}

\noindent
which would allow us to determine the fermion shift functions if we knew the
gauge parameters $\Lambda^{I}$. In order to determine the $\Lambda^{I}$s we
have to close the supersymmetry algebra on the 1-forms. In these expressions
and in those that will follow, we use the shorthand notation

\begin{equation}
a^{I} \equiv \xi^{\mu}A^{I}{}_{\mu}\, ,\hspace{.5cm}
b^{i}{}_{\mu} \equiv \xi^{\nu}B^{i}{}_{\nu\mu}\, ,\hspace{.5cm}
c_{\mu\nu} \equiv \xi^{\rho}C_{\rho\mu\nu}\, ,\hspace{.5cm}
\mathrm{etc.} 
\end{equation}


\subsection{The 1-forms $A^{I}$}

The next step in this procedure is to consider the 1-forms that we just
introduced to construct covariant derivatives for the scalars.


\subsubsection{The 2-form field strengths $F^{I}$}

The gauge transformations of the 1-forms are given in Eq.~(\ref{eq:deltaAI})
and we first need to determine their covariant field strengths. A general
result of the embedding-tensor formalism tells us that we need to introduce
2-form potentials in the covariant field strengths. In this case only have the
$SL(2,\mathbb{R})$ doublet $B^{i}$ at our disposal and, therefore, the 2-form
field strengths have the form

\begin{equation}
\label{eq:FI}
F^{I} = dA^{I} +\tfrac{1}{2}X_{JK}{}^{I}A^{J}\wedge A^{K}
+Z^{I}{}_{i}B^{i}\, ,  
\end{equation}

\noindent
where $X_{JK}{}^{I}$ has been defined in Eq.~(\ref{eq:XIJKXIjk}) and
$Z^{I}{}_{i}$ is precisely the deformation tensor we introduced in
Eq.~(\ref{eq:deltaAI}). $F^{I}$ will transform covariantly under
Eq.~(\ref{eq:deltaAI}) if simultaneously the 2-forms $B^{i}$ transform
according to

\begin{equation}
\label{eq:deltaBi}
\delta_{\Lambda}B^{i}
= 
-\mathfrak{D}\Lambda^{i} 
-2h_{IJ}{}^{i}\left[\Lambda^{I}F^{J} 
+\tfrac{1}{2}A^{I}\wedge \delta_{\Lambda}A^{J}\right] 
+Z^{i}\Lambda\, ,
\end{equation}

\noindent
where $h_{IJ}{}^{i}$ and $Z^{i}$ are two possible new deformation tensors the
first of which must satisfy the constraint

\begin{equation}
\label{eq:constraint4}
X_{(JK)}{}^{I} +Z^{I}{}_{i}h_{JK}{}^{i} =  0\, ,
\end{equation}

\noindent
while $Z^{i}$ must satisfy the orthogonality constraint

\begin{equation}
\label{eq:constraint6}
Z^{I}{}_{i}Z^{i} = 0\, .
\end{equation}

\noindent
Both of them must satisfy the constraints that express their gauge invariance:

\begin{eqnarray}
\label{eq:constraint7}
X_{I\, j}{}^{i}h_{JK}{}^{j} -2X_{I(J}{}^{L}h_{K)L}{}^{i} & = & 0\, ,
\\
& & \nonumber \\
\label{eq:constraint9}
X_{I}Z^{i}- X_{I\, j}{}^{i}Z^{j} & = & 0\, ,
\end{eqnarray}

\noindent
where 

\begin{equation}
X_{I} \equiv \vartheta_{I}{}^{A}T^{(1)}_{A}\, .
\end{equation}


\subsubsection{Closure of the supersymmetry algebra on the 
1-forms $A^{I}$}

We assume, as we are doing with all the bosons, that the supersymmetry
transformations of the 1-forms of the theory are not deformed by the gauging,
so they take the form

\begin{eqnarray}
\delta_{\epsilon}A^{0}{}_{\mu}
& = & 
\tfrac{i}{2}e^{\frac{2}{\sqrt{7}}\varphi}\bar{\epsilon}
\left(\psi_{\mu} -\tfrac{i}{\sqrt{7}}\gamma_{\mu}\tilde{\lambda}^{*}\right) 
+\mathrm{h.c.}
\, ,
\\ 
& & \nonumber \\ 
\delta_{\epsilon}A^{\mathbf{1}}{}_{\mu}
& = & 
\tfrac{i}{2}\tau^{*}e^{-\frac{3}{2\sqrt{7}}\varphi+\frac{1}{2}\phi}
\left(
\bar{\epsilon}^{*}\psi_{\mu} 
-\tfrac{i}{4}\bar{\epsilon}\gamma_{\mu}\lambda
+\tfrac{3i}{4\sqrt{7}}\bar{\epsilon}^{*}\gamma_{\mu}\tilde{\lambda}^{*}
\right)
+\mathrm{h.c.}
\, ,
\\ 
& & \nonumber \\ 
\delta_{\epsilon}A^{\mathbf{2}}{}_{\mu}
& = & 
\tfrac{i}{2}e^{-\frac{3}{2\sqrt{7}}\varphi+\frac{1}{2}\phi}
\left(
\bar{\epsilon}^{*}\psi_{\mu} 
-\tfrac{i}{4}\bar{\epsilon}\gamma_{\mu}\lambda
+\tfrac{3i}{4\sqrt{7}}\bar{\epsilon}^{*}\gamma_{\mu}\tilde{\lambda}^{*}
\right)
+\mathrm{h.c.}
\end{eqnarray}

\noindent
The commutator of two of them gives, to lowest order in fermions,


\begin{equation}
  \begin{array}{rcl}
\left[\delta_{\epsilon_{1}},\delta_{\epsilon_{2}}\right]A^{0}{}_{\mu}
& = &   
\xi^{\nu}F^{0}{}_{\nu\mu} 
-\mathfrak{D}_{\mu}\left(e^{\frac{2}{\sqrt{7}} \varphi}b\right)
+\tfrac{2}{\sqrt{7}}
e^{\frac{2}{\sqrt{7}}\varphi}
\left\{
\left[\Re{\rm e}(\tilde{h})-\sqrt{7}\, \Im{\rm m}(f)\right]\xi_{\mu}
\right.
 \\
& &  \\
& & 
\left.
+\left[\Re{\rm e}(\tilde{g})-\sqrt{7}\, \Im{\rm m}(k)\right]\sigma_{\mu}
+\left[\Im{\rm m}(\tilde{g})-\sqrt{7}\, \Re{\rm e}(k)\right]\rho_{\mu}
\right\}
\, ,
\end{array}
\end{equation}

\begin{equation}
  \begin{array}{rcl}
\left[\delta_{\epsilon_{1}},\delta_{\epsilon_{2}}\right]A^{\mathbf{1}}{}_{\mu}
& = &   
\xi^{\nu}F^{\mathbf{1}}{}_{\nu\mu} 
-\partial_{\mu}
\left[
e^{-\frac{3}{2\sqrt{7}}\varphi+\frac{1}{2}\phi}(\chi d +e^{-\phi}c)
\right]
 \\
& &  \\
& & 
\hspace{-3cm}
-A^{I}{}_{\mu}
\left[
(\tfrac{1}{2}\vartheta_{I}{}^{1}-\tfrac{3}{4}\vartheta_{I}{}^{5})
e^{-\frac{3}{2\sqrt{7}}\varphi+\frac{1}{2}\phi}(\chi d +e^{-\phi}c)
+
\tfrac{1}{2}(\vartheta_{I}{}^{2}+\vartheta_{I}{}^{3})
e^{-\frac{3}{2\sqrt{7}}\varphi\frac{1}{2}\phi}d
\right]
 \\
& & \\
& & 
\hspace{-3cm}
-2e^{-\frac{3}{2\sqrt{7}}\varphi\frac{1}{2}\phi}\left\{
\chi
\left[
\Im{\rm m}(k)+\tfrac{3}{4\sqrt{7}}\Re{\rm e}(\tilde{g}) -\tfrac{1}{4}\Re{\rm e}(g)
\right]
+e^{-\phi}
\left[
-\Re{\rm e}(k)-\tfrac{3}{4\sqrt{7}}\Im{\rm m}(\tilde{g}) -\tfrac{1}{4}\Im{\rm m}(g)
\right]
\right\}\xi_{\mu}
 \\
& &  \\
& & 
\hspace{-3cm}
-2e^{-\frac{3}{2\sqrt{7}}\varphi\frac{1}{2}\phi}
\left\{
\chi
\left[
-\Re{\rm e}(f) -\tfrac{3}{4\sqrt{7}}\Im{\rm m}(\tilde{h}) 
+\tfrac{1}{4}\Im{\rm m}(h)
\right]
+
e^{-\phi}
\left[
-\Im{\rm m}(f) -\tfrac{3}{4\sqrt{7}}\Re{\rm e}(\tilde{h}) 
-\tfrac{1}{4}\Re{\rm e}(h)
\right]
\right\}\rho_{\mu}
 \\
& &  \\
& & 
\hspace{-3cm}
-2e^{-\frac{3}{2\sqrt{7}}\varphi\frac{1}{2}\phi}
\left\{
\chi
\left[
\Im{\rm m}(f)+\tfrac{3}{4\sqrt{7}}\Re{\rm e}(\tilde{h}) 
-\tfrac{1}{4}\Re{\rm e}(h)
\right]
+e^{-\phi}
\left[
-\Re{\rm e}(f)-\tfrac{3}{4\sqrt{7}}\Im{\rm m}(\tilde{h}) 
-\tfrac{1}{4}\Im{\rm m}(h)
\right]
\right\}
\sigma_{\mu}\, ,
\end{array}
\end{equation}

\noindent
and

\begin{equation}
  \begin{array}{rcl}
\left[\delta_{\epsilon_{1}},\delta_{\epsilon_{2}}\right]A^{\mathbf{2}}{}_{\mu}
& = &   
\xi^{\nu}F^{\mathbf{2}}{}_{\nu\mu} 
-\partial_{\mu}
\left(
e^{-\frac{3}{2\sqrt{7}}\varphi+\frac{1}{2}\phi}d
\right)
 \\
& &  \\
& & 
-A^{I}{}_{\mu}
\left[
\tfrac{1}{2}(\vartheta_{I}{}^{2}-\vartheta_{I}{}^{3})
e^{-\frac{3}{2\sqrt{7}}\varphi+\frac{1}{2}\phi}(\chi d +e^{-\phi}c)
-\tfrac{1}{2}\vartheta_{I}{}^{1}
e^{-\frac{3}{2\sqrt{7}}\varphi+\frac{1}{2}\phi}d
\right]
 \\
& &  \\
& & 
-2e^{-\frac{3}{2\sqrt{7}}\varphi+\frac{1}{2}\phi}
\left[
\Im{\rm m}(k)+\tfrac{3}{4\sqrt{7}}\Re{\rm e}(\tilde{g}) -\tfrac{1}{4}\Re{\rm e}(g)
\right]\xi_{\mu}
 \\
& &  \\
& & 
-2e^{-\frac{3}{2\sqrt{7}}\varphi+\frac{1}{2}\phi}
\left[
-\Re{\rm e}(f) -\tfrac{3}{4\sqrt{7}}\Im{\rm m}(\tilde{h}) 
+\tfrac{1}{4}\Im{\rm m}(h)
\right]
\rho_{\mu}
 \\
& &  \\
& & 
-2e^{-\frac{3}{2\sqrt{7}}\varphi+\frac{1}{2}\phi}
\left[
\Im{\rm m}(f)+\tfrac{3}{4\sqrt{7}}\Re{\rm e}(\tilde{h}) 
-\tfrac{1}{4}\Re{\rm e}(h)
\right]
\sigma_{\mu}\, ,
\end{array}
\end{equation}

\noindent
where $\sigma_{\mu}$ and $\rho_{\mu}$ are spinor bilinears defined in
Appendix~\ref{d9:sec-bilinears}.

The closure of the local supersymmetry algebra requires the commutators to
take the form

\begin{equation}
\left[\delta_{\epsilon_{1}},\delta_{\epsilon_{2}}\right]A^{I}{}_{\mu}
 =    
\pounds_{\xi}A^{I}{}_{\mu}
-\mathfrak{D}_{\mu}\Lambda^{I} +Z^{I}{}_{i}\Lambda^{i}{}_{\mu}\, ,
\end{equation}

\noindent
which will only happen if gauge parameters $\Lambda^{I}$ are given by 

\begin{equation}
\begin{array}{rcl}
\label{eq:LambdaI}
\Lambda^{0}
& = & 
a^{0}  +e^{\frac{2}{\sqrt{7}}\varphi}b\, ,
\\
& &  \\
\Lambda^{\mathbf{1}}
& = & 
a^{\mathbf{1}}
+e^{-\frac{3}{2\sqrt{7}}\varphi+\frac{1}{2}\phi}(\chi d +e^{-\phi}c)\, ,
\\
& &  \\
\Lambda^{\mathbf{2}}
& = & 
a^{\mathbf{2}} +e^{-\frac{3}{2\sqrt{7}}\varphi+\frac{1}{2}\phi}d\, ,\\
\end{array}
\end{equation}

\noindent
and the 1-form gauge parameters $\Lambda^{i}{}_{\mu}$ satisfy the relations

\begin{eqnarray}
\left[\Re{\rm e}(\tilde{h})-\sqrt{7}\, \Im{\rm m}(f)\right]\xi_{\mu}
+\left[\Re{\rm e}(\tilde{g})-\sqrt{7}\, \Im{\rm m}(k)\right]\sigma_{\mu}
+\left[\Im{\rm m}(\tilde{g})-\sqrt{7}\, \Re{\rm e}(k)\right]\rho_{\mu}
& &   
\nonumber \\
& & \nonumber \\
=
\tfrac{\sqrt{7}}{2}e^{-\frac{2}{\sqrt{7}}\varphi}Z^{0}{}_{i}
\left[
\Lambda^{i}{}_{\mu}
-(b^{i}{}_{\mu} -h_{IJ}{}^{i}a^{I}A^{J}{}_{\mu})
\right]\, ,
\label{eq:1-formshifts1}
& &  \\
& & \nonumber \\
\left\{
\chi
\left[
\Im{\rm m}(k)+\tfrac{3}{4\sqrt{7}}\Re{\rm e}(\tilde{g}) -\tfrac{1}{4}\Re{\rm e}(g)
\right]
+e^{-\phi}
\left[
-\Re{\rm e}(k)-\tfrac{3}{4\sqrt{7}}\Im{\rm m}(\tilde{g}) -\tfrac{1}{4}\Im{\rm m}(g)
\right]
\right\}\xi_{\mu}
\nonumber \\
& & \nonumber \\
\hspace{-3cm}
+
\left\{
\chi
\left[
-\Re{\rm e}(f) -\tfrac{3}{4\sqrt{7}}\Im{\rm m}(\tilde{h}) 
+\tfrac{1}{4}\Im{\rm m}(h)
\right]
+
e^{-\phi}
\left[
-\Im{\rm m}(f) -\tfrac{3}{4\sqrt{7}}\Re{\rm e}(\tilde{h}) 
-\tfrac{1}{4}\Re{\rm e}(h)
\right]
\right\}\rho_{\mu}
& & \nonumber \\
& & \nonumber \\
\hspace{-3cm}
+\left\{
\chi
\left[
\Im{\rm m}(f)+\tfrac{3}{4\sqrt{7}}\Re{\rm e}(\tilde{h}) 
-\tfrac{1}{4}\Re{\rm e}(h)
\right]
+e^{-\phi}
\left[
-\Re{\rm e}(f)-\tfrac{3}{4\sqrt{7}}\Im{\rm m}(\tilde{h}) 
-\tfrac{1}{4}\Im{\rm m}(h)
\right]
\right\}
\sigma_{\mu}\, ,
& & \nonumber \\
& & \nonumber \\
 = 
-\tfrac{1}{2}e^{+\frac{3}{2\sqrt{7}}\varphi-\frac{1}{2}\phi}Z^{\mathbf{1}}{}_{i}
\left[
\Lambda^{i}{}_{\mu}
-(b^{i}{}_{\mu} -h_{IJ}{}^{i}a^{I}A^{J}{}_{\mu})
\right]\, , 
& & \\
& & \nonumber \\
\left[
\Im{\rm m}(k)+\tfrac{3}{4\sqrt{7}}\Re{\rm e}(\tilde{g}) -\tfrac{1}{4}\Re{\rm e}(g)
\right]
\xi_{\mu}
+
\left[
-\Re{\rm e}(f) -\tfrac{3}{4\sqrt{7}}\Im{\rm m}(\tilde{h}) 
+\tfrac{1}{4}\Im{\rm m}(h)
\right]
\rho_{\mu}
& & \nonumber \\
& & \nonumber \\
\hspace{-3cm}
+\left[
\Im{\rm m}(f)+\tfrac{3}{4\sqrt{7}}\Re{\rm e}(\tilde{h}) 
-\tfrac{1}{4}\Re{\rm e}(h)
\right]
\sigma_{\mu}\, ,
& & \nonumber \\
& & \nonumber \\
 = 
-\tfrac{1}{2}e^{+\frac{3}{2\sqrt{7}}\varphi-\frac{1}{2}\phi}Z^{\mathbf{2}}{}_{i}
\left[
\Lambda^{i}{}_{\mu}
-(b^{i}{}_{\mu} -h_{IJ}{}^{i}a^{I}A^{J}{}_{\mu})
\right]\, .
\label{eq:1-formshifts2}
\end{eqnarray}

\noindent
Using the values of the parameters $\Lambda^{I}$ that we just have determined
in the relations Eqs.~(\ref{eq:relation1}) and (\ref{eq:relation2}) we can
determine some of the fermions shifts:

\begin{eqnarray}
\Re{\rm e}(\tilde{h})
& = & 
\vartheta_{0}{}^{A}k_{A}{}^{\varphi} e^{\frac{2}{\sqrt{7}}\varphi}\, ,
\\
& & \nonumber \\
\tilde{g}
& = & 
(\vartheta_{\mathbf{1}}{}^{A}\tau^{*} +\vartheta_{\mathbf{2}}{}^{A})
k_{A}{}^{\varphi} e^{-\frac{3}{2\sqrt{7}}\varphi+\tfrac{1}{2}\phi}\, ,
\\
& & \nonumber \\
h
& = &
i\vartheta_{0}{}^{A}k_{A}{}^{\tau} e^{\frac{2}{\sqrt{7}}\varphi+\phi}\, ,
\\
& & \nonumber \\
g
& = &
\vartheta_{\mathbf{1}}{}^{A}
k_{A}{}^{\tau} e^{-\frac{3}{2\sqrt{7}}\varphi+\frac{1}{2}\phi}\, .
\end{eqnarray}

As a matter of fact, $g$ is overdetermined: we get two different expression
for it that give the same value if and only if

\begin{equation}
(\vartheta_{\mathbf{1}}{}^{A}\tau +\vartheta_{\mathbf{2}}{}^{A})k_{A}{}^{\tau}  
= 
0\, ,
\end{equation}

\noindent
which, upon use of the explicit expressions of the holomorphic Killing vectors
$k_{A}{}^{\tau}$ in Section~\ref{sec-global}, leads to the following linear
constraints on the components of the embedding tensor:

\begin{equation}
\label{eq:linearconstraints1}
\begin{array}{rcl}
\vartheta_{\mathbf{2}}{}^{2}  +\vartheta_{\mathbf{2}}{}^{3}
& = & 0\, ,
\\
& & \\
\vartheta_{\mathbf{1}}{}^{2}  +\vartheta_{\mathbf{1}}{}^{3} +2\vartheta_{\mathbf{2}}{}^{1}
-\tfrac{3}{2}\vartheta_{\mathbf{2}}{}^{5}
& = & 0\, ,
\\
& &  \\
\vartheta_{\mathbf{2}}{}^{2}  -\vartheta_{\mathbf{2}}{}^{3} -2\vartheta_{\mathbf{1}}{}^{1}
+\tfrac{3}{2}\vartheta_{\mathbf{1}}{}^{5}
& = & 0\, ,
\\
& &  \\
\vartheta_{\mathbf{1}}{}^{2}  -\vartheta_{\mathbf{1}}{}^{3}
& = & 0\, .
\end{array}
\end{equation}

These constraints allow us to express 4 of the 15 components of the embedding
tensor in terms of the remaining 11, but we are only going to do this after we
take into account the constraints that we are going to find in the closure of
the local supersymmetry algebra on the doublet of 2-forms $B^{i}$.

The values of $g,h.\tilde{g},\tilde{h}$ and the above constraints are
compatible with those of the primary deformations found in
Ref.~\cite{Bergshoeff:2002nv}.


\subsection{The 2-forms $B^{i}$}

In the previous subsection we have introduced a doublet of 2-forms $B^{i}$
with given gauge transformations to construct the 2-form field strengths
$F^{I}$. We now have to construct their covariant field strengths and check
the closure of the local supersymmetry algebra on them.


\subsubsection{The 3-form field strengths $H^{i}$}

In general we need to introduce 3-form potentials to construct the covariant
3-form field strengths and, since in maximal 9-dimensional supergravity, we
only have $C$ at our disposal, the 3-form field strengths will be given by

\begin{equation}
\label{eq:Hi}
H^{i}=
\mathfrak{D}B^{i} -h_{IJ}{}^{i}A^{I}\wedge dA^{J} 
-\tfrac{1}{3} X_{[IJ}{}^{L}h_{K]L}{}^{i}A^{IJK} +Z^{i}C\, ,
\end{equation}

\noindent
and they transform covariantly under the gauge transformations of the 1- and
2-forms that we have previously determined provided if the 3-form $C$
transforms as

\begin{equation}
\label{eq:deltaC}
\delta_{\Lambda}C = -\mathfrak{D}\Lambda
+g_{Ii}
\left[
-\Lambda^{I}H^{i} -F^{I}\wedge \Lambda^{i} +\delta_{\Lambda}A^{I}\wedge B^{i}
-\tfrac{1}{3}h_{JK}{}^{i}A^{IJ}\wedge \delta_{\Lambda}A^{K}
\right] 
+Z\tilde{\Lambda}\, .
\end{equation}

\noindent
where $g_{Ii}$ and $Z$ are two possible new deformation parameters. $g_{Ii}$
must satisfy the constraint

\begin{equation}
\label{eq:constraint5}
2h_{IJ}{}^{i}Z^{J}{}_{j}  +X_{I\, j}{}^{i} +Z^{i}g_{Ij}
=  0\, ,
\end{equation}

\noindent
while $Z$ must satisfy the orthogonality constraint

\begin{equation}
\label{eq:constraint10}  
Z^{i}Z  =  0\, .  
\end{equation}

\noindent
Both must by gauge-invariant, which implies the constraints

\begin{eqnarray}
\label{eq:constraint14}  
X_{IJ}{}^{L}g_{Li}+X_{I\, i}{}^{j}g_{Jj}-X_{I}g_{Ji} & = & 0\, ,
\\
& & \nonumber \\
\label{eq:constraint18}
(X_{I}-\tilde{X}_{I})Z & = & 0\, ,
\end{eqnarray}

\noindent
where 

\begin{equation}
\tilde{X}_{I} \equiv \vartheta_{I}{}^{A}T^{(\tilde{1})}_{A}\, .  
\end{equation}

Using the constraints obeyed by the deformation parameters and the explicit
form of the 2-form field strengths $F^{I}$ we can rewrite the 3-form field
strengths in the useful form

\begin{equation}
H^{i} =
\mathfrak{D}B^{i} -h_{IJ}{}^{i}A^{I}\wedge F^{J} 
+\tfrac{1}{6} X_{[IJ}{}^{L}h_{K]L}{}^{i}A^{IJK}
-\tfrac{1}{2}X_{Ij}{}^{i}A^{I}\wedge B^{j} 
+Z^{i}(C -\tfrac{1}{2}g_{Ij}A^{I}\wedge B^{j})\, .
\end{equation}


\subsubsection{Closure of the supersymmetry algebra on the 2-forms $B^{i}$}

In the undeformed theory, the supersymmetry transformation rules for the
2-forms are

\begin{eqnarray}
\delta_{\epsilon}B^{1}
& = &
\tau^{*} e^{\frac{1}{2\sqrt{7}}\varphi +\frac{1}{2}\phi} 
\left[ 
\bar{\epsilon}^{*}\gamma_{[\mu}\psi_{\nu]} 
-\tfrac{i}{8}\bar{\epsilon}\gamma_{\mu\nu}\lambda 
-\tfrac{i}{8\sqrt{7}}\bar{\epsilon}^{*}\gamma_{\mu\nu}\tilde{\lambda}^{*}
\right]   
\nonumber \\
& & \nonumber \\
& & 
-\delta^{1}{}_{\mathbf{i}}\left(A^{0}{}_{[\mu|}\delta_{\epsilon}A^{\mathbf{i}}{}_{|\nu]}
+A^{\mathbf{i}}{}_{[\mu|}\delta_{\epsilon}A^{0}{}_{|\nu]}\right)
+\mathrm{h.c.}\, ,
\\
& & \nonumber \\
\delta_{\epsilon}B^{2}
& = &
e^{\frac{1}{2\sqrt{7}}\varphi +\frac{1}{2}\phi} 
\left[ 
\bar{\epsilon}^{*}\gamma_{[\mu}\psi_{\nu]} 
-\tfrac{i}{8}\bar{\epsilon}\gamma_{\mu\nu}\lambda 
-\tfrac{i}{8\sqrt{7}}\bar{\epsilon}^{*}\gamma_{\mu\nu}\tilde{\lambda}^{*}
\right]   
\nonumber \\
& & \nonumber \\
& & 
-\delta^{2}{}_{\mathbf{i}}
\left(A^{0}{}_{[\mu|}\delta_{\epsilon}A^{\mathbf{i}}{}_{|\nu]}
+A^{\mathbf{i}}{}_{[\mu|}\delta_{\epsilon}A^{0}{}_{|\nu]}\right)
+\mathrm{h.c.}\, .
\end{eqnarray}

\noindent
The last terms in both transformations are associated to the presence of
derivatives of $A^{\mathbf{1}}$ and $A^{\mathbf{2}}$ in the field strengths of
$B^{1}$ and $B^{2}$ in the undeformed theory (see
Eq.~(\ref{eq:Hiundeformed})). In the deformed theory, the terms $-(A^{0}\wedge
dA^{\mathbf{i}} +A^{\mathbf{i}}\wedge dA^{0})$ are replaced by more general
couplings $-h_{IJ}{}^{i}A^{I}\wedge dA^{J}$ and, therefore, it would be natural to
replace the last terms in $\delta_{\epsilon}B^{i}{}_{\mu\nu}$ by

\begin{equation}
-2h_{IJ}{}^{i}A^{I}{}_{[\mu|}\delta_{\epsilon}A^{J}{}_{|\nu]}\, .  
\end{equation}

In the commutator of two supersymmetry transformations on the 2-forms, these
terms give the right contributions to the terms
$-2h_{IJ}{}^{i}\Lambda^{I}F^{J}$ of the gauge transformations (see
Eq.~(\ref{eq:deltaBi})). However, these terms must receive other contributions
in order to be complete and it turns out that the only terms of the form
$-2h_{IJ}{}^{i}\Lambda^{I}F^{J}$ that can be completed are precisely those of
the undeformed theory, which correspond to

\begin{equation}
\label{eq:valueofhIJi}
h_{\mathbf{i}0}{}^{j} = -\tfrac{1}{2}\delta_{\mathbf{i}}{}^{j}\, .  
\end{equation}

In order to get more general $h_{IJ}{}^{i}$s it would be necessary to deform
the fermions' supersymmetry rules, something we will not do here. Furthermore,
the structure of the Chern-Simons terms of the field strengths is usually
determined by the closure of the supersymmetry algebra at higher orders in
fermions and it is highly unlikely that a more general structure of the
Chern-Simons terms will be allowed by supersymmetry. Therefore, from now on,
we will set $h_{IJ}{}^{i}$ to the above value and we will set the values of
the deformation tensors in the Chern-Simons terms of the higher-rank field
strengths, to the values of the undeformed theory. Using the above value of
$h_{IJ}{}^{i}$ in the constraints in which it occurs will help us to solve
them, sometimes completely, as we will see.  Nevertheless, we will keep using
the notation $h_{IJ}{}^{i}$ for convenience.
 
Using the identity

\begin{equation}
  \begin{array}{rcl}
\xi^{\rho}H^{i}{}_{\rho\mu\nu}
-2h_{IJ}{}^{i}A^{I}{}_{\mu}\pounds_{\xi}A^{J}{}_{\nu}
& = &
\pounds_{\xi}B^{i}{}_{\mu\nu}
-2\mathfrak{D}_{[\mu|}(b^{i}{}_{|\nu]} 
-h_{IJ}{}^{i}a^{I}A^{J}{}_{|\nu]})]
\\
& & \\
& & 
-2h_{IJ}{}^{i}a^{I}F^{J}{}_{\mu\nu}  
\\
& & \\
& & 
+Z^{i}\left( c_{\mu\nu} -g_{Ij}a^{I} B^{j}{}_{\mu\nu} 
+\tfrac{2}{3} g_{Jj}h_{IK}{}^{j}a^{I}A^{JK}{}_{\mu\nu]}\right)\, ,
\end{array}
\end{equation}

\noindent
we find that the local supersymmetry algebra closes on the $B^{i}$s in the
expected form (to lowest order in fermions) 

\begin{equation}
\left[\delta_{\epsilon_{1}},\delta_{\epsilon_{2}}\right]B^{i}{}_{\mu\nu}
 =    
\pounds_{\xi}B^{i}{}_{\mu\nu}
+\delta_{\Lambda}B^{i}{}_{\mu\nu}\, ,
\end{equation}

\noindent
where $\delta_{\Lambda}B^{i}{}_{\mu\nu}$ is the gauge transformation given in
Eq.~(\ref{eq:deltaBi}) in which the 0-form gauge parameters $\Lambda^{I}$ are
as in Eqs.~(\ref{eq:LambdaI}), the 1-form gauge parameters
$\Lambda^{i}{}_{\mu}$ are given by 

\begin{equation}
\Lambda^{i}{}_{\mu}
= 
\lambda^{i}{}_{\mu}
+b^{i}{}_{\mu}-h_{IJ}{}^{i}a^{I}A^{J}{}_{\mu}\, ,  
\end{equation}

\noindent
where

\begin{equation}
\begin{array}{rcl}
\lambda^{1}{}_{\mu}
& \equiv & 
e^{\frac{1}{2\sqrt{7}}\varphi+\frac{1}{2}\phi} (\chi\sigma_{\mu}
-e^{-\phi}\rho_{\mu})\, ,\\
& & \\
\lambda^{2}{}_{\mu}
& \equiv & 
e^{\frac{1}{2\sqrt{7}}\varphi} \sigma_{\mu}
\, ,\\    
\end{array}
\end{equation}

\noindent
and the shift term is given by

\begin{eqnarray}
Z^{1}\left[\Lambda_{\mu\nu} 
-\left( c_{\mu\nu} -g_{Ij}a^{I} B^{j}{}_{\mu\nu} 
+\tfrac{2}{3} g_{Jj}h_{IK}{}^{j}a^{I}A^{JK}{}_{\mu\nu}\right)\right]  
& &  \nonumber \\
& & \nonumber \\
&  &
\hspace{-5cm} 
= e^{\frac{1}{2\sqrt{7}}\varphi+\frac{1}{2}\phi}
\left[
\left(
\tfrac{1}{2} \Im {\rm m} (g) 
-4 \Re{\rm e} (k)
+\tfrac{1}{2\sqrt{7}} \Im {\rm m} (\tilde{g}) 
\right)\chi
\right.
\nonumber \\
& & \nonumber \\
& &
\hspace{-5cm} 
\left.
-
\left(
\tfrac{1}{2} \Re {\rm e} (g) 
+4 \Im{\rm m} (k)
-\tfrac{1}{2\sqrt{7}} \Re {\rm e} (\tilde{g}) 
\right)e^{-\phi}
\right]\xi_{\mu\nu}\, ,
\label{eq:2formshift1}
\\
& & \nonumber \\
Z^{2}\left[\Lambda_{\mu\nu} 
-\left( c_{\mu\nu} -g_{Ij}a^{I} B^{j}{}_{\mu\nu} 
-\tfrac{2}{3} g_{Jj}h_{IK}{}^{j}a^{I}A^{JK}{}_{\mu\nu}\right)\right]  
& &  \nonumber \\
& & \nonumber \\
&  &
\hspace{-5cm} 
= e^{\frac{1}{2\sqrt{7}}\varphi+\frac{1}{2}\phi}
\left(
\tfrac{1}{2} \Im {\rm m} (g) 
-4 \Re{\rm e} (k)
+\tfrac{1}{2\sqrt{7}} \Im {\rm m} (\tilde{g}) 
\right)\xi_{\mu\nu}\, .
\label{eq:2formshift2}
\end{eqnarray}

Now, let us analyze the constraints that involve $h_{IJ}{}^{i}$. From those
that only involve the embedding tensor we find seven linear constraints that
imply those in Eqs.~(\ref{eq:linearconstraints1}) and that can be used to
eliminate seven components of the embedding tensor:

\begin{equation}
  \begin{array}{rclrclrcl}
\vartheta_{\mathbf{2}}{}^{1} & = & 0\, ,\,\,\,  &
\vartheta_{\mathbf{1}}{}^{2}& = & \tfrac{3}{4}\vartheta_{\mathbf{2}}{}^{5}\, ,\,\,\,  & 
\vartheta_{\mathbf{1}}{}^{3}& = & \tfrac{3}{4}\vartheta_{\mathbf{2}}{}^{5}\, ,\,\,\,  \\
 & & & & & & & & \\
 \vartheta_{\mathbf{1}}{}^{1}& = & \tfrac{3}{2}\vartheta_{\mathbf{1}}{}^{5}\, ,\,\,\, & 
\vartheta_{\mathbf{2}}{}^{2}& = & \tfrac{3}{4}\vartheta_{\mathbf{1}}{}^{5}\, ,\,\,\,\hspace{.5cm}  &
\vartheta_{\mathbf{2}}{}^{3}& = & -\tfrac{3}{4}\vartheta_{\mathbf{1}}{}^{5}\, ,\,\,\,  \\
 & & & & & & & & \\
\vartheta_{0}{}^{4}& = & -\tfrac{1}{6}\vartheta_{0}{}^{5}\, ,\hspace{.5cm}  & 
& & & & & \\
\end{array}
\end{equation}

\noindent
leaving the eight components (a triplet of $SL(2,\mathbb{R})$ in the upper
component, a singlet and two doublets of $SL(2,\mathbb{R})$ in the lower
components)

\begin{equation}
\vartheta_{0}{}^{m}\, ,\,\,\, m=1,2,3\, ,\,\,\,
\vartheta_{0}{}^{5}\, ,\,\,\,
\vartheta_{\mathbf{i}}{}^{4}\, ,\,\,\,
\vartheta_{\mathbf{i}}{}^{5}\, ,\,\,\, \mathbf{i}=\mathbf{1},\mathbf{2}\, ,
\end{equation}

\noindent
as the only independent ones.  These components correspond to the eight
deformation parameters of the primary deformations studied in
Ref.~\cite{Bergshoeff:2002nv}. More precisely, the relation between them are

\begin{equation}
  \begin{array}{rclrclrcl}
\vartheta_{0}{}^{m} & = &  m_{m}\, ,\,\,\, (m=1,2,3)\hspace{.5cm}& 
\vartheta_{\mathbf{1}}{}^{4} & = & -m_{11}\, ,\hspace{.5cm}& 
\vartheta_{\mathbf{1}}{}^{5} & = & \tilde{m}_{4}\, ,\,\,\,
\\
& & & & & & & & \\
\vartheta_{0}{}^{5} & = & -\tfrac{16}{3}m_{\rm IIB}\, ,\,\,\,&
\vartheta_{\mathbf{2}}{}^{4} & = & m_{\rm IIA}\, ,\,\,\,& 
\vartheta_{\mathbf{2}}{}^{5} & = & m_{4}\, . 
\end{array}
\end{equation}

From the constraints that relate $h_{IJ}{}^{i}$ to $Z^{I}{}_{i},Z^{i}$ and
$g_{Ii}$ we can determine all these tensors, up to a constant $\zeta$,  in
terms of the independent components of the embedding tensor:

\begin{equation}
\label{eq:valuesofsometensors}
  \begin{array}{rclrcl}
Z^{\mathbf{i}}{}_{j} 
& = & 
\vartheta_{0}{}^{m}(T_{m})_{\mathbf{j}}{}^{\mathbf{i}} 
-\tfrac{3}{4}\vartheta_{0}{}^{5}\delta_{j}{}^{1}\delta_{1}{}^{\mathbf{i}}\,
,\,\,\, &
Z^{0}{}_{i} & = & 3\vartheta_{\mathbf{i}}{}^{4}
+\tfrac{1}{2}\vartheta_{\mathbf{i}}{}^{5}\, ,\\ 
& & & & & \\
g_{0i} & = & 0\, ,\,\,\, &
g_{\mathbf{i}j} & = & \varepsilon_{\mathbf{i}j}\, .\,\,\, \\
\end{array}
\end{equation}
 
The constant $\zeta$ is the coefficient of a Chern-Simons term in the 4-form
field strength and, therefore, will be completely determined by supersymmetry.

Finally, using all these results in
Eqs.~(\ref{eq:1-formshifts1}-\ref{eq:1-formshifts2}) we find 

\begin{eqnarray}
k
& = 
& -\tfrac{9i}{14} e^{-\frac{3}{2\sqrt{7}}\varphi +\frac{1}{2}\phi}
(\vartheta_{\mathbf{1}}{}^{4}\tau +\vartheta_{\mathbf{2}}{}^{4})\, ,  
\\
& & \nonumber \\
\Im{\rm m} (f) 
& = & 
\tfrac{3}{28} \vartheta_{0}{}^{5}e^{\frac{2}{\sqrt{7}}\varphi}\, ,
\\
& & \nonumber \\
\Re{\rm e} (f) +\tfrac{3}{4\sqrt{7}}\Im {\rm m}(\tilde{h}) 
& = & 
\tfrac{1}{4}e^{\frac{2}{\sqrt{7}}\varphi +\phi} 
\left\{
\tfrac{1}{2}(\vartheta_{0}{}^{2}+\vartheta_{0}{}^{3})
+(\vartheta_{0}{}^{1}-\tfrac{3}{4}\vartheta_{0}{}^{5}) \chi
\right.
\nonumber \\
& & \nonumber \\
& & 
\left.
-\tfrac{1}{2}(\vartheta_{0}{}^{2}-\vartheta_{0}{}^{3})|\tau|^{2}
\right\}\, ,
\end{eqnarray}

\noindent
which determines almost completely all the fermion shifts. We find that, in
order to determine completely $\Re{\rm e} (f)$ and $\Im {\rm m}(\tilde{h})$,
separately, one must study the closure of the supersymmetry algebra on the
fermions of the theory or on the bosons at higher order in fermions. The
result is  

\begin{eqnarray}
\Re{\rm e} (f)
& = & 
\tfrac{1}{14}e^{\frac{2}{\sqrt{7}}\varphi}\vartheta_{0}{}^{m}\mathcal{P}_{m}\, ,
\\
& & \nonumber \\
\Im {\rm m}(\tilde{h})  
& = & 
\tfrac{4}{\sqrt{7}}e^{\frac{2}{\sqrt{7}}\varphi}\vartheta_{0}{}^{m}\mathcal{P}_{m}\, .
\end{eqnarray}

All these results are collected in Appendix~\ref{d9:sec-final}.


\subsection{The 3-form $C$}

In the next step we are going to consider the last of the fundamental,
electric $p$-forms of the theory, the 3-form $C$, whose gauge transformation
is given in Eq.~(\ref{eq:deltaC}).


\subsubsection{The 4-form field strength $G$}

The 4-form field strength $G$ is given by 

\begin{equation}
\label{eq:G}
G = \mathfrak{D}C -g_{Ii}\left(F^{I}-\tfrac{1}{2}Z^I{}_jB^j\right)\wedge B^{i} 
-\tfrac{1}{3} h_{IK}{}^{i}g_{Ji}A^{IJ}\wedge dA^{K}  +Z\tilde{C}\, ,
\end{equation}

\noindent
and it is covariant under general gauge transformations provided that the 4-form
$\tilde{C}$ transforms as

\begin{equation}
\label{eq:deltatildeC}
\begin{array}{rcl}
\delta_{\Lambda}\tilde{C}
& = &  
-\mathfrak{D}\tilde{\Lambda} 
-\tilde{g}_{I}\left[\Lambda^{I}G +C\wedge\delta_{\Lambda} A^{I} 
+F^{I}\wedge\Lambda +\tfrac{1}{12}g_{Ji}h_{KL}{}^{i}A^{IJK}\wedge
\delta_{\Lambda} A^{L}\right]
\\
& & \\
& & 
-\tilde{g}_{ij}[2H^{i}\wedge \Lambda^{j}-B^{i}\wedge\delta_{\Lambda} B^{j}
+2h_{IJ}{}^{i}B^{j}\wedge A^{I}\wedge\delta_{\Lambda} A^{J}]
\\
& & \\
& & 
-\tilde{g}_{IJK}\left[3\Lambda^{I} F^{JK} +2(F^{I}-Z^{I}{}_{i}B^{i})\wedge
  A^{J}\wedge\delta_{\Lambda} A^{K}
  -\tfrac{1}{4}X_{LM}{}^{J}A^{ILM}\wedge\delta_{\Lambda} A^{K}\right]
\\
& & \\
& & 
+Z^{i}\tilde{\Lambda}_{i}\, ,
\end{array}
\end{equation}

\noindent
where the new deformation tensors that we have introduced,
$\tilde{g}_{I},\tilde{g}_{ij}=-\tilde{g}_{ji}$ and
$\tilde{g}_{IJK}=\tilde{g}_{(IJK)}$, are subject to the constraints

\begin{eqnarray}
\label{eq:constraint11}  
g_{I[i} Z^{I}{}_{j]} +Z \tilde{g}_{ij} & = & 0\, ,\\
& & \nonumber \\
\label{eq:constraint12}  
X_{I}+g_{Ii}Z^{i}+Z\tilde{g}_{I} & = & 0\, ,\\
& & \nonumber \\
\label{eq:constraint13}  
h_{(IJ}{}^{i}g_{K)i}-Z \tilde{g}_{IJK} & = & 0\, ,
\end{eqnarray}

\noindent
plus the constraints that express the gauge invariance of the new deformation
parameters 

\begin{eqnarray}
\label{eq:constraint15}  
\tilde{X}_{I}\tilde{g}_{J} - X_{I\, J}{}^{K}\tilde{g}_{K}
& = & 0\, ,\\
& & \nonumber \\
\label{eq:constraint16}  
\tilde{X}_{I}\tilde{g}_{ij} -2X_{I\, [i|}{}^{k}\tilde{g}_{k|j]}
& = & 0\, ,\\
& & \nonumber \\
\label{eq:constraint17}  
\tilde{X}_{I}\tilde{g}_{JKL} -3X_{I\, (J}{}^{M}\tilde{g}_{KL)M}
& = & 0\, .  
\end{eqnarray}


\subsubsection{Closure of the supersymmetry algebra on the 3-form $C$}

Taking into account the form of $\delta_{\epsilon}C_{\mu\nu\rho}$ in the
undeformed case and the form of the field strength $G$, we arrive at the
following Ansatz for the supersymmetry transformation of the 3-form $C$:

\begin{equation}
\delta_{\epsilon}C_{\mu\nu\rho} 
= 
-\tfrac{3}{2} e^{-\frac{1}{\sqrt{7}}\varphi} \bar{\epsilon} \gamma_{[\mu\nu}
\left(\psi_{\rho]} +\tfrac{i}{6\sqrt{7}}\tilde{\lambda}^{*}
\right)   
+\mathrm{h.c.}
+ 3\delta_{\epsilon}A^{I}{}_{[\mu|}
\left(
g_{Ii}B^{i}{}_{|\nu\rho]}
+\tfrac{2}{3} h_{IJ}{}^{i} g_{Ki} A^{JK}{}_{|\nu\rho]}
\right)\, .
\end{equation}

\noindent
The last two terms are written in terms of the tensors $g_{Ii}$ and
$h_{IJ}{}^{i}$. In the undeformed theory these tensors have values which are
determined by supersymmetry (at orders in fermions higher than we are
considering here) and that cannot be changed in the deformed theory, as we
already discussed when we considered the 2-forms for $h_{IJ}{}^{i}$. Thus,
$h_{IJ}{}^{i}$ is given by Eq.~(\ref{eq:valueofhIJi}) and $g_{Ii}$ is given by
Eqs.~(\ref{eq:valuesofsometensors}) with $\zeta=+1$

Using the identity

\begin{equation}
\begin{array}{rcl}
\xi^{\sigma}G_{\sigma\mu\nu\rho} 
+3\pounds_{\xi}A^{I}{}_{[\mu|}
\left[g_{Ii} B^{i}{}_{|\nu\rho]}
+\tfrac{2}{3} h_{IJ}{}^{i}g_{Ki} A^{JK}{}_{|\nu\rho]}
\right] 
& = & 
\\
& & \\
& & \hspace{-7cm}
=\pounds_{\xi}C_{\mu\nu\rho}   
-3 \mathfrak{D}_{[\mu|}
\left[
\left( c_{|\nu\rho]} -g_{Ij}a^{I} B^{j}{}_{|\nu\rho]} 
+\tfrac{2}{3} g_{Jj}h_{IK}{}^{j}a^{I}A^{JK}{}_{|\nu\rho]}\right)\right]  
\\
& & \\
& & \hspace{-7cm}
+g_{Ii}
\left[
-a^{I} H^{i}{}_{\mu\nu\rho} 
-3F^{I}{}_{[\mu\nu|}(b^{i}{}_{|\rho]} -h_{JK}{}^{i}a^{J}A^{K}{}_{|\rho]}) 
\right]
\\
& & \\
& & \hspace{-7cm}
+Z
\left\{
\tilde{c}_{\mu\nu\rho} -\tilde{g}_{I}a^{I} C_{\mu\nu\rho} 
+3\tilde{g}_{ij}B^{i}{}_{[\mu\nu|}(b^{j}{}_{|\rho]}
-h_{JK}{}^{j} a^{J}A^{K}{}_{\rho})
-12\tilde{g}_{IJK}a^{I}A^{J}{}_{[\mu}\partial_{\nu}A^{K}{}_{\rho]} 
\right.
\\
& & \\
& & \hspace{-7cm}
\left.
+3h_{IJ}{}^{i}\tilde{g}_{ij}a^{I}A^{J}{}_{[\mu}B^{j}{}_{\nu\rho]}
-\tfrac{1}{4}
\left(
h_{IJ}{}^{i}g_{Ki}\tilde{g}_{L} +3 X_{JK}{}^{M}\tilde{g}_{ILM}
\right)a^{I}A^{JKL}{}_{\mu\nu\rho}
\right\}\, ,
\end{array}
\end{equation}

\noindent
one can see that the local supersymmetry algebra  closes into a general
coordinate transformation plus a gauge transformation of $C$ of the form  
Eq.~(\ref{eq:deltaC}) with

\begin{equation}
\label{eq:Lambda}
\Lambda_{\mu\nu} 
= e^{\frac{1}{\sqrt{7}}\varphi} \xi_{\mu\nu}
+\left( c_{\mu\nu} -g_{Ij}a^{I} B^{j}{}_{\mu\nu} 
-\tfrac{2}{3} g_{Jj}h_{IK}{}^{j}a^{I}A^{JK}{}_{\mu\nu}\right)\, ,
\end{equation}

\noindent
and with the identification

\begin{equation}
\begin{array}{rcl}
Z \left\{
\tilde{\Lambda}_{\mu\nu\rho} -\tilde{c}_{\mu\nu\rho} 
+\tilde{g}_{I}a^{I}C_{\mu\nu\rho} +3\tilde{g}_{ij} B^{i}{}_{[\mu\nu|}
\left(b^{j}{}_{|\rho]} -h_{JK}{}^{j}a^{J}A^{K}{}_{|\rho]} \right)
-12\tilde{g}_{IJK}a^{I}A^{J}{}_{[\mu}\partial_{\nu}A^{K}{}_{\rho]}
\right.
& & \\
& & \\
\left.
-3\tilde{g}_{ij} h_{IJ}{}^{i}a^{I}A^{J}{}_{[\mu}B^{j}{}_{\nu\rho]}
+\tfrac{1}{4} \left(\tilde{g}_{L}g_{Ki}h_{IJ}{}^{i}
+3\tilde{g}_{ILN}X_{JK}{}^{N} \right) a^{I}A^{JKL}{}_{\mu\nu\rho}
\right\}
& & \\
& & \\
= 
6 e^{-\frac{1}{\sqrt{7}}\varphi} 
\left[ 
\Im \mathrm{m} (f) +\tfrac{1}{6\sqrt{7}} \Re\mathrm{e}(\tilde{h})
\right] \zeta_{\mu\nu\rho}\, .   \\
\end{array}
\end{equation}

\noindent
Comparing Eq.~(\ref{eq:Lambda}) with Eqs.~(\ref{eq:2formshift1}) and
(\ref{eq:2formshift2}) we find that 

\begin{equation}
Z^{1}   =  X_{\mathbf{2}} =
3\vartheta_{\mathbf{2}}{}^{4} -\tfrac{1}{4}\vartheta_{\mathbf{2}}{}^{5}\, ,
\hspace{1cm}
Z^{2} = -X_{\mathbf{1}} =  
-3\vartheta_{\mathbf{1}}{}^{4}+\tfrac{1}{4}\vartheta_{\mathbf{1}}{}^{5} \, .
\end{equation}

To make further progress it is convenient to compute the 5-form $\tilde{G}$
since it will contain the tensors
$\tilde{g}_{I},\tilde{g}_{ij},\tilde{g}_{IJK}$ that appear in the above
expression. These tensors cannot be deformed (just as it happens with
$h_{IJ}{}^{i}$) and their values can be found by comparing the general form of
$\tilde{G}$ with the value found by duality, Eq.~(\ref{eq:tildeGundeformed}).

The generic form of the magnetic 5-form field strength $\tilde{G}$ is

\begin{equation}
\label{eq:tildeG}
\begin{array}{rcl}
\tilde{G} 
& = & 
\mathfrak{D}\tilde{C}
-\tilde{g}_{J}\left[
(F^{J}-Z^{J}{}_{j}B^{j})\wedge C
+\tfrac{1}{12}g_{Kj}h_{MN}{}^{j}A^{JKM}\wedge dA^{N}\right]
\\
& & \\
& & 
+2\tilde{g}_{ij}\left(H^{i}-\tfrac{1}{2}\mathfrak{D}B^{i}\right)\wedge B^{j}
-\tilde{g}_{JKL}\left(A^{J}\wedge dA^{KL}
+\tfrac{3}{4}X_{MN}{}^{L}A^{JMN}\wedge dA^{K}\right)
\\
& & \\
& & 
+Z^{i}\tilde{B}_{i}\, ,\\
\end{array}
\end{equation}

\noindent
and comparing this generic expression with Eq.~(\ref{eq:tildeGundeformed}) we
find that 

\begin{equation}
\tilde{g}_{I} = -\delta_{I}{}^{0}\, ,
\hspace{1cm}
\tilde{g}_{ij} = -\tfrac{1}{2}\varepsilon_{ij}\, ,
\hspace{1cm}
\tilde{g}_{IJK} = 0\, .  
\end{equation}

\noindent
Plugging these values into the constraints that involve $Z$
Eqs.~(\ref{eq:constraint10}),(\ref{eq:constraint18}), and
(\ref{eq:constraint11}-\ref{eq:constraint13}) we find that it must be related
to $\vartheta_{0}{}^{5}$ by 

\begin{equation}
Z = -\tfrac{3}{4} \vartheta_{0}{}^{5}\, ,  
\end{equation}

\noindent
and that $\vartheta_{0}{}^{5}$ must satisfy the two doublets of quadratic constraints

\begin{align}
\vartheta_{\mathbf{i}}{}^{4}\vartheta_{0}{}^{5}  
& = 
0\, ,
\\
& \nonumber \\
\vartheta_{\mathbf{i}}{}^{5}\vartheta_{0}{}^{5}  
& = 
0\, .
\end{align}

\noindent
Plugging our results into all the other constraints between deformation
tensors, we find that all of them are satisfied provided that the quadratic
constraints

\begin{align}
\varepsilon^{\mathbf{ij}}\vartheta_{\mathbf{i}}{}^{4}\vartheta_{\mathbf{j}}{}^{5}
& = 0\, ,
\\
& \nonumber \\
\vartheta_{0}{}^{m}
\left(12\vartheta_{\mathbf{i}}{}^{4}+5\vartheta_{\mathbf{i}}{}^{5} \right)
& = 0\, ,
\\
& \nonumber \\
\vartheta_{\mathbf{j}}{}^{4} \left(\vartheta_{0}^{m} T_{m}
\right)_{\mathbf{i}}{}^{\mathbf{j}}
& = 0\, , 
\end{align}

\noindent
are also satisfied. This set of irreducible quadratic constraints that cannot
be used to solve some deformation parameters in terms of the rest in an
analytic form, and to which the 9-form potentials of the theory may be
associated as explained in Section~\ref{sec-magnetic} is one of our main
results.


\section{Summary of results and discussion}
\label{sec-summary}

In the previous section we have constructed order by order in the rank of the
$p$-forms the supersymmetric tensor hierarchy of maximal 9-dimensional
supergravity, up to $p=3$, which covers all the fundamental fields of the
theory.
 
As it usually happens in all maximal supergravity theories, all the
deformation parameters can be expressed in terms of components of the
embedding tensor. Furthermore, we have shown that gauge invariance and local
supersymmetry allow for one triplet, two doublets and one singlet of
independent components of the embedding tensor

\begin{equation}
\label{eq:independent}
\vartheta_{0}{}^{m}\, ,\,\,\, m=1,2,3\, ,\,\,\,
\vartheta_{0}{}^{5}\, ,\,\,\,
\vartheta_{\mathbf{i}}{}^{4}\, ,\,\,\,
\vartheta_{\mathbf{i}}{}^{5}\, ,\,\,\, \mathbf{i}=\mathbf{1},\mathbf{2}\, .
\end{equation}

\noindent 
They can be identified with the deformation parameters studied in
Ref.~\cite{Bergshoeff:2002nv}:

\begin{equation}
  \begin{array}{rclrclrcl}
\vartheta_{0}{}^{m} & = &  m_{m}\, ,\,\,\, (m=1,2,3)\hspace{.5cm}& 
\vartheta_{\mathbf{1}}{}^{4} & = & -m_{11}\, ,\hspace{.5cm}& 
\vartheta_{\mathbf{1}}{}^{5} & = & \tilde{m}_{4}\, ,\,\,\,
\\
& & & & & & & & \\
\vartheta_{0}{}^{5} & = & -\tfrac{16}{3}m_{\rm IIB}\, ,\,\,\,&
\vartheta_{\mathbf{2}}{}^{4} & = & m_{\rm IIA}\, ,\,\,\,& 
\vartheta_{\mathbf{2}}{}^{5} & = & m_{4}\, . 
\end{array}
\end{equation}

\noindent
This proves, on the one hand, that no more deformations are possible and, on
the other hand, that all the deformations of maximal 9-dimensional
supergravity have a higher-dimensional origin, as shown in
Ref.~\cite{Bergshoeff:2002nv}.

Furthermore, we have also shown that it is not possible to give non-zero
values to all the deformation parameters at the same time, since they
must satisfy the quadratic constraints

\begin{align}
\label{eq:irreduciblequadraticconstraints}
\vartheta_{0}{}^{m}
\left(12\vartheta_{\mathbf{i}}{}^{4}+5\vartheta_{\mathbf{i}}{}^{5} \right)
& \equiv \mathcal{Q}^{m}{}_{\mathbf{i}}=0\, ,
\\
& \nonumber \\
\vartheta_{\mathbf{i}}{}^{4}\vartheta_{0}{}^{5}  
& \equiv \mathcal{Q}^{4}{}_{\mathbf{i}}= 0\, ,
\\
& \nonumber \\
\vartheta_{\mathbf{i}}{}^{5}\vartheta_{0}{}^{5}  
& \equiv \mathcal{Q}^{5}{}_{\mathbf{i}} = 0\, ,
\\
& \nonumber \\
\vartheta_{\mathbf{j}}{}^{4} \left(\vartheta_{0}^{m} T_{m}
\right)_{\mathbf{i}}{}^{\mathbf{j}}
& \equiv \mathcal{Q}_{\mathbf{i}} = 0\, , 
\\
& \nonumber \\
\varepsilon^{\mathbf{ij}}\vartheta_{\mathbf{i}}{}^{4}\vartheta_{\mathbf{j}}{}^{5}
& \equiv \mathcal{Q} = 0\, ,
\end{align}

\noindent
all of which are related to gauge invariance. 

Using these results, we can now apply the arguments developed in
Section~\ref{sec-magnetic} to relate the number of symmetries (Noether
currents), deformation parameters, and quadratic constraints to the numbers
(and symmetry properties) of 7-, 8- and 9-forms of the theory. Our results can
be compared with those presented in Ref.~\cite{Bergshoeff:2010xc} (Table~6)
and Ref.~\cite{Bergshoeff:2011zk} (Table~3) and found from $E_{11}$ level
decomposition.

\begin{table}[t!]
    \centering
\begin{tabular}{cccccc}
\hline 
$\mathbb{R}^{+}$ & 
$j_{1}$ & 
$j_{2}-j_{3}$ & 
$j_{2}+j_{3}$ & 
$j_{4}$ &  $j_{5}$  
\\
\hline 
\hline 
$\alpha$ & 
$0$ & $0$ & $0$ &
$0$ & $0$  
\\
$\beta$ &
$0$ & $+3/4$ &  $-3/4$  & 
$0$ & $0$
\\
$\gamma$ &
$0$ & $-2$ & $+2$ &
$0$ & $0$ 
\\
$\delta$ & 
$0$ & $0$ & $0$  &
$0$ & $0$
\\
\hline 
\end{tabular}
{\it   \caption{Weights of the Noether currents}
    \label{tab:noetherweights}
    }
\end{table}

Associated to the symmetry group of the equations of motion of the theory,
$SL(2,\mathbb{R})\times \mathbb{R}^{2}$ there are 5 Noether currents $j_{A}$
that fit into one triplet and two singlets of $SL(2,\mathbb{R})$ and are
explicitly given in Appendix~\ref{d9:sec-noether}. Their weights are given in
Table~\ref{tab:noetherweights}. They can be dualized as explained in
Section~\ref{sec-magnetic} into a triplet and two singlets of 7-forms
$\tilde{A}_{(7)}$ whose weights are given in
Table~\ref{tab:789formweights}. In
Refs.~\cite{Bergshoeff:2010xc,Bergshoeff:2011zk} the $\beta$ rescaling has not
been considered. As mentioned before, it corresponds to the so-called trombone
symmetry which may not survive to higher-derivative string corrections. The
associated 7-form singlet $\tilde{A}^{5}_{(7)}$ does not appear in their
analysis. The weights assigned in those references to the fields correspond to
one third of the weight of the $\alpha$ rescaling in our conventions.

\begin{table}[t!]
\centering
\hspace{-1cm}
\begin{tabular}{ccccccc}
\hline 
$\mathbb{R}^{+}$ & 
$\vartheta_{0}{}^{1}$ & 
$\vartheta_{0}{}^{2}-\vartheta_{0}{}^{3}$ & 
$\vartheta_{0}{}^{2}+\vartheta_{0}{}^{3}$ & 
$\vartheta_{\mathbf{1}}{}^{4},\vartheta_{\mathbf{1}}{}^{5}$ &  
$\vartheta_{\mathbf{1}}{}^{4},\vartheta_{\mathbf{2}}{}^{5}$ &
$\vartheta_{0}{}^{5}$ 
\\
\hline\hline 
$\alpha$ & 
$-3$ & $-3$ & $-3$ &
$0$ & $0$  & 
$-3$
\\
$\beta$ &
$-1/2$ & $-5/4$ &  $1/4$  & 
$3/4$ & $0$ & 
$-1/2$
\\
$\gamma$ &
$0$ & $2$ & $-2$ &
$-1$ & $1$ &
$0$ 
\\
$\delta$ & 
$0$ & $0$ & $0$  &
$-2$ & $-2$ &
$0$
\\
\hline 
\end{tabular}
{\it   \caption{Weights of the embedding tensor components}
    \label{tab:embeddingtensorweights}
    }
\end{table}

Associated to each of the $SL(2,\mathbb{R})$ multiplets of independent
embedding-tensor components there is a dual multiplet of 8-forms
$\tilde{A}_{(8)}$ (i.e.~one triplet, two doublets and one singlet) whose
weights are given in Table~\ref{tab:789formweights}. The doublet and singlet
associated to the gauging of the trombone symmetry using the doublet and
singlet of 1-forms are missing in
Refs.~\cite{Bergshoeff:2010xc,Bergshoeff:2011zk}, but the rest of the 8-forms
and their weights are in perfect agreement with those obtained from $E_{11}$.
Given the amount of work that it takes to determine which are the independent
components of the embedding tensor allowed by supersymmetry, this is a quite
non-trivial test of the consistency of the $E_{11}$ and the embedding-tensor
approaches.

\begin{table}[t!]
\centering
{\footnotesize
\begin{tabular}{cccccccccccc}
\hline
$\mathbb{R}^{+}$ & 
$\mathcal{Q}_{\mathbf{1}}{}^{1}$ & 
$\mathcal{Q}_{\mathbf{2}}{}^{1}$ & 
$\mathcal{Q}_{\mathbf{1}}{}^{2-3}$ & 
$\mathcal{Q}_{\mathbf{2}}{}^{2-3}$ & 
$\mathcal{Q}_{\mathbf{1}}{}^{2+3}$ & 
$\mathcal{Q}_{\mathbf{2}}{}^{2+3}$ & 
$\mathcal{Q}_{\mathbf{1}}{}^{4},\mathcal{Q}_{\mathbf{1}}{}^{5}$ &  
$\mathcal{Q}_{\mathbf{2}}{}^{4},\mathcal{Q}_{\mathbf{2}}{}^{5}$ &  
$\mathcal{Q}_{\mathbf{1}}$ &
$\mathcal{Q}_{\mathbf{2}}$ &
$\mathcal{Q}$
\\
\hline 
\hline 
$\alpha$ & 
$-3$ & $-3$ & $-3$ &
$-3$ & $-3$ & $-3$ &
$-3$ & $-3$  & 
$-3$ & $-3$ &
0
\\
$\beta$ &
$1/4$ & $-1/2$ & $-1/2$ & $-5/4$ & $1$ & $1/4$ & 
$1/4$ & $-1/2$ & 
$1/4$ & $-1/2$ & 
$3/4$
\\
$\gamma$ &
$-1$ & $1$ & $1$ & $3$ & $-3$ & $-1$ &
$-1$ & $1$ &
$-1$ & $1$&
$0$
\\
$\delta$ & 
$-2$ & $-2$ & $-2$  &
$-2$ & $-2$ & $-2$  &
$-2$ & $-2$ &
$-2$ & 
$-2$ & 
$-4$
\\
\hline 
\end{tabular}}
{\it   \caption{Weights of quadratic constraints components.}
    \label{tab:qqweights}
    }
\end{table}

Finally, associated to each of the quadratic constraints that the components
of the embedding tensor must satisfy
$\mathcal{Q}_{\mathbf{i}}{}^{m},\mathcal{Q}_{\mathbf{i}}{}^{4},
\mathcal{Q}_{\mathbf{i}}{}^{5},\mathcal{Q}_{\mathbf{i}},\mathcal{Q}$ there is
a 9-form potential $\tilde{A}_{(9)}$. The weights of these potentials are
given in Table~\ref{tab:789formweights}. If we set to zero the
embedding-tensor components associated to the trombone symmetry
$\vartheta_{A}{}^{5}$, the only constraints which are not automatically solved
are

\begin{equation}
\mathcal{Q}_{\mathbf{i}}{}^{m} = 12
\vartheta_{0}{}^{m}\vartheta_{\mathbf{i}}{}^{4}=0\, ,
\hspace{1cm}
\mathcal{Q}_{\mathbf{i}}=
\vartheta_{\mathbf{j}}{}^{4} \left(\vartheta_{0}^{m} T_{m}
\right)_{\mathbf{i}}{}^{\mathbf{j}}
=0\, .   
\end{equation}

The first of these constraints can be decomposed into a quadruplet and a
doublet: rewriting $\mathcal{Q}_{\mathbf{i}}{}^{m}$ in the equivalent form

\begin{equation}
\mathcal{Q}_{\mathbf{i}(\mathbf{j}\mathbf{k})}= 
\vartheta_{\mathbf{i}}{}^{4} 
\left(\vartheta_{0}^{m} T_{m}
\right)_{\mathbf{j}}{}^{\mathbf{l}}\varepsilon_{\mathbf{kl}}\, ,  
\end{equation}

\noindent
the quadruplet corresponds to the completely symmetric part
$\mathcal{Q}_{(\mathbf{i}\mathbf{j}\mathbf{k})}$ and the doublet to

\begin{equation}
\varepsilon^{\mathbf{jk}}\mathcal{Q}_{\mathbf{j}(\mathbf{k}\mathbf{i})}= -
\mathcal{Q}_{\mathbf{i}}\, ,
\end{equation}

\noindent
which is precisely the other doublet. Therefore, we get the quadruplet and one
doublet of 9-forms with weight $4$ under $\alpha/3$, while one more doublet is
found in Refs.~\cite{Bergshoeff:2010xc,Bergshoeff:2011zk} .


\begin{table}[t!]
    \centering
\begin{tabular}{ccccccccccccc}
\hline 
$\mathbb{R}^{+}$ & 
$\tilde{A}_{(7)}^{m}$ & 
$\tilde{A}_{(7)}^{4}$ &  
$\tilde{A}_{(7)}^{5}$ &
$\tilde{A}_{(8)}^{m}$ & 
$\tilde{A}_{(8)}^{4\, \mathbf{i}}$ &  
$\tilde{A}_{(8)}^{5\, \mathbf{i}}$ &
$\tilde{A}_{(8)}^{4}$ &
$\tilde{A}_{(9)\, m}^{\mathbf{i}}$ &
$\tilde{A}_{(9)\, 4}^{\mathbf{i}}$ &
$\tilde{A}_{(9)\, 5}^{\mathbf{i}}$ &
$\tilde{A}_{(9)}^{\mathbf{i}}$ &
$\tilde{A}_{(9)}$ 
\\
\hline 
\hline 
$\alpha$ & 
$9$ & $9$ & $9$  &
$12$ & $9$ & $9$ & $12$ 
& $12$ & $12$ & $12$  & $12$ &  $9$
\\
$\delta$ & 
$8$ & $8$ & $8$ & $8$ & $2$ & $2$ & $8$ & 
$10$ & $10$ & $10$ & $10$ & $12$  
\\
\hline
\end{tabular}
{\it   \caption{Weights of the 7-, 8- and 9-form fields.}
    \label{tab:789formweights}
    }
\end{table}

This situation is similar to the one encountered in the $N=2$ theories in
$d=4,5,6$ dimensions \cite{Huebscher:2010ib}. In those cases, the
Ka\v{c}-Moody (here $E_{11}$) approach predicts one doublet of $d$-form
potentials more than the embedding-tensor formalism
\cite{Kleinschmidt:2008jj}. However, it can be seen that taking the undeformed
limit of the results obtained in the embedding-tensor formalism, one
additional doublet of $d$-forms arises because some St\"uckelberg shifts
proportional to deformation tensors that could be used to eliminate them, now
vanish. Furthermore, the local supersymmetry algebra closes on them as
independent fields.

By analogy with what happens in the $N=2$ theories in $d=4,5,6$ dimensions,
the same mechanism can make our results compatible with those of the $E_{11}$
approach (up to the trombone symmetry): we expect the existence of two
independent doublets of 9-forms in the undeformed theory but we also expect
new St\"uckelberg transformations in the deformed theory such that one a
combination of them is independent and the supersymmetry algebra closes.

This possibility (and the exclusion of any further 9-forms) can only be proven
by the direct exploration of all the possible candidates to 9-form
supersymmetry transformation rules, to all orders in fermions, something that
lies outside the boundaries of this work.


\section{Concluding remarks}
\label{sec-conclusions}

In this chapter we have applied the embedding-tensor formalism to the study of
the most general deformations (\textit{i.e.}~gaugings and massive
deformations) of maximal 9-dimensional supergravity. We have used the complete
global $SL(2,\mathbb{R})\times \mathbb{R}^{2}$ symmetry of its equations of
motion, which includes the so-called \textit{trombone symmetry}. We have found
the constraints that the deformation parameters must satisfy in order to
preserve both gauge and supersymmetry invariance (the latter imposed through
the closure of the local supersymmetry algebra to lowest order in
fermions). We have used most of the constraints to express some components of
the deformation tensors in terms of a few components of the embedding tensor
which we take to be independent and which are given in
Eq.~(\ref{eq:independent}). At that point we have started making contact with
the results of Ref.~\cite{Bergshoeff:2002nv}, since those independent
components are precisely the 8 possible deformations identified there. All of
them have a higher-dimensional origin discussed in detail in
Ref.~\cite{Bergshoeff:2002nv}. The field strengths, gauge transformations and
supersymmetry transformations of the deformed theory, written in terms of the
independent deformation tensors, are collected in Appendix~\ref{d9:sec-final}.

The 8 independent deformation tensors are still subject to quadratic
constraints, given in Eq.~(\ref{eq:irreduciblequadraticconstraints}), but
those constraints cannot be used to express analytically some of them in terms
of the rest, and, therefore, we must keep the 8 deformation parameters and
we must enforce these irreducible quadratic constraints. 

In Section~\ref{sec-summary} we have used our knowledge of the global
symmetries (and corresponding Noether 1-forms), the independent deformation
tensors and the irreducible quadratic constraints of the theory, together with
the general arguments of Section~\ref{sec-magnetic} to determine the possible
7-, 8- and 9-forms of the theory (Table~\ref{tab:789formweights}), which are
dual to the Noether currents, independent deformation tensors and irreducible
quadratic constraints. We have compared this spectrum of higher-rank forms
with the results of Refs.~\cite{Bergshoeff:2010xc,Bergshoeff:2011zk}, based on
$E_{11}$ level decomposition. We have found that, in the sector unrelated to
the trombone symmetry, which was excluded from that analysis, the
embedding-tensor formalism predicts one doublet of 9-forms less than the
$E_{11}$ approach. However, both predictions are not contradictory: the extra
doublet of 9-forms may not survive the deformations on which the
embedding-tensor formalism is built: new 9-form St\"uckelberg shifts
proportional to the deformation parameters may occur that can be used to
eliminate it so only one combination of the two 9-form doublets survives. This
mechanism is present in the $N=2$ $d=4,5,6$ theories \cite{Huebscher:2010ib},
although the physics behind it is a bit mysterious.

We can conclude that we have satisfactorily identified the extended field
content (the tensor hierarchy) of maximal 9-dimensional supergravity and,
furthermore, that all the higher-rank fields have an interpretation in terms
of symmetries and gaugings. This situation is in contrast with our
understanding of the extended field content of the maximal 10-dimensional
supergravities ($N=2A,B$) for which the $E_{11}$ approach can be used to get a
prediction of the higher-rank forms (which turns out to be correct
\cite{Bergshoeff:2005ac,Bergshoeff:2006qw,Bergshoeff:2010mv}) but the
embedding-tensor approach apparently cannot be used\footnote{In the $N=2B$
  case there are no 1-forms to be used as gauge fields and in the $N=2A$ case
  the only 1-form available is not invariant under the only rescaling symmetry
  available.} for this end. This seems to preclude an interpretation for the
9- and 10-form fields in terms of symmetries and gaugings\footnote{The 8-form
  fields are dual to the Noether currents of the global symmetries.}, at least
if we insist in the standard construction of the tensor hierarchy that starts
with the gauging of global symmetries. Perhaps a more general point of view is
necessary.




\def\beq{\begin{equation}}
\def\eeq{\end{equation}}
\def\beqa{\begin{eqnarray}}
\def\eeqa{\end{eqnarray}}

\def\a{{\alpha}}
\def\b{{\beta}}
\def\g{{\gamma}}
\def\d{{\delta}}
\def\var{{\varepsilon}}

\def\bfone{\relax{\rm 1\kern-.35em 1}}
\def\dop{{\rm d}\hskip -1pt}


\newcommand{\cM}{{\cal M}}
\newcommand{\cN}{{\cal N}}
\newcommand{\cG}{{\cal G}}
\newcommand{\cL}{{\cal L}}
\newcommand{\cV}{{\cal V}}
\newcommand{\cK}{{\cal K}}
\newcommand{\cW}{{\cal W}}
\newcommand{\mZ}{\mathbb{Z}}
\newcommand{\mI}{\mathbb{I}}
\newcommand{\mR}{\mathbb{R}}
\newcommand{\nH}{n_{\rm H}}
\newcommand{\nV}{n_{\rm V}}
\newcommand{\bP}{\mathbb P}
\newcommand{\bQ}{\mathbb Q}
\newcommand{\bZ}{\mathbb Z}
\newcommand{\fg}{{\mathfrak g}}

\newcommand{\eins}{\mbox{$1 \hspace{-1.0mm} \text{l}$}}
\newcommand{\ft}[2]{{\textstyle\frac{#1}{#2}}}

\newcommand {\Rbar} {{\mbox{\rm $\mbox{I}\!\mbox{R}$}}}
\newcommand {\Hbar} {{\mbox{\rm $\mbox{I}\!\mbox{H}$}}}
\newcommand {\Cbar}
            {\mathord{\setlength{\unitlength}{1em}
             \begin{picture}(0.6,0.7)(-0.1,0)
                \put(-0.1,0){\rm C}
                \thicklines
                \put(0.2,0.05){\line(0,1){0.55}}
             \end {picture}}}

\newcommand{\eqn}[1]{(\ref{#1})}

\newenvironment{matr}[1]{\left[ \begin{array}{{#1}}}{\end{array} \right]}

\newcommand{\llceil}{{|\!\!|\!\!\lceil}}
\newcommand{\rrfloor}{{\rfloor\!\!|\!\!|}}

\newcommand{\be}{\begin{equation}}
\newcommand{\ee}{\end{equation}}
\newcommand{\ben}{\begin{displaymath}}
\newcommand{\een}{\end{displaymath}}
\newcommand{\bea}{\begin{eqnarray}}
\newcommand{\eea}{\end{eqnarray}}
\newcommand{\nn}{\nonumber}
\newcommand{\non}{\nonumber\\}
\newcommand{\bean}{\begin{eqnarray*}}
\newcommand{\eean}{\end{eqnarray*}}

\newcommand{\mathon}{\mathversion{bold}}
\newcommand{\mathoff}{\mathversion{normal}}

\newcommand{\fp}{f^{\text{\tiny{(\!+\!)}}}}
\newcommand{\fm}{f^{\text{\tiny{(--)}}}}
\newcommand{\fo}{f^{\text{\tiny{(1)}}}}
\newcommand{\ftw}{f^{\text{\tiny{(2)}}}}
\newcommand{\fth}{f^{\text{\tiny{(3)}}}}

\newcommand{\Fp}{F^{\text{\tiny{(\!+\!)}}}}
\newcommand{\Fm}{F^{\text{\tiny{(--)}}}}
\newcommand{\Fo}{F^{\text{\tiny{(1)}}}}
\newcommand{\Ftw}{F^{\text{\tiny{(2)}}}}
\newcommand{\Fth}{F^{\text{\tiny{(3)}}}}

\newcommand{\remark}[1]{{\bf [#1]}\marginpar[\hfill ${\bf \Longrightarrow}$]{${\bf \Longleftarrow}$} }


\chapter{DFT and Duality orbits of non-geometric fluxes}
\label{ch:duality_orbits}



After having studied gauged supergravities as deformations of the ungauged theories, we will study the gaugings that arise from dimensional reductions of higher-dimensional supergravities. The existence of a mismatch between the catalog of gaugings (probably obtained with the help of the embedding tensor formalism) and the ones that arise from compactification has motivated the formulation of theories that include T-duality as a true symmetry. We will study of double field theory (DFT), one of these T-duality proposals, is able to reproduce the whole set of gaugings that the embedding tensor formalism supplies.

\section{Introduction}
\label{sec:introduction}

In the context of half-maximal  \cite{Schon:2006kz}
and maximal \cite{ deWit:2007mt}  supergravities, not only
does supersymmetry tightly organize the ungauged theory, but also it
strictly determines the set of possible deformations (\emph{i.e.}
gaugings). 

When compactifying heterotic, type II or eleven-dimensional supergravity 
on a given background, one obtains lower-dimensional effective theories
whose features depend on the fluxes included in the compactification
procedure and, in particular, on the amount of supersymmetry
preserved by the chosen background. When some supersymmetry is
preserved during the compactification, the effective theories under 
consideration are then gauged supergravities. Compactification can be
considered then a way of ``deforming'' supergravities.

As we have seen in Chapter \ref{ch:gauged_sugra}, the  embedding tensor
formalism enable us to formally describe all the possible
deformations in a single universal formulation, which therefore
completely restores duality covariance. Not all the
deformations obtained in this way 
have a clear higher-dimensional origin, in the sense
that they can be obtained by means of a certain compactification of
ten or eleven dimensional supergravity.

One of the most interesting open problems concerning flux
compactifications is to reproduce, by means of a suitable flux
configuration, a given lower-dimensional gauged supergravity theory.
Although this was done in particular cases (see for example
\cite{Roest:2009dq, Dall'Agata:2009gv}), an exhaustive analysis
remains to be done. This is due to fact that, on the one hand we
lack a classification of the possible gauging configurations allowed
in gauged supergravities and, on the other hand, only a limited set
of compactification scenarios are known. Typically, to go beyond the
simplest setups one appeals to dualities. The paradigmatic example
\cite{Shelton:2005cf} starts by applying T-dualities to a simple
toroidal background with a non-trivial two-form generating a single
$H_{abc}$ flux. By T-dualizing this setup, one can construct a chain
of T-dualities leading to new backgrounds (like twisted-tori or
T-folds) and generating new (dual) fluxes, like the so-called
${Q_a}^{bc}$ and $R^{abc}$. It is precisely by following duality
covariance arguments in the lower-dimensional effective description
that non-geometric fluxes \cite{Shelton:2005cf} were first
introduced in order to explain the mismatch between particular flux
compactifications and generic gauged supergravities.

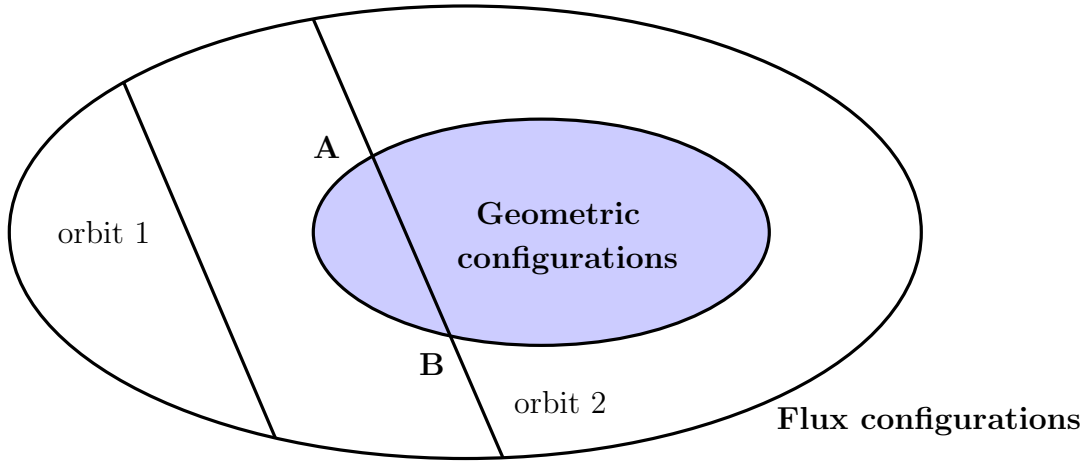
\begin{figure}[ph!]
\begin{center}
\begin{tikzpicture}[scale=0.5,>=latex']
        \path[P_3] \ellip;
        \path[P_4] \ellipp;
        \draw[very thick] (-9,4) -- (-5,-5.45);
        \draw[very thick] (-4,5.65) -- (1,-6);
        \path (-3,2.25) node[left] {\textbf{A}}
            (-1.5,-3.5) node [right] {\textbf{B}};
        \path (-8,0) node[left] {orbit 1}
            (1,-4.5) node [right] {orbit 2};
        \path (16.5,-5) node[left] {\textbf{Flux configurations}}
            (0,.5) node [right] {\textbf{Geometric}};
        \path (-.5,-.75) node[right] {\textbf{configurations}};
\end{tikzpicture}
{\it \caption{ The space of flux configurations sliced into
duality orbits (vertical lines). Moving along a given orbit
corresponds to applying dualities to a certain flux configuration
and hence it does not imply any physical changes in the
lower-dimensional effective description. Geometric fluxes only
constitute a subset of the full configuration space. Given an orbit,
the physically relevant question is whether (orbit 2 between A and
B) or not (orbit 1) this intersects the geometric subspace. We refer
to a given point in an orbit as a {\it
representative}.}
\label{pic:orbits}
}
\end{center}
\end{figure}

Here we would like to emphasize that all these (a priori) different
T-duality connected flux configurations by definition lie in the
same orbit of gaugings,  and therefore give rise to the same
lower-dimensional physics. In order to obtain a different gauged
supergravity, one should consider more general configurations of
fluxes, involving for example combinations of geometric and
non-geometric fluxes, that can never be T-dualized to a frame in
which the non-geometric fluxes vanish. For the sake of clarity, we
depict this concept in Figure~\ref{pic:orbits}.

Non-geometric fluxes are
the inevitable consequence of string dualities, and only a theory
which promotes such dualities to symmetries could have a
chance to describe them together with geometric fluxes and to understand their origin in a unified way. From the
viewpoint of the lower-dimensional effective theory, it turns out
that half-maximal and maximal gauged supergravities give
descriptions which are explicitly covariant with respect to T- and
U-duality respectively. This is schematically depicted in table~
\ref{dualities}, even though only restricted to the cases we will address in this work.

\begin{table}[t!]
\renewcommand{\arraystretch}{1.25}
\begin{center}
\scalebox{0.85}[0.85]{
\begin{tabular}{ccc}
\hline
$D$ & T-duality & U-duality \\
\hline \hline
$9$ &  O($1,1$) & $\mathbb{R}^{+}\,\times\,$SL($2$)\\

$8$ &  O($2,2)\,=\,$SL($2)\,\times\,$SL($2$) & SL($2)\,\times\,$SL($3$)\\

$7$ &  O($3,3)\,=\,$SL($4$) & SL($5$)\\
\hline
\end{tabular}
}
\end{center}
{\it \caption{The various T- and U-duality groups in $D>6$. These turn
out to coincide with the global symmetry groups of half-maximal and
maximal supergravities respectively.} 
\label{dualities}
}
\end{table}

In recent years, a new proposal (DFT) aiming to promote T-duality to a
fundamental symmetry in field theory has received increasing interest. 
It is named Double Field Theory (DFT)
\cite{Hull:2009mi} since T-duality invariance requires
a doubling of the spacetime coordinates, by supplementing them with
dual coordinates associated to the stringy winding modes, whose dynamics can
become important in the compactified theory. Recently it has been
pointed out how to obtain gaugings of $\mathcal{N}=D=4$ supergravity
by means of twisted double torus reductions of DFT
\cite{Aldazabal:2011nj, Geissbuhler:2011mx}, even though at that
stage, the so-called \emph{weak} and \emph{strong} constraints imposed for
consistency of DFT represented a further restriction that prevented
one from describing the most general gaugings that solve the Quadratic
Constraints (QC) of gauged supergravity.

Subsequently, an indication has been
given that gauge consistency of DFT does not need the weak and strong
constraints \cite{Grana:2012rr}. Following this direction, we could 
wonder whether
relaxing these constraints can provide a higher-dimensional origin
for all gaugings of extended supergravity in DFT. 

The aim of our work will be to assess to what extent DFT can 
improve our description of
non-geometric fluxes by giving a higher-dimensional origin to orbits
which do not follow from standard supergravity compactifications. 
We will call such orbits of gaugings {\it non-geometric} 
(in Figure \ref{pic:orbits} they are represented by orbit 1).

As a starting point for this investigation, we will address the
problem in the context of maximal and half-maximal gauged
supergravities in seven dimensions and higher, where the global
symmetry groups are  small enough to allow for a general
classification of orbits, without needing to consider truncated
sectors. 
We will show that in the half-maximal supergravities in
seven and higher-dimensions, where the classifications of orbits can
be done exhaustively, {\it all} the orbits (including geometric and
non-geometric) admit an uplift to DFT, through Scherk-Schwarz (SS)
\cite{Scherk:1979zr} compactifications on appropriate backgrounds.
We provide explicit backgrounds for every orbit, and discuss their
(un)doubled nature. 
The result is that truly doubled DFT provides the appropriate framework 
to deal with orbits that can not be obtained from supergravity. 
In contrast, in maximal supergravities in eight and higher-dimensions, all
orbits are geometric and hence can be obtained without resorting to DFT.

The chapter is organized as follows.  
In Sections 
\ref{labelflux} and 
\ref{labeltduality}
we present a general introduction to
flux compactifications
and different methods used in $T$-duality covariant constructions.

In Section~\ref{ch:duality_orbits0}, we introduce and motive DFT emphasizing in the aspect of its SS compactification and its connection with gauged supergravities.
%
%
We will explicitly show how the gaugings in the
effective theory are related to the compactification ansatz, in
order to make a link with the results of the following sections. In
Section~\ref{sec:U_Dualitites} we present the classification of
consistent gaugings in maximal supergravity in terms of U-duality
orbits. 
In particular, 
we work out the $D=9$ and $D=8$ orbits. 
In both cases we are able to show that all the duality orbits have a
geometric origin in compactifications of ten dimensional
supergravity. In Section~\ref{sec:T_Dualitites} we classify the
consistent gaugings in half-maximal supergravity in terms of
T-duality orbits. 
In particular, 
we work out the $D=8$ and $D=7$ orbits.
Here we encounter the first orbits lacking a geometric
higher-dimensional origin. 
We show that such orbits do follow from dimensional reductions of DFT. 
Finally, our conclusions are
presented in Section~\ref{sec:conclusions}. 
We defer a number of technical
details on gauge algebras and 't Hooft symbols to the Appendix \ref{app:duality_orbits}.

\section{Flux compactification: a primer}
\label{labelflux}

\subsection{Geometric fluxes}

Let us briefly introduce the geometric fluxes origin from 
Scherk-Schwarz (SS) compactifications of supergravities. 
We will closely follow the references 
\cite{Kaloper:1999yr,Aldazabal:2013sca,Roest:2004pk}.

Let us consider the common ``NSNS'' bosonic sector of supergravity, 
spanned by a $D$-dimensional metric $\hat g_{\hat\mu\hat\nu}$, 
a 2-form field $\hat B_{\hat\mu\hat\nu}$ and a dilaton $\hat\phi$.
This sector is shared by all the superstring-derived theories.
Its effective action in the ``string frame'' is 
given by (\emph{cf}. \eqref{eq:effective_ST})
\begin{align}
S=\frac{g^2}{16\pi G_N^{(d)}}\int d^D x \sqrt{|g|}e^{-2\phi}\left[
	R
	-4(\partial\phi)^2
	+\frac{1}{2\cdot 3!} H_{\mu\nu\rho}H^{\mu\nu\rho}
	\right]
\, .
\label{eq:effective_ST2}
\end{align}
This action can arise as part of the low-energy effective action of the 
bosonic string with $D = 26$ 
or froms the ``common sector'' of the heterotic or type II superstrings 
in $D = 10$ dimensions.
All of the fields will depend on  $D$ space-time coordinates. 
The standard Kaluza-Klein dimensional reduction of the 
previous action  on a $n$-torus $T^n$ would give a theory in 
$d=D-n$ dimensions with a manifest $O(n, n)$ global invariance and a  $U(1)^{2n}$ gauge symmetry.
 The scalar fields would take values in the coset 
$O(n, n)/O(n)\times O(n)$.
If one is  interested in gauged supergravities that arise as 
deformations of this theory,
one possibility of deformations consists of promoting
a $2n$-dimensional subgroup $G_0\subset O(n, n)$ to 
a local symmetry gauged by the vector fields already existing
on it using the embedding tensor formalism (see previous 
chapters).  

Another possibility of deformation is to perform a more
general dimensional reduction. 
Let us for this purpose  split the coordinates as follows:
\begin{align}
x^{\hat\mu}&=(x^\mu,y^m)\, ,
\end{align}
where $y^m$, $m=1,\ldots,n$ are compact space directions and $x^\mu$, $\mu=1,\ldots,d$ are spacetime directions. The fields must be decomposed into representations of the symmetry group of the lower-dimensional theory,
\begin{align}
\hat g_{\hat\mu\hat\nu}
&=
\left(\begin{array}{cc}
\hat g_{\mu\nu}+\hat g_{pq} \hat A^p{}_\mu \hat A^q{}_\nu & \hat A^p{}_\mu \hat g_{p m}
\\
\hat g_{mp}\hat A^p{}_\nu & \hat g_{\mu\nu}
\end{array}\right)
\, ,
\\
\hat b_{\hat\mu\hat\nu}
&=
\left(\begin{array}{cc}
\hat b_{\mu\nu}-\frac{1}{2}\left(\hat A^p{}_\mu \hat V_{p\nu}-\hat A^p{}_{\nu} \hat V_{p\mu}\right)+\hat A^p{}_\mu \hat A^q{}_\nu \hat b_{pq} & \hat V_{n\mu}-\hat b_{np} \hat A^p{}_\mu
\\
-\hat V_{m\nu}+\hat b_{mp} \hat A^p{}_\nu & \hat b_{\mu\nu}
\end{array}\right)	
\, ,
\end{align}
where $\hat A^m{}_\mu$ and $\hat V_{m\mu}$ are vector fields and $\hat g_{mn}$ and $\hat b_{mn}$ are symmetric and antisymmetric 
scalar 
matrices, respectively.
In principle, all the fields in the matrices above depend on both 
$(x^\mu,y^m)$ coordinates.

A reduction ansatz, expressing the dependence of these $D$-dimensional 
fields on the effective fields that will live in $d$ dimensions (unhatted)
is necessary.
We can assume an ansatz in which these fields do not depend on the compact coordinates, as for example:
\begin{align}
\hat g_{\mu\nu}&=g_{\mu\nu} (x)\, ,&\hat g_{mn}&=u^a{}_m(y) u^b{}_n(y) g_{ab}(x)\, ,\nonumber\\
\hat b_{\mu\nu}&=b_{\mu\nu} (x)\, ,&\hat b_{mn}&=u^a{}_m(y) u^b{}_n(y) b_{ab}(x)
+v_{mn}(y)\, ,\nonumber\\
\hat A^m{}_\mu &=u_a{}^m (y)A^a{}_\mu (x)\, ,&\hat V_{m\mu}&=
u^a{}_m(y) V_{a\mu}(x)\, ,\nonumber\\
\hat \phi&=\phi(x)\, .
\label{eq:red_ansatz_fluxes}
\end{align}
Thus, we are left with a $d$-dimensional metric and a 2-form plus $2n$ 
vector fields, $A^a{}_\mu$ and $V_{a\mu}$, and $n^2+1$ scalar fields 
$(g_{ab},b_{ab},\phi)$. The $y$-dependent elements $u^a{}_m(y)$ 
and $v_{mn}(y)$ carry the deformation of the compactified manifold, and 
they have to combine in such a way that there is not $y$-dependence in 
the effective action.
\footnote{We note the formal similarity of the $u^a{}_m$ quantities with 
a `vielbein'.}

The SS reduction of the gauge transformation parameters implies new contributions to the gauge transformations of the effective fields. For a detailed discussion, we refer to \cite{Kaloper:1999yr,Roest:2004pk}. Schematically, if we have a $D$-dimensional gauge parameter
\begin{align}
\hat \lambda^{\hat \mu}
&=
(\epsilon^\mu,\Lambda^m)
\, ,
\end{align}
and an arbitrary vector field of the type
\begin{align}
\hat V^{\hat \mu}
&=
(V^\mu(x),u_a{}^m(y)V^a(x))
\, ,
\end{align}
the effective Lie derivative gets modified. Namely, if
\begin{align}
\mathcal{\hat L}_{\hat\lambda}\hat V^{\hat\mu}
&=
\hat\lambda^{\hat\nu}\partial_{\hat \nu}\hat V^{\hat\mu}
-\hat V^{\hat\nu}\partial_{\hat \nu}\hat \lambda^{\hat\mu}
\end{align}
is the $D$-dimensional Lie derivative, the (unhatted) effective Lie derivative results
\begin{align}
\mathcal{L}_\epsilon V^a
&=
\mathcal{\hat  L}_{\epsilon} V^a
+f_{bc}{}^a\Lambda^b V^c
\, ,
\end{align}
where
\begin{align}
f_{ab}{}^c
&=
u_a{}^m\partial_m u_b{}^n u^c{}_n
-u_b{}^m\partial_m u_a{}^n u^c{}_n
\, .
\end{align}
These structure constants are 
known as \emph{metric fluxes}, due to the role that $u$ plays on the 
definition of the lower-dimensional metric.

Inspired by  $O(n,n)$, we can rearrange the fields and gauge parameters 
into $O(n,n)$ multiplets,
\begin{align}
\xi&=(\epsilon_\mu, \epsilon^\mu,\Lambda^A)\, ,\nonumber\\
\Lambda^A&=(\lambda_a,\lambda^a)\, ,\nonumber\\
A^A{}_\mu 
&=
(V_{a\mu},A^a{}_\mu)
\, ,
\nonumber
\\
M_{AB}
&=
\left(\begin{array}{cc}
g_{ab} & -g^{ac}b_{cb}
\\
b_{ac}g^{cb} & g_{ab}-b_{ac}g^{cd}b_{db}
\end{array}\right)
\, ,
\end{align}
where indices $A,B=1,\ldots,2n$ are raised and lowered by means of the metric
\begin{align}
\eta_{AB}
&=
\left(\begin{array}{cc}
0 & \delta^a{}_b
\\
\delta_a{}^b & 0
\end{array}\right)
\, .
\end{align}
The gauge transformations of the effective fields result modified and their dependence on the compact manifold is reflected in the structure constants $f_{ABC}$,
\begin{align}
\delta_\xi g_{\mu\nu}
&=
\mathcal{L}_\epsilon g_{\mu\nu}
\, ,
\nonumber\\
\delta_\xi b_{\mu\nu}
&=
\mathcal{L}_\epsilon b_{\mu\nu}
+(\partial_\mu\epsilon_\nu-\partial_\nu\epsilon_\mu)
\, ,
\nonumber\\
\delta_\xi A^A{}_\mu
&=
\mathcal{L}_\epsilon A^A{}_\mu
-\partial_\mu\Lambda^A
+f_{BC}{}^A\Lambda^B A^C{}_\mu
\, ,
\nonumber\\
\delta_\xi M_{AB}
&=
\mathcal{L}_\epsilon M_{AB}
+f_{AC}{}^D\Lambda^C M_{DB}
+f_{BC}{}^D\Lambda^C M_{AD}
\, .
\end{align}
The structure constants $f_{ABC}$ have non-vanishing components
\begin{align}
f_{ab}{}^c
&=
u_a{}^m\partial_m u_b{}^n u^c{}_n
-u_b{}^m\partial_m u_a{}^n u^c{}_n
\, ,
\nonumber\\
f_{abc}
&=
3(
	\partial_{[a}v_{bc]}
	+f_{[ab}{}^dv_{c]d}
	)
	\, ,
\end{align}
so that $f_a{}^{bc}=f^{abc}=0$.

Substituting the ansatz \eqref{eq:red_ansatz_fluxes} 
into the $D$-dimensional SUGRA bosonic action 
\eqref{eq:effective_ST2}, 
we have the effective d-dimensional action
\begin{multline}
S=\int d^dx \sqrt{|g|}e^{-2\phi}\left(
	R
	+4(\partial\phi)^2
	-\frac{1}{4}M_{AB}F^{A\mu\nu}F^B{}_{\mu\nu}
	-\frac{1}{12}G_{\mu\nu\rho}G^{\mu\nu\rho}
\right.
\\
\left.
	+\frac{1}{8}D_\mu M_{AB}D^\mu M^{AB}
	-V
	\right)
\, ,
\label{actiond}
\end{multline}
where
\begin{align}
F^{A}{}_{\mu\nu}
&=
2\partial_{[\mu} A^A{}_{\nu]}
-f_{BC}{}^A A^B{}_\mu A^C{}_\nu
\, ,
\\
G_{\mu\nu\rho}
&=
3\partial_{[\mu} b_{\nu\rho]}
-f_{ABC} A^A{}_\mu A^B{}_\nu A^C{}_\rho
+3\partial_{[\mu}A^A{}_\nu A_{|A|\rho]}
\, ,
\end{align}
and the covariant derivative is
\begin{align}
D_\mu M_{AB}
&=
\partial_\mu M_{AB}
-f_{AD}{}^C A^D{}_\mu M_{CB}
-f_{BD}{}^C A^D{}_\mu M_{AC}
\, .
\end{align}
In addition, a scalar potential naturally arises. This is given by the expression
\begin{align}
V
&=
\frac{1}{4}f_{DA}{}^Cf_{CB}{}^D M^{AB}
+\frac{1}{12}f_{AC}{}^E f_{BD}{}^F M^{AB} M^{CD} M_{EF}
+\frac{1}{6}f_{ABC}f^{ABC}
\, .
\end{align}
The structure constants that have appeared as a consequence of the 
dimensional reduction of the $2$-form and  metric fields 
are called \emph{geometric fluxes} due to their geometrical reduction 
origin. 
In the literature, they are also denoted as
\begin{align}
H_{abc} 
&\equiv
f_{abc}
\, ,
&
\omega_{ab}{}^c
&\equiv
f_{ab}{}^c
\, .
\label{notationfluxes1}
\end{align}
This is a notation which we will frequently use in the following sections.

\subsubsection{Beyond geometric fluxes}

If in the 
ansatz \eqref{eq:red_ansatz_fluxes}, 
we choose
\begin{align}
g_{ab}(x)&=
\delta_{ab}
\, ,
&
b_{ab}(x)
&=
0
\, ,
\end{align}
the twist matrices $u$ and $v$ can be understood as the background fields associated to the vielbein and the 2-form that live in the compact space. 
Since T duality exchanges metric and 2-form components by means of the 
Buscher's rules \eqref{eq:Buscher_2AB},
these geometric fluxes can be transformed into each other as well. 
Let us study a simple setting of these fluxes to see explicitly how 
this applies \cite{Shelton:2005cf}.

Let us consider a compactification on a 3-torus with a 
non-trivial 2-form, \emph{e.g.}
\begin{align}
{\hat g}_{mn}
&=
\delta_{mn}
\, ,
&
b_{23}
&=
C y^1
\, ,
\end{align}
whose associated twist matrices are
\begin{align}
u_m{}^a
&=
\delta_m{}^a
\, ,
&
v_{23}
&=
Cy^1
\, .
\end{align}
The corresponding fluxes are 
\begin{align}
H_{123}
&=
C
\, ,
&
\omega_{12}{}^3
&=
\omega_{23}{}^1
=
\omega_{31}{}^2
=
0
\, .
\end{align}
Since these backgrounds enjoy isometries in the $y^2$ and $y^3$ directions, we can perform T duality transformations on these directions. So, applying \eqref{eq:Buscher_2AB}, we get certain $g_{mn}$ and $b_{mn}$,
\begin{align}
ds^2
&=
g_{mn}dy^mdy^n
=
(dy^1)^2
+(dy^2)^2
+(dy^3+Cy^1dy^2)^2
\, ,
&
b_{mn}
&=
0
\, ,
\end{align}
which imply the following fluxes:
\begin{align}
H_{123}
&=
\omega_{23}{}^1
=
\omega_{31}{}^2
=
0
\, ,
&
\omega_{12}{}^3
&=
C
\, .
\end{align}
By simple inspection, we notice that these fluxes still can be T-dualized in the direction $y^2$. Again, using the Buscher's rules, they transform into 
\begin{align}
ds^2
&=
g_{mn}dy^mdy^n
=
(dy^1)^2
+\frac{1}{1+(Cy^1)^2}\left[(dy^2)^2+(dy^3)^2\right]
\, ,
&
b_{23}
&=
-\frac{Cy^1}{1+(Cy^1)^2}
\, .
\end{align}
The non-vanishing component of the 2-form is associated to a new flux, which in the literature is called $Q_1{}^{23}$. 

Symbolically we have built T-duality 
transformations, such that
\begin{align}
H_{abc} \overset{T_c}{\longleftrightarrow}\omega_{ab}{}^c
\overset{T_b}{\longleftrightarrow}Q_a{}^{bc}
\, 
\label{eq:duality_chain}
\end{align}
The first T-duality transformation $T_c$ relates 
the \emph{metric} and \emph{gauge fluxes}. 
The second one, $T_b$, produces the so-called $Q$ fluxes, 
which describe locally geometric backgrounds despite of 
not being globally well-defined.

A last T-duality transformation, $T_a$ in the diagram below, 
would generate the $R$ fluxes. and 
since there are no isometries in the $y^1$ direction, there 
does not exist even a local description for these 
background fluxes.
\begin{align}
H_{abc} \overset{T_c}{\longleftrightarrow}\omega_{ab}{}^c
\overset{T_b}{\longleftrightarrow}Q_a{}^{bc}
\overset{T_a}{\longleftrightarrow}R^{abc}\, 
\label{eq:duality_chain2}
\end{align}

Thus, T-duality would allow to transform a single 
non-geometric flux into a geometric one. However, a 
configuration of both geometric and non-geometric fluxes simultaneously
turned on such that T-duality  is not 
capable of converting all the non-geometric fluxes into 
geometric ones is a special situation. This kind of 
setting is called 
\emph{duality orbit of a non-geometric flux} and is 
treated in Section \ref{ch:duality_orbits}, where we 
explicitly show that a standard SS reduction is not able 
to reproduce it. This fact turns out to wonder whether we 
need extra ingredients in our compactification procedure 
to get these additional fluxes.

We realize that T duality is going to be crucial in 
the development and inclusion of these non-geometric 
backgrounds. Indeed, the way in that these fluxes have 
emerged suggests a new framework in which T-duality 
becomes a true symmetry of the genuine theory, instead 
of appearing after the compactification.


\section{T-duality covariant constructions}
\label{labeltduality}


Several approaches have been developed to solve the problem 
of getting 
non-geometric fluxes in a natural and covariant formalism. 
We can distinguish three different trends. 
The first one is the \emph{doubled geometry}, in which the 
local charts or patches 
that define the background geometry are slightly modified. 
Another possibility is the 
so-called \emph{generalized complex geometry}, 
which is defined on a manifold whose bundle structure is extended to 
include new elements. Finally, there exists 
the \emph{double field theory} formalism, which suggests the doubling 
of spatial coordinates, associating the new ones to their 
corresponding dual winding modes.

Despite of the different approaches under which these theories are 
built, their aim is the same: to be able to host T-duality as 
a global symmetry by construction.

\subsection{Doubled Geometry}

The distinctive characteristic of 
doubled geometry is that given a manifold 
the group of 
transition functions between overlapping coordinate charts 
is generalized   to include, in addition to diffeomorphisms and 
gauge transformations, duality transformations. When 
these duality transformations are  T-duality transition 
functions the manifold equipped with the extra structure
is named a  {\it T-fold} \cite{Dabholkar:2005ve,Hull:2004in}.\footnote{Similarly, U-folds with U-duality transition functions,
or mirror-folds with mirror symmetry transition functions
can be defined. Locally,
 T-folds or U-folds require each
coordinate chart or patch to be the product of a torus 
with some open set, 
while a mirror-folds have a Calabi-Yau fibration.}

In \cite{Hull:2006va}, $O(n,n)$ duality twist reductions 
have been 
performed by making use of this T-fold structure. Later on, dimensional 
reduction over twisted doubled tori were performed to include 
non-geometric fluxes configurations 
in \cite{Dall'Agata:2007sr,Hull:2009sg,ReidEdwards:2009nu}. 

\subsection{Generalized Complex Geometry}


The starting point of this approach consists  of a modification of the
tangent bundle structure associated to the manifold. 
The main idea is the treatment of the tangent and cotangent space at 
the same level, without distinguishing them. 
In its original formulation \cite{Hitchin:2004ut},  a new generalized 
tangent bundle is 
constructed by the direct sum of both spaces,
\begin{align}
X+\xi\in TM_n \oplus T^*M_n
\, .
\end{align}
The elements of such a space 
 are formal sums of a vector field and a one-form.


This generalized bundle induces a natural metric $\mathcal{I}$,
\begin{align}
\mathcal{I}(X+\xi,Y+\eta)
&\equiv
\frac{1}{2}(\imath_Y\xi+\imath_X\eta)
\, ,
\end{align}
where $\imath_Y\xi\equiv Y^m\xi_m$. In the coordinate basis $(\partial_m,dx^m)$, the metric is realized by the matrix
\begin{align}
\mathcal{I}
&=
\frac{1}{2}
\left(\begin{array}{cc}
0 & \mathds{1}_n
\\
\mathds{1}_n & 0
\end{array}\right)
\, .
\end{align}

Thus, a \emph{generalized almost-complex structure} on this bundle is 
defined as an endomorphism $J$,
\begin{align}
J:TM_n\oplus T^*M_n\longrightarrow TM_n\oplus T^*M_n
\end{align}
such that  
$$J^2=-\mathds{1}_{2n}$$ 
and 
$$J^T\mathcal{I} J = \mathcal{I}.$$

Following the parallelism of an almost-complex structure, 
a generalized Lie bracket can be defined. 
This is the so-called \emph{Courant bracket},
which is defined as
\begin{align}
[X+\xi,Y+\eta]_C
&=
[X,Y]
+\mathcal{L}_X\eta
-\mathcal{L}_Y\xi
-\frac{1}{2}d(\imath_X\eta-\imath_Y\xi)
\, .
\end{align}
A {\it generalized complex structure} is a generalized almost complex 
structure such that the space of smooth bundle sections  is 
closed under the Courant bracket.
This bracket is also defined in double field theory, as we will see in the next section.
Interesting monographs dealing with generalized complex 
geometry are for example \cite{Berman:2010is,Berman:2011pe,Berman:2011kg,Berman:2011jh,Berman:2012vc,Berman:2013eva}.

\subsection{Double field theory}

Double field theory (DFT) can be understood as a T duality invariant 
formulation of  string theory and supergravity. 
That is, it contains T duality as a symmetry of the theory by construction. 
From the stringy point of view this is achieved by doubling the spacetime 
coordinates and associating the winding modes of the strings to the 
new dual coordinates that are required to be introduced to have T-duality as a symmetry. 
Its original version was developed to describe the dynamics of 
closed strings on tori \cite{Hull:2009mi}. 
However, due to the successful development of a background independent 
version \cite{Hohm:2010jy}, it was  quickly used to perform SS reductions 
over different manifolds. These dimensional compactifications done in 
a DFT scenario allowed to obtain the gaugings associated to the electric 
sector of $N=4$ $D=4$ supergravity, thus establishing
a relation between  DFT and non-geometric fluxes.

We will show here a brief introduction to 
the main features of DFT and its relation with gauged supergravities. 
Let us introduce the necessary ingredients of DFT and some notation. 
For a $D$-dimensional spacetime with $d$ non-compact spacetime coordinates 
and $n$ compact dimensions ($D=d+n$), the fields depend on coordinates 
\begin{align}
X^M&=(\tilde x_i,x^i)=(\tilde x_\mu,\tilde y_m,x^\mu,y^m)\, ,
\end{align}
where $M=1,\ldots,2D$ is an $O(D,D)$ index. 
The $2D$ coordinates can be split into the genuine $D$ spacetime 
coordinates $x^i$ and their dual coordinates, $\tilde x_i$. 
In addition, the $i$ index can be split into extended and compact 
coordinates, $i=\{\mu,m\}$, where $\mu=1,\ldots,d$ represents extended 
coordinates and $m=1,\ldots,n$ runs over the compactified coordinates.

Any fields and gauge parameters of DFT is supposed to be annihilated by 
the differential operator
\begin{align}
\partial_i\tilde \partial^i \Phi&=0\, .
\label{eq:levelmatching_constraint}
\end{align}
Where $\Phi$ denotes any field or gauge parameter of the theory.
This is the DFT \emph{weak constraint} (WC).\footnote{In its original stringy formulation, DFT was restricted to satisfy the level matching condition
\begin{align}
L_0-\bar L_0&=0\, ,
\end{align}
arising for closed string theory. This condition translates to the WC.}
A background independent action is constructed \cite{Hohm:2010jy} under 
a stronger restriction: \eqref{eq:levelmatching_constraint} 
must 
hold not only for any field or gauge parameter, but for any product of 
them. This is the so-called \emph{strong constraint} (SC). 
If we define a generalized field $\mathcal{E}_{ij}$ in terms 
of the metric and the 2-form,
\begin{align}
\mathcal{E}_{ij}
&\equiv
g_{ij}+b_{ij}
\, ,
\end{align}
and a T-duality invariant scalar field $d$,\footnote{The context should 
be enough to clarify the difference between the dimension `$d$' and the scalar field `$d$'}
\begin{align}
e^{-2d}
&\equiv
\sqrt{|g|}e^{-2\phi}
\, ,
\end{align}
the background independent action is given by
\begin{multline}
S
=
\int d^{d}x d^{d}\tilde x e^{-2d}\left[
	-\frac{1}{4}g^{ik} g^{jl} \mathcal{D}^p \mathcal{E}_{kl}\mathcal{D}_p \mathcal{E}_{ij}
	+\frac{1}{4}g^{kl}\left(
		\mathcal{D}^j \mathcal{E}_{ik}\mathcal{D}^i \mathcal{E}_{jl}
		+\mathcal{\tilde D}^j \mathcal{E}_{ki} \mathcal{\tilde D}^i \mathcal{E}_{lj}
		\right)
\right.
\\
\left.
	+\mathcal{D}^i d \mathcal{\tilde D}^j \mathcal{E}_{ij}
	+\mathcal{\tilde D}^i d \mathcal{D}^j \mathcal{E}_{ji}
	\right]
\, .
\end{multline}
The derivative operators $\mathcal{D}^i$, $\mathcal{\tilde D}^i$ are defined as
\begin{align}
\mathcal{D}_i
&\equiv
\frac{\partial}{\partial x^i}
-\mathcal{E}_{ik}\frac{\partial}{\partial \tilde x_k}
\, ,
&
\mathcal{\tilde D}_i
&\equiv
\frac{\partial}{\partial x^i}
+\mathcal{E}_{ik}\frac{\partial}{\partial \tilde x_k}
\, .
\end{align}

This action is invariant under the $O(D,D)$ T-duality group, which acts on the fields as follows:
\begin{align}
\mathcal{E}'(X')
&=
\frac{a\mathcal{E}(X)+b}{c\mathcal{E}(X)+d}
\, ,
&
d'(X')
&=
d(X)
\, ,
&
X'
&=
hX
\, ,
\end{align}
where $h$ is
\begin{align}
h
&=
\left(\begin{array}{cc}
a & b
\\
c& d
\end{array}\right)
\in O(D,D)
\, ,
&
h^T\eta h
&=
\eta
\text{  with  }
\eta
=
\left(\begin{array}{cc}
0 & \mathds{1}_D
\\
\mathds{1}_D & 0
\end{array}
\right)
\, .
\end{align}
This action can be rewritten in terms of the so-called \emph{generalized metric}, $\mathcal{H}_{MN}$. This is a $2D\times 2D$ symmetric matrix constructed from the $D\times D$ matrices $g_{ij}$ and $b_{ij}$, with the remarkable property that it transforms as an $O(D,D)$ tensor,
\begin{align}
\mathcal{H}
&=
\left(\begin{array}{cc}
g^{ij} & -g^{ik}b_{kj}
\\
b_{ik}g^{kj} & g_{ij} - b_{ik}g^{kl}b_{lj}
\end{array}\right)
\, .
\end{align}
Under $h\in O(D,D)$ transformations, the fields transform as
\begin{align}
\mathcal{H}_{MN}(X)
&\rightarrow
h_M{}^P h_N{}^Q \mathcal{H}_{PQ}(h X)
\, ,
&
d(X)
&\rightarrow
d(h X)
\, ,
\end{align}
For cases in which $h$ corresponds to a T-duality transformation, it reproduces the corresponding Buscher's rules \eqref{eq:Buscher_2AB} for $\{g_{ij},b_{ij},\phi\}$. In fact, it has been shown that these transformation rules allow the possibility of performing a T-duality transformation in non-isometric directions \cite{Hull:2004in,Hull:2006va,Hull:2007jy,Hull:2009sg}. Then, in terms of this generalized metric formulation, the original action 
is rewritten as
\begin{multline}
S
=
\int
d^dx d^d\tilde x e^{-2d}
\left(
	\frac{1}{8}\mathcal{H}^{MN}\partial_M\mathcal{H}^{PQ}\partial_N\mathcal{H}_{PQ}
	-\frac{1}{2}\mathcal{H}^{MN}\partial_N\mathcal{H}^{PQ}\partial_Q\mathcal{H}_{MP}
\right.
\\
	-2\partial_M d \partial_N \mathcal{H}^{MN}
	+4\mathcal{H}^{MN}\partial_M d \partial_N d
	\bigg)
\, .
\label{eq:DFT_action_H}
\end{multline}
Gauge invariance of the action and the closure of the algebra of DFT happens upon the weak (WC) and strong (SC) versions of 
\eqref{eq:levelmatching_constraint}, 
which in $O(D,D)$ indices are rewritten, respectively, as
\begin{align}
\partial_M \partial^M A
&=
0
\, ,
&
\partial_M A \partial^M B
&=
0
\, ,
\label{eq:SC1}
\end{align}
where $A$, $B$, again refers to any field and/or gauge parameter. 
Gauge transformations of the fields $\{\mathcal{H},d\}$ are driven by the transformation rules of $\mathcal{E}_{ij}$,
\begin{align}
\delta_\xi \mathcal{H}^{MN}
&=
\xi^P\partial_P \mathcal{H}^{MN}
+(\partial^M\xi_P-\partial_P\xi^M)\mathcal{H}^{PN}
+(\partial^N\xi_P-\partial_P\xi^N)\mathcal{H}^{MP}
\, ,
\nonumber
\\
\delta \xi d
&=
\xi^M\partial_M d
-\frac{1}{2}\partial_M \xi^M
\, .
\end{align}
This motivated the definition of a generalized Lie derivative $\mathcal{\hat L}_\xi$ such that, for an arbitrary $O(D,D)$ tensor $V^M{}_N$, 
\begin{align}
\mathcal{\hat L}_\xi V^{M}{}_N
&=
\xi^P\partial_P V^M{}_N
+(\partial^M\xi_P-\partial_P\xi^M)V^P{}_N
-(\partial_N\xi^P-\partial^P\xi_N)V^M{}_P
\, .
\end{align}
Then, the field transformations are rewritten as
\begin{align}
\delta_\xi \mathcal{H}^{MN}
&=
\mathcal{\hat L}_\xi \mathcal{H}^{MN}
\, ,
\\
\delta_\xi d
&=
\mathcal{\hat L}_\xi d
\, .
\end{align}
Upon the SC constrain, this generalized Lie derivative (and thus the gauge transformations) close under the 
 \emph{Courant} or \emph{C-bracket} (equivalent to the 
one defined in the previous section),
\begin{align}
\left[\xi_1,\xi_2\right]_C{}^M
&=
2\xi_{[1}^M\partial_N\xi_{2]}^M
-\xi^N_{[1}\partial^M\xi_{2]N}
\, .
\end{align}
The SC results essential in this DFT development in this way. 
However, some deficiencies to this formulation 
arose \cite{Geissbuhler:2011mx,Aldazabal:2011nj}. 
In these works, some, but not all of the gaugings of $N=4$ $D=4$ SUGRA 
were obtained by SS reductions of DFT. 
Indeed, the gaugings associated to non-geometric fluxes could be 
geometrized by performing suitable T-duality transformations as the ones 
shown before. 
This, together with the presence of the constraint \eqref{eq:SC1}, led to 
think about a new reformulation of DFT in which the SC \eqref{eq:SC1} 
would be somehow relaxed and genuine non-geometric fluxes would 
be captured as consequence.

In ref. \cite{Grana:2012rr}, DFT was formulated without imposing any 
constraint at the very beginning.
Without this constraint the requirements of gauge invariance of the 
action, the closure of the generalized Lie derivatives and the generalized 
Jacobi identities are not automatically satisfied.
 When the SS compactification is performed on the theory,  
it is shown that \eqref{eq:SC1} is indeed a sufficient but not a 
necessary condition for the consistency of the theory. 
In particular, they find a less restrictive condition under which 
the 3 previous requirements are fulfilled. 
These relaxed constraints are
\begin{align}
\partial_M\partial^M \hat A&=0\, ,&
\partial_M \hat A \partial^M \hat B&=0\, ,
\label{eq:SC2}
\end{align}
where $\hat A$, $\hat B$ denote any effective (that is, living in the 
lower-dimensional theory) 
field and/or gauge parameter. 
That is, while \eqref{eq:SC1} is required not only for the 
lower-dimensional fields but also for the fields of the higher-dimensional theory, the 
new constraints \eqref{eq:SC2} are only imposed on fields living in the 
lower-dimensional theory. Moreover, not only the 3 consistency 
constraints (gauge invariant action, closure of the gauge transformations, 
Courant-like Jacobi identities) are satisfied, but an additional term, which is killed by \eqref{eq:SC1}, can be this time added to the action,
\begin{align}
\int
d^dxd^d\tilde x e^{-2d}\frac{1}{2}\partial_M\mathcal{E}^a{}_P\partial^M\mathcal{E}^b{}_QS_{ab}\eta^{PQ}
\, .
\end{align}
Actually, this term becomes crucial for matching fluxes and gaugings, as 
we will verify in the following sections.

\section{Duality orbits of non-geometric fluxes}
\label{ch:duality_orbits0}

As we have mentioned in the last section, compactifications in duality covariant constructions such as
generalized geometry and double field theory have proven to be suitable
frameworks to reproduce gauged supergravities containing
non-geometric fluxes. However, it is a priori unclear whether these
approaches only provide a reformulation of old results, or also
contain new physics. To address this question, we classify the T-
and U-duality orbits of gaugings of (half-)maximal supergravities in
dimensions seven and higher. It turns out that all orbits have a
geometric supergravity origin in the maximal case, while there are
non-geometric orbits in the half-maximal case. We show how the
latter are obtained from compactifications of double field theory. Some technical material used in the development of this chapter can be found in Appendix \ref{appendix_A}. The results of this chapter were first obtained in refs. \cite{deRoo:2011fa,FernandezMelgarejo:2011wx,Dibitetto:2012rk}.

\subsection{Orbits from double field theory}
\label{sec:DFT}

While toroidal compactifications of DFT lead to half-maximal
ungauged supergravities, SS compactifications on more general double
spaces are effectively described by gauged supergravities like the
ones we will analyze in the next sections. If the internal space is
restricted in such a way that there always exists a frame without
dual coordinate dependence, the only orbits allowed in the effective
theory are those admitting representatives that can be obtained from
 compactifications of ten dimensional supergravity. This is not the most general case, and we will show that some orbits require
the compact space to be truly doubled, capturing information of both momentum and winding modes.

Recently in ref.~\cite{Grana:2012rr}, a new set of solutions to the
constraints for DFT has been found. For these solutions the internal
dependence of the fields is not dynamical, but fixed. The
constraints of DFT restrict the dynamical external space to be
undoubled, but allows for a doubling of the internal coordinates as
long as the QC for the gaugings are satisfied. Interestingly, these
are exactly the constraints needed for consistency of gauged
supergravity, so there is a priori no impediment to uplift any orbit
to DFT in this situation. In fact, in the following sections we show
that all the orbits in half-maximal $D = 7,8$ gauged supergravities
can be reached from twisted double tori compactifications of DFT.

\subsubsection{DFT and (half-)maximal gauged supergravities}
\label{subsec:GDFT}

%

In the SS procedure, the coordinates $X^M$ are split into  external directions $\mathbb{X}=(\tilde x_i,x^i)$ and compact internal
$\mathbb{Y}=(\tilde y_i,y^i)$ coordinates.  The former set contains pairs of O$(D,D)$
dual coordinates, while the latter one contains pairs of O$(n,n)$ dual
coordinates, with $d = D + n$. This means that if a given coordinate
is external (internal), its dual must also be external (internal),
so the effective theory is formally a (gauged) DFT. The SS procedure
is then defined in terms of a reduction ansatz, that specifies the
dependence of the fields in $(\mathbb{X},\mathbb{Y})$ \be {\cal
H}_{MN} (\mathbb{X},\mathbb{Y}) = U(\mathbb{Y})^A{}_M \ \widehat
{\cal H}(\mathbb{X})_{AB}\ U(\mathbb{Y})^B{}_N
 \ , \ \ \ \ \ d(\mathbb{X},\mathbb{Y}) = \widehat d(\mathbb{X}) + \lambda(\mathbb{Y})\ .\ee
Here the hatted fields $\widehat {\cal H}$ and $\widehat{d}$  are
the dynamical fields in the effective theory, parametrizing
perturbations around the background, which is defined by
$U(\mathbb{Y})$ and $\lambda(\mathbb{Y})$. The matrix $U$ is
referred to as the \emph{twist matrix}, and must be an element of
O$(n,n)$. It contains a DFT T-duality index $M$, and another index
$A$ corresponding to the T-duality group of the effective theory.
When DFT is evaluated on the reduction ansatz, the twists generate
the gaugings of the effective theory \bea f_{ABC} &=& 3 \eta_{D[A}\
(U^{-1})^M{}_B (U^{-1})^N{}_{C]} \partial_M U^D{}_M \
,\label{f_from_U_}
\\
\xi_A &=& \partial_M (U^{-1})^M{}_A - 2 (U^{-1})^M{}_A \partial_M
\lambda \ ,\label{f_from_U}\eea
where $f_{ABC}$ and $\xi_{A}$ build the generalized structure
constants of the gauge group in the lower-dimensional theory.

Although $U$ and $\lambda$ are $\mathbb{Y}$ dependent quantities,
the gaugings are forced to be constants in order to eliminate the
$\mathbb{Y}$ dependence from the lower dimensional theory.
 When the external-internal splitting is performed, namely $d = D + n$, the dynamical fields are written in terms of their components which are a $D$-dimensional metric, a $D$-dimensional $2$-form, $2n$ $D$-dimensional vectors and $n^2$ scalars. These are the degrees of freedom of half-maximal supergravities. Since these fields are contracted with the gaugings, one must make sure that after the splitting the gaugings have vanishing Lorentzian indices, and this is achieved by stating that the twist matrix is only non-trivial in the internal directions. Therefore, although formally everything is covariantly written in terms of O$(d,d)$ indices $A,B,C,...$, the global symmetry group is actually broken to O$(n,n)$. We will not explicitly show how this splitting takes place, and refer to \cite{Aldazabal:2011nj} for more details. In this work,  for the sake of simplicity, we will restrict to the case $\xi_A  =0$, which should be viewed as a constraint for $\lambda$. Also we will restrict to $O(n,n)$ global symmetry groups, without additional vector fields.

There are two possible known ways to restrict the fields and gauge
parameters in DFT, such that the action is gauge invariant and the
gauge algebra closes. On the one hand, the  weak and strong
constraints can be imposed, which in this context they read as 
\be
\partial_M \partial^M A = 0\ , \ \ \ \ \partial_M A\ \partial^M B =
0\ ,
\ee 
where $A$ and $B$ generically denote products  of
(derivatives of) fields and gauge parameters. When this is the case,
one can argue \cite{Hohm:2010jy} that there is always a frame in
which the fields do not depend on the dual coordinates. On the other
hand, in the SS compactification scenario, it is enough to impose
the weak and strong constraints only on the external space
(\emph{i.e.}, on hatted quantities) \be
\partial_M \partial^M \widehat A = 0\ , \ \ \ \ \partial_M \widehat A\ \partial^M \widehat B = 0
\ ,\ee and impose QC for the gaugings \be f_{E[AB} f^E{}_{C]D}
= 0 \ .\ee This second option is more natural for our purposes,
since these constraints exactly coincide with those of half-maximal
gauged supergravities\footnote{We are working under the assumption that the structure constants not only specify the gauging, but all couplings of the theory. Reproducing the correct structure constants therefore implies reproducing the full theory correctly, as has been proven in $D=4$ and $D=10$ \cite{Aldazabal:2011nj, Geissbuhler:2011mx, Hohm:2011ex,
 *Hohm:2010xe}.} (which are undoubled theories in the external
space, and contain gaugings satisfying the QC).

Notice that if a given $U$ produces a solution to the QC, any T-dual
$U$ will also. Therefore, it is natural to define the notion of {\it
twist orbits}  as the sets of twist matrices connected through
T-duality transformations. If a representative of a twist orbit
generates a representative of an orbit of gaugings, one can claim
that the twist orbit will generate the entire orbit of gaugings.
Also, notice that if a twist matrix satisfies the weak and strong
constraints, any representative of its orbit will, so one can define
the notions of undoubled and truly doubled twist orbits.

\subsubsection*{Non-geometry VS weak and strong constraint violation}

Any half-maximal supergravity can be uplifted to the maximal theory
whenever the following constraint holds\footnote{$D=4$ half-maximal
supergravity is slightly different because its global symmetry group
features an extra SL($2$) factor; for full details, see \cite{Aldazabal:2011yz, Dibitetto:2011eu}.}
\be f_{ABC}\,f^{ABC}\,=\,0\ . \label{Extra_f}\ee
This constraint plays the role of an orthogonality condition between geometric and non-geometric fluxes.
Interestingly, the constraint \eqref{Extra_f} evaluated in terms of
the twist matrix $U$ and $\lambda$ can be rewritten as follows (by taking
relations \eqref{f_from_U_} and \eqref{f_from_U} into account)
\be
f_{ABC}\,f^{ABC}\,=\,-3\,\partial_{D}{U^{A}}_{P}\,\partial^{D}{\left(U^{-1}\right)^{P}}_{A}-24\,\partial_{D}\lambda\,\partial^{D}\lambda\,+\,24\,\partial_{D}\partial^{D}\lambda\
. \label{Max_VS_Geom}\ee
The RHS of this equation is zero whenever the background defined by
$U$ and $\lambda$ satisfies the weak and strong constraints. This immediately
implies that any background satisfying weak and strong constraints
defines a gauging which is upliftable to the maximal theory.
Conversely, if an orbit of gaugings in half-maximal supergravity
does not satisfy the extra constraint \eqref{Extra_f}, the RHS of
this equation must be non-vanishing, and then the strong and weak
constraint must be relaxed. In conclusion, the orbits of
half-maximal supergravity that do not obey the QC of the maximal
theory require truly doubled twist orbits, and are therefore
genuinely non-geometric. This point provides a concrete criterion to
label these orbits as non-geometric. Also, notice that these orbits
will never be captured by non-geometric flux configurations obtained
by T-dualizing a geometric background\footnote{However, we would like to
stress that, in general, it is not true that an orbit satisfying the
QC constraints of maximal supergravity (\ref{Extra_f}) is
necessarily generated by an undoubled twist orbit. An example can be found at the end of Section \ref{sec:T_Dualitites}.}.

For the sake of clarity, let us briefly review the definitions that we use. A twist orbit is non-geometric if it doesn't satisfy the weak/strong constraint, and geometric if it does. Therefore, the notion of geometry that we consider is local, and we will not worry about global issues (given that the twist matrix is taken to be an element of the global symmetry group, the transition functions between coordinate patches are automatically elements of $O(n,n)$). On the other hand an orbit of gaugings is geometric if it contains a representative that can be obtained from 10 dimensional supergravity (or equivalently from a geometric twist orbit), and it is non-geometric if it does not satisfy the constraints of maximal supergravity.

We have now described all the necessary ingredients to formally relate dimensional reductions of DFT and the orbits
of half-maximal gauged supergravities. In particular, in what follows we will:
\begin{enumerate}
\item Provide a classification of all the orbits of gaugings in maximal and half-maximal supergravities in $D\geq7$.
\item Explore mechanisms to generate orbits of gaugings from twists, satisfying
\begin{itemize}
\item $U(\mathbb{Y}) \in \textrm{O}(n,n)$
\item Constant $f_{ABC}$
\item  $f_{E[AB} f^E{}_{C]D} = 0$
\end{itemize}
\item Show that in the half-maximal theories all the orbits of gaugings  can be obtained from twist orbits in DFT.
\item Show that in the half-maximal theories the orbits that satisfy the QC of maximal supergravity  admit a representative with a higher-dimensional supergravity origin. For these we provide concrete realizations in terms of undoubled backgrounds in DFT. Instead, the orbits that fail to satisfy (\ref{Extra_f}) require, as we argued, truly doubled twist orbits for which we also provide concrete examples.
\item Show that  there is a degeneracy in the space of twist orbits giving rise to the same orbit of gaugings. Interestingly, in some cases a given orbit can be obtained either from undoubled or truly doubled twist orbits.
\end{enumerate}

In the next sections we will classify all the orbits in (half-)maximal $D\geq 7$ supergravities, and provide the half-maximal ones with concrete uplifts to DFT, explicitly proving the above points.

\subsubsection{Parametrizations of the duality twists}
\label{subsec:twist_matrices}

Here we would like to introduce some notation that will turn out to
be useful in the uplift of orbits to DFT. We start by noting the
double internal  coordinates as $\mathbb{Y}^A = (\tilde y_a, y^a)$
with $a = 1,...,n$. As we saw, the SS compactification of DFT is
defined by the twists $U(\mathbb{Y})$ and $\lambda(\mathbb{Y})$. The
duality twist $U(\mathbb{Y})$ is not generic, but forced to be an
element of O($n,n$), so we should provide suitable
parametrizations. One option is the  {\it light-cone}
parametrization, where the metric of the (internal) global symmetry
group is taken to be of the form 
 \be \eta_{AB} =
\begin{pmatrix} 0 & \mathds{1}_n \\ \mathds{1}_n & 0\end{pmatrix}\ . \ee The most
general form of the twist matrix is then given by \be U(\mathbb{Y}) =
\left(\begin{matrix} e & 0 \\ 0 & e^{-T}\end{matrix}\right)
\,\left(\begin{matrix} \mathds{1}_n & 0 \\ -B &
\mathds{1}_n\end{matrix}\right)\, \left(\begin{matrix} \mathds{1}_n & \beta \\
0 & \mathds{1}_n\end{matrix}\right)\ , \ee with $e \in
\textrm{GL}(n)$ and $B$ and $\beta$ are generic $n\times n$
antisymmetric matrices. When $\beta = 0$, $e = e(y^a)$ and $B =
B(y^a)$, the matrix $e$ can be interpreted as a $n$-dimensional
internal vielbein and $B$ as a background $2$-form for the
$n$-dimensional internal Kalb-Ramond field $b$. Whenever the
background is of this form, we will refer to it as geometric (notice
that this still does not determine completely the background, which
receives deformations from scalar fluctuations). In this case the
gaugings take the simple form \bea
f_{abc} &=& 3 (e^{-1})^\alpha{}_{[a}(e^{-1})^\beta{}_{b}(e^{-1})^\gamma{}_{c]} \partial_{[\alpha} B_{\beta\gamma]}\ , \nn\\
f^a{}_{bc} &=& 2 (e^{-1})^\beta{}_{[b}(e^{-1})^\gamma{}_{c]}\partial_{\beta} e^a{}_{\gamma }\ , \nn\\
f^{ab}{}_c &=& f^{abc} = 0 \ .\eea

If we also turn on a  $\beta (y^a)$, the relation of $e$, $B$ and
$\beta$ with the internal $g$ and $b$ is less trivial, and typically
the background will be globally well defined up to O$(n,n)$
transformations mixing the metric and the two-form (this is typically called a T-fold). In this case,
we refer to the background as locally geometric but globally
non-geometric, and this situation formally allows for non-vanishing
$f^{ab}{}_c$ and $f^{abc}$. Finally, if the twist matrix is a
function of $\tilde y_a$, we refer to the background as locally
non-geometric. Notice however, that if it satisfies the weak and
strong constraints, one would always be able to rotate it to a frame
in which it is locally geometric, and would therefore belong to an
undoubled orbit.

Alternatively, one could also define the  {\it Cartesian}
parametrization of the twist matrix, by taking
 the metric of the (internal) global symmetry group to be of the form
\be \eta_{AB} = \begin{pmatrix} \mathds{1}_n & 0 \\ 0 &
-\mathds{1}_n\end{pmatrix} \ .\ee This formulation is related to the
light-cone parametrization through a SO$(2n)$ transformation, that
must also rotate the coordinates. In this case the relation between
the components of the twist matrix and the internal $g$ and $b$ is
non-trivial. We will consider the O($n,n$) twist
matrix to contain a smaller O($n-1,n-1$) matrix in the directions
$(y^2,...,y^n,\tilde y_2,...,\tilde y_n)$ fibred over the flat
directions $(y^1,\tilde y_1)$. We have seen that this typically
leads to constant gaugings.

Of course these are not the most general parametrizations and
ansatz, but they will serve our purposes of uplifting all the orbits
of half-maximal supergravity to DFT. Interesting works on how to
generate gaugings from twists are \cite{Dall'Agata:2007sr,
*Andriot:2009fp}.

\subsection{U-duality orbits of maximal supergravities}
\label{sec:U_Dualitites}

Following the previous discussion of DFT and its relevance for generating duality orbits, we turn to the actual classification of these. In particular, we start with orbits under U-duality of gaugings of maximal supergravity. Moreover, we will demonstrate that all such orbits do have a higher-dimensional supergravity origin.

Starting with the highest dimension for  maximal supergravity,
$D=11$, no known deformation is possible here. Moreover, in $D=10$
maximal supergravities, the only possible deformation occurs in what
is known as massive IIA supergravity\footnote{Throughout this section
we will not consider the trombone gaugings giving rise to theories
without an action principle, as discussed in
\emph{e.g.}~\cite{Howe:1997qt, Bergshoeff:2002nv, LeDiffon:2008sh,
LeDiffon:2011wt}.} \cite{Romans:1985tz}. It consists of a
St\"uckelberg-like way of giving a mass to the 2-form $B_{2}$.
Therefore, such a deformation cannot be interpreted as a gauging.
The string theory origin of this so-called Romans' mass parameter is
nowadays well understood as arising from D8-branes
\cite{Polchinski:1995mt}. Furthermore, its DFT uplift has been
constructed in ref.~\cite{Hohm:2011cp}. Naturally, the structure of
possible orbits becomes richer when going to lower dimensions. In what follows we will perform
the explicit classification in dimensions nine and eight.

\subsubsection{Orbits and origin of the $D=9$ maximal case}
\label{subsec:Max9}

\subsubsection*{Maximal $D=9$ gauged supergravity}

The maximal (ungauged) supergravity in $D=9$ \cite{Gates:1984kr} can
be obtained by reducing either massless type IIA or type IIB
supergravity in ten dimensions on a circle. The global symmetry
group of this theory is taken here to be
\be G_{0}\,=\,\mathbb{R}^{+}\,\times\,\textrm{SL}(2)\ . \notag\ee
Note that $G_{0}$ is the global symmetry of the action and hence it
is realized off-shell, whereas the on-shell symmetry has an extra
$\mathbb{R}^{+}$ with respect to which the Lagrangian has a
non-trivial scaling weight. This is normally referred to as the
\emph{trombone symmetry}. As a consequence, the on-shell symmetry
contains three independent rescalings \cite{Bergshoeff:2002nv,
Roest:2004pk}, which we summarize in Table~\ref{rescalings}.
\begin{table}[h!]
\begin{center}
\scalebox{1}[1]{
\begin{tabular}{ c  c  c  c  c  c  c  c  c  c c  c  c c}
\hline
\textrm{ID} & $e_{\mu}^{\phantom{\mu}a}$ & $A_{\mu}$ &  $A_{\mu}{}^{1}$ & $A_{\mu}{}^{2}$ & $B_{\mu\nu}{}^{1}$ & $B_{\mu\nu}{}^{2}$ & $C_{\mu\nu\rho}$ & $e^{\varphi}$ & $\chi$ & $e^{\phi}$ & $\psi_{\mu}$ & $\lambda\,,\,\tilde{\lambda}$ & $\mathcal{L}$  \\[1mm]
\hline \hline
$\alpha$ & $\frac{9}{7}$ & $3$ & $0$ & $0$ & $3$ & $3$ & $3$ & $\frac{6}{\sqrt{7}}$ & $0$ & $0$ & $\frac{9}{14}$ & $-\frac{9}{14}$ & $9$ \\[1mm]

$\beta$ & $0$ & $\frac{1}{2}$ & $-\frac{3}{4}$ & $0$ & $-\frac{1}{4}$ & $\frac{1}{2}$ & $-\frac{1}{4}$ & $\frac{\sqrt{7}}{4}$ & $-\frac{3}{4}$ & $\frac{3}{4}$ & $0$ & $0$ & $0$ \\[1mm]

$\gamma$ & $0$ & $0$ & $1$ & $-1$ & $1$ & $-1$ & $0$ & $0$ & $2$ & $-2$ & $0$ & $0$ & $0$ \\[1mm]

$\delta$ & $\frac{8}{7}$ & $0$ & $2$ & $2$ & $2$ & $2$ & $4$
& $-\frac{4}{\sqrt{7}}$ & $0$ & $0$ & $\frac{4}{7}$ & $-\frac{4}{7}$
&
$8$ \\[1mm] \hline
\end{tabular}
}
\end{center}
{\it \caption{The scaling weights of the nine-dimensional fields. As
already anticipated, only three rescalings are independent since
they are subject to the following constraint:
$8\alpha-48\beta-18\gamma-9\delta=0$. As the scaling weight of the
Lagrangian $\mathcal{L}$ shows, $\beta$ and $\gamma$ belong to the
off-shell symmetries, whereas $\alpha$ and $\delta$ can be combined
into a trombone symmetry and an off-shell
symmetry.}
\label{rescalings}}
\end{table}
The full field content consists of the following objects
(see also Chapter \ref{ch:gaugings_d9} for more details) which
arrange themselves into irreducible representations of
$\mathbb{R}^{+}\,\times\,\textrm{SL}(2)$:
\be \textrm{9D :}\qquad \underbrace{e_{\mu}^{\phantom{\mu}a}\,,\,A_{\mu}\,,\,A_{\mu}{}^{i}\,,\,B_{\mu\nu}{}^{i}\,,\,C_{\mu\nu\rho}\,,\,\varphi\,,\,\tau=\,\chi\,+\,i\,e^{-\phi}}_{\textrm{bosonic dof's}}\,\,\,;\,\underbrace{\psi_\mu\,,\,\lambda\,,\,\tilde{\lambda}}_{\textrm{fermionic dof's}}\ ,\label{fields_9D} \ee%
where $\mu,\nu,\ldots$ denote 9-dimensional curved spacetime,
$a,b,\ldots$ 9-dimensional flat spacetime and $i,j,\cdots$
fundamental $SL(2)$ indices respectively.

The general deformations of this theory have been studied in detail in 
Chapter \ref{ch:gaugings_d9} 
(see also ref.~\cite{FernandezMelgarejo:2011wx}), where both embedding
tensor deformations and gaugings of the trombone symmetry have been
considered. For the present scope we shall restrict ourselves to the
first ones. The latter ones would correspond to the additional mass
parameters $m_{\textrm{IIB}}$ and $(m_{11},m_{\textrm{IIA}})$ in
refs~\cite{Bergshoeff:2002nv, FernandezMelgarejo:2011wx}, which give
rise to theories without an action principle.

The vectors of the theory $\{A_{\mu}\,,\,{A_{\mu}}^{i}\}$ transform
in the $V^\prime\,=\,\textbf{1}_{(+4)}\,\oplus\,\textbf{2}_{(-3)}\,$
of $\mathbb{R}^{+}\,\times\,\textrm{SL}(2)\,$, where the
$\mathbb{R}^{+}$ scaling weights are included as well\footnote{The
$\mathbb{R}^{+}$ factor in the global symmetry is precisely the
combination
$\left(\frac{4}{3}\,\alpha\,-\,\frac{3}{2}\,\delta\right)\,$ of the
different rescalings introduced in ref.~\cite{Bergshoeff:2002nv}.}.
The resulting embedding tensor deformations live in the following
tensor product
\be \mathfrak{g}_{0}\otimes V=\textbf{1}_{(-4)}\,\oplus\, 2\,\cdot\,\textbf{2}_{(+3)}\,\oplus\,\textbf{3}_{(-4)}\,\oplus\,\textbf{4}_{(+3)}\ . \ee
The Linear Constraint (LC) projects out the $\textbf{4}_{(+3)}$, the
$\textbf{1}_{(-4)}$ and one copy of the $\textbf{2}_{(+3)}$ since
they would give rise to inconsistent deformations. As a consequence,
the consistent gaugings are parametrized by embedding tensor
components in the $\textbf{2}_{(+3)}\,\oplus\,\textbf{3}_{(-4)}$. We
will denote these allowed deformations by $\theta^{i}$ and
$\kappa^{(ij)}$.

The closure of the gauge algebra and the antisymmetry of the brackets impose the following Quadratic Constraints (QC)
\bea
\epsilon _{ij}\,\theta^{i}\,\kappa^{jk} &=&0\ ,\qquad\qquad
\textbf{2}_{(-1)}\label{quadratic constraints in 9D1}\\
\theta^{(i}\,\kappa^{jk)} &=&0\ .\qquad\qquad\,
\textbf{4}_{(-1)}\label{quadratic constraints in 9D2}
\eea

\subsubsection*{The $\mathbb{R}^{+}\,\times\,$SL($2$) orbits of
solutions to the QC}

The QC \eqref{quadratic constraints in 9D1} and \eqref{quadratic
constraints in 9D2} turns out to be very simple to solve; after
finding all the solutions, we studied the duality orbits,
\emph{i.e.} classes of those solutions which are connected via a
duality transformation. The resulting orbits of consistent gaugings
in this case are presented in Table~\ref{orbits_max9}.

\begin{table}[t!]
\begin{center}
\begin{tabular}{ c  c  c  c }
\hline
\textrm{ID} & $\theta^{i}$ & ${\kappa}^{ij}$ &  gauging \\[1mm]
\hline \hline
$1$ & \multirow{3}{*}{$(0,0)$} &  diag($1,1$) &  SO($2$) \\[1mm]
 $2$ & & diag($1,-1$) & SO($1,1$) \\[1mm]
 $3$ & & diag($1,0$) & $\mathbb{R}^{+}_{\gamma}$ \\[1mm]
\hline  $4$ & $(1,0)$ & diag($0,0$) & $\mathbb{R}^{+}_{\beta}$ \\[1mm]
\hline
\end{tabular}
\end{center}
{\it \caption{All the U-duality orbits of consistent gaugings in maximal
supergravity in $D=9$. For each of them, the simplest representative
is given. The subscripts $\beta$ and $\gamma$ refer to the
rescalings summarized in Table~\protect\ref{rescalings}.}
\label{orbits_max9}
}
\end{table}

\subsubsection*{Higher-dimensional geometric origin}

The four different orbits of maximal $D=9$ theory have the following higher-dimensional origin in terms of geometric compactifications  \cite{Bergshoeff:2002mb}:

\begin{itemize}

\item \textbf{Orbits 1 -- 3:}  These come from reductions of type IIB supergravity on a circle with an SL($2$) twist.

\item \textbf{Orbit 4:}  This can be obtained from a reduction of type IIA supergravity on a circle with the inclusion of an $\mathbb{R}^{+}_{\beta}$ twist.

\end{itemize}

\subsubsection{Orbits and origin of the $D=8$ maximal case}
\label{subsec:Max8}

\subsubsection*{Maximal $D=8$ gauged supergravity}

The maximal (ungauged) supergravity in $D=8$ \cite{Salam:1984ft} can
be obtained by reducing eleven-dimensional supergravity on a $T^3$.
The global symmetry group of this theory is
\be G_{0}\,=\,\textrm{SL}(2)\,\times\,\textrm{SL}(3)\ . \notag\ee
The full field content consists of the following objects which
arrange themselves into irrep's of
$\textrm{SL}(2)\,\times\,\textrm{SL}(3)$:
\be \textrm{8D :}\qquad \underbrace{e_{\mu}^{\phantom{\mu}a}\,,\,A_{\mu}{}^{\alpha m}\,,\,B_{\mu\nu m}\,,\,C_{\mu\nu\rho}\,,\,L_{m}^{\phantom{m}I}\,,\,\phi\,,\,\chi}_{\textrm{bosonic dof's}}\,\,\,;\,\underbrace{\psi_\mu\,,\,\chi_I}_{\textrm{fermionic dof's}}\ ,\label{fields_8D} \ee%
where $\mu,\nu,\cdots$ denote eight-dimensional curved spacetime,
$a,b,\cdots$ eight-dimensional flat spacetime, $m,n,\cdots$
fundamental SL($3$), $I,J,\cdots$ fundamental SO($3$) and
$\alpha,\beta,\cdots$ fundamental SL($2$) indices respectively. The
six vector fields $A_{\mu}{}^{\alpha m}$ in \eqref{fields_8D}
transform in the $V'=\left( \textbf{2},\textbf{3}^\prime\right)$.
There are eleven group generators, which can be expressed in the
adjoint representation $\mathfrak{g}_{0}$.

The embedding tensor $\Theta $ then lives in the representation
$\mathfrak{g} _{0}\,\otimes\,V$, which can be decomposed into
irreducible representations as
\be \mathfrak{g}_{0}\otimes V=2\,\cdot\left(
\textbf{2},\textbf{3}\right) \oplus \left(
\textbf{2},\textbf{6}^\prime\right) \oplus  \left(
\textbf{2},\textbf{15}\right) \oplus \left(
\textbf{4},\textbf{3}\right)\,. \ee
The LC restricts the embedding tensor to the
$\left( \textbf{2},\textbf{3}\right) \oplus \left(
\textbf{2},\textbf{6}^\prime\right) $ \cite{Weidner:2006rp}. It is
worth noticing that there are two copies of the $\left(
\textbf{2},\textbf{3} \right) $ irrep in the above composition; the
LC imposes a relation between them
\cite{Samtleben:2008pe}. This shows that, for consistency, gauging
some SL($2$) generators implies the necessity of gauging some
SL($3$) generators as well. Let us denote the allowed embedding
tensor irrep's by $\xi _{\alpha m}$ and $f_{\alpha}{}^{(mn)}$
respectively.

The quadratic constraints (QC) then read \cite{Dani:2008,
deRoo:2011fa}
\bea
\epsilon ^{\alpha\beta}\,\xi _{\alpha p}\xi _{\beta q} &=&0
\text{ ,}\qquad\qquad
\left( \textbf{1},\textbf{3}^\prime\right)\label{quadratic constraints in 8D1}\\%
f_{(\alpha}{}^{np}\xi _{\beta)p} &=&0 \text{ ,} \qquad\qquad \left(
\textbf{3},\textbf{3}^\prime\right)\label{quadratic constraints in 8D2}\\%
\epsilon ^{\alpha\beta}\left(\epsilon
_{mqr}f_{\alpha}{}^{qn}f_{\beta}{}^{rp}+f_{\alpha}{}^{np}\xi _{\beta
m}\right)
&=&0\text{ .} \qquad\left( \textbf{1},\textbf{3}^\prime\right)\oplus\left( \textbf{1},\textbf{15}\right)\label{quadratic constraints in 8D3} %
\eea
Any solution to the QC \eqref{quadratic constraints in 8D1},
\eqref{quadratic constraints in 8D2} and \eqref{quadratic
constraints in 8D3} specifies a consistent gauging of a subgroup of
SL($2)\,\times\,$SL($3$) where the corresponding generators are
given by
\bea \label{gauge_gen_D=8} {\left(X_{\alpha
m}\right)_{\beta}}^{\gamma} &=& \delta_{\alpha}^{\gamma}\,\xi_{\beta
m}\,-\,\frac{1}{2}\,\delta_{\beta}^{\gamma}\,\xi_{\alpha m}\ ,\\
{\left(X_{\alpha m}\right)_{n}}^{p} &=&
\epsilon_{mnq}\,{f_{\alpha}}^{qp}\,-\,\frac{3}{4}\,\left(\delta_{m}^{p}\,\xi_{\alpha
n}\,-\,\frac{1}{3}\,\delta_{n}^{p}\,\xi_{\alpha m}\right)\ . \eea

\subsubsection*{The SL($2$)$\,\times\,$SL($3$) orbits of solutions to
the QC}

We exploited an algebraic geometry tool called the
Gianni-Trager-Zacharias (GTZ) algorithm \cite{GTZ}. This algorithm
has been computationally implemented by the \textsc{\,Singular\,}
project \cite{DGPS} and it consists in the primary decomposition of
ideals of polynomials. After finding all the solutions to the QC by
means of the algorithm mentioned above, one has to group together
all the solutions which are connected through a duality
transformation, thus obtaining a classification of such solutions in
terms of duality orbits. The resulting orbits of consistent
gaugings\footnote{Recently, also the possible vacua of the different
theories have been analyzed \cite{deRoo:2011fa}. It was found that
only  {\bf orbit 3} has maximally symmetric vacua.} in this case are
presented in Table~\ref{orbits_max8}.

\begin{table}[t!]
\begin{center}
\begin{tabular}{ c  c  c  c  c  c }
\hline
\textrm{ID} & ${f_{+}}^{mn}$ & ${f_{-}}^{mn}$ & $\xi_{+m}$ & $\xi_{-m}$ &  gauging \\[1mm]
\hline \hline
$1$ & diag($1,1,1$) & \multirow{5}{*}{$\textrm{diag}(0,0,0)$} & \multirow{5}{*}{$(0,0,0)$} & \multirow{5}{*}{$(0,0,0)$} &  SO($3$) \\[1mm]
 $2$ & diag($1,1,-1$) & & & &  SO($2,1$) \\[1mm]
 $3$ & diag($1,1,0$) & & & &  ISO($2$) \\[1mm]
 $4$ & diag($1,-1,0$) & & & &  ISO($1,1$) \\[1mm]
 $5$ & diag($1,0,0$) & & & &  CSO($1,0,2$) \\[1mm]
\hline $6$ & diag($0,0,0$) & diag($0,0,0$) & $(1,0,0)$ &
$(0,0,0)$ & Solv$_{2}\,\times\,$Solv$_{3}$\\[1mm]
\hline $7$ & diag($1,1,0$) &
\multirow{3}{*}{$\textrm{diag}(0,0,0)$} & \multirow{3}{*}{$(0,0,1)$}
& \multirow{3}{*}{$(0,0,0)$} & \multirow{3}{*}{Solv$_{2}\,\times\,$Solv$_{3}$}\\[1mm]
 $8$ & diag($1,-1,0$) & & & &   \\[1mm]
 $9$ & diag($1,0,0$) & & & &   \\[1mm]
\hline $10$ & diag($1,-1,0$) & \scalebox{0.7}[0.7]{$\left(\begin{array}{ccc}1 & 1 & 0\\
1 & 1 & 0\\ 0 & 0 & 0\end{array}\right)$} & $\frac{2}{9}(0,0,1)$ &
$(0,0,0)$ & Solv$_{2}\,\times\,$SO$(2)\,\ltimes\,$Nil$_{3}(2)$\\[1mm]

\hline
\end{tabular}
\end{center}
{\it \caption{All the U-duality orbits of consistent gaugings in maximal
supergravity in $D=8$. For each of them, the simplest representative
is given. We denote by  Solv$_{2}\,\subset\,$SL($2$) and
Solv$_{3}\,\subset\,$SL($3$) a solvable algebra of dimension 2 and 3
respectively. To be more precise, Solv$_{2}$ identifies the Borel
subgroup of SL($2$) consisting of $2\times 2$ upper-triangular
matrices. Solv$_{3}$, instead, is a Bianchi type V algebra.}
\label{orbits_max8}
}
\end{table}

\subsubsection*{Higher-dimensional geometric origin}

\begin{itemize}
\item \textbf{Orbits 1 -- 5:}  These stem from reductions of eleven-dimensional supergravity on a three-dimensional group manifold of type  A in the Bianchi classification \cite{Bergshoeff:2003ri}. The special case in orbit 1 corresponds to a reduction over an SO($3$) group manifold and it was already studied in ref.~\cite{Salam:1984ft}.

\item \textbf{Orbit 6:}  This can be obtained from a reduction of maximal nine-dimensional supergravity on a circle with the inclusion of an  $\mathbb{R}^{+}$ twist inside the global symmetry group.

\item \textbf{Orbits 7 -- 9:}  These can come from the same reduction from $D=9$ but upon inclusion of a more general $\mathbb{R}^{+}\,\times\,\textrm{SL}(2)$ twist.

\item \textbf{Orbit 10:}  This orbit seems at first sight more complicated to be obtained from a dimensional reduction owing to its non-trivial SL($2$) angles. Nevertheless, it turns out that one can land on this orbit by compactifying type IIB supergravity on a circle with an SL($2$) twist and then further reducing on another circle with $\mathbb{R}^{+}\,\times\,\textrm{SL}(2)$ twist given by the residual little group leaving invariant the intermediate nine-dimensional deformation.
\end{itemize}

\subsection*{Remarks on the $D=7$ maximal case}
\label{subsec:Max7}

The general deformations of the maximal theory in $D=7$ are
constructed and presented in full detail in
ref.~\cite{Samtleben:2005bp}. For the present aim we only summarize
here a few relevant facts.

The global symmetry group of the theory is SL($5$). The vector
fields $A_{\mu}{}^{MN}=A_{\mu}{}^{[MN]}$ transform in the
\textbf{10}$^\prime$ of SL($5$), where we denote by $M$ a
fundamental SL($5$) index. The embedding tensor $\Theta$ takes
values in the following irreducible components
\be
\textbf{10}\otimes\textbf{24}\,=\,\textbf{10}\oplus\textbf{15}\oplus\textbf{40}^\prime\oplus
\textbf{175}\,.\ee
The LC restricts the embedding tensor to the $
\textbf{15}\,\oplus\,\textbf{40}^\prime$, which can be parametrized
by the following objects
\be Y_{(MN)}\,,\qquad\textrm{and}\qquad
Z^{[MN],P}\quad\textrm{with}\quad Z^{[MN,P]}=0\ . \ee
The generators of the gauge algebra can be written as follows
\be
{\left(X_{MN}\right)_P}^Q\,=\,\delta_{[M}^Q\,Y_{N]P}\,-\,2\,\epsilon_{MNPRS}\,Z^{RS,Q}\,,
\label{gen_max}\ee
or, identically, if one wants to express them in the $\textbf{10}$,
\be
{\left(X_{MN}\right)_{PQ}}^{RS}\,=\,2\,{\left(X_{MN}\right)_{[P}}^{[R}\,\,\delta_{Q]}^{S]}\,.
\label{gen_max10}\ee
The closure of the gauge algebra and the antisymmetry of the
brackets imply the following QC
\be
Y_{MQ}\,Z^{QN,P}\,+\,2\,\epsilon_{MRSTU}\,Z^{RS,N}\,Z^{TU,P}\,=\,0\
, \label{QC_max7}\ee
which have different irreducible pieces in the
$\textbf{5}^\prime\,\oplus\,\textbf{45}^\prime\,\oplus\,\textbf{70}^\prime$.
Unfortunately, in this case, both the embedding tensor deformations
and the quadratic constraints reach a level of complexity that makes an exhaustive and general analysis difficult.
 Such analysis lies beyond the scope of our work.

\subsection{T-duality orbits of half-maximal supergravities}
\label{sec:T_Dualitites}

After the previous section on maximal supergravities, we turn our
attention to theories with half-maximal supersymmetry. In
particular, in this section we will classify the orbits under
T-duality of all gaugings of half-maximal supergravity. We will only
consider the theories with duality groups $\mathbb{R}^+ \times
\textrm{SO}(d,d)$ in $D= 10-d$, which places a restriction on the
number of vector multiplets. For these theories we will classify all
duality orbits, and find a number of non-geometric orbits.
Furthermore, we demonstrate that double field theory does yield a
higher-dimensional origin for all of them.

Starting from $D=10$ half-maximal supergravity without vector
multiplets, it can be seen that there is no freedom to deform this
theory, rendering this case trivial. In $D=9$, instead, we have the
possibility of performing an Abelian gauging inside
$\mathbb{R}^{+}\,\times\,$SO($1,1$), which will depend on one
deformation parameter. However, this is precisely the parameter that
one expects to generate by means of a twisted reduction from $D=10$.
This immediately tells us that non-geometric fluxes do not yet
appear in this theory. In order to find the first non-trivial case,
we will have to consider the $D=8$ case.

\subsubsection{Orbits and origin of the $D=8$ half-maximal case}
\label{subsec:Half_Max8}

\subsubsection*{Half-maximal $D=8$ gauged supergravity}

Half-maximal supergravity in $D=8$ is related to the maximal theory
analyzed in the previous section by means of a $\mathbb{Z}_{2}$
truncation. The action of such a $\mathbb{Z}_{2}$ breaks
$\textrm{SL}(2)\times\textrm{SL}(3)$ into
$\mathbb{R}^{+}\times\textrm{SL}(2)\times\textrm{SL}(2)$, where
$\textrm{SL}(2)\times\textrm{SL}(2)=\textrm{O}(2,2)$ can be
interpreted as the T-duality group in $D=8$ as shown in
Table~\ref{dualities}. The embedding of
$\mathbb{R}^{+}\times\textrm{SL}(2)$ inside SL($3$) is unique and it
determines the following branching of the fundamental representation
\bea
\textbf{3}\,\,&\longrightarrow&\,\,\textbf{1}_{(+2)}\,\oplus\,\textbf{2}_{(-1)}\ ,\notag\\
m\,&\longrightarrow&\,\,(\bullet\,,\,i)\ ,\notag \eea
where the $\mathbb{R}^{+}$ direction labeled by $\bullet$ is parity
even, whereas $i$ is parity odd, such as the other SL($2$) index
$\alpha$. In the following we will omit all the $\mathbb{R}^{+}$ weights since they do not play any role in the truncation.

The embedding tensor of the maximal theory splits in the following
way
\bea
(\textbf{2},\textbf{3})\,\,&\longrightarrow&\,\,\xcancel{(\textbf{2},\textbf{1})}\,\oplus\,(\textbf{2},\textbf{2})\ ,\notag\\
(\textbf{2},\textbf{6}^{\prime})\,\,&\longrightarrow&\,\,\xcancel{(\textbf{2},\textbf{1})}\,\oplus\,(\textbf{2},\textbf{2})\,\oplus\,\xcancel{(\textbf{2},\textbf{3})}\
,\notag \eea
where all the crossed irrep's are projected out because of
$\mathbb{Z}_2$ parity. This implies that the consistent embedding
tensor deformations of the half-maximal theory can be described by
two objects which are doublets with respect to both SL($2$)'s. Let
us denote them by $a_{\alpha i}$ and $b_{\alpha i}$. This statement
is in perfect agreement with the Ka\v{c}-Moody analysis performed in
ref.~\cite{Bergshoeff:2007vb}. The explicit way of embedding
$a_{\alpha i}$ and $b_{\alpha i}$ inside $\xi _{\alpha m}$ and
$f_{\alpha}{}^{mn}$ is given by
\bea
{f_{\alpha}}^{i\bullet}&=&{f_{\alpha}}^{\bullet i}\,=\,\epsilon^{ij}\,a_{\alpha j}\ , \label{ET_Half_Max81}\\[2mm]
\xi _{\alpha i}&=&4\,b _{\alpha i}\ .\label{ET_Half_Max82} \eea

The QC given in \eqref{quadratic constraints in 8D1},
\eqref{quadratic constraints in 8D2} and \eqref{quadratic
constraints in 8D3} are decomposed according to the following
branching
\bea
(\textbf{1},\textbf{3}^{\prime})\,\,&\longrightarrow&\,\,(\textbf{1},\textbf{1})\,\oplus\,\xcancel{(\textbf{1},\textbf{2})}\ ,\notag\\
(\textbf{3},\textbf{3}^{\prime})\,\,&\longrightarrow&\,\,(\textbf{3},\textbf{1})\,\oplus\,\xcancel{(\textbf{3},\textbf{2})}\ ,\notag\\
(\textbf{1},\textbf{15})\,\,&\longrightarrow&\,\,(\textbf{1},\textbf{1})\,\oplus\,\xcancel{2\,\cdot\,(\textbf{1},\textbf{2})}\,\oplus\,2\,\cdot\,(\textbf{1},\textbf{3})\,\oplus\,\xcancel{(\textbf{1},\textbf{4})}\
.\notag\eea
As a consequence, one expects the set of $\mathbb{Z}_{2}$ even QC to
consist of 3 singlets, a $(\textbf{3},\textbf{1})$ and 2 copies of
the $(\textbf{1},\textbf{3})$. By plugging \eqref{ET_Half_Max81} and
\eqref{ET_Half_Max82} into \eqref{quadratic constraints in 8D1},
\eqref{quadratic constraints in 8D2} and \eqref{quadratic
constraints in 8D3}, one finds
\bea \epsilon ^{\alpha\beta}\,\epsilon^{ij}\,b_{\alpha i}\,b_{\beta
j}&=&0\ ,\qquad\qquad
\left(\textbf{1},\textbf{1}\right)\label{QC_Half_Tot_81}\\
\epsilon ^{\alpha\beta}\,\epsilon^{ij}\,a_{\alpha i}\,b_{\beta
j}&=&0\ ,\qquad\qquad
\left(\textbf{1},\textbf{1}\right)\label{QC_Half_Tot_82}\\
\epsilon ^{\alpha\beta}\,\epsilon^{ij}\,a_{\alpha i}\,a_{\beta
j}&=&0\ ,\qquad\qquad
\left(\textbf{1},\textbf{1}\right)\label{QC_Half_Tot_83}\\
\epsilon^{ij}\,a_{(\alpha i}\,b_{\beta) j}&=&0\ ,\qquad\qquad
\left(\textbf{3},\textbf{1}\right)\label{QC_Half_Tot_84}\\
\epsilon ^{\alpha\beta}\,a_{\alpha (i}\,b_{\beta j)}&=&0\
.\qquad\qquad
\left(\textbf{1},\textbf{3}\right)\label{QC_Half_Tot_85} \eea
With respect to what we expected from group theory, we seem to be
finding a $(\textbf{1},\textbf{3})$ less amongst the even QC. This
could be due to the fact that $\mathbb{Z}_{2}$ even QC can be
sourced by quadratic expressions in the odd embedding tensor
components that we truncated away. After the procedure of turning
off all of them, the two $(\textbf{1},\textbf{3})$'s probably
collapse to the same constraint or one of them vanishes directly.

The above set of QC characterizes the consistent gaugings of the
half-maximal theory which are liftable to the maximal theory, and
hence they are more restrictive than the pure consistency
requirements of the half-maximal theory. In order to single out only
these we need to write down the expression of the gauge generators
and impose the closure of the algebra. The gauge generators in the
$(\textbf{2},\textbf{2})$ read
\be \label{Gen_Half_max8} {\left(X_{\alpha i}\right)_{\beta
j}}^{\gamma k} =
\frac{1}{2}\,\delta^{\gamma}_{\beta}\,\epsilon_{ij}\,\epsilon^{kl}\,a_{\alpha
l} \,+\, \delta^{\gamma}_{\alpha}\,\delta^{k}_{j}\,b_{\beta i} \,-\,
\frac{3}{2}\,\delta^{\gamma}_{\beta}\,\delta^{k}_{i}\,b_{\alpha j}
\,+\,
\frac{1}{2}\,\delta^{\gamma}_{\beta}\,\delta^{k}_{j}\,b_{\alpha i}
\,+\, \epsilon_{\alpha \beta}\,\epsilon^{\gamma
\delta}\,\delta^{k}_{j}\,b_{\delta i}\ . \ee
The closure of the algebra generated by \eqref{Gen_Half_max8}
implies the following QC
\bea \epsilon ^{\alpha\beta}\,\epsilon^{ij}\,\left(a_{\alpha
i}\,a_{\beta j}\,-\,b_{\alpha i}\,b_{\beta j}\right)&=&0\
,\qquad\qquad
\left(\textbf{1},\textbf{1}\right)\label{QC_Half_81}\\
\epsilon ^{\alpha\beta}\,\epsilon^{ij}\,\left(a_{\alpha i}\,b_{\beta
j}\,+\,b_{\alpha i}\,b_{\beta j}\right)&=&0\ ,\qquad\qquad
\left(\textbf{1},\textbf{1}\right)\label{QC_Half_82}\\
\epsilon^{ij}\,a_{(\alpha i}\,b_{\beta) j}&=&0\ ,\qquad\qquad
\left(\textbf{3},\textbf{1}\right)\label{QC_Half_83}\\
\epsilon ^{\alpha\beta}\,a_{\alpha (i}\,b_{\beta j)}&=&0\
.\qquad\qquad \left(\textbf{1},\textbf{3}\right)\label{QC_Half_84}
\eea

To facilitate the mapping of gaugings $a_{\alpha i}$ and $b_{\alpha i}$ with the more familiar $f_{ABC}$ and $\xi_A$ in the DFT language, we have written a special section in the appendix \ref{appendix_B}. The mapping is explicitly given in \eqref{X2f_D=8}.

\subsubsection*{The O($2,2$) orbits of solutions to the QC}

After solving the QC given in \eqref{QC_Half_81},
\eqref{QC_Half_82}, \eqref{QC_Half_83} and \eqref{QC_Half_84} again
with the aid of \textsc{\,Singular\,}, we find a 1-parameter
family of T-duality orbits plus two discrete ones. The results are
all collected in Table~\ref{orbits_half_max8}.

\begin{table}[t!]
\begin{center}
\scalebox{1}[1]{
\begin{tabular}{ c  c  c  c }
\hline
\textrm{ID} & $a_{\alpha i}$ & $b_{\alpha i}$ & gauging \\[1mm]
\hline \hline
$1$ & diag($\,\cos\alpha,0$) & diag($\,\sin\alpha,0$) &  Solv$_{2}\,\times\,$SO($1,1$) \\[1mm]
\hline 
$2$ & diag($1,1$) & diag($-1,-1$) &  \multirow{2}{*}{SL$(2)\,\times\,$SO($1,1$)} \\[1mm]
$3$ & diag($1,-1$) & diag($-1,1$) &  \\[1mm]
\hline
\end{tabular}
}
\end{center}
{\it \caption{All the T-duality orbits of consistent gaugings in
half-maximal supergravity in $D=8$. For each of them, the simplest
representative is given. Solv$_{2}$ refers again to the solvable
subgroup of SL($2$) as already explained in the caption of
Table~\ref{orbits_max8}.} 
\label{orbits_half_max8}
}
\end{table}

\subsubsection*{Higher-dimensional geometric origin}

The possible higher-dimensional origin of the three different orbits is as follows:

\begin{itemize}

\item \textbf{Orbit 1:} This orbit can be obtained by performing a
two-step reduction of type I supergravity. In the first step, by
reducing a circle, we can generate an
$\mathbb{R}^{+}\,\times\,$SO($1,1$) gauging of half-maximal $D=9$
supergravity. Subsequently, we reduce such a theory again on a
circle with the inclusion of a new twist commuting with the previous
deformation. Also, these orbits include a non-trivial $\xi_A$ gauging, so we will not address it from a DFT perspective.

\item \textbf{Orbits 2 -- 3:} These do not seem to have any obvious geometric
higher-dimensional origin in supergravity. In fact, they do not
satisfy the extra constraints \eqref{Extra_f}, so one can only hope
to reproduce them from truly doubled twist orbits in DFT.

\end{itemize}
Therefore we find that, while the half-maximal orbits in $D = 9$ all have a known geometric higher-dimensional origin, this is not the case for the latter two orbits in $D = 8$. We have finally detected the first signals of
non-geometric orbits.

\subsubsection*{Higher-dimensional DFT origin}

As mentioned, the {\bf orbits 2} and {\bf 3} lack of a clear
higher-dimensional origin. Here we would like to provide a
particular twist matrix giving rise to these gaugings. We chose to
start in the Cartesian framework, and propose the following form for
the SO$(2,2)$ twist matrix \be U = \begin{pmatrix} 1& 0 & 0 & 0 \\ 0
&
\cosh (m\,y^1 + n \, \tilde y_1) & 0 & \sinh (m\,y^1 + n \, \tilde y_1) \\
0& 0& 1& 0\\ 0 & \sinh (m\,y^1 + n \, \tilde y_1) & 0 & \cosh (m\,y^1
+ n \, \tilde y_1) \end{pmatrix}\ .  \vspace{2mm}\ee

\noindent This is in fact an element of $\textrm{SO}(1,1)$ lying in the
directions ($\tilde y_2, y^2$), fibred over the double torus
($\tilde y_1 , y^1 $). Here, the coordinates are written in the
Cartesian formulation, so we must rotate this in order to make
contact with the light-cone case.

For this twist matrix, the weak and strong constraints in the
light-cone formulation read $(m+n) (m-n) = 0$, while the QC are
always satisfied. The gaugings are constant, and when written in
terms of $a_{\alpha i}$ and $b_{\alpha i}$ we find \be a_{\alpha i}
= - b_{\alpha i } ={\rm diag} \left(-\frac{m+n}{2\,\sqrt{2}} ,\
\frac{m-n}{2\,\sqrt{2}}\right)\ , \ee so {\bf orbit 2} is obtained
by choosing $m = 0$, $n = -2\,\sqrt{2}$, and {\bf orbit 3} by
choosing $m = - 2\,\sqrt{2}$, $n = 0$. Notice that in both cases the
twist orbit is truly doubled, so we find the first example of an
orbit of gaugings without a clear supergravity origin, that finds an
uplift to DFT in a truly doubled background.

\subsubsection{Orbits and origin of the $D=7$ half-maximal case}
\label{subsec:Half_Max7}

\subsubsection*{Half-maximal $D=7$ gauged supergravity}

A subset of half-maximal gauged supergravities is obtained from the
maximal theory introduced in Section~\ref{subsec:Max7} by means of a
$\mathbb{Z}_2$ truncation. Thus, we will in this section perform
this truncation and carry out the orbit analysis in the half-maximal
theory. As we already argued before, this case is not only simpler,
but also much more insightful from the point of view of
understanding T-duality in gauged supergravities and its relation to
DFT.

The action of our $\mathbb{Z}_2$ breaks\footnote{The $\mathbb{Z}_2$
element with respect to which we are truncating is the following
USp($4)\,=\,$SO($5$) element \be
\alpha\,=\,\left(\begin{array}{cc}\mathds{1}_{2} & 0\\ 0 &
-\mathds{1}_{2}\end{array}\right) \notag\ee projecting out half of
the supercharges.} SL($5$) into $\mathbb{R}^+\,\times\,$SL($4$). Its
embedding inside SL($5$) is unique and it is such that the
fundamental representation splits as follows
\be
\textbf{5}\,\,\longrightarrow\,\,\textbf{1}_{(+4)}\,\oplus\,\textbf{4}_{(-1)}\,.
\ee
After introducing the following notation for the indices in the $\mathbb{R}^+$ and in the SL($4$) directions
\be M\,\,\longrightarrow\,\,(\,\diamond\,,\,m)\,, \ee
we assign an even parity to the $\diamond$ direction and odd parity
to $m$ directions.

The embedding tensor of the maximal theory splits according to
\bea
\textbf{15}&\longrightarrow & \textbf{1}\,\oplus\,\xcancel{\textbf{4}}\,\oplus\,\textbf{10}\,,\\[2mm]
\textbf{40}^\prime&\longrightarrow &\xcancel{\textbf{4}^\prime}\,\oplus\,\textbf{6}\,\oplus\,\textbf{10}^\prime\,\oplus\,\xcancel{\textbf{20}}\,,
\eea
where again, as in Section~\ref{subsec:Half_Max8}, all the crossed
irrep's are projected out because of $\mathbb{Z}_2$ parity. This
implies that the embedding tensor of the half-maximal theory lives
in the
$\textbf{1}\,\oplus\,\textbf{6}\,\oplus\,\textbf{10}\,\oplus\,\textbf{10}^\prime$
and hence it is described by the following objects
\be
\theta\,\,,\,\,\xi_{[mn]}\,\,,\,\,M_{(mn)}\,\,,\,\,\tilde{M}^{(mn)}\,.
\label{Theta_half} \ee
This set of deformations agrees with the decomposition
$\textrm{D}_8^{+++}\,\rightarrow\,\textrm{A}_3\,\times\,\textrm{A}_6$
given in ref.~\cite{Bergshoeff:2007vb}. The objects in
\eqref{Theta_half} are embedded in $Y$ and $Z$ in the following way
\bea
Y_{\diamond\,\diamond}&=&\theta\,, \label{ExprY1}\\[2mm]
Y_{mn}&=&\frac{1}{2}\,M_{mn}\,,\label{ExprY2}\\[3mm]
Z^{mn,\,\diamond}&=&\frac{1}{8}\,\xi^{mn}\,,\label{ExprZ1}\\[2mm]
Z^{m\,\diamond,n}&=&-Z^{\diamond\,m,n}\,=\,\frac{1}{16}\,\tilde{M}^{mn}\,+\,\frac{1}{16}\,\xi^{mn}\,,\label{ExprZ2}
\eea
where for convenience we defined $\xi^{mn}\,=\, \frac{1}{2}\,\epsilon^{mnpq}\,\xi_{pq}$.

Now we will obtain the expression of the gauge generators of the
half-maximal theory by plugging the expressions \eqref{ExprY1} --
\eqref{ExprZ2} into \eqref{gen_max}. We find
\be
{\left(X_{mn}\right)_p}^q\,=\,\frac{1}{2}\,\delta_{[m}^q\,M_{n]p}\,-\,\frac{1}{4}\,\epsilon_{mnpr}\,\left(\tilde{M}\,+\,
\xi\right)^{rq}\, \,, \label{gen_half-max}\ee
which extends the expression given in  ref.~\cite{Roest:2009tt} by
adding an antisymmetric part to $\tilde{M}$ proportional to $\xi$.
Note that the $\xi$ term is also the only one responsible for the
trace of the gauge generators which has to be non-vanishing in order
to account for $\mathbb{R}^+$ gaugings.

The presence of such a term in the expression \eqref{gen_half-max}
has another consequence: the associated structure constants that one
writes by expressing the generators in the $\textbf{6}$
${\left(X_{mn}\right)_{pq}}^{rs}$ will not be automatically
antisymmetric in the exchange between $mn$ and $pq$. This implies
the necessity of imposing the antisymmetry by means of some extra
QC\footnote{The QC which
ensure the antisymmetry of the gauge brackets are given by \\
${\left(X_{mn}\right)_{pq}}^{rs}\,X_{rs}\,+\,(mn\,\leftrightarrow\,pq)\,=\,0$,
where $X$ is given in an arbitrary representation.}.

The QC of the maximal theory are branched into
\bea
\textbf{5}^\prime&\longrightarrow & \textbf{1}\,\oplus\,\xcancel{\textbf{4}^\prime}\,,\\[2mm]
\textbf{45}^\prime&\longrightarrow &\xcancel{\textbf{4}}\,\oplus\,\textbf{6}\,\oplus\,\textbf{15}\,\oplus\,\xcancel{\textbf{20}}\,,\\[2mm]
\textbf{70}^\prime&\longrightarrow &\textbf{1}\,\oplus\,\xcancel{\textbf{4}}\,\oplus\,\xcancel{\textbf{4}^\prime}\,\oplus\,\textbf{10}^\prime\,\oplus\,\textbf{15}\,\oplus\,\xcancel{\textbf{36}^\prime}\,.
\eea
By substituting the expressions \eqref{ExprY1} -- \eqref{ExprZ2}
into the QC \eqref{QC_max7}, one finds
\bea
\theta\,\xi_{mn}&=&0\,,\qquad \,\,\,\,\,\,(\textbf{6}) \label{theta_xi}\\[2mm]
\left(\tilde{M}^{mp}\,+\,  \xi^{mp}\right)\, M_{pq}&=&0\,,\qquad ( \textbf{1}\,\oplus\,\textbf{15})\label{Q_Qtilde}\\[2mm]
M_{mp}\,\xi^{pn}\,-\,\xi_{mp}\,\left(\tilde{M}^{pn}\,+\,\xi^{pn}\right)&=&0\,,\qquad ( \textbf{1}\,\oplus\,\textbf{15})\label{Q_xi}\\[2mm]
\theta\,\tilde{M}^{mn}&=&0\,.\qquad \,\,\,\,\,(\textbf{10}^\prime)
\label{theta_Qtilde} \eea
Based on the Ka\v{c}-Moody analysis performed in
ref.~\cite{Bergshoeff:2007vb}, the QC constraints of the
half-maximal theory should only impose conditions living in the
$\textbf{1}\,\oplus\,\textbf{6}\,\oplus\,\textbf{15}\,\oplus\,\textbf{15}$.
The problem is then determining which constraint in the $\textbf{1}$
is already required by the half-maximal theory and which is not.

By looking more carefully at the constraints \eqref{theta_xi} --
\eqref{theta_Qtilde}, we realize that the traceless part of
\eqref{Q_Qtilde} exactly corresponds to the Jacobi identities that
one gets from the closure of the algebra spanned by the generators
\eqref{gen_half-max}, whereas the full \eqref{Q_xi} has to be
imposed to ensure antisymmetry of the gauge brackets. Since there is
only one constraint in the  $\textbf{6}$, we do not have ambiguities
there\footnote{We would like to stress that the parameter $\theta$
within the half-maximal theory is a consistent deformation, but it
does not correspond to any gauging and hence QC involving it cannot
be derived as Jacobi identities or other consistency constraints
coming from the gauge algebra.}.

We are now able to write down the set of QC of the half-maximal theory:
\bea
\theta\,\xi_{mn}&=&0\,,\qquad \,(\textbf{6}) \label{QC1}\\[2mm]
\left(\tilde{M}^{mp}\,+\,  \xi^{mp}\right)\, M_{pq}\,-\,\frac{1}{4}\,\left(\tilde{M}^{np}\,M_{np}\right)\,\delta_q^m&=&0\,,\qquad (\textbf{15})\label{QC2}\\[2mm]
M_{mp}\,\xi^{pn}\,+\,\xi_{mp}\,\tilde{M}^{pn}&=&0\,,\qquad (\textbf{15})\label{QC3}\\[2mm]
\epsilon^{mnpq}\,\xi_{mn}\,\xi_{pq}&=&0\,.\qquad \,\,\,(
\textbf{1})\label{QC4} \eea
We are not really able to confirm whether (\ref{QC1}) is part of the
QC of the half-maximal theory, in the sense that there appears a
top-form in the \textbf{6} from the $\textrm{D}_{8}^{+++}$
decomposition but it could either be a tadpole or a QC. This will
however not affect our further discussion, in that we only consider
orbits of gaugings in which $\theta=0$. The extra QC required in
order for the gauging to admit an uplift to maximal supergravity are
\bea
\tilde{M}^{mn}\,M_{mn}&=&0\,,\qquad \,(\textbf{1}) \label{extra1}\\[2mm]
\theta\,\tilde{M}^{mn}&=&0\,.\qquad (\textbf{10}^\prime)
\label{extra2} \eea

\subsubsection*{The O($3,3$) orbits of solutions to the QC in the
$\textbf{10}\,\oplus\,\textbf{10}^\prime$}

The aim of this section is to solve the constraints summarized in
\eqref{QC1}, \eqref{QC2}, \eqref{QC3} and \eqref{QC4}. We will start
by considering the case of gaugings only involving the
$\textbf{10}\,\oplus\,\textbf{10}^\prime$. This restriction is
motivated by flux compactification, as we will try to argue later
on.

The only non-trivial QC are the following
\be
\tilde{M}^{mp}\,M_{pn}-\frac{1}{4}\left(\tilde{M}^{pq}\,M_{pq}\right)\,\delta^m_n\,=\,0\,,
\label{QCQQtilde}\ee
which basically implies that the matrix product between $M$ and
$\tilde{M}$, which in principle lives in the
$\textbf{1}\,\oplus\,\textbf{15}$, has to be pure trace. We made use
of a GL($4$) transformation in order to reduce $M$ to pure
signature; as a consequence, the QC \eqref{QCQQtilde} imply that
$\tilde{M}$ is diagonal as well \cite{Dibitetto:2010rg}. This
results in a set of eleven 1-parameter orbits\footnote{We would like
to point out that the extra discrete generator $\eta$ of O($3,3$)
makes sure that, given a certain gauging with $M$ and $\tilde{M}$,
it lies in the same orbit as its partner with the role of $M$ and
$-\tilde{M}$ interchanged.} of solutions to the QC which are given
in Table~\ref{orbits_halfmax7}.

\begin{table}[t!]
\begin{center}
\scalebox{.93}[1]{
\begin{tabular}{ c  c  c  c  c }
\hline
\textrm{ID} & $M_{mn}/\,\cos\alpha\,$ & $\tilde{M}^{mn}/\,\sin\alpha\,$ & range of $\alpha$ & gauging \\[1mm]
\hline \hline
$1$ & diag($1,1,1,1$) & diag($1,1,1,1$) & $-\frac{\pi}{4}\,<\,\alpha\,\le\,\frac{\pi}{4}$ & $\left\{\begin{array}{cc}\textrm{SO}($4$)\ , & \alpha\,\ne\,\frac{\pi}{4}\ ,\\ \textrm{SO}(3)\ , & \alpha\,=\,\frac{\pi}{4}\ .\end{array}\right.$\\[4mm]

$2$ & diag($1,1,1,-1$) & diag($1,1,1,-1$) & $-\frac{\pi}{4}\,<\,\alpha\,\le\,\frac{\pi}{4}$ & SO($3,1$)\\[1mm]

$3$ & diag($1,1,-1,-1$) & diag($1,1,-1,-1$) & $-\frac{\pi}{4}\,<\,\alpha\,\le\,\frac{\pi}{4}$ & $\left\{\begin{array}{cc}\textrm{SO}($2,2$)\ , & \alpha\,\ne\,\frac{\pi}{4}\ ,\\ \textrm{SO}(2,1)\ , & \alpha\,=\,\frac{\pi}{4}\ .\end{array}\right.$\\[2mm]
\hline 
$4$ & diag($1,1,1,0$) & diag($0,0,0,1$) & $-\frac{\pi}{2}\,<\,\alpha\,<\,\frac{\pi}{2}$ & ISO($3$)\\[1mm]

$5$ & diag($1,1,-1,0$) & diag($0,0,0,1$) & $-\frac{\pi}{2}\,<\,\alpha\,<\,\frac{\pi}{2}$ & ISO($2,1$)\\[1mm]
\hline
$6$ & diag($1,1,0,0$) & diag($0,0,1,1$) & $-\frac{\pi}{4}\,<\,\alpha\,\le\,\frac{\pi}{4}$ & $\left\{\begin{array}{cc}\textrm{CSO}(2,0,2)\ , & \alpha\,\ne\,\frac{\pi}{4}\ ,\\ \mathfrak{f}_{1}\quad(\textrm{Solv}_{6}) \ , & \alpha\,=\,\frac{\pi}{4}\ .\end{array}\right.$\\[4mm]

$7$ & diag($1,1,0,0$) & diag($0,0,1,-1$) & $-\frac{\pi}{2}\,<\,\alpha\,<\,\frac{\pi}{2}$ & $\left\{\begin{array}{cc}\textrm{CSO}(2,0,2)\ , & |\alpha|\,<\,\frac{\pi}{4}\ ,\\ \textrm{CSO}(1,1,2)\ , & |\alpha|\,>\,\frac{\pi}{4}\ ,\\ \mathfrak{g}_{0}\quad(\textrm{Solv}_{6}) \ , & |\alpha|\,=\,\frac{\pi}{4}\ .\end{array}\right.$\\[4mm]

$8$ & diag($1,1,0,0$) & diag($0,0,0,1$) & $-\frac{\pi}{2}\,<\,\alpha\,<\,\frac{\pi}{2}$ & $\mathfrak{h}_{1}\quad(\textrm{Solv}_{6})$\\[1mm]

$9$ & diag($1,-1,0,0$) & diag($0,0,1,-1$) & $-\frac{\pi}{4}\,<\,\alpha\,\le\,\frac{\pi}{4}$ & $\left\{\begin{array}{cc}\textrm{CSO}(1,1,2)\ , & \alpha\,\ne\,\frac{\pi}{4}\ ,\\ \mathfrak{f}_{2}\quad(\textrm{Solv}_{6}) \ , & \alpha\,=\,\frac{\pi}{4}\ .\end{array}\right.$\\[4mm]

$10$ & diag($1,-1,0,0$) & diag($0,0,0,1$) & $-\frac{\pi}{2}\,<\,\alpha\,<\,\frac{\pi}{2}$ & $\mathfrak{h}_{2}\quad(\textrm{Solv}_{6})$\\[1mm]
 \hline 
 $11$ & diag($1,0,0,0$) & diag($0,0,0,1$) & $-\frac{\pi}{4}\,<\,\alpha\,\le\,\frac{\pi}{4}$ & $\left\{\begin{array}{cc}\mathfrak{l}\quad(\textrm{Nil}_{6}(3)\,)\ , & \alpha\,\ne\,0\ ,\\
\textrm{CSO}(1,0,3)\ , &
\alpha\,=\,0\ .\end{array}\right.$\\[4mm]
\hline
\end{tabular}
}
\end{center}
{\it \caption{All the T-duality orbits of consistent gaugings in
half-maximal supergravity in $D=7$. Any value of $\,\alpha\,$
parametrizes inequivalent orbits. More details about the
non-semisimple gauge algebras $\mathfrak{f}_{1}$,
$\mathfrak{f}_{2}$, $\mathfrak{h}_{1}$, $\mathfrak{h}_{2}$,
$\mathfrak{g}_{0}$ and $\mathfrak{l}$ are given in
appendix~\ref{appendix_A}.} 
\label{orbits_halfmax7}
}
\end{table}

As we will see later, some of these consistent gaugings in general
include non-zero non-geometric fluxes, but at least in some of these
cases one will be able to dualize the given configuration to a
perfectly geometric background.

\subsubsection*{Higher-dimensional geometric origin}

Ten-dimensional heterotic string theory compactified on a $T^3$
gives rise to a half-maximal supergravity in $D=7$ where the
SL($4$)$\,=\,$SO($3,3$) factor in the global symmetry of this theory
can be interpreted as the T-duality group. The set of generalized
fluxes which can be turned on here is given by
\be \left\{f_{abc},\,{f_{ab}}^c,\,{f_a}^{bc},\,f^{abc}\right\}\equiv \left\{H_{abc},\,{\omega_{ab}}^c,\,{Q_a}^{bc},\,R^{abc}\right\}\
, \label{Fluxes}\ee
where $a,b,c\,=\,1,2,3$.

These are exactly the objects that one obtains by decomposing a
three-form of SO($3,3$) with respect to its GL($3$) subgroup. The
number of independent components of the above fluxes (including
traces of $\omega$ and $Q$) amounts to $1+9+9+1\,=\,20$, which is
the number of independent components of a three-form of SO($3,3$).
Nevertheless, the three-form representation is not irreducible since
the Hodge duality operator in 3+3 dimensions squares to 1. This
implies that one can always decompose it in a self-dual (SD) and
anti-self-dual (ASD) part
\be \textbf{10}\,\oplus\,\textbf{10}^\prime\quad\textrm{of
SL}(4)\quad\longleftrightarrow\quad\textbf{10}_{\textrm{SD}}\,\oplus\,\textbf{10}_{\textrm{ASD}}\quad\textrm{of
SO}(3,3)\ , \ee
such that the matching between the embedding tensor deformations
$(M_{mn},\,\tilde{M}^{mn})$ and the generalized fluxes given in
\eqref{Fluxes} now perfectly works. The explicit mapping between
vectors of SO($3,3$) expressed in light-cone coordinates and
two-forms of SL($4$) can be worked out by means of the SO($3,3$) 't
Hooft symbols $\left(G_A\right)^{mn}$ (see
Appendix~\ref{appendix_B}). This gives rise to the following
dictionary between the $M$ and $\tilde{M}$-components and the fluxes
given in \eqref{Fluxes}
\be
M\,=\,\textrm{diag}\,\left(H_{123},\,{Q_1}^{23},\,{Q_2}^{31},\,{Q_3}^{12}\right)\
,\quad
\tilde{M}\,=\,\textrm{diag}\,\left(R^{123},\,{\omega_{23}}^1,\,{\omega_{31}}^2,\,{\omega_{12}}^3\right)\
.\label{dictionary}\ee

The QC given in equations \eqref{QC1}-\eqref{QC4} enjoy a symmetry
in the exchange
\be (M,\,\xi)\,\overset{\eta}{\leftrightarrow}\,(-\tilde{M},\,-\xi)\
. \label{triple_duality}\ee
The discrete $\mathbb{Z}_2$ transformation $\eta$ corresponds to the
following O($3,3$) element with determinant $-1$
\be \eta\,=\,\left(
\begin{array}{cc}
0 & \mathds{1}_3\\
\mathds{1}_3 & 0
\end{array}\right)\ ,
\label{eta}\ee
which can be interpreted as a triple T-duality exchanging the three
compact coordinates $y^{a}$ with the corresponding winding
coordinates $\tilde{y}_{a}$ in the language of DFT.

Now we have all the elements to analyze the higher dimensional origin
of the orbits classified in Table~\ref{orbits_halfmax7}.

\begin{itemize}

\item \textbf{Orbits 1 -- 3:}  These gaugings are non-geometric for
every $\alpha\ne 0$; for $\alpha =0$, they correspond to coset
reductions of heterotic string theory. See \emph{e.g.} the $S^{3}$
compactification in ref.~\cite{Cvetic:2000dm} giving rise to the
SO($4$) gauging. This theory was previously obtained in
ref.~\cite{Salam:1983fa} as $\mathcal{N}=2$ truncation of a maximal
supergravity in $D=7$.

\item \textbf{Orbits 4 -- 5:} For any value of $\alpha$ we can
always dualize these representatives to the one obtained by means of a
twisted $T^{3}$ reduction with $H$ and $\omega$ fluxes.

\item \textbf{Orbits 6 -- 7:} For any $\alpha\ne 0$ these orbits could be obtained from supergravity compactifications on locally-geometric T-folds, whereas for $\alpha=0$ it falls again in a special
case of the reductions described for orbits 4 and 5.

\item \textbf{Orbits 8 -- 11:} For any value of $\alpha$, these
orbits always contain a geometric representative involving less
general $H$ and $\omega$ fluxes.

\end{itemize}

To summarize, in the half-maximal $D=7$ case, we encounter a number
of orbits which do not have an obvious higher-dimensional origin. To
be more precise, these are orbits 1, 2 and 3 for $\alpha\ne 0$. The
challenge in the next subsection will be to establish what DFT can
do for us in order to give these orbits a higher-dimensional origin.
Again, before reading the following subsections we refer to the
Section~\ref{subsec:twist_matrices} for a discussion of what we mean
by light-cone and Cartesian formulations.

\subsubsection*{Higher-dimensional DFT origin}

First of all we would like to show here how to capture the gaugings that only
involve (up to duality rotations) fluxes $H_{abc}$ and
${\omega_{ab}}^{c}$. For this, we start from the light-cone
formulation, and propose the following Ansatz for a {\it globally geometric twist} (involving $e$ and $B$ and physical coordinates $y$)
\bea
e &=& \begin{pmatrix}1 & 0& \frac{\omega_1}{\omega_3} \sin (\omega_1\,\omega_3\,y^2) \\ 0 & \cos(\omega_2\,\omega_3\,y^1) & -\frac{\omega_2}{\omega_3} \cos (\omega_1\,\omega_3\,y^2) \sin(\omega_2\,\omega_3\,y^1) \\ 0 & \frac{\omega_3}{\omega_2} \sin(\omega_2\,\omega_3\,y^1) & \cos (\omega_1\,\omega_3\,y^2) \cos(\omega_2\,\omega_3\,y^1)\end{pmatrix}\ ,\\
B&=& \begin{pmatrix}0& 0 & 0\\ 0& 0& H\,y^1\,\cos
(\omega_1\,\omega_3\,y^2) \\ 0 & -H\,y^1\,\cos
(\omega_1\,\omega_3\,y^2)& 0\end{pmatrix} \ , \\
\lambda &=& - \frac 12 \log(\cos (\omega_1\omega_3 y^2)) \ .\eea
This is far from being the most general ansatz, but it serves our
purposes of reaching a large family of geometric orbits. The
parameters $\omega_i$ can be real, vanishing or imaginary, since $U$
is real and well-behaved in these cases. The QC, weak and strong
constraints are all automatically satisfied, and the gaugings read
\be M = {\rm diag} (H\ ,\ 0\ ,\ 0\ ,\ 0) \ ,\ \ \ \ \tilde M = {\rm
diag} (0\ , \ \omega_1^2\ ,\ \omega_2^2\ ,\ \omega_3^2)\ . \ee From
here, by choosing appropriate values of the parameters the {\bf
orbits 4, 5, 8, 10} and {\bf 11} can be obtained. Indeed these are
geometric as they only involve gauge and (geo)metric fluxes.

Secondly, in order to address the remaining orbits, we consider an SO($2,2$) twist $U_{4}$ embedded in O($3,3$) in the
following way
\be
U =
\begin{pmatrix}1 & 0 & 0 & 0\\ 0& A & 0 & B\\ 0& 0 & 1 & 0 \\ 0& C &
0 & D
\end{pmatrix}\ , \ \ \ \ \ \ U_{4} = \begin{pmatrix}A& B \\ C &
D\end{pmatrix}\ ,  \ \ \ \ \ \ \lambda =0\ . \ee
This situation is analog to the SO$(1,1)$ twist considered in the $D
= 8$ case, but with a more general twist. Working in the Cartesian formulation, one can define the generators
and elements of SO$(2,2)$ as \be [t_{IJ}]_K{}^L = \delta^L_{[I}
\eta_{J]K}\ , \ \ \ \ \  U_4 = \exp\left(t_{IJ} \phi^{IJ}\right)\ ,
\ee where the rotations are generated by $t_{12}$ and $t_{34}$, and
the boosts by the other generators. Also, we take $\phi^{IJ} =
\alpha^{IJ} y^1 + \beta^{IJ} \tilde y_1$ to be linear.

From the above $\textrm{SO}(2,2)$ duality element one can reproduce the following orbits employing a {\it locally geometric twist} (including $e$, $B$ and $\beta$ but only depending on $y$, usually referred to as a T-fold):
~
\begin{itemize}

\item {\bf Orbit 6}  can be obtained by taking
\be {\bf (6)} \ \ \ \alpha^{12} = - \beta^{12} = - \frac
1{\sqrt{2}}\,(\cos \alpha + \sin \alpha)   \ , \ \ \ \alpha^{34} =
-\beta^{34}= - \frac 1{\sqrt{2}}\,(\cos \alpha + \sin \alpha)\
.\nn\ee
and all other vanishing.

\item \textbf{Orbits 7} and {\textbf 9}  can be obtained by the following particular identifications
\be
\begin{array}{lclclc}
\phi^{14} = \phi^{23} & , & \phi^{12} = \phi^{34} & \textrm{and} &
\phi^{13} = \phi^{24} & .\end{array}\nn\ee
\be {\bf (7)} \ \ \ \alpha^{14} = - \beta^{14} = -
\frac{1}{\sqrt{2}}\,\sin \alpha \ , \ \ \ \alpha^{12} = - \beta^{12}
= - \frac{1}{\sqrt{2}}\,\cos \alpha \ , \ \ \ \alpha^{13} =
\beta^{13} =0 \ ,\nn\ee
\be {\bf (9)} \ \ \ \alpha^{14} = - \beta^{14} = -
\frac{1}{\sqrt{2}}\,\sin \alpha \ , \ \ \ \alpha^{12} = \beta^{12} =
0 \ , \ \ \ \alpha^{13} = \beta^{13} = - \frac{1}{\sqrt{2}}\,\cos
\alpha \ .\nn\ee
\end{itemize}
All these backgrounds satisfy both the weak and the strong
constraints  and hence they admit a locally geometric
description. This is in agreement with the fact that the simplest
representative of \textbf{orbits 6, 7} and {\textbf 9} given in
Table~\ref{orbits_halfmax7} contains $H$, $\omega$ and $Q$ fluxes
but no $R$ flux.

Finally, one can employ the same SO($2,2$) duality elements with
different identifications to generate the remaining orbits with a
{\it non-geometric twist} (involving both $y$ and $\tilde y$
coordinates): ~
\begin{itemize}

\item \textbf{Orbits 1, 3}  can be again obtained by considering an
SO($2)\,\times\,$SO($2$) twist  with arbitrary $\phi^{12}$ and
$\phi^{34}$:
\be {\bf (1)} \ \ \ \alpha^{12} = - 2\,\sqrt{2}\,(\cos \alpha + \sin
\alpha) \ , \ \ \ \beta^{34} = 2\,\sqrt{2}\,(\cos \alpha - \sin
\alpha)\ , \ \ \ \alpha^{34} = \beta^{12} = 0 \ ,\nn\ee
\be {\bf (3)} \ \ \ \alpha^{34} = - 2\,\sqrt{2}\,(\cos \alpha + \sin
\alpha) \ , \ \ \ \beta^{12} = 2\,\sqrt{2}\,(\cos \alpha - \sin
\alpha)\ , \ \ \ \alpha^{12} = \beta^{34} = 0 \ .\nn\ee
\item \textbf{Orbit 2} can be obtained by means of a different
SO($2,2$) twist built out of the two rotations and two boosts
subject to the following identification
\be
\begin{array}{lcl}
\phi^{14} = \phi^{23} & \textrm{, } & \phi^{12} = \phi^{34}\ .
\end{array}
\ee
\be {\bf (2)} \ \ \ \alpha^{14} = \beta^{12} = \frac
1{\sqrt{2}}\,(\cos \alpha - \sin \alpha) \ , \ \ \ \alpha^{12} = -
\beta^{14} = - \frac 1{\sqrt{2}}\,(\cos \alpha + \sin \alpha)\ .
\nn\ee
\end{itemize}
These backgrounds violate both the weak and the strong constraints for $\alpha \neq 0$.
This implies that these backgrounds are truly doubled and they do
not even admit a locally geometric description.

Finally, let us also give an example of degeneracy in twist orbits-space reproducing the same orbit of gaugings. The following twist
\be
\phi^{12} = \phi^{13} \ , \ \ \ \phi^{34} = \phi^{24} \ , \ \ \ \phi^{23} = \phi^{14} = 0
\ee
\be {\bf (6)} \ \ \ \alpha^{13} = - \frac{1}{\sqrt{2}}(\cos \alpha + \sin
\alpha) \ , \ \ \ \beta^{24} = \frac{1}{\sqrt{2}}(\cos \alpha - \sin
\alpha)\ , \ \ \ \alpha^{24} = \beta^{13} = 0 \ ,\nn\ee
also reproduces the {\bf orbit 6}, but in this case through a non-geometric twist. What happens in this case is that although the twist matrix does not satisfy the weak/strong constraints, the contractions in (\ref{Max_VS_Geom}) cancel.

\section{Concluding remarks}
\label{sec:conclusions}

In the research summarized in this chapter 
we have addressed  the notion of  non-geometry, by classifying the 
explicit orbits of consistent gaugings of different supergravity 
theories, and considering the possible higher-dimensional origins of 
these. 
The results turn out to be fundamentally different for the cases of 
U-duality orbits of maximal supergravities, and T-duality orbits of 
half-maximal theories.

In the former case we have managed to explicitly classify all U-duality 
orbits in dimensions $8 \leq D \leq 11$. This led to zero, one, four and 
ten discrete orbits in dimensions $D=11, 10,9$ and $8$, respectively, with different associated gauge groups. Remarkably, we have found that all of these orbits have a higher-dimensional origin via some geometric compactification, be it twisted reductions or compactifications on group manifolds or coset spaces. In our parlance, we have therefore found that all U-duality orbits are geometric. The structure of U-duality orbits is therefore dramatically different from the sketch of Figure \ref{pic:orbits} in the introduction. Although a full classification of all orbits in lower-dimensional cases becomes increasingly cumbersome, we are not aware of any examples that are known to be non-geometric. It could therefore hold in full generality that all U-duality orbits are necessarily geometric.

This is certainly not the case for T-duality orbits of gaugings of half-maximal supergravities. In this case, we have provided the explicit classification in dimensions $7 \leq D \leq 10$ (where in $D=7$ we have only included three-form fluxes). The numbers of distinct families of orbits in this case are zero, one, three and eleven in dimensions $D=10,9,8$ and $7$, respectively, which includes both discrete and one-parameter orbits. A number of these orbits do not have a higher-dimensional origin in terms of a geometric compactification. Such cases are {\bf orbits 2} and {\bf 3} in $D=8$ and {\bf orbits 1, 2} and {\bf 3} in $D=7$ for $\alpha\neq 0$. Indeed, these are exactly the orbits that do not admit an uplift to the maximal theory. As proven in Section~\ref{subsec:GDFT}, all such orbits necessarily violate the weak and/or strong constraints, and therefore need truly doubled backgrounds. Thus, the structure of T-duality orbits is very reminiscent of Figure \ref{pic:orbits} in the introduction. Given the complications that already arise in these simpler higher-dimensional variants, one can anticipate that the situation will be similar in four-dimensional half-maximal supergravity.

Fortunately, the formalism of double field theory seems tailor-made to
generate additional T-duality orbits of half-maximal supergravity. Building on the recent generalization of the definition of
double field theory \cite{Grana:2012rr}, we have demonstrated that
all T-duality orbits, including the non-geometric ones in $D=7,8$,
can be generated by a twisted reduction of double field theory. We
have explicitly provided duality twists for all orbits. For locally-geometric orbits the twists only depend on the physical
coordinates $y$, while for the non-geometric orbits these
necessarily also include $\tilde y$. Again, based on our exhaustive
analysis in higher-dimensions, one could conjecture that also in
lower-dimensional theories, all T-duality orbits follow from this
generalized notion of double field theory.

At this point we would like to stress once more that a given orbit of 
gaugings can be generated from different twist orbits. Therefore, there 
is a degeneracy in the space of twist orbits giving rise to a particular 
orbit of gaugings. 
Interestingly, as it is the case of {\bf orbit 6} in $D=7$  for instance, 
one might find two different twist orbits reproducing the same orbit of 
gaugings, one  violating weak and strong constraints, the other one 
satisfying both. 
Our notion of a locally
geometric orbit of gaugings is related to the existence of at least one 
undoubled background giving rise to it. However, this ambiguity seems to 
be peculiar of gaugings containing $Q$ flux. 
These can, in principle, be independently obtained by
either adding a $\beta$ but no $\tilde{y}$ dependence 
(locally geometric choice, usually called T-fold), or by including 
non-trivial $\tilde{y}$
dependence but no $\beta$ (non-geometric choice) \cite{Aldazabal:2011nj}.

Another remarkable degeneracy occurs for the case of semi-simple
gaugings, corresponding to {\bf orbits 1 -- 3} in $D=7$. For the
special case of $\alpha = 0$, we have two possible ways of
generating such orbits from higher-dimensions: either a coset
reduction over a sphere or analytic continuations thereof, or a
duality twist involving non-geometric coordinate dependence.
Therefore $d$-dimensional coset reductions seem to be equivalent to
$2d$-dimensional twisted torus reductions (with the latter in fact
being more general, as it leads to all values of $\alpha$).
Considering the complications that generally arise in proving the
consistency of coset reductions, this is a remarkable reformulation
that would be interesting to understand in more detail. Furthermore,
when extending the notion of double field theory to type II and
M-theory, this relation could also shed new light on the consistency
of the notoriously difficult four-, five- and seven-sphere
reductions of these theories.

Our results mainly focus on Scherk-Schwarz compactifications leading to gauged supergravities with vanishing $\xi_M$ fluxes. In addition, we have restricted to the NSNS sector and ignored $\alpha'$-effects.
Also, we stress once again that relaxing the strong and weak constraints
is crucial in part of our analysis. If we kept the weak constraint, typically the Jacobi identities would lead to backgrounds satisfying also the strong constraint\ \cite{Grana:2012rr}.
However, from a purely (double) field theoretical analysis the weak constraint is not necessary. 
A sigma model analysis beyond tori would help us to clarify the 
relation between DFT without the weak and strong constraints and 
string field theory on more general backgrounds. 
\renewcommand{\sp}[2]{\left\langle #1 | #2 \right\rangle}
\newcommand{\F}{\mathcal{F}}
\newcommand{\X}{\mathcal{X}}
\newcommand{\D}{\mathcal{D}}
\newcommand{\V}{{V}}
\newcommand{\K}{\mathcal{K}}
\newcommand{\R}{\mbox{$\mathcal{R}$}}
\newcommand{\I}{\mbox{$\mathcal{I}$}}
\newcommand{\M}{\mathcal{M}}
\newcommand{\Z}{\mathcal{Z}}
\newcommand{\Q}{\mathcal{Q}}
\newcommand{\G}{\mathcal{G}}
\renewcommand{\L}{\mathcal{L}}
\renewcommand{\H}{\mathcal{H}}
\newcommand{\SSzinf}{\mathcal{\hat Z_\infty^2}}
\newcommand{\Pp}{\text{P}_+}
\newcommand{\Pm}{\text{P}_-}
\newcommand{\Ppm}{\text{P}_\pm}
\newcommand{\Pmp}{\text{P}_\mp}
\newcommand{\sigmaad}{\sigma^\dagger}
\newcommand{\pf}[1]{\mathrm{Pf}\left(#1\right)}
\newcommand{\vecr}{\mathbf{x}}
\newcommand{\rot}{\mathbf{rot}}
\newcommand{\ssigmaad}{{\cal S}^\dagger}
\newcommand{\ssigma}{{\cal S}}
\newcommand{\atot}{\mbox{$A_\infty$}}
\newcommand{\ah}{\mbox{$S_h$}}

\chapter{Studies on $N=2$ extremal multicenter black holes}
\label{ch:bh}

In this chapter we present a systematic study of extremal, stationary,
multicenter black hole solutions in ungauged four dimensional
Einstein-Maxwell  $N=2$ supergravity theories minimally coupled to  scalars, i.e. theories with quadratic prepotentials.

We  show how it is possible to derive in a systematic and
straightforward way a fully analytic, explicit description
of the multicenter black holes, the attractor mechanism
and their properties  making an intensive use of
the  matrices ($\ssigma_N,\ssigma_F$) and
their adjoints with respect to the symplectic product
(to be defined here).
The compatibility   of these matrices with respect
to the symplectic product makes possible the definition of
an associated inner product for which these matrices are
unitary.  This unitarity suggests the decomposition of
the $2 n_v+2$ dimensional
($n_v$ the number of vector multiplets)
symplectic space into a subspace generated by the center charge vectors $q_a$, and their associated vectors $\ssigmaad q_a$ and
its orthogonal complement subspace.

In particular, this decomposition results useful for
understanding some questions related to multicenter black
holes, as the entropy increasing effects in the fragmentation
of a single center black holes into two o more centers, or
the extremality of the solutions by simple considerations of
the dimensions of each subspace.
The results presented here are easily extendable to
general prepotentials or even theories without them.

The study we are going to developed in this chapter is based on the work done in \cite{Fernandez-Melgarejo:2013ksa}.f

\section{Introduction}

In this chapter, we are interested in general, stationary, multicenter
black hole solutions in ungauged four dimensional $N=2$ supergravity
theories coupled to an arbitrary number of $N=2$ vector multiplets.
The action of the theory can be determined, in the framework of
special geometry, in terms of a  holomorphic
section $\Omega$ of the scalar manifold.
The set of field equations  and Bianchi identities
associated to the action  is invariant under the group
of symplectic transformations
$Sp(2n_v+2,\mathbb{R})$.
This group acts linearly on the section $\Omega$, becoming this
a symplectic vector which can be written as
$\Omega=(X^I,F_I),$ with $ I=(0,n_v)$.

Black hole solutions in $N=2$ $D=4$ supergravity have been
 extensively studied for a long term by now.\footnote{See, for example, refs. \cite{Ferrara:1995ih,Ferrara:1996um,Gibbons:1996af,Shmakova:1996nz,Ferrara:1997tw,
 Behrndt:1997ny,Sabra:1997dh,Sabra:1997kq,Bellucci:2007ds,Ceresole:2010hq,Andrianopoli:1996vr,Galli:2011fq}. Multicenter black holes have been treated in refs.
\cite{Liu:2000ah,Bellorin:2006xr,Bobev:2009zz,Bena:2009en,Bena:2008wt,Anninos:2011vn,Manschot:2011xc}.}

The values of the $n_v$ scalar fields constitute the moduli space of the theory.
A distinctive feature of many of these theories is that the, possibly
disconnected, black hole horizon acts as an attractor for the scalar
fields present in the spectrum. The values of the moduli at any of
the  horizon components does not depend on their asymptotic values,
but only on the symplectic vector of charge assigned to that horizon
component \cite{Ferrara:1996um,Bellucci:2007ds}.
The embedding of the duality group of the moduli space into the
symplectic group $Sp(2n_v+2,\mathbb{R})$ establishes, in general, a relation between
the upper and lower components of $\Omega$, $F_I=F_I(X^J)$.
In some cases, $F_I$ is the derivative of a single function,
the prepotential $F=F(X^J)$. The choice of a particular embedding
determines the full Lagrangian of the theory and
whether a prepotential exists \cite{Sabra:1996xg,Sabra:1996kw}.

In this chapter, we focus in general quadratic prepotentials.
These theories include the simplest examples of special
K\"ahler homogeneous manifolds, the
\begin{align}
\mathbb{C}P^n\equiv \frac{SU(1,n)}{U(1)\times SU(n)}
\end{align}
case.

%
These models correspond to Maxwell-Einstein
$N=2$ supergravities  minimally coupled to $n_v$ vector multiplets.
They lead to phenomenologically
interesting $N = 1$ minimally coupled supergravities
\cite{Ferrara:2012qp}.
Theories derived from particular examples of these quadratic prepotentials have been studied in detail.
The case $n_v=1$ corresponds to the
$SU(1,1)/U(1)$ axion-dilaton black hole
(see for example
\cite{Sabra:1996bk,Behrndt:1997fq} or
\cite{Ferrara:2010cw}) with
prepotential $ F=-i X^0X^1$.

The  aim of this study 
is the  explicit, detailed study of stationary
multicenter black hole solutions with any number of scalar fields,
the study of the properties of the bosonic field solutions
 and their global and local properties.
For this purpose, we make a systematic use of, some previously
well-known objects of the theory, the \emph{stabilization matrices},
and some new ones, their symplectic ``adjoints''.
These stabilization matrices,
named $\ssigma_F,\ssigma_N$ along this work, are related
to the vector kinetic matrix and the matrix of
second derivatives of
the prepotential. They are real $Sp(2 n_v+2,\mathbb{R})$ matrices,
isometries of the symplectic quadratic form,
connecting   the
real and imaginary parts of the  special geometry sections.
Their adjoints with respect to the symplectic product
$\ssigmaad_N,\ssigmaad_F$ are defined and shown
to lie inside the Lie algebra of the isometry group.
They are such that (for any $\ssigma=\ssigma_F,\ssigma_N$) $\ssigma+\ssigmaad=0$.
This property, together with $\ssigma^2=-\mathds{1}$, make these matrices
unitary with respect to the symplectic
product, $\ssigma \ssigmaad=\mathds{1}$.
We show how it is possible to derive or rederive again in a
systematic and straightforward way a
fully analytic, explicit description of the multicenter black holes
and their properties (attractor mechanism, central charge,
horizon areas, masses, \ldots) making an intensive use of
these stabilization matrices, their adjoints and the
algebraic properties of both.

As we will show, the properties of these matrices, specially their
symplectic unitarity property, suggest  the convenience of the
 separation of the $(2 n_v+2)$-dimensional symplectic space into a $2 n_a$ dimensional subspace
generated by the $n_a$ center charges $q_a$ and their associated
vectors $\ssigmaad q_a$ (or $\ssigma q_a$) and its orthogonal
complement subspace (possibly of dimension zero depending on the
number and on the linear dependency of center charge vectors).
For quadratic prepotentials, this separation into
``charge-longitudinal'' and ``transversal'' subspaces can be made
global by choosing $\ssigma=\ssigma_F$. A similar, but local,
scalar dependent separation  can be advantageously considered also
for generic prepotentials, or even theories without them.
The projection of any symplectic vector appearing in the theory (for example a subset of the charge vectors themselves or vectors
characterizing the black hole ansatz at infinity) in terms of these
new bases appears as a promising technique. The use of this
projection allows, in particular, the understanding of questions as
entropy increasing effects in the fragmentation of a single center
black holes into two o more centers, or the extremality of the
solutions, in terms, for example, of simple considerations of the
dimensions of each of the charge-longitudinal and transversal subspaces.

 Although we have focused in the study of minimally coupled
theories with quadratic prepotentials, the main techniques,
properties and expressions presented are extendable to
theories governed by general prepotentials or even theories without
them. 

This study is organized as follows. In Section \ref{ssection1}, we present a brief introduction of the Reissner-Nordstr\"om black hole and the concept of extremality. Section \ref{ssection2} treats the attractor mechanism. In  Section \ref{ssection3},
 we present some well-known basic aspects of $N=2$ $D=4$ supergravity theories and their formulation in terms
of special and symplectic geometry.
In Section \ref{ssection4},  we first introduce the matrices
$\ssigma_{N,F}$, stressing some of their known properties and deriving new ones. We also construct projective operators (as well as their corresponding symplectic adjoints) based on these matrices.
After the consideration of the attractor mechanism in terms of these
projectors, we enter in a full explicit description of multicenter
black hole solutions, their horizons and their asymptotic properties. This is done in Sections \ref{ssection5} and \ref{ssection6}.
We finally present Section \ref{ssection7}, which contains a summary and discussion of our work, as well as an outlook on further proposals.


\section{Reissner-Nordstr\"om: a window to extremality}
\label{ssection1}

In this section, we review some elementary properties of the
Reissner-Nordstr\"om black hole, laying stress on its extremal case.

Our starting point is the Einstein-Maxwell action
in 4 dimensions,
\begin{align}
 \label{eq:EM_action}
\mathcal{L}
&= \int d^4 x  \sqrt{g} \left (R -\frac{1}{4}F_{\mu\nu}F^{\mu\nu}\right )
\, ,
\end{align}
which allows charged black holes as solutions.
For the sake of simplicity, we  consider  a
 static and spherically symmetric metric ansatz.
The most general solution of the field equations
 satisfying these requirements is
\begin{align}
\label{eq:RN_ansatz2}
ds^2&=-e^{2U(r)}dt^2+e^{-2U(r)}\left[dr^2+r^2d\Omega^2\right]\, ,
\end{align}
where $d\Omega^2=d\theta+\sin^2\theta d\phi^2$ and $U(r)$ is the warp factor.
Imposing the same symmetry conditions on the Maxwell field,
the field strength 2-form is restricted to be
\begin{align}
F&=P\sin\theta d\theta\wedge d\phi+Q dt\wedge \frac{1}{r^2}dr
\, ,
\end{align}
where the constants $P$ and $Q$ can be interpreted as
 the magnetic and electric charges, respectively.
Solving the field equations  derived
from \eqref{eq:EM_action}, we get that the {\em Reissner-Nordstr\"om
metric} is given by \eqref{eq:RN_ansatz2} with warp factor
\begin{align}
e^{2U(r)}
&=1-\frac{2M}{r}+\frac{P^2+Q^2}{r^2}\, .
\end{align}
This solution enjoys a singularity at $r=0$ hidden by the horizons at $r=r_\pm$, which appear when the metric element vanishes ($e^{2U(r_\pm)}=0$),
\begin{align}
r_\pm&=M\pm\sqrt{M^2-(P^2+Q^2)}\, .
\end{align}
Both $r_\pm$ values are real when $M^2\ge P^2+Q^2$.
We define so-called \emph{extremality parameter} $c$ as
the
\begin{align}
c&=r_+-r_-=\sqrt{M^2-(P^2+Q^2)}\, ,
\end{align}
The properties of the solution depend on the values of
the mass and the electromagnetic charges.
For $c>0$ then the solution describes a {\em non-extremal }
black hole with two, interior and exterior, horizons.
Its surface gravity $\kappa_S$
and exterior horizon area are, respectively, given by
\begin{align}
A &= 4\pi (c+M)^2
\, ,
\\
\kappa_S &= \frac{4\pi c}{A}
\, .
\end{align}
The two horizons coincide
when  $c=0$ or, equivalently, when
\begin{align}
M^2
&=
P^2+Q^2
\, .
\end{align}
In this case,  the surface gravity
vanishes and the horizon area is given exclusively in
terms of the charge,
\begin{align}
A &= 4\pi (P^2+Q^2)^2
\, .
\end{align}
The kind of black hole that results for $c=0$ is called \emph{extremal black hole}.
In the case $c<0$, the event horizons disappear and the singularity
at $r=0$ becomes a naked singularity.
The Schwarzschild black hole and the Minkowski spacetime are
special cases for, respectively, $M>0, P=Q=0$ and
$M=P=Q=0$ values.

Let us focus on the \emph{extremal} case. By introducing a radial coordinate $v=r-M$, the metric can be expressed as
\begin{align}
\label{eq:RN_ansatz}
ds^2&=-\left(1+\frac{M}{v}\right)^{-2} dt^2+
\left(1+\frac{M}{v}\right)^{2}\left[dv^2+v^2d\Omega^2\right]
\, .
\end{align}
The horizon is now at $v=0$ and
the near-horizon metric for $v\rightarrow0$ is
\begin{align}
ds_{NH}^2&=-\frac{v^2}{M^2}dt^2+\frac{M^2}{v^2}dv^2+M^2d\Omega^2\, .
\label{eq611}
\end{align}
Defining a new coordinate $z= M^2/v$, this metric is rewritten as
\begin{align}
ds^2&=\frac{M^2}{z^2}\left(-dt^2+dz^2\right)+M^2d\Omega^2\, .
\end{align}
This line element describes the direct product of two manifolds, $AdS_2\times S^2$, where the $AdS$ scale $L$ and the radius of the sphere $r_S$ coincide, $L=r_S=M$.
This metric is the so-called \emph{Robinson-Bertotti metric} \cite{Bertotti:1959pf,Robinson:1959ev}.


The Reissner-Nordstr\"om solution can be expressed in
\emph{isotropic coordinates}. Under the transformation
\begin{align}
r
&=
\rho \left[\left(1+\tfrac{M}{2\rho }\right)^2 - \left(\tfrac{e}{2\rho }\right)^2 \right]
\, ,
\end{align}
where $e^2\equiv P^2+Q^2$, the metric element becomes
\begin{align}
ds^2 = -H_1(\rho)^2dt^2 +H_2(\rho)^2\left(d\rho ^2 + \rho ^2 d\theta ^2 + \rho ^2 \sin^2 \theta d\phi ^2 \right)
\, .
\end{align}
The functions $H_1(\rho)$ and $H_2(\rho)$ are
\begin{align}
H_1(\rho)
&=
 \frac{1-\left(\frac{M}{2\rho }\right)^2 +\left(\frac{e}{2\rho }\right)^2 }{1+\frac{M}{\rho}+\left(\frac{M}{2\rho}\right)^2 - \left(\frac{e}{2\rho}\right)^2 }
\, ,
\\
H_2(\rho)
&=
1+\frac{M}{\rho}+\left(\frac{M}{2\rho}\right)^2 - \left(\frac{e}{2\rho}\right)^2
\, ,
\end{align}

In the extremal limit $c=0$, or $M^2=P^2+Q^2$,
the coordinate transformation becomes linear,
$r=\rho+M$,
 and the metric results
\begin{align}
ds^2
&=
-H^{-2} dt^2+H^2 d{\mathbf x}^2
\, ,
\end{align}
where the factor
\begin{align}
H(\rho)=1+\frac{M}{\rho}
\end{align}
is an elementary solution of a Laplace
equation (in the 3-space $(\rho,\theta,\phi), \Delta_3H=0$).
This is the \emph{extremal Reissner-Nordstrom metric in isotropic form}.
It is not casual that the factor $H$ is a harmonic function,
since it can been understood as a special case of a more
general
family of solutions of Einstein-Maxwell action without
  spherical symmetry that we review on continuation.

The \emph{Majumdar-Papapetrou solution}
\cite{Majumdar:1947eu,Papaetrou:1947ib} can be considered
a non-spherical
generalization of the Reissner-Nordstr\"om solution.
Let us take the metric element
\begin{align}
ds^2&=-H^{-2}(\vecr ) dt^2+H^2(\vecr) d{\vecr}^2\, .
\end{align}
The
Einstein-Maxwell equations of motion
reduce to the following  equation of motion for the warp factor $H(\vecr)$
\begin{align}
\Delta_3 H&=0\, ,
\label{delta3}
\end{align}
with $\Delta_3$ the 3-dimensional Laplacian. The electromagnetic
field strength, solution to the Einstein-Maxwell equations, results
\begin{align}
F_{ti}&=-\cos\theta\partial_i H^{-1}
\, ,
\nonumber\\
F_{ij}&=\sin\theta\epsilon_{ijk}\partial^k H
\, .
\end{align}

The Laplace equation \eqref{delta3} is satisfied,
in particular, by  harmonic functions with an
(arbitrary) number of point singularities,
\begin{align}
H&\equiv e^{-U}=1+\sum_{n} \frac{m_n}{|\vecr-\vecr_n|}.
\label{eq119bb}
\end{align}
The charges $e_i$ inside a closed surface surrounding each
point singularity may be identified by computing the
flux of the electromagnetic field through the aforementioned surface.
They result equal to the residues $m_n$ at any of the
singularities of the function \eqref{eq119bb},
\begin{align}
e_i^2&=m_i^2\, ,
\end{align}
These types of geometries, with $m_n\geq 0,\forall n$,
have event horizons with spherical topology and
represent genuine black hole solutions
(in fact, they are
the only Majumdar-Papapetrou solutions\footnote{Not only that: also the only IWP
 solutions with that property.} with this property
\cite{Hartle:1972ya}).
In the case of a single point singularity solution, one recovers
the extremal Reissner-Nordstr\"om black hole.
In the general case,
they  can be seen as  an arbitrary, static,
configuration of
 single Reissner-Nordstr\"om black holes.
These \emph{multicenter}, static,  black hole solutions
are in static equilibrium with the gravitational and
the electrostatic forces cancelling each other.\footnote{Newtonian point charged particles can remain
in static equilibrium if all the charges share the same
sign and satisfy
$| e_i|=m_i$, no matter how arranged they are
\cite{Hartle:1972ya}.}
The solution can be seen as a simple example of BPS
configuration. In fact, this type of solutions
emerge as a BPS solutions
of $N=2$ supergravity \cite{Gibbons:1982fy,Mohaupt:2000mj}.


A further generalization to \emph{stationary} solutions
is possible. A class of stationary solutions
of the pure Einstein-Maxwell equations
are given by
\cite{Israel:1972vx,Perjes:1971gv}
\begin{align}
ds^2
&=
-(H\bar{H})^{-1} (dt+\omega)^2+H\bar{H} d{\vecr}^2
\, ,
\end{align}
where $\omega$ is a purely
spatial 1-form $\omega=\omega_i dx^i$ and
 $H=H(\vecr)$ is any complex solution to the
3-dimensional Laplace equation
\begin{align}
\Delta_3 H
&=0
\, .
\end{align}
The term
 $\omega$ is given by the equation
\begin{align}
\star_3 d\omega&= 2 \im{ \bar{H} d H}
\, .
\label{eq687b}
\end{align}
A integrability condition for this equation is given by
the complex Laplace equation above.
In particular, if $H$ is real (or purely imaginary) then
$d\omega=0$ and we can write, by the Poincar\'e
lemma, $\omega=d\lambda$.
If we do a translation on the time coordinate, $t\to t+\lambda$,  we  recover the
Majumdar-Papapetrou class of solutions.
These are the so-called \emph{Israel-Wilson-Perj\'es (IWP) solutions.}
Similarly, as we have seen before, particular solutions of
the Laplace equation with a finite number of
point singularities are given by
\begin{align}
H&\equiv e^{-U}=1+\sum_{n} \frac{m_n}{|\vecr-\vecr_n|}
\, ,
\label{eq255}
\end{align}
but in this case the parameters $m_n$ and $\vecr_n$ are allowed
to be complex.

For example \cite{Israel:1972vx}, the Kerr-Newman solution
with $M^2 = Q^2$  corresponds to a simple case of
\eqref{eq255} for the values  $n=1$, $m_1=M$ and $\vecr_1=(0,0,i a)$,
where $a, M\in\mathbb{R}$, the former related to the
form $\omega$.
One can consider a generalized solution with similarly defined real $m_n$ and arbitrary complex parameters $\vecr_n$.
The resulting metric will represent the field of a set of
arbitrarily spinning, charged Kerr-like particles in
neutral equilibrium.
The single point source  solution has a naked singularity
and no horizons. Superposition of a number of solutions
also generally results in naked singularities except in
some special cases for some concrete configuration of the
parameters. However, it can be shown that in this case, the
solution becomes static and reduces to the
Majumdar-Papapetrou class \cite{Hartle:1972ya}.

Another simple case of the IWP solution  includes a
single point singularity at $\vecr_1=0$ and $m_1=M+i N$,
it is given by ($r=| \vecr|$) \cite{Hartle:1972ya}
\begin{align}
H&=1+\frac{M+i N}{ r}.
\end{align}
The
1-form $\omega$
is given by (up to an additive constant)
\begin{align}
\omega
&=
\frac{2N}{r}\frac{\cos\theta-1}{\sin\theta} d\phi
\, .
\end{align}
This is proportional to $N$, so for $N=0$ we recover the static
case.
Let us take the generic case, with $n$ complex
quantities $m_i=M_i+i N_i$ and arbitrary real
parameters $\vecr_i$. At large distances the corresponding
function $H$ is given by
\begin{align}
H&
\sim
1+\sum_{n}\frac{ M_n+i N_n}{r}
\, .
\label{eq255b}
\end{align}
We note that for $r\to\infty$, the behavior
becomes that of a  Majumdar-Papapetrou solution if the
imaginary part of \eqref{eq255b} is zero, namely,
if the following condition is satisfied
\begin{align}
\sum_n N_n
&=
0
\, .
\end{align}

The only IWP solutions with point singularities which
represent black hole solutions
are those for which all the
imaginary parts, $N_i$, are null.
That is, they are nothing but the
Majumdar-Papapetrou static solutions
\cite{Chrusciel:2005ve} (see also \cite{Hartle:1972ya}).


\section{Black holes in SUGRA and the attractor mechanism}
\label{ssection2}

We will study now black hole solutions in
 gravity theories that contain gauge and scalar fields
(as for example it happens  in supergravity).
An important mechanism appears in these theories,
the {\em  attractor mechanism}.
This phenomenon was originally discovered for
BPS extremal black holes in $N=2$ supergravity
theories \cite{Ferrara:1995ih,Ferrara:1996dd,Ferrara:1997tw}.
The flow of the scalar fields towards the horizon exhibits the
feature of a gradient flow towards a fixed point,
which, in the supergravity case,
 is the minimum of a function related to the central
charge of the SUSY algebra.
Among other properties, a basic feature of the
attractor mechanism  is that the ADM mass
is minimized, for fixed values of the conserved charges
carried by the black hole,
when the scalar fields  are constant (they
take their attractor values through the spacetime).

In the last years, the attractor mechanism has been
investigated for extremal black holes
in non-supersymmetric
theories,  in theories beyond GR as $D=5$ Gauss-Bonnet gravity
 \cite{Anber:2007gk}, as well as for non-BPS extremal solutions
in $N\ge 1$ supersymmetric theories.

A generic Lagrangian describing
the bosonic sector of
$D=4$ supergravity
coupled to scalars and $n_v$ vector multiplets  is of the form
\begin{align}
\label{eq:N2_action}
S
&= \int d^4x e\left (
R\star\mathds{1}
-\frac{1}{2}g_{ij}(\phi)d \phi^i\wedge \star d\phi^j
+\frac{1}{4}\mathcal{I}_{\Lambda\Sigma}(\phi)F^\Lambda\wedge
F^{\Sigma}
+\frac{1}{4}\mathcal{R}_{\Lambda\Sigma}(\phi)F^\Lambda\wedge
\star F^{\Sigma}
\right )
\, ,
\end{align}
where $g_{ij}(\phi)$ is the metric of the scalar
$\sigma$-model,
and $\mathcal{I}_{\Lambda\Sigma}$ and $\mathcal{R}_{\Lambda\Sigma}$ describe the  couplings terms of the vector fields. In particular ,
$\mathcal{R}_{\Lambda\Sigma}$ is the generalization of
the $\theta$-angle terms in presence
of scalar and vector fields. We assume that there is no-scalar potential.

We are interested here in finding single center, static,
extremal,
spherically symmetric and charged black hole solutions.
In addition, we assume asymptotical flatness.
A suitable ansatz for such requirements is of the form
\begin{align}
\label{eq:RN_generic_metric}
ds^2&=-e^{2U}dt^2
+e^{-2U}\left(
	\frac{c^4}{\sinh^4(cz)}dz^2
	+\frac{c^2}{\sinh^2(cz)}d\Omega^2
	\right)
\, ,
\end{align}
where the constant $c$ is an  extremality parameter.
The $z$ coordinate
 runs from $z=-\infty$ (horizon) to $z=0$ (spatial infinity).
The unknown function $U=U(z)$ is such that
$\exp(-2U(z\to 0))=\exp(-2 M z)\to 1$
(asymptotic flatness).
At $z\to -\infty$ we require, in order to ensure a finite
horizon area ($c\not =0$),
\begin{eqnarray}
U(z\to -\infty)&=& c z
\, ,\nonumber\\
U'(z\to -\infty)&=& c
\, .
\end{eqnarray}

In the extremal limit $c\to 0$, we recover the metric
\begin{align}
\label{eq:RN_generic_metric2}
ds^2&=-e^{2U}dt^2
+e^{-2U}\left(\frac{1}{z^2}dz^2+\frac{1}{z^2}d\Omega^2\right)
\, .
\end{align}
In this case, the condition of having a finite horizon area
implies the boundary condition for the regime $z\to-\infty$,
\begin{eqnarray}
\exp\left ( -2 U\right )
&\to & \frac{A}{4\pi} z^2
\, .
\end{eqnarray}

A similar static, spherically
symmetric ansatz can be introduced
for the gauge fields.
Due to the structure of the
 couplings in \eqref{eq:N2_action} and
the dependence of $\mathcal{R}_{\Sigma\Lambda}$ and $\mathcal{I}_{\Sigma\Lambda}$ on the
moduli,  the Bianchi identities are
\begin{align}
dF^\Lambda&=0\, ,\nonumber\\
dG_\Lambda&=d\left(
	\mathcal{R}_{\Lambda\Sigma}F^\Sigma
	-\mathcal{I}_{\Lambda\Sigma}\star F^\Sigma
	\right)=0
\, ,
\end{align}
where the second equation defines the quantity $G$.
This set of equations remains invariant when performing a
symplectic rotation of the field strengths
 \cite{Gaillard:1981rj}.

Electric and magnetic conserved  charges can be defined
in terms of the field strengths and their duals,
\begin{align}
\frac{1}{4\pi}\int F^\Lambda&=
p^\Lambda
\, ,
&
\frac{1}{4\pi}\int G_\Lambda
&=
q_\Lambda\, .
\end{align}

We can introduce the pair of potentials $(A^\Sigma,A_\Sigma)$,
corresponding to the symplectic vector of 2-forms
 $(F^\Sigma,G_\Sigma)$ with the required symmetry:
\begin{align}
A^\Lambda
&=
\chi^\Lambda(r)dt-p^\Lambda\cos\theta d\phi
\, ,
\nonumber
\\
A_\Lambda
&=
\psi_\Lambda(r)dt
-q_\Lambda\cos\theta d\phi
\, .
\label{eq:A_ansatz}
\end{align}
The electric-magnetic duality relation imposes
the constraint 
\begin{align}
\chi'^\Lambda
&=
e^{2U}\mathcal{I}^{\Lambda\Sigma}(q_\Sigma-\mathcal{R}_{\Sigma\Gamma}p^\Gamma)
\, .
\end{align}

At this point, we can write the Einstein field equations
for the metric and the gauge field ansatze, \eqref{eq:RN_generic_metric} and \eqref{eq:A_ansatz} respectively. The equations of motion for the gauge fields may then be directly
solved.
 The  equations of motion for metric and scalar fields
simplify to the equations \cite{Ferrara:1997tw}
\begin{align}
 U''-e^{2U}V_{BH}&=0
 \, ,
\label{eq:eoms_attractor1}
\\
(U')^2+\frac{1}{2}g_{ij}{\phi^i}'{\phi^j}'-e^{2U}V_{BH}-c^2&=0
\, ,
\label{eq:eoms_attractor2}
\\
{\phi^i}''+\Gamma_{jk}{}^i{\phi^j}'{\phi^k}'-e^{2U}g^{ij}\partial_j V_{BH}&=0 \, ,
\label{eq:eoms_attractor3}
\end{align}
where $V_{BH}$ and $\Gamma_{jk}{}^i$ are scalar functions.

The non-linear system of second order differential
equations
\eqref{eq:eoms_attractor1}-\eqref{eq:eoms_attractor3}
is complemented by
the  asymptotic  boundary conditions for the metric
at infinity (flatness) and at the horizon
(finite area condition for $c\to 0$)
and, in principle,
by two initial or boundary conditions for each of the
scalar fields.
However, only one of these two theoretically possible conditions
for each of the scalars survives. This is due to the properties of this non-linear system,
the existence of the first order
constraint equation \eqref{eq:eoms_attractor2}
and the requirement of everywhere
regularity of the solutions.
In fact, in the extremal case
$c\to 0$ the value of the scalars and their
first derivatives will be fixed at the horizon.
Only the values of the scalars at infinity will remain as  free parameters
of the theory.
We will see more details in what follows.

The quantity $V_{BH}$ is the \emph{black hole potential},
which encodes the terms of the energy momentum tensor
corresponding to the vector fields that appear in
the Lagrangian \cite{Ferrara:1997tw}.
It can be written as
\begin{align}
V_{BH}&=-\frac{1}{2}Q^T\mathcal{M}Q\, ,
\end{align}
where $\mathcal{M}$ is a scalar-dependent matrix
(see Sections \ref{ssection3} and \ref{ssection4})
and $Q$ is a symplectic charge vector,
\begin{align}
Q
&=
\left(\begin{array}{c}
	p^\Lambda
	\\
	q_\Lambda
\end{array}\right)
\, .
\end{align}
By making use of the scalar matrix $\mathcal{M}$, we can rewrite the gauge field strengths in a covariant way as
\begin{align}
\left(\begin{array}{c}
	F
	\\
	G
\end{array}\right)
&=
e^{2U}
\tilde \Omega
\mathcal{M}
\left(\begin{array}{c}
	p^\Lambda
	\\
	q_\Lambda
\end{array}\right) dt\wedge dz
-\left(\begin{array}{c}
	p^\Lambda
	\\
	q_\Lambda
\end{array}\right)\sin\theta d\theta\wedge d\phi
\, ,
\end{align}
where $\tilde \Omega$ is the symplectic metric
\begin{align}
\tilde \Omega
&=
\left(\begin{array}{cc}
	0 & -\mathds{1}_{n_v}
	\\
	\mathds{1}_{n_v} & 0
\end{array}\right)
\, .
\end{align}


At this point, let us consider the possibility of
deriving the field equations above from an
effective Lagrangian. Let us take the following Lagrangian
\begin{align}
\label{eq:effective_action_attractor}
\mathcal{L}&=(U')^2+\frac{1}{2}g_{ij}{\phi^i}'{\phi^j}'
+e^{2U}V_{BH}(\phi)+c^2\, ,
\end{align}
which depends on the ``fields'' $U(z)$, $\phi(z)$.
 The Lagrangian does not explicitly depend on $z$, Noether's theorem implies that the effective `energy'
\begin{align}
\label{eq:constraint_attractor}
\mathcal{E}&\equiv
(U')^2+\frac{1}{2}g_{ij}{\phi^i}'{\phi^j}'-e^{2U}V_{BH}-c^2
\end{align}
is a constant.
The  Euler-Lagrange equations corresponding to the fields $U(z),\phi^i(z)$
agree with the equations
\eqref{eq:eoms_attractor1}
and
\eqref{eq:eoms_attractor3}, respectively.
However, the condition  \eqref{eq:eoms_attractor2}
has to be implemented by hand, as an additional constraint,
\begin{align}
\mathcal{E}
&=
0
\, .
\label{eq:constraint_attractorE}
\end{align}
Thus, this effective action plus the constraint $\mathcal{E}=0$ is equivalent to the system of differential equations.

From the behavior of the
constraint equation (\ref{eq:constraint_attractorE})
at spatial
infinity
(considering that $\exp(-2U(z\to 0))=\exp(-2 M z)\to 1$, we
get the following constraint between the black hole mass $M$,
the scalar charges\footnote{The scalar charges $\Sigma^i$ of the black hole
are defined by (at spatial infinity, $z\to 0$)
\begin{align*}
\phi^i
&=\phi^i_\infty+z \frac{\Sigma^i}{r}+\mathcal{O}(z^2)
\, .
\end{align*}
} 
and the potential at infinity:
\begin{align}
\label{eq:constraint_attractor2}
M^2+\frac{1}{2}g_{ij}{\Sigma^i}{\Sigma^j}'-V_{BH}(\phi^i_\infty)&=c^2
\, .
\end{align}


The equations
\eqref{eq:eoms_attractor1}-\eqref{eq:eoms_attractor3}
can be solved  by constant values
$\phi^i (z)={\tilde\phi}^i=\phi^i_\infty$
for the  scalar fields. This is possible
 if these values represent a critical point of the
effective potential, \emph{i.e.} if
\begin{align}
\label{eq:minimization_vBH2}
\partial_i V_{BH}({\tilde\phi}^{i},q,p)&=0
\, .
\end{align}
The black hole charges are the only parameters that
appear in \eqref{eq:minimization_vBH2}. Thus
the extremal points will be solved in terms of them,
\begin{align}
{\tilde\phi}^i&={\tilde\phi}^i(p,q)\, .
\end{align}
The value of BH potential at the
minimum is a constant given by
\begin{align}
{\tilde V}_{BH}&=V_{BH}({\tilde\phi}^i(p,q),p,q)\, .
\end{align}
The equations
 \eqref{eq:eoms_attractor1} and \eqref{eq:eoms_attractor2}
for the warp factor at the horizon can be directly solved
giving
\begin{align}
U''(z)&=(U'(z))^2\quad ,\\
U(z)&=-\log\left( r_H z\right)\, ,
\end{align}
where $r_H=\sqrt{{\tilde V}_{BH}}$.
This implies that the black hole entropy is
\begin{align}
S_{BH}&=\frac{A}{4}=\pi {\tilde V}_{BH}(q,p)\, .
\end{align}
%
If  we take the extremal case\footnote{Extremal black hole solutions in which
the scalar fields take
constant values are usually
 called {\em double-extreme} black holes.} $c=0$,
from equation (\ref{eq:constraint_attractor2}),
we get\footnote{Here, we assume the vanishing of the scalar charges, $\Sigma^i=0$.}
\begin{align}
\label{eq:constraint_attractor3}
M^2&=V_{BH}(\phi^i_\infty)= {\tilde V}_{BH}(q,p)\, .
\end{align}


Next, we will study general, non-constant scalar solutions
focusing in the extremal $c=0$ case.
The critical values of the effective black hole potential
represent possible attractor values for the moduli
scalars (provided positivity of the Hessian).
The moduli and their derivatives  will have the same value at the horizon, whilst their asymptotic values may be varied freely.
We will see how, in this general case, the same
 universal properties (as the  attractor
mechanism and  the area of extremal black holes)  can be
deduced only by demanding a regular behaviour of the geometry and the moduli near the horizon.

If the scalar fields and their derivatives do not blow up near
the horizon, the following
asymptotic expression
is valid for $z\to\infty$
\begin{eqnarray}
\phi^i &=& \tilde\phi^i_h+ \frac{a^i}{z}+\mathcal{O}(z^2)
\, .
\end{eqnarray}
By making use of this series and of the relation (for the
extremal $c\to 0$ case)
$ \exp(-2U)_h=\tfrac{z^2 A}{4\pi}$, the differential equations
\eqref{eq:eoms_attractor1}-\eqref{eq:eoms_attractor3} near the horizon
become, at leading orders,
\begin{align}
\frac{1}{z^2}-\frac{4\pi}{A z^2}V_{BH,h}
 &=0
 \, ,
 \\
\frac{1}{z^2}+ \frac{a_1}{z^4}-\frac{4\pi}{A z^2}V_{BH,h}
 +\mathcal{O}\left(\frac{1}{z^5}\right)
&=0
\, ,
\\
\frac{a_2}{z^3}+ \frac{a_3}{z^4}-\frac{4\pi}{A z^2}\left
(g^{ij}\partial_j V_{BH}\right)_h
 +\mathcal{O}\left(\frac{1}{z^5}\right)
&=0
\, ,
\end{align}
or, equivalently,
\begin{align}
\label{eq:eoms_attractor4}
1-\frac{4\pi}{A}V_{BH,h}&=0\, ,\\
a_1+z^2 \left(1-\frac{4\pi z^2}{A }V_{BH,h}\right)
 +\mathcal{O}\left(\frac{1}{z}\right)&=0\, ,\label{eq:eoms_attractor4b}\\
a_2 z+ a_3-z^2 \frac{4\pi}{A}\left
(g^{ij}\partial_j V_{BH}\right)_h
 +\mathcal{O}\left(\frac{1}{z}\right)
 &=0
 \, .
\label{eq:eoms_attractor4c}
\end{align}
with $a_1,a_2,a_3$ are arbitrary constants.
If we compare the coefficients order by order in $1/z$, we obtain that, in order to the solutions not blow up at the horizon,
the following conditions should be imposed:\footnote{It is assumed that the matrix
$g^{ij}(\phi)$ is invertible at the horizon.}
\begin{align}
A&=4\pi V_{BH,h}\, ,\\
\phi^i(z\to-\infty)&=0\, ,\\
\left (\frac{\partial V_{BH}}{\partial \phi^i}\right)_h&=0\, .
\end{align}
These conditions show
that the area of the horizon of extremal
black holes coincides with the area of the horizon of double-extremal black holes with the same values of charges
and is given by the value of $V_{BH}$ (\emph{cf.} \eqref{eq:eoms_attractor4}).
Moreover, the entropy of the black hole,  related
to the area of the horizon, will be determined by the charges.\footnote{For quantized charges this means, in particular,
that the entropy does not depend on any
continuous parameter}
In addition, we see
that the values of the moduli at the horizon can be considered
as free initial conditions, since they are given by the minimization
of the effective black hole potential
(\ref{eq:eoms_attractor4c}). That is,
the horizon is an attractor point
\cite{Ferrara:1995ih,Ferrara:1996dd,Ferrara:1996um,Strominger:1996kf}.



\subsubsection{Supergravity central charge and flow equations}

Let us consider in more detail the special case of $N=2$
supergravity, for which the scalar manifold is a special K\"ahler
manifold.
For $N=2$ theories, special geometry can be used and the
expressions are somehow simplified.\footnote{We will closely follow the notation of \cite{Ceresole:2007wx}.}
The black hole potential is given by
\begin{align}
V_{BH}&=|Z|^2+4g^{ij}\partial_i|Z|\bar\partial_j|Z|\, ,
\end{align}
where $Z$ is the central charge of the $N=2$ SUSY algebra. This central charge is defined
by the special K\"ahler geometry, as it is discussed in the next section. 
Thus, the Lagrangian \eqref{eq:effective_action_attractor} and the constraint \eqref{eq:constraint_attractor} become
\begin{align}
\mathcal{L}&=(U')^2+g_{ij}{\phi^i}'{\phi^j}'+e^{2U}(|Z|^2+4g^{ij}\partial_i|Z|\bar\partial_j|Z|)\, ,\\
(U')^2+g_{ij}{\phi^i}'{\phi^j}'&=e^{2U}(|Z|^2+4g^{ij}\partial_i|Z|\bar\partial_j|Z|)
\, .
\end{align}
The energy constraint is an equality between two different sums of squares  with the same weight $e^{2U}$. So, a possible
ansatz for the solution would be
\begin{align}
U'&=\pm e^U|Z|\, ,\nonumber\\
{\phi^i}'&=\pm 2e^{U} g^{i\bar \jmath}\bar\partial_{\bar \jmath}|Z|
\, .
\end{align}
It can be checked that this ansatz is also a solution for
the equations of motion
\eqref{eq:eoms_attractor1}-\eqref{eq:eoms_attractor3}
when the same signs are chosen.
Thus, this is a reduction of the original second order
system to a first order system  governed by $|Z|$.
Due to asymptotic flatness arguments, the physical sign is
fixed and the equations become
\begin{align}
\label{eq:BPS_flow}
U'&=-e^U|Z|\, ,\\
{\phi^i}'&=- 2e^{U} g^{i\bar \jmath}\bar\partial_{\bar \jmath}|Z|
\, . \label{eq:BPS_flowb}
\end{align}
These same first order equations can be obtained
 by analyzing the Killing spinor equations for the theory.
The conditions for the gravitino and gaugino supersymmetry
transformations
\begin{eqnarray}
\delta\psi_\mu^A&=&0 \, ,\\
\delta \lambda_A^i&=&0\, .
\end{eqnarray}
are equivalent to \eqref{eq:BPS_flow} and \eqref{eq:BPS_flowb}, respectively.\footnote{Actually, the Killing spinor equation for gauginos
implies a new first order equation for a phase factor.
However, it is also related to the K\"ahler connection and
once the flow equations
\eqref{eq:BPS_flow}-\eqref{eq:BPS_flowb}
are fulfilled,
this additional equation is automatically satisfied,
showing that the phase factor is not an independent quantity.}

By evaluating the  equations
\eqref{eq:BPS_flow}-\eqref{eq:BPS_flowb}
at infinity and at the horizon,
 similarly as in the previous section, we  infer that the central charge fully determines the
solution. The fixed values of the scalars at horizon are
given by the minimization condition
\begin{align}
\label{eq:BPS_attractor_cond}
\partial_i|Z|_h&=0\, ,
\end{align}
whose  critical points are also a critical point for the
black hole potential.
Solutions  corresponding to  a critical point describes a
supersymmetric extremal black hole.
The central charge at the horizon is fixed
in terms of the discrete charges
\begin{align}
|Z|_h&=|Z\left (p,q,\phi_h^i(p,q)\right )|_h\, .
\end{align}

The special K\"ahler nature of the scalar manifold
guarantees that the second derivative of
the central charge is such that
\begin{align}
\partial_i\bar\partial_{\bar \jmath}|Z|&=
g_{i\bar \jmath}|Z| >0\, ,
\end{align}
\emph{i.e.} all the critical points are minima of the central charge.
No matter what the values of the scalars are at infinity,
they will be driven towards the minimum of the central charge.
This constitutes an attractor behaviour.

The extremality condition for the central charge was brought
to a purely algebraic and equivalent form in
\cite{hep-th/9602136,hep-th/9603090,hep-th/9610105,hep-th/9610157,hep-th/9612076}
under the condition that the special geometry is not singular.

\section{$N=2$ $D=4$ SUGRA and Special K\"ahler geometry}
\label{ssection3}

The field content of the
$N=2$ supergravity theory coupled to  vector multiplets
consists of
\begin{align}
\left\{
	e_\mu{}^a \, ,
	A_\mu{}^I \, ,
	z^\alpha \, ,
	\psi_\mu{}^r \, ,
	\lambda_r{}^\alpha
	\right\}
\, ,
\end{align}
with $\alpha=1,\ldots, n_v$, and $I=0,\ldots,n_v$.
The theory
also contains some hypermultiplets, which can be safely  taken
as constant or neglected (further
details can be found in \cite{Bellorin:2006xr}, whose notation and concepts we generally adopt).
The   bosonic $N=2$ action  can be written as
\begin{align}
S&=\int_{M(4d)} R\star 1
+\G_{\alpha\bar\beta} d z^\alpha\wedge \star d\bar z^{\bar\beta}
+  F^I\wedge  G_I
\, .
\end{align}
The fields $F^I,G_I$ are not independent.
Whilst $F^I$  is given by $F^I=d A^I$, $G_I$ is a set of combinations
of the $F^I$ and their Hodge duals,
\begin{align}
G_I=a_{IJ} F^I+b_{IJ}\star F^I
\, ,
\end{align}
with scalar-dependent coefficients $a_{IJ}$ and $b_{IJ}$.

Abelian charges with
respect the $U(1)^{n_v+1}$ local symmetry of the
theory are defined by means of the integrals of the gauge field strengths.
The total charges of the geometry  are
\begin{align}
q
&\equiv (p^I,q_I)
\equiv  \frac{1}{2\pi^2}\int_{S_\infty} ( F^I, G_I)
\, .
\label{eq22}
\end{align}
Similar charges can be defined for specific finite regions.

The theory is defined, in the special geometry formalism,
by the introduction of some projective scalar coordinates $X^I$,
as for example,  `special' projective  coordinates
$ z^\alpha\equiv X^I/X^0$.  By introducing a covariantly holomorphic section of a symplectic bundle, $V$, we are able to arrange $2n_v$ quantities that transform as a vector under symplectic transformations at any point of the manifold. $V$ has the following structure
\begin{align}
V&=V(z,{\bar z}) \equiv \sp{V^I}{V_I}
\, ,
\end{align}
and satisfies the following identities:\footnote{We
choose a basis such that $\omega = \left( \begin{array}{cc}  0 & -\mathds{1}_{n_v} \\ \mathds{1}_{n_v} & 0 \end{array} \right)$.}
\begin{align}
\sp{V}{\bar V}
&\equiv V^t {\cal \omega}V\equiv {\bar V}^I V_I-V^I {\bar V}_I
=-i
\, .
\end{align}

The scalar kinetic term metric is given by
\begin{align}
\G_{\alpha\bar\beta}= \partial_\alpha \partial_{\bar\beta}\K
\, ,
\end{align}
where the K\"ahler potential $\K$ is defined by the relations
$V=\exp(-{\cal K}/2) \Omega$ being $\Omega \equiv (X^I, F_I)$
a holomorphic section
and
\begin{align}
e^{-{\cal K}}
&=
i \left ({\bar X}^I F_I- X^I F_I\right ) =
i \sp{\Omega}{\bar \Omega}.
\label{eq697a}
\end{align}
In $N=2$ theories, the central charge
$Z$  can be expressed as
a  linear function on the charge space:
\begin{align}
\label{eq26}
 Z(z^\alpha,q)\equiv\sp{V}{q}=e^{\K/2} \left(p^I F_I-q_I X^I\right)
 \, .
\end{align}

The
embedding of the isometry group of the scalar manifold
metric $\G_{\alpha\bar\beta}$,
into the symplectic group
fixes, through the K\"ahler potential $\K$, a
functional relation between the lower and upper parts of
$V$ and $\Omega$ \cite{hep-th/9611210,hep-th/9608075},
\begin{align}
F_I&=F_I(X^I)
\, ,
\nonumber
\\
V_I&=V_I(V^I)
\, .
\label{eq74b}
\end{align}

There always exists a symplectic frame under which
 the theory can be described in terms of a single holomorphic
function,
the \emph{prepotential} $F(X)$. It is a second degree homogeneous function
on the projective scalar  coordinates $X^I$, such that
$F_I(X)=\partial_I F (X) $.
For simplicity, we will assume the existence of such prepotential
along this study although the results will not depend on such
existence. Using the notation $F_{IJ}= \partial_I\partial_J F$,
the lower and upper components of $\Omega$  are related by
\begin{align}
F_I &= F_{IJ} X^J
\, .
\end{align}

The lower and upper components of $V$  are related by
a field dependent matrix $N_{IJ}$, which is determined by the
special geometry relations
\cite{Ferrara:1996um}
\begin{eqnarray}
V_I &=& N_{IJ} V^J,\quad \label{eq24}\\
D_{\bar\imath} {\bar V}_I &=& N_{IJ} D_{\bar \imath} {\bar V}^J.
\label{eq24b}
\end{eqnarray}

The matrix $N$,
which also fixes the vector couplings $(a_{IJ},b_{IJ})$ in the action,
can be related  to $F_{IJ}$ \cite{Ceresole:1995ca}
by
\begin{align}
N_{IJ}={\bar F}_{IJ}+ T_I T_J
\, ,
\end{align}
where the quantities $T_I$ are proportional to
the projector of the graviphoton, whose flux defines the
$N=2$ central charge \cite{Ceresole:1995ca}.
For our purposes, it is convenient to  write this relation
between the $N_{IJ}$ and $F_{IJ}$ quantities as
\begin{align}
N_{IJ} &\equiv F_{IJ}+N_{IJ}^\perp
\nonumber
\\
&=
F_{IJ}- 2i \im{F_{IJ}}+2 i\frac{ \im{F_{IK}} L^K \im{F_{JQ}} L^Q}
{L^P\im{F_{PQ}} L^Q}
\, ,
 \end{align}
where we have decomposed
the matrix $N_{IJ}$
into   ``longitudinal''  (the $F_{IJ}$ themselves)
and  ``transversal'' parts  ($N_{IJ}^\perp$).
The perpendicular term (defined by the expression above)
  annihilates $L^I$, or any multiple of it,
\begin{eqnarray}
N_{IJ}^\perp (\alpha L^J) &=& 0.
 \end{eqnarray}
From this, \eqref{eq24}
can be written as
\begin{align}
V_I & =N_{IJ} L^J = \left (F_{IJ}+ N_{IJ}^\perp \right) L^J
\nonumber
\\
              &=         F_{IJ}  L^J
\, .
\label{eq786}
 \end{align}
Thus, the upper and lower components of $V$ and $\Omega$
are connected by the same matrix $F_{IJ}$.

The existence of  functional dependencies
 among the upper and lower components of the
vectors $V$ or  $\Omega$
 imply further relations
between their respective   real and imaginary parts.
They are related by symplectic
matrices $\ssigma(N),\ssigma(F)\in Sp(2 n_v+2,\mathbb{R}) $ which are
respectively associated to the quantities $N_{IJ},F_{IJ}$ as follows:
\begin{align}
\re{\Omega}&=\ssigma(F)\im{\Omega}
\, ,
\label{eq234a}
\\
\re{V}&=\ssigma(N)\im{V}=\ssigma(F)\im{V}
\, .
\label{eq234}
\end{align}
The last expression is obtained by means of the relation (\ref{eq786}). These same relations \eqref{eq234a}-\eqref{eq234}
are valid for any complex multiple of $\Omega$ or $V$.
It is straightforward to  show, for example, that for
 any $\lambda\in\mathbb{C}$, we have
\begin{align}
\re{ \lambda V}
&=
\ssigma(N)\im{ \lambda V}
=
\ssigma(F)\im {\lambda V}
\, .
\label{eq122}
\end{align}


%
The matrix  $\ssigma (F)$ is of the form \cite{Ceresole:1995ca}
\begin{align}
\mathcal{S}(F) &=
\left( \begin{array}{cc}
1 & -\re {F_{IJ}}^t \\
0 & 1
\end{array} \right)
\left( \begin{array}{cc}
\im {F_{IJ}}& 0 \\
0 & \im {F_{IJ}}^{-1}
\end{array} \right)
\left( \begin{array}{cc}
1 & 0 \\
-\re{F_{IJ}}   & 1\\
\end{array} \right)
\, .
\label{eq680bb}
\end{align}
Similarly, the same result applies for $\mathcal{S}(N)$ with $F_{IJ}\to N_{IJ}$.\footnote{The matrix $\ssigma_N$ is related to $\mathcal{M}$, the matrix that appears in the black hole effective potential $V_{BH}=-\tfrac{1}{2} q^t{\cal M}q $,  by
$\ssigma(N){\cal\omega}={\cal M}$.}


In $N=2$ theories, the matrix $\ssigma(N)$ always exhibits a moduli dependence \cite{Ferrara:2012qp}.
However, this is not the case for $\ssigma(F)$.
We will  focus in this work
on the particular case of theories with
quadratic prepotentials,\footnote{Or, equivalently, $D=4$ theories with
  U-duality groups of ``degenerate type $E_7$'' \protect\cite{Ferrara:2012qp}.}
\begin{align}
 F(X) &= \frac{1}{2} F_{IJ} X^IX^J
 \, ,
\end{align}
where  $F_{IJ}$ is a complex, constant, symmetric matrix.
Then, the corresponding matrix $\ssigma(F)$
is a field-independent,  constant matrix.
We can assume that $\re{F_{IJ}}=0$  and $\im{F_{IJ}}$ is
 negative definite.
In what follows, we will use the notation $\ssigma\equiv \ssigma(F)$.
The condition $e^{-\K}>0$ and the expression 
\eqref{eq697a}
imply a restriction on the prepotential. We will write
this restriction in a convenient form in 
section \eqref{ssection4} in terms of the positivity of 
a quadratic form.


\subsubsection{General supersymmetric stationary  solutions}

The most general stationary (time independent) 4-dimensional metric compatible with supersymmetry can be
written in the IWP form
\cite{Israel:1972vx,Perjes:1971gv,Sabra:1997yd},
\begin{align}
 ds^2&=e^{2 U}(dt+\omega)^2-e^{-2 U} d{\vecr}^2.
\label{eq211}
\end{align}

Supersymmetric $N=2$ supergravity solutions can be constructed
systematically following well-established methods
\cite{Bellorin:2006xr}.
The 1-form $\omega$ and the function $e^{-2U}$ are related in
these theories to the K\"ahler potential and
connection, $\K,Q$ \cite{Sabra:1997yd}.
We demand
 asymptotic flatness, $e^{-2 U}\to 1$
 together with $\omega\to 0$
for $|\vecr|\to\infty$
BPS field equation solutions for the action above (for example, quantities that appear in the metric, as $e^{-2U}$ or $\omega$) can be written
in terms of the following real symplectic vectors $\R$ and $\I$
\begin{align}
\R&=\frac{1}{\sqrt 2}\re{\frac{V}{ X}} \, ,
\\
\I&=\frac{1}{\sqrt 2}\im{ \frac{V}{ X}}\, .
\end{align}
$X$ is an arbitrary complex function of space coordinates
 such that $1/X$ is harmonic.
The $2 n_v+2$ components of $\I$ and $\R$ are real harmonic functions
in $\mathbb{R}^3$.
There is an algebraic
 relation between $\R$ and $\I$ and
the solutions can be written in terms only of the  vector $\I$.
Due to the relations \eqref{eq234a}-\eqref{eq234} and
\eqref{eq122}, we can write the following
\emph{stabilization equation}
\begin{align}
\R &= \ssigma \I
\, .
\label{eq888}
\end{align}

In practice,
specific solutions are determined by giving a
particular, suitable, ansatz for the symplectic vector $\I$ as
a function of the spacetime coordinates.


Using these symplectic vectors we rewrite the
only independent metric component as
\begin{align}
e^{-2 U}
&=
e^{-\K}= \frac{1}{2 | X|^2}
\nonumber
\\
               &=  \sp{\R}{\I}=\sp{\ssigma\I}{\I}
\, .
\label{eq215}
\end{align}
Similarly,  the time independent $3$-dimensional 1-form  $\omega=\omega_i dx^i $ satisfies the equation
\begin{align}
 d\omega=2\sp{\I}{\star_3 d\I}
 \, ,
\end{align}
where $\star_3$ is the Hodge dual on flat $R^3$, together with
 the integrability condition
\begin{align}
 \sp{\I}{\Delta\I}&=0
 \, .
\label{eq221}
\end{align}
The asymptotic flatness condition implies
\begin{align}
 \sp{\R_\infty}{\I_\infty}&=\sp{\ssigma\I_\infty}{\I_\infty}=1.
\label{eq689a}
\end{align}

The gauge field equations of motion and Bianchi identities
 can be directly solved  in terms of spatially
dependent harmonic functions \cite{Bellorin:2006xr}.
The modulus of the central charge function defined in
\eqref{eq26} can be written,  taking into account
\eqref{eq215}, as
\begin{eqnarray}
| Z(q)|^2 e^{- 2U} &=&| \sp{\R}{q}|^2+ | \sp{ \I}{q}|^2.
\end{eqnarray}
At spatial infinity, assuming asymptotic flatness
\eqref{eq689a}, we arrive to
\begin{eqnarray}
| Z_\infty(q)|^2   &=&| \sp{\R_\infty}{q}|^2+ | \sp{ \I_\infty}{q}|^2.
\label{eq218b}
\end{eqnarray}

The, assumed time independent, $n_v$ complex scalar fields
$z^\alpha$ solutions to the field equations, are given in
this formalism by
\begin{eqnarray}
 z^\alpha&=&\frac{\Omega^\alpha}{\Omega^0}=\frac{V^\alpha}{\V^0}=
\frac{\R^\alpha+i\I^\alpha}{\R^0+i\I^0}
\, .
\label{eq226}
\end{eqnarray}
This is, in general, a formal expression as the $\I$ or $\R$ quantities
may be scalar dependent.\footnote{Even for a non scalar dependent
ansatz $\I$, the matrix $\ssigma$ is, in general, scalar dependent.}

These scalar fields can, in principle, take any values
($z_\infty$) at infinity. These values will
appear as free parameters in the ansatz that we give for $\I$.
Nevertheless, according to the attractor mechanism,
the moduli adjust themselves  at some fixed points.


We are interested in this work in extremal, single- or
multi-center black hole-type solutions  determined by
an $\I$ ansatz with point-like singularities of the form
\begin{align}
 \I&=\I_{\infty}+\sum_{a}\frac{q_a}{|\vecr-\vecr_a|}
 \, ,
\label{eq216}
\end{align}
where $a=1,\ldots,n_a$ being the number $n_a$ arbitrary and
$q_a=(p_a{}^I, q_{aI})$ and $\I_\infty$ real, constant,
symplectic vectors.

For this kind of solutions, the quantities
$\I_\infty$ are related to the values at infinity of
the moduli while the ``charge'' vectors $q_a$ are
related to their values at the fixed points.
The fixed values of the scalars, $z(\vecr)\to z(\vecr_a)=z_{f}^a$, are the solutions
of the following
\emph{attractor equations}
\cite{Shmakova:1996nz,Ferrara:1995ih,Ferrara:1996um}:
\begin{align}
q^a=\re{2i\bar Z(z_{f}^a) V(z_f^a)}
\, .
\label{eq224}
\end{align}
The prepotential performs its influence throughout $V$ and $Z$ (\emph{cf.} \eqref{eq26}).
The scalar attractor values are independent of their
asymptotic values and only depend on the discrete charges
$ z^a_{f}=z^a_{f}(q_I)$.

Single center black hole solutions are known to exist for all
regions of the moduli scalars at infinity, under very mild
conditions on the charge vector. In the multicenter case,
for fixed charge vectors, not all the positions $\vecr_a$
 in the ansatz \eqref{eq216} are allowed.
The integrability condition (\ref{eq221}) imposes
necessary conditions on the relative positions and on the moduli
at spatial infinity (through $\I_\infty$)
for the existence of a solution.
In this framework, a particular black hole solution is
completely determined by a triplet of charge vectors,
distances and values of the moduli at
infinity $(q_a,\vecr_a,z^\alpha_\infty)$.

\section{The stabilization matrix and the attractor equations}
\label{ssection4}
\label{sec:stabilizationequations}
Let us consider now the attractor equations (\ref{eq224}),
in more detail. We will use the properties of the
stabilization matrix $\ssigma$ to solve them in a purely
algebraic way to obtain some properties and give some explicit
expressions for the scalars at the fixed points.

For this purpose, we first establish some well-known
properties of  $\ssigma_N,\ssigma\equiv\ssigma_F$ and
define new matrices: some projector operators associated
to them and their respective symplectic adjoints.

It can be shown by explicit computation that the real
symplectic  matrices
$\ssigma_N,\ssigma\equiv\ssigma_F\in Sp(2n_v+2,\mathbb{R})$ defined by
\eqref{eq234a}-\eqref{eq234} and 
whose explicit expressions are given by \eqref{eq680bb},
  satisfy the relations
\begin{eqnarray}
\ssigma_N^2=\ssigma_F^2&=&-\mathds{1}
\, .
\label{eq235}
\end{eqnarray}
From this, it is possible as well as convenient
 to define  the projector operators
for the matrix $\ssigma$ (similarly for $\ssigma_N$) as 
\begin{align}
 \Ppm = \frac{1\pm i\ssigma}{2}
 \, .
\end{align}
They satisfy the following straightforward properties
\begin{align}
 \Ppm^2       &=   \Ppm\, , \nonumber\\
 \ssigma \Ppm &= \mp i  \Ppm\, , \nonumber \\
 (\Ppm )^*    & =\Pmp\, , \label{eq312}
\end{align}
and, for $X,Y$ arbitrary real vectors,
\begin{align}
 \Ppm X    & =\Ppm Y\,\quad \Rightarrow\, X=Y.  
\label{eq312cc}
\end{align}

According to \eqref{eq312}, the  $\Ppm$ are the
projectors into the eigenspaces of the matrix $\ssigma$.
The symplectic space $W$ can be decomposed into
eigenspaces of the matrix $\ssigma$:
\begin{align}
 W &= W^+\oplus W^-\, ,
\end{align}
where $W^{\pm}=\Ppm  W$.
Complex conjugation interchanges the $W^+$ and $W^-$ subspaces, $(\Ppm )^*=\Pmp$, so that both subspaces are isomorphic to each other.

We note that we can rewrite a stabilization relation similar to \eqref{eq122} in a different way with the help of
these projection operators $\Ppm$.
For any vector $V\in W$ for which there is a relation between
its real and imaginary parts of the form
$\re{V}=\ssigma\im{V}$, we have the relations
 (for an arbitrary $\lambda\in \mathbb{C}$)
 \begin{align}
\lambda V&=  \re{\lambda V}+i \im{\lambda V} 
\nonumber
\\
&=   2 i \Pm\im{\lambda V}
\nonumber
\\
&=  2  \Pm\re{\lambda V}
\, .
\label{eq41}
\end{align}
Thus, the full vector $V$ can be reconstructed by applying one of the projectors either from its real or imaginary
part.
We see that such vectors are fully contained in the
subspace $W^-$ or, equivalently, they are
eigenvectors of $\ssigma$
 \begin{align}
\ssigma V&= 2 i\ssigma   P_-\im { V}=  2  P_-\im { V}\nonumber \\
         &=    i V
\, .
\label{eq41cc}
\end{align}


We will find convenient to define the adjoint operator of the
matrix $\ssigma$,  $\ssigmaad$, with respect to the symplectic
bilinear product  such that, for any  vectors $A,B$,
\begin{equation}
 \sp{\ssigma A}{B}=\sp{A}{\ssigmaad B}.
\end{equation}
A straightforward computation shows that  $\ssigmaad$ is given by
\begin{equation}
 \ssigmaad=-\Omega \ssigma^t\Omega
 \, .
\end{equation}
Under the assumption of a symmetric $F_{IJ}$ matrix, it is given by
\begin{align}
 \ssigmaad &= -\ssigma
 \, .
\label{eq1000}
\end{align}
In summary, the matrix $\ssigma$ is skew-adjoint with respect
to $\omega$ and its square is $\ssigma^2=-\mathds{1}$.
It fulfills an ``unitarity'' condition $\ssigmaad \ssigma=\mathds{1}$.

In mathematical terms, $\ssigma$ defines an
(almost) complex structure on the symplectic space.
This complex structure preserves the symplectic bilinear form,
the matrix $\ssigma$ is an isometry of the symplectic space,
\begin{eqnarray}
 \sp{\ssigma A}{\ssigma B}&=&\sp{A}{B}
 \, .
\end{eqnarray}
From \eqref{eq1000}, we see that $\ssigma$ is an
element of the symplectic Lie algebra $\mathfrak{sp}(2 n_v+2)$.

Moreover, the bilinear form defined by
\begin{align}
g(X,Y)\equiv \sp{\ssigma X}{Y}
\end{align}
is symmetric. This can be easily seen:
\begin{align}
 g(X,Y)&= \sp{\ssigma X}{Y}=\sp{Y}{\ssigmaad X}=\sp{\ssigma Y}{X}
 \nonumber\\
 &= g(Y,X)
 \, .
\label{eq1123b}
\end{align}


We will apply these properties to the study of the attractor
equations. In general, the matrices
$\ssigma_N,\ssigma_F$ are scalar dependent.
Only one of them, $\ssigma_F$, is constant,
in the case of quadratic prepotentials.
Let us write $\ssigma_N^f=\ssigma_N(z=z_f)$
$\ssigma_F^f=\ssigma_F(z=z_f)$ for the matrices
evaluated at (anyone of) the fixed points. Let us use the sub/superindex $f$ to denote any quantity at the fixed points. For instance, $Z^f\equiv Z(z^\alpha_f)$ or $V^f\equiv V(z^\alpha_f)$.
If we multiply both sides of
\eqref{eq224} by $\ssigma_N^f=\ssigma_N(z=z_f)$,
 we arrive to
\begin{align}
\ssigma_N^{f} q^a&=
\ssigma_N^{f}\re{2i\bar Z^f V^f}=\ssigma_F^{f}\re{2i\bar Z^f V^f}
=\ssigma q^a
\, ,
\label{eq323}
\end{align}
where we have used the properties \eqref{eq234} and (\ref{eq122}).\footnote{Following \cite{Ceresole:1995ca}, we note that
$V_{BH}=| Z_i|^2+| Z|^2=-\tfrac{1}{2} q^t \ssigma(N)\omega q$
 and
$
| Z_i|^2-| Z|^2=\frac{1}{2} q^t \ssigma(F)\omega q
$.
At the
fixed points, we have $Z_i=0$, so that
$| Z|^2=-\frac{1}{2} q^t \ssigma_N \omega q=-\frac{1}{2} q^t\ssigma_F\omega  q $.
This last equation is satisfied by a solution of \eqref{eq224}.}

The  attractor equations can be written yet in another
alternative way. By using
(\ref{eq41}) and (\ref{eq224}), we can write
\begin{align}
i\bar Z_{f} V_{f}
&= 2 \Pm i\bar Z^{f} V^{f}
\nonumber\\
&= \Pm q,
\label{eq48}
\end{align}
or  its conjugate equation
\begin{eqnarray}
-i Z^{f} \bar V^{f}&=& \Pp q
\, .
\label{eq311zz}
\end{eqnarray}
That is, the attractor equations simply
equal (a multiple of) the vector $V$ (which, as we have seen above lies
on the subspace $W^-$) with the part of the charge
vector which lies on such a subspace.

From \eqref{eq48}-\eqref{eq311zz}, by taking symplectic
products, we obtain
\begin{eqnarray}
|Z_{f}|^2 \sp{ V_{f}}{\bar V_{f}}
      &=&\sp{\Pm q}{\Pp q}=\sp{q}{\Pp q}\nonumber\\
      &=& -\frac{i}{2} \sp{\ssigma q}{ q}
      \, .
\end{eqnarray}
If we insert the constraint $\sp{V}{\bar V}=-i$,
we arrive in a straightforward and purely algebraic way
 to the  well known formula
\begin{eqnarray}
 |Z_\text{f}|^2&=& \frac{1}{2}\sp{\ssigma q}{q}
 \, ,
\label{eq412}
\end{eqnarray}
which relates the absolute value
of the central charge at any fixed point to a quadratic
expression on the charge.
It is obvious from \eqref{eq412} that the
 positivity of the quadratic form $g(q,q)=\sp{\ssigma q}{q}$
(at least locally at  all  the fixed points)
is  a necessary consistency condition for the
existence of solutions to the attractor mechanism.

Moreover the mathematical 
consistency condition $e^{-\K}>0$ 
can be written as (\emph{cf}.\eqref{eq697a}) 
\begin{eqnarray}
e^{-\K} &=& i \sp{\Omega}{\bar\Omega}=2 
\sp{\re{\Omega}}{\im{\Omega}}\nonumber\\
 &=&2 \sp{\ssigma\im{\Omega}}{\im{\Omega}}>0.
\label{eq6150b}
\end{eqnarray}
This last equation is automatically 
satisfied for  a definite positive  
quadratic form $g$ at any point.

Positivity (which is physically imposed by \eqref{eq412}
and \eqref{eq6150b})
and symmetry (demanded by \eqref{eq1123b}), implies that the
bilinear form $g(X,Y)$ is an inner product.
In addition, a hermitian form $h$ can be defined from it and
from the symplectic form. We define
\begin{eqnarray}
h(X,Y) &=& \sp{\ssigma X}{Y}+i \sp{X}{Y}
\, ,
\end{eqnarray}
which can be written in terms of the
projection operators $\Ppm$ as
\begin{eqnarray}
h(X,Y) &=& 2 i \sp{\Pm X}{Y}\nonumber\\
       &=& 2 i \sp{\Pm X}{\Pp Y}.
\end{eqnarray}
The three defined structures $\{g, \omega, \ssigma\}$
 form a compatible triple, each structure can be specified
by the two others.\footnote{It seems natural to think that
 the existence of these structures
can be traced back to the scalar manifold being K\"ahler.}


Let us address now to the problem of obtaining the values of the
moduli at the fixed points and at infinity.
 The values of the scalar fields at the fixed points
can  be computed by an explicit expression,
which involves only the matrix $\ssigma_F$.
The fixed values of the $n_v$ complex scalars
$z^\alpha_f(q)$
(at a generic fixed point with charge $q$)
are given, using the expressions (\ref{eq226}) and (\ref{eq48}), by
\begin{align}
z^\alpha_f(q)
&=\frac{(\ssigma \I)^\alpha+i\I^\alpha}{(\ssigma\I)^0+i\I^0}
=\frac{\left((\ssigma +i\mathds{1})\I\right)^\alpha}{\left((\ssigma+i\mathds{1})\I\right)^0}
\nonumber\\
&=\frac{(\Pm q)^\alpha}{(\Pm q)^0}
\, .
\label{eq416}
\end{align}
That is, the fixed values of the scalars are given in terms of the projection
of the charges into the ei\-gen\-spa\-ces of the
matrix $\ssigma$.
For quadratic prepotentials, where this matrix is a constant,
 this is a complete,
explicit solution of the attractor equations.

The values of the $n_v$ complex scalars at spatial infinity,
$|\vecr|\rightarrow\infty$ are given by (using
again (\ref{eq226}) and
defining $\I_\infty=\lim_{|\vecr|\to\infty} \I$, we are not
assuming any particular ansatz for $\I$ at this moment)
\begin{align}
 z^\alpha_\infty
&=\lim_{|\vecr|\rightarrow\infty} \frac{(\Pm \I)^\alpha}
{(\Pm\I)^0}=\frac{(\Pm \I_\infty)^\alpha}{(\Pm \I_\infty)^0}
\, .
\label{eq3434}
\end{align}
According to this formula, the `moduli' $z^\alpha_\infty$
are simple rational functions of the $2 n_v+2$ real constant
components of $\I_\infty$. They are thus independent of the
fixed attractor values \eqref{eq416} (at least for a vector 
 $\I$ with only point like singularities,
as \eqref{eq216}).

We note that  the expression (\ref{eq3434})  is
formally identical to the expression (\ref{eq416}), since both give the moduli values at a fixed point in terms of the charges,
where the roles of $\I_\infty$ and $q$ are exchanged.
It is suggesting then to  write an ``effective
attractor equation'' at infinity, where the  r\^ole of the
center charge  is played by the vector $\I_\infty$. That is,
the scalar solutions of the equation
\begin{align}
\I_\infty=\left.\re{2i\bar Z V}\right|_{\infty}
\, ,
\end{align}
are those precisely given  by (\ref{eq3434}).



One can extract some algebraic relations
for the vectors $\I_\infty$ and $q^a$ and the
equations
\eqref{eq416}-\eqref{eq3434}
in specific cases,
for example for solutions with constant scalars.
Let us assume a one-scalar theory and
$z_f=z_\infty (\not =0) $.
In this case, equations
\eqref{eq416}-\eqref{eq3434}
 imply the projective equality
($\lambda\in\mathbb{R}$ an arbitrary  non-zero, constant)
\begin{eqnarray}
\Pm \I_\infty&=& \lambda \Pm q\, .
 \end{eqnarray}
which, due to  \eqref{eq312cc}, implies 
\begin{eqnarray}
 \I_\infty&=& \lambda  q \, .
 \end{eqnarray}
In addition, the asymptotic flatness condition
\eqref{eq689a} implies
\begin{eqnarray}
\lambda^2&=& \frac{1}{\sp{\ssigma q}{q}}=\frac{1}{2| Z_{f}|^2}
\, .
\end{eqnarray}
The consistency of the last equation is assured by the
positivity of the quadratic form $\sp{\ssigma q}{q}$.
Thus, we can finally arrive to a characterization of the
$\I_\infty$ parameters in the case of constant scalar solutions
\begin{align}
\I_\infty &= \pm \frac{q}{\surd\sp{\ssigma q}{q}}\, .
\label{eq118b}
\end{align}
Similar arguments can be stated in the multicenter case.


Let us finish this section with some qualitative remarks.
We have arrived to the expressions \eqref{eq416}-\eqref{eq3434}
which can be written,  in terms of the projective
 complex, vector $\Omega=(X^I,F_I)$, as
\begin{eqnarray}
\Omega_{fix} &=& P_- q\, , \nonumber\\
\Omega_{\infty} &=& P_- \I_\infty\, . \label{eq445}
\end{eqnarray}
We could have predicted these expressions a priori:\footnote{Extending  arguments presented in \cite{Mohaupt:2008gt} (and references therein).}
if SUSY  solutions are uniquely determined by the
symplectic real vectors $q_a$, then
the also symplectic but complex vector $\Omega=(X^I,F_I)$
must be related to these vectors in a linear way,
respecting symplectic covariance at the same time.
Moreover, the symplectic sections $\Omega$ (or $V$)
lie on the subspace $W^-$, one eigenspace of the
stabilization matrix $\ssigma$.
The only  possibility for such a relation
would be the expressions  in \eqref{eq445},
where precisely appear the projections of
$q$ or $\I_\infty$ into such subspace.
These expressions, evaluated at the points of maximal symmetry
(the horizon and  infinity), are   equivalent forms of
the  standard horizon attractor equations  and the
generalized attractor equation at infinity presented here.


\section{Complete solutions for quadratic prepotentials}
\label{ssection5}

We have got some general results in the previous section
without using a concrete form  for the solutions, for \I.
In this section we will make use of
the ansatz \eqref{eq216} for theories with quadratic
prepotentials to obtain a full characterization of the
solutions.

Let us insert the ansatz \eqref{eq216} into the general
expression for the complex scalars, \eqref{eq226}.
The values for the time independent $n_v$ complex scalar
solutions to the field equations are explicitly given by
\begin{align}
z^\alpha(\vecr)&=
\frac{(\Pm \I)^\alpha}{(\Pm \I)^0}=\frac{(\Pm \I_\infty)^\alpha+\sum_a \frac{(\Pm q_a)^\alpha}{|\vecr-\vecr_a|}}{(\Pm \I_\infty)^0+\sum_a\frac{(\Pm q_a)^0}{|\vecr-\vecr_a|}}
\, .
\label{eq514}
\end{align}
This equation 
is a simple rational expression
for the value of the scalar fields in the whole space.
The fields and their derivatives are regular everywhere,
including the fixed points
(there could be singularities for special charge configurations
which make zero the denominator of \eqref{eq514}).

The expression (\ref{eq514})
 interpolates between the values at the fixed points and at infinity. After some simple manipulations, it can be
written as
\begin{align}
z^\alpha(\vecr)
&=c_\infty^\alpha(\vecr) z^\alpha_\infty +c_a^\alpha(\vecr) z^\alpha_{a,f}
\, ,
\label{eq514s}
\end{align}
where
$c_\infty^\alpha(\vecr)$ and $c_a^\alpha(\vecr)$
are spatial dependent complex functions such that
\begin{align}
c_\infty^\alpha(\vecr)+c_a^\alpha(\vecr)
&=
1
\, ,
\nonumber\\
c_\infty^\alpha(\infty)
&=
1
\, ,
\nonumber\\
c_\infty^\alpha(\vecr_a)
&=
0
\, ,
\nonumber\\
\lim_{\vecr\to\vecr_b}c_a^\alpha(\vecr)&=\delta_{ab}
\, .
\end{align}
For a single center solution, we note that
if $z_\infty^\alpha=z_f^\alpha$ then
the scalar fields are constant in all the space.
%

It is straightforward to see that the attractor mechanism
is automatically fulfilled by the ansatz (\ref{eq216}).
The value of $z^\alpha$ at any center $\vecr_a$ is given,
by taking the corresponding limit in \eqref{eq514}, by
\begin{align}
z^\alpha(\vecr_a)
&=\frac{(\Pm q_a)^\alpha}{(\Pm q_a)^0}=z^\alpha_f(q_a)
\, ,
\end{align}
where, after the second equality, we have used the
fixed point expression \eqref{eq416},
which is a direct consequence
of the attractor equations.

On the other hand, the solution at the spatial infinity
 recovers spherical symmetry. Again, taking limits, we have
(with $|\vecr|\equiv r$)
\begin{align}
z^\alpha( r\to\infty)
&=
\frac{r (\Pm \I_\infty)^\alpha+\sum_a (\Pm q_a)^\alpha}
{r(\Pm \I_\infty)^0+\sum_a(\Pm q_a)^0}
\nonumber\\
&= (1-c^\alpha(r))z^\alpha_\infty +c^\alpha(r)z^\alpha_f(Q)
\, ,
\label{eq514b}
\end{align}
where $z_{f}(Q)$ is the fixed point scalar value which
would correspond, according to the attractor equations,
 to a total charge $Q=\sum_a q_a$.
The asymptotically interpolating functions appearing
above are
\begin{align}
c^\alpha(r)&=\frac{1}{1+\frac{r}{r_0^\alpha}}
\, ,
\end{align}
with the (non-zero) scale parameter
\begin{align}
r_0^\alpha
&= \frac{\sum_a(\Pm q_a)^0}{\sum_a (\Pm \I_\infty)^0}
\, .
\end{align}
They  are such that
\begin{align}
c^\alpha(0)
&=
1
\, ,
\nonumber\\
c^\alpha(\infty)
&=
0
\, .
\end{align}

The \emph{scalar charges} $\Sigma^\alpha$ associated to
the scalar fields can be simply defined by the
asymptotical series
\begin{align}
z^\alpha( r\to\infty)&=z^\alpha_\infty
+ \frac{\Sigma^\alpha}{r}
+\mathcal{O}\left(\frac{1}{r^2}\right)
\, .
\end{align}
Expanding (\ref{eq514b}), we have
\begin{align}
z^\alpha( r\to\infty)&=z^\alpha_\infty
+ \frac{r_0^\alpha(  z^\alpha_f(Q)-z^\alpha_\infty)}{r}
+\mathcal{O}\left(\frac{1}{r^2}\right)
\, ,
\end{align}
and thus the scalar charges are given by
\begin{eqnarray}
\Sigma^\alpha&=&r_0^\alpha\left(  z^\alpha_f(Q)-z^\alpha_\infty\right)
\, .
\label{eq47}
\end{eqnarray}
In the special case of a single center solution,
the expression (\ref{eq47}) is in agreement
with the well known fact that the scalar charges vanish
for double extremal black holes.
In the multicenter case, from this formula we infer a
similar result: the scalar charges vanish if
\begin{eqnarray}
z_\infty^\alpha&=&z_{f}^\alpha (Q)
\, ,
\label{eq618bb}
\end{eqnarray}
where $Q$ is the sum of the individual charges.
Obviously, in this case this does not mean that the scalars
are constant in all the space. Thus the conditions \eqref{eq618bb} could be considered a convenient generalization of
double extremal solutions in the multicenter case.
By taking into account the considerations of the previous
section, \eqref{eq118b}, a candidate vector $\I_\infty$
corresponding to such a solution would be of the form
\begin{eqnarray}
\I_\infty&=& \pm\frac{Q}{\sqrt{ \sp{\ssigma Q}{Q}}}
\, ,
\label{eq6179cc}
\end{eqnarray}
whereas the scalar fields would be given at any point by
\begin{align}
z^\alpha(\vecr)&=c_\infty^\alpha(\vecr) z^\alpha_f(Q)
+c_a^\alpha(\vecr) z^\alpha_{ f}(q_a)
\, .
\end{align}
%
%

The integrability condition for the equation determining 
$\omega$ are, 
for any charge center $q_b$ \cite{Bellorin:2006xr},
\begin{align}
\sp{\I_\infty}{q_b}+\sum_a\frac{\sp{q_a}{q_b}}{r_{ab}} &=0
\, ,
\end{align}
where $r_{ab}=| \vecr_a-\vecr_b |$. 
The solutions for these equations give the possible intercenter positions. 

Let us see the consequence of the integrability equations 
for a double extremal two center configuration. In this 
case, if $\I_\infty=\lambda Q$, we have
\begin{align}
0&=\lambda \sp{Q}{q_1}+\frac{\sp{q_1}{q_2}}{r_{12}}
\nonumber\\
 &=\lambda \sp{q_2}{q_1}+\frac{\sp{q_1}{q_2}}{r_{12}} 
 \nonumber\\
 &=\sp{q_2}{q_1}\left (\lambda-\frac{1}{r_{12}}\right). 
\end{align}
Comparing this last equation with
\eqref{eq6179cc}
we conclude that the double extremal 
intercenter distance is given by
\begin{eqnarray}
\left.r_{12}\right|_{doubl. extrm}&=&  \sp{\ssigma Q}{Q}
\, .
\end{eqnarray}

\subsection{Near horizon and infinity geometry}

Let us now study the  gravitational field.
The   metric has the form given by (\ref{eq211}),
with the asymptotic flatness conditions
$ -g_{rr}=\sp{\R_\infty}{\I_\infty}=1$ and
$\omega(\vecr\to\infty)\to 0$.
For point-like sources, as those represented by the
ansatz (\ref{eq216}),
the  compatibility equation (\ref{eq221}) takes the form
(see, for example \cite{Bellorin:2006xr})
\begin{eqnarray}
N  & \equiv& \sum_a\sp{\I_\infty}{q_a}=\sp{\I_\infty}{Q}=0
\, .
\label{eq49}
\end{eqnarray}
 An explicit computation
of the total field strength shows  that (\ref{eq49})
 is equivalent to the requirement of absence of NUT charges:
only  after imposing the condition $N=0$, the overall integral
of the $(F^I,G_I)$ field strengths at infinity
 is equal to $Q=\sum q_a$.
Another consequence of the condition $N=0$, which can
be checked by direct computation from \eqref{eq687b}, is that
the 1-form $\omega$  evaluated at each horizon of the multicenter solution is the same and is equal to its value at spatial infinity, which can be taken to be zero.

Let us write  a more  explicit expression for the $g_{rr}$
component at any space point.
We can write, using the `stabilization equation' \eqref{eq888}
and the ansatz \eqref{eq216}, the expression
\begin{eqnarray}
\sp{\R}{\I}&=&
\sp{\ssigma\I_\infty+\sum_a\frac{\ssigma q_a}{|\vecr-\vecr_a|}}{\I_\infty+\sum_b\frac{q_b}{|\vecr-\vecr_b|}}\nonumber\\
&=&1
+\sum_b\frac{1}{|\vecr-\vecr_b|}
\left (\sp{\ssigma\I_\infty}{q_b}+\sp{\ssigma q_b}{\I_\infty}\right )
+
\sum_{a,b}\frac{\sp{\ssigma q_a}{q_b}}{|\vecr-\vecr_a||\vecr-\vecr_b|}
\nonumber\\
&=&1+2\sum_b\frac{\sp{\ssigma\I_\infty}{q_b}}{|\vecr-\vecr_b|}
+\sum_{a,b}\frac{\sp{\ssigma q_a}{q_b}}{|\vecr-\vecr_a||\vecr-\vecr_b|}
\, ,
\label{eq222}
\end{eqnarray}
where we have  used  the property
$\ssigma^\dagger=-\ssigma$ and the
asymptotic flatness
condition $\sp{\ssigma \I_\infty}{\I_\infty}=1$.
We  introduce now the quantities
\begin{align}
 M_a        &\equiv \sp{\ssigma \I_\infty}{q_a}\, , \label{eq432}\\
 A_{ab} &\equiv \sp{\ssigma q_a}{q_b}\, , \label{eq433}
\end{align}
where $A_{ab}$ is symmetric in its indices
due to the property \eqref{eq1123b}.

With these definitions, we can finally write the
expression for the metric element as
\begin{align}
-g_{rr}&=\sp{\R}{\I}
\nonumber\\
&=
1+2\sum_b\frac{M_b}{|\vecr-\vecr_b|}
+\sum_{a,b}\frac{  A_{ab}}{|\vecr-\vecr_a||\vecr-\vecr_b|}
\, .
\label{eq415}
\end{align}
If the metric element (\ref{eq415}) describes a
black hole,  then the right part should be kept
 always positive  and  finite for any finite $|\vecr|$.
\footnote{Also consider that $-g_{rr}\sim e^{-\K}$.}
Its positivity is  ensured as long as
the ``mass'' $M_a$ and ``area'' $A_{ab}$ parameters
are positive. But less strict conditions can be imposed,
for the positivity it is sufficient that the
matrix $(A_{ab})$ is (semi-) definite positive.
This is guaranteed by the fact that this matrix is
the Gram matrix of a set of (linearly independent or not)
vectors $q_a$  with the \emph{inner product} $g$
(see discussion in Section \ref{ssection4}).\footnote{If the vectors $q_a$ are not linearly independent,
then $(A_{ab})$ is only semi definite positive, positive
definite otherwise.}
\footnote{From the mathematical point of view 
 it is not neccessary that  the $M_b$ are  all positive. 
The quadratic form $-g_{rr}(y)=1+2 \Sigma M_a y_a+\Sigma A_{ab} y_a y_b $ is strictly convex, and hence has a global
minimum as long as $A_{ab}$ is positive definite.
Positivity of this minimum ($-g_{rr}>0$) is guaranteed
in this case at least if
$ M_a (A^{-1})_{ab} M_b<1$.}


\subsubsection*{Behaviour at fixed points and at infinity}

We will define new quantities the mass
$M_{ADM}$ and $\atot$ from the behaviour of the metric
at infinity.
At spatial infinity $|\vecr|\to \infty$, $\frac{1}{| \vecr-\vecr_a|}\to 1/r$, the metric element \eqref{eq415}
becomes  spherically symmetric:
\begin{align}
-g_{rr}
&=1+\frac{2\sum_a M_a}{r}+ \frac{ \sum_{ab}A_{ab}}{r^2}+\mathcal{O}\left(\frac{1}{r^3}\right)
\nonumber\\
&\equiv
1+\frac{2M_{ADM}}{r}+ \frac{ \atot}{r^2}+\mathcal{O}\left(\frac{1}{r^3}\right)
\, .
\end{align}
The second equation defines $M_{ADM}$ and \atot.
Comparing both expressions and using
\eqref{eq432}, \eqref{eq433} and $Q=\sum_a q_a$, we have
\begin{align}
M_{ADM} &= \sum_a M_a=\sp{\ssigma I_\infty}{Q}\, ,
\label{eq529}\\
\atot &= \sum_{ab} A_{ab} =\sp{\ssigma Q}{Q}\, .
\label{eq530}
\end{align}
The expression for the central charge at
infinity, (\ref{eq218b}),
becomes then
\begin{align}
| Z_\infty |^2 &= M_{ADM}^2+N^2\, ,
\label{eq531b}
\end{align}
where $N$ is defined by
(\ref{eq49}).
The compatibility
condition $N=0$  is equivalent to the saturation
of a BPS condition
\begin{align}
| Z_\infty |^2 &= M_{ADM}^2 =| \sp{\ssigma \I_\infty}{Q}|^2
\, .
\label{eq531bc}
\end{align}

The $M_{ADM}$ quantity, opposed to  $\atot$,
depends on the scalar values at infinity
through the implicit dependence of $\I_\infty$ on them.
These can take arbitrary values there, or at least can be
chosen in a continuous range.
In the single center case, for any given charge vector, one
can obtain a certain particular solution by setting the
scalar fields to constant
values ($z_{f}^\alpha=z^\alpha_\infty$), giving this the minimal possible $M_{ADM}$ mass
\cite{Gibbons:1996af}.
For multicenter solutions and generic non-trivial
charge vectors, it is not possible to have constant scalar
fields. Nevertheless, we can still proceed to the minimization of
$M_{ADM}(z_\infty^\alpha)$,
\begin{align}
\left.\frac{\partial M_{ADM}}{\partial z_\infty^\alpha}\right|_{z_{\infty,min}} &=0\, ,
\end{align}
with respect to the scalar fields at infinity for a given
configuration. On view of the relation (\ref{eq47}),
if this minimum coincides with $z^\alpha_\infty=z_{f}(Q)$,
we would have full analogy with the one center case.

That is indeed the case. We can show that, for a
given configuration of charges,
\begin{align}
| Z_\infty|^2&=M_{ADM}^2\le \sp{\ssigma Q}{Q} =\atot
\, ,
\label{eq6139bb}
\end{align}
the equality appears at the minimum
$\left(z_\infty^\alpha\right)_{min}=z_{f}(Q)$.
The proof is a simple application of the
fact that the bilinear form $\sp{\ssigma X}{Y}$
is an inner product.
If we apply  the Cauchy-Schwartz inequality to
\eqref{eq531bc} and then the asymptotical flatness
condition, we have
\begin{align}
 M_{ADM}^2 &= |\sp{\ssigma \I_\infty}{Q}|^2
\nonumber \\
  & \le \sp{\ssigma \I_\infty}{\I_\infty}  \sp{\ssigma Q}{Q}
 =   \sp{\ssigma Q}{Q}
 \, .
\label{eq6139cc}
\end{align}
The saturation of the inequality (\ref{eq6139cc}) happens
when $\I_\infty$ is of the form
\begin{eqnarray}
  \I_\infty &=& \lambda Q
  \, ,
\end{eqnarray}
where $\lambda\in \mathbb{R}$.
Such $\I_\infty$ trivially satisfies the $N=0$ condition,
$\sp{\I_\infty}{Q}=\sp{Q}{Q}=0$ and therefore
the values of the scalar fields at infinity
are given by
$z_\infty^\alpha=z_{f}(Q)$. Thus, this configuration is
a multicenter generalization of the double extremal solutions.

Let us proceed now to the study of the geometry near the centers.
For $\vecr\to \vecr_a$ the metric element
given by  (\ref{eq415}) becomes spherically symmetric.
Moreover, it can be shown that,
by fixing additive integration constants,
we can take $\omega_a=\omega(\vecr\to\vecr_a)=0$
at the same time that $\omega_\infty=\omega(\vecr\to\infty)=0$.
As a consequence, the metric at any of the horizon components
with charge $q_a$
approaches an $AdS_2\times S^2$ metric of the form
\begin{align}
ds^2
&= \frac{r^2}{\sp{\ssigma q_a}{q_a}} dt^2- \frac{\sp{\ssigma q_a}{q_a}}{r^2} d{\vecr}^2
\, .
\label{eq425}
\end{align}
This is a Robinson-Bertotti metric, of the form
\eqref{eq611}. 
Positivity of $-g_{rr}$ in any of the fixed point limits
is ensured if we request that, for all the center charges 
$q_a$
\begin{eqnarray}
\sp{\ssigma q_a}{q_b}&>&0.
\label{eq6194cc}
\end{eqnarray}
The parameter $M_{RB}$ appearing there satisfies a
charge extremal condition of the form
\begin{align}
M_{RB} &=\sp{\ssigma q_a}{q_a}
\, .
\end{align}

The near horizon geometry is thus,
completely determined in terms of the individual
horizon areas $\ah_{,a}=\sp{\ssigma q_a}{q_a}$.
The horizon area $\ah$ is the sum
of the areas of its disconnected parts
\begin{align}
\ah=\sum_a \ah_{,a} &= \sum_a \sp{\ssigma q_a}{q_a}\nonumber\\
                    &=2 \sum_a | Z_{f,a}|^2
\, .
\end{align}
This expression can be compared with the
area corresponding to a single center black hole with
the same total charge
$Q=\sum_a q_a$, which is given by
$\ah(q=Q)=\sp{\ssigma Q}{Q}$.


\section{Other properties: charge vector expansions}
\label{ssection6}

Given generic real charge vectors $(q_1,q_2,\ldots,q_{n})$ 
one can define a  subspace of $W$ generated 
by eigenvectors of the matrix $\ssigma$ associated 
to the center charges, directly of the form
\begin{align}
B(q_n)\equiv \text{Span}(\Ppm q_1,\ldots,\Ppm q_n)\, ,
\end{align}
or, equivalently, in the slightly modified basis
\begin{align}
B(q_n)\equiv \text{Span}(q_1,\ldots,q_n,\ssigma q_1,\ldots,\ssigma q_n)
\, .
\end{align}
In particular, we can consider the subspace $B(q_{na})$
generated by the $n_a$ pairs ($q_a$, $\ssigma q_a$)
of center charges, whose dimension is, in general,
$\dim B(q_{na})\leq 2 n_a$.
The dimension  of the orthogonal complement to this
space, $B(q_{na})^\perp$,
\emph{i.e.} those vectors $s$ such
that $\sp{q}{s}=\sp{\ssigma q}{s}=0$
is, generically, $\dim B(q_{na})^\perp= 2 (n_v-n_a)+2$.\footnote{Or, $B(q_{na})^\perp$ is defined as the
set of vectors $s$ such that
$h(s,q)=0$ for all $q\in (q_{na})$, where $h$ is the hermitian
inner product defined in Section \ref{ssection4}.}
This dimension  is zero for one scalar, one
center black holes ($n_v=0,n_a=1$).
The set of vectors ($q_a$, $\ssigma q_a$)
may form themselves a (maybe overcomplete) basis for the
$(2 n_v+2)$ symplectic space. Otherwise, they
can be extended with as many other vectors
as necessary to complete such a basis.
Naturally, other bases are possible or convenient, for example
bases including linear combinations of the charge vectors,
the total charge vector $Q$, $\I_\infty$, etc.

We will use several expansions of different quantities
in such  a basis formed by charge and extra vectors,
to get different results.
In a first illustrative case, we will get a bound on the
black hole areas $\ah,\atot$.
In the second place,
by decomposition of   the $\I_\infty$ vector,
we will study different properties. In particular, we will see how the extremality of the solutions imposes
strong conditions on such extra vectors.


\subsection{A bound on $\ah$}

The relation between the asymptotic ``area'' $\atot$ and
the multicenter horizon area, or horizon entropy $\ah$, is simply
\begin{align}
\atot&=\sp{\ssigma Q}{Q}= \sum_{a,b} \sp{\ssigma q_a}{q_b}
\nonumber\\
&= \ah+2\sum_{a< b} \sp{\ssigma q_a}{q_b}
\, .
\label{eq429}
\end{align}
Taking into account the 
positivity of the quantities $\sp{\ssigma q_a}{q_b}$, \emph{cf.} \eqref{eq6194cc}, we arrive to 
\begin{eqnarray}
\atot-\ah&\geq&0\, .
\end{eqnarray}
For one center solution we always have $\atot=\ah$.
For the case of two centers, for example, with
charges $q_{1,2}$ the difference is
\footnote{
This is in agreement with \cite{Ferrara:2010cw} where
it has been shown  that for quadratic prepotentials, the
single center BPS extremal black hole area with
charge $Q=q_1+q_2$ is always larger
than the corresponding two-center area 
\begin{eqnarray}
\ah(Q=q_1+q_2)&\geq&\ah_{,1}+\ah_{,2}\, ,
\end{eqnarray}
or, into account that $\atot$ is also the area of
the equivalent single black hole with the same total
charge $\atot_{q1,q2}=S_h(Q=q_1+q_2)$
\begin{eqnarray}
\atot-\ah&=& 2 \sp{\ssigma q_1}{q_2}\ge 0.
\end{eqnarray}
}
\begin{eqnarray}
\atot-\ah&=& 2 \sp{\ssigma q_1}{q_2}>0\, .
\label{eq430}
\end{eqnarray}
We can use  a combination of Cauchy-Schwartz and Jensen
inequalities applied
to the scalar product which appear in the
last equation to  write   the expression
\begin{align}
2| \sp{\ssigma q_1}{q_2}|
&\le
2\sqrt{ \sp{\ssigma q_1}{q_1} \sp{\ssigma q_2}{q_2}}
\nonumber\\
&\le
 \sp{\ssigma q_1}{q_1}+ \sp{\ssigma q_2}{q_2}
=\ah
\, .
\end{align}
Then, we  arrive to the bound
\begin{align}
0\le \atot-\ah=& \ah
\, ,
\end{align}
or, equivalently,\footnote{For the multicenter case, using only
\eqref{eq429}
and applying Cauchy-Schwartz and the triangle inequalities,
we get the slightly less restrictive bound
\begin{align*}
| \atot-\ah|&\le \ah\, .
\end{align*}
}
\begin{align}
\ah \le \atot &\le 2 \ah
\, .
\label{eq430cc}
\end{align}

\subsection{Decomposition of $\I_\infty$ and double extremality}

We will decompose now the vector $\I$ into a basis of
charge and extra vectors. For the sake of simplicity we
will discuss here the case of
a single center solution and an arbitrary number of scalars.
We will see, in
particular, how the extremality of the solutions imposes
strong conditions on such extra vectors.
In addition, we will show, using this decomposition,
the double extremality of the black hole solutions
 for quadratic prepotentials.

Let us decompose the vector $\I_\infty$ in the following way
(with  $\sp{\ssigma q}{q}\neq0$)
\begin{eqnarray}
\I_\infty
&=&\alpha q+\beta\ssigma q+\gamma s +\epsilon\ssigma s\, ,
\label{eq51}
\end{eqnarray}
where $\alpha,\beta,\gamma,\epsilon\in\mathbb{R}$
and $s$ is an arbitrary but fixed, normalized vector
such that $s\in B(q_{na},\ssigma q_{na})^\perp$, \emph{i.e.}
\begin{align}
 \sp{s}{q}        &=\sp{s}{\ssigma q}=0\, ,\nonumber\\
 \sp{\ssigma s}{s}&=1\, .
\end{align}
Such vector $s$ can  be always determined by a modified
 Gram-Schmidt procedure for a given pair of
vectors $(q,\ssigmaad q)$.
By projecting the relation (\ref{eq51}) over any of the
individual vectors $q,\ssigma q$,
we get
\begin{eqnarray}
\sp{\I_\infty }{q} &=& \beta \sp{\ssigma q}{q}\, ,\nonumber\\
\sp{\I_\infty }{\ssigma q} &=& -\alpha \sp{\ssigma q}{q}\, .
\end{eqnarray}
Using the expressions \eqref{eq49},\eqref{eq529} and \eqref{eq530},
we can rewrite these last two expressions respectively as
\begin{eqnarray}
N&=& \beta \atot\, ,\nonumber
\\
M_{ADM}&=& -\alpha \atot \, ,\label{eq57}
\end{eqnarray}
from where we read the values of
the $\alpha,\beta$ coefficients
in terms of some other, more physical, parameters.
The condition $N=0$ implies that $\beta=0$, hence the
$\I_\infty$ vector does not contain any component in the
``$\ssigma q$'' direction.

Let us consider now the asymptotic flatness condition
and apply the ansatz (\ref{eq51}) for $\I_\infty$,
but without imposing at this moment the $N=0$ condition.
We have,
using the $\alpha,\beta$ values, the definition
$\Delta^2\equiv (\gamma^2+\epsilon^2)$ and \eqref{eq531b},
the expression
\begin{align}
1
&=
\sp{\ssigma\I_\infty}{\I_\infty}
\nonumber\\
&=
\left(  \alpha^2+\beta^2\right)\sp{\ssigma q}{q}
+\left(  \gamma^2+\epsilon^2\right)\sp{\ssigma s}{s}\nonumber\\
&=
\frac{M_{ADM}^2+N^2}{\atot^2}\sp{\ssigma q}{q}+\Delta^2
\,  ,
\end{align}
or, equivalently,
\begin{align}
| Z_\infty |^2=M_{ADM}^2+N^2
&=\sp{\ssigma q}{q}(1-\Delta^2)
\, .
\end{align}
The BPS condition
$| Z_\infty |=M_{ADM}=\sp{\ssigma q}{q}$ is only fulfilled
if $N=0$ (in concordance with \eqref{eq531bc})
and the additional condition $\Delta=0$. The parameter
$\Delta$ is an ``extremality'' parameter.

The vanishing of these quantities can be directly
seen  by
imposing extremality in the metric elements,
 by requesting extremal RN black hole type metric or,
$-g_{rr}\sim f^2$ with $f$ an spatially harmonic function.
The metric component $g_{rr}$ is
\begin{align}
-g_{rr}
&=  
1+\frac{2M_{ADM}}{r}+\frac{\sp{\ssigma q}{q}}{r^2}
\nonumber\\
&=
1+\frac{2M_{ADM}}{r}+\frac{(M_{ADM}^2+N^2)/(1-\Delta^2)}{r^2}
\nonumber\\
&=
\left (1+\frac{M_{ADM}}{r}\right )^2+\frac{1}{r^2}
\frac{1}{1-\Delta^2}
\left(M_{ADM}^2\Delta^2+ N^2 \right)
\, .
\end{align}
The metric element is of the form
$-g_{rr}\sim f^2$ with $f$ an spatially harmonic function
if and only if the second part of the previous expression is
zero, that is, if and only if
\begin{align}
M_{ADM}^2\Delta^2+ N^2
&=0
\, .
\end{align}
Thus, the conditions
$N=0$ and $\Delta=0$
(which is equivalent to $\gamma=\epsilon=0$ in \eqref{eq51})
are necessary conditions to
recover an extremal RN black hole type metric.
In this case, the central charge at infinity is
\begin{align}
| Z_\infty |^2 &= M_{ADM}^2 =\sp{\ssigma q}{q}\, .\label{eq616bc}
\end{align}

We see that the vanishing of the non-extremality
parameter $\Delta$ is equivalent to
require that   $\I_\infty$ is fully contained in the
subspace $\text{Span}(q,\ssigma q)$, whereas the condition
$N=0$ further restricts it to be proportional to  the
vector charge $\I_\infty=q/M_{ADM}$.
In this case, after imposing the conditions $N=\Delta=0$,
we can finally write
\begin{align}
\I  =& \frac{q}{M_{ADM}}\left (1+ \frac{M_{ADM}}{r}\right)
\, .
\label{eq567}
\end{align}
As a consequence of having $\I_\infty=q/M_{ADM}$
the scalar fields $z^\alpha$ are
constant everywhere
and equal to their
values at the fixed point (see \eqref{eq3434} and the discussion
in Section \ref{ssection5}).
It might be interesting to remark that in this expression
the ``unphysical'' vector $\I$ appears written in terms of 
the physical quantities $q$ and $M_{ADM}$ which can be input 
by hand from the beginning.

\section{Summary and concluding remarks}
\label{ssection7}

We have presented a systematic study of general, stationary,
multicenter black hole solutions in ungauged four dimensional Einstein-Maxwell  $N=2$ supergravity theories minimally
coupled to  scalars, \emph{i.e.}
theories with quadratic prepotentials.
An important part of our analysis  has been based on the
  matrices  $\ssigma_F,\ssigma_N$ and their symplectic adjoints.
These  matrices are isometries of the
symplectic bilinear form. Their adjoints with respect to
the symplectic product
$\ssigmaad_N,\ssigmaad_F$, which fulfills the
property $\ssigma^2=-\mathds{1}$,  are  shown to lay inside the
Lie algebra of the isometry group, they are such
that $\ssigma+\ssigmaad=0$. They are  ``unitary'',
$\ssigma \ssigmaad=\mathds{1}$, with respect to the symplectic product.
Inner  products, $g,h$, are defined.
The three defined structures $(g, \omega, \ssigma)$
 form a compatible triple, each structure can be specified
by the two others.
The symplectic $2n_v+2$ dimensional space $W$ is decomposed
into eigenspaces of the matrix $\ssigma$. Projection
operators over these subspaces are considered.

Using the properties of these matrices, it is shown
in particular that symplectic vectors (for which a
\emph{stabilization} equation relating their imaginary
and real parts, $\re{X}=\ssigma\im{X}$ is valid) are
inside the subspace $W^-$, one  eigenspace of  the
matrix $\ssigma$.

We derive using pure algebraic properties,
some alternative expressions for the attractor
equations, \eqref{eq48} or \eqref{eq311zz}. In
this form, the attractor equations simply
equal (a multiple of) the vector $V$,
 which  lies on the subspace $W^-$,  with the part
of the charge  vector which lies on such a subspace.
We show some
properties of the central charge modulus which
can be expressed as a norm of a charge vector induced
by the inner product $g$.

Similarly, the values of the scalars at the fixed points
and at infinity
are given by  explicit expressions, \eqref{eq416} and \eqref{eq3434}, respectively.
By these formulas, the 
values of the scalar fields at the fixed points and at infinity are given  in terms of the projection of the
charges into the ei\-gen\-spa\-ces of the
matrix $\ssigma$.

Supposing a generic multicenter  ansatz, \eqref{eq216},
(which depends on  the center charges $q_a$ and the value at
infinity $\I_\infty$) and a new form of the attractor equations,
we have  derived, or rederived in a simple  way, different
relations.
The   scalar field solutions are explicitly given by
\eqref{eq514}-\eqref{eq514s}. In particular, we study some
properties of configurations for which
$z_\infty^\alpha=z_{f}^\alpha (Q)$. For these configurations,
the scalar charges vanish, \emph{cf.} \eqref{eq47}, and the
 vector $\I_\infty$ is  of the form
 \begin{align}
 \I_\infty
 &=
 \frac{Q}{\sqrt{ \sp{\ssigma Q}{Q}}}
\, .
 \end{align}
In fact,  the vanishing of the
scalar charges  is shown to be equivalent to the vanishing
of the quantities $(z_\infty^\alpha - z_{f}^\alpha (Q))$.
This is in close
analogy with the single center case, in which the vanishing of the
scalar charges is a necessary and sufficient condition for
the double extremality of the black hole \cite{Gibbons:1996af}.

The study of the near horizon and infinity geometry of
the black hole lead us to the consideration of  the area-like
quantities
$A_{ab}=\sp{\ssigma q_a}{q_b}$ and $\atot=\sum_{ab} A_{ab}$,
in addition to the horizon areas
$\ah_{,a}=\sp{\ssigma q_a}{q_a}$.
The metric element is written as (\emph{cf.} \eqref{eq415})
\begin{align}
-g_{rr}
&=
\sp{\R}{\I}
\nonumber\\
&=
1+2\sum_b\frac{M_b}{|\vecr-\vecr_b|}
+\sum_{a,b}\frac{  A_{ab}}{|\vecr-\vecr_a||\vecr-\vecr_b|}
\, ,
\end{align}
which is  positive  and  finite for any finite $|\vecr|$
if the matrix $(A_{ab})$ is  definite positive.
This is guaranteed by the fact that this matrix is
the Gram matrix of a set of (linearly independent)
vectors $q_a$  with the \emph{inner product} $g$.

We  proceed to the minimization of
$M_{ADM}(z_\infty^\alpha)$
($M_{ADM}^2= \sp{\ssigma \I_\infty}{Q}$),
\begin{align}
\left.\frac{\partial M_{ADM}}{\partial z_\infty^\alpha}\right|_{z_{min}} &= 0
\, ,
\end{align}
with respect to the scalars at infinity for a given
charge configuration. We can show that, for a
given charge configuration, we have
\begin{align}
| Z_\infty|^2
&=M_{ADM}^2\le \sp{\ssigma Q}{Q} =\atot
\, ,
\end{align}
where the equality appears at the minimum
$\left(z_\infty^\alpha\right)_{min}=z_{f}(Q)$.
The proof of this relation
is a simple application of the
 Cauchy-Schwartz inequality to
the inner product $\sp{\ssigma X}{Y}$ and the
asymptotic flatness condition.

The near horizon geometry is
completely determined in terms of the individual
horizon areas $\ah_{,a}=\sp{\ssigma q_a}{q_a}$.
The total horizon area $\ah$ is the sum
of the areas of its disconnected parts
\begin{align}
\ah
&=
\sum_a \ah_{,a}
=
\sum_a \sp{\ssigma q_a}{q_a}
\nonumber\\
&
=
2 \sum_a | Z_{f,a}|^2
\, .
\end{align}

We  use expansions of different quantities
in terms  of symplectic charge and extra vectors
to get a series of different results. We get, for example
different   bounds of the  quantities $\ah,\atot$.
For the case of two centers  with
charges $q_{1,2}$ the relation between both quantities is
given by
$\atot-\ah= 2 \sp{\ssigma q_1}{q_2}$, \emph{cf.} \eqref{eq430}.
Using some simple general arguments, we arrive to the bound
\begin{eqnarray}
\ah \le \atot &\le& 2 \ah
\, .
\end{eqnarray}


Finally, we have studied  diverse properties and given
some explicit expression of the quantity $\I_\infty$ by
expanding this vector in a certain symplectic basis of the
form \eqref{eq51},
\begin{eqnarray}
\I_\infty
&=&\alpha q+\beta\ssigma q+\gamma s +\epsilon\ssigma s
\, .
\end{eqnarray}
We arrive to the expression for the central charge
\begin{align}
| Z_\infty |^2
&=
M_{ADM}^2+N^2
=
\sp{\ssigma q}{q}(1-\Delta^2)
\, ,
\end{align}
where $\Delta^2=\gamma^2+\epsilon^2$.
The  condition
$| Z_\infty |=M_{ADM}=\sp{\ssigma q}{q}$ is
 fulfilled
if $N=0$ and $\Delta=0$.
The vanishing of
parameter $\Delta$ is equivalent to
demanding  $\I_\infty$ to be fully contained in the
subspace $\text{Span}(q,\ssigma q)$.
We finally arrive to an explicit expression for the
solution ansatz $\I$, which for this case results
\begin{eqnarray}
\I  &=& \frac{q}{M_{ADM}}\left (1+ \frac{M_{ADM}}{r}\right)
\, .
\end{eqnarray}
As a consequence of having $\I_\infty=q/M_{ADM}$,
the scalar fields $z^\alpha$ are constant everywhere
and equal to their values at the fixed point.
In this expression
the `unphysical' vector $\I$ is written in terms of 
the physical quantities $q$ and $M_{ADM}$, which can be input 
by hand from the beginning.

The projection of any symplectic vector that appears in the
theory (for example, a subset of the charge vectors
themselves or vectors characterizing the black hole ansatz
at infinity) in terms of these new bases might be of general
interest.
The use of this projection, as it has been shown here, allows the understanding of questions as the entropy effects in the fragmentation of a single center black hole into a multicenter one. It also simplifies the study of the extremality of the solutions in terms, for example, of simple dimensional considerations of each of the charge-longitudinal and transversal subspaces.


In this study, we have focused on minimal
coupling theories with quadratic prepotentials.
It is of interest to study  to which extent, and which
modifications are needed, to apply the main
techniques, properties and expressions presented here
 to the study of extremal and non extremal solutions in
theories with general prepotentials (where the
matrix $\ssigma$ is not constant)
or even theories without them.

\chapter{Conclusions and prospects}
\label{ch:conclusions}
\markboth{6. Conclusions and prospects}{6. Conclusions and prospects}

This work comprises an analysis of diverse theoretical
topics of supergravity with three well differentiated parts:
first, the study of
gauged supergravities in higher (9D) dimensions within
the embedding tensor formalism.
The second part addresses the study of  maximal and half-maximal gauged supergravities in $D=9,8,7$.
By using the double field theory formalism, we classify which ones have a higher-dimensional geometric origin or, otherwise, are
obtained by means of a generalized Scherk-Scharwz reduction of DFT, in which the dual coordinates have a crucial importance.
Finally, extremal multicenter black hole solutions have been
considered in the context of some specific $N=2$ supergravity theories, emphasazing on those coming from special geometry
and quadratic  prepotentials.
We present full conclusions at the end of any of the
three parts and we refer to them. Here we collect a summary of these conclusions.

The first part treats the study and classification
 of maximal gauged supergravities in $d=9$ by means of  the embedding tensor formalism.
This formalism is   a covariant tool to generate all possible
gauged supergravities from a basic given theory.
It scans along all the possible combinations of the global symmetry generators catching all the gaugings allowed by the global symmetry that the ungauged theory enjoys.
Maximal $D=9$ supergravity is a feasible example on which perform this analysis due to its relatively simple field content
and group structure.

We have applied the embedding-tensor formalism to the study of
the most general deformations (\textit{i.e.}~gaugings and massive
deformations) of maximal 9-dimensional supergravity. We have used the complete
global $SL(2,\mathbb{R})\times \mathbb{R}^{2}$ symmetry of its equations of
motion, which includes the so-called \textit{trombone symmetry}. We have found
the constraints that the deformation parameters must satisfy in order to
preserve both gauge and supersymmetry invariance (the latter imposed through
the closure of the local supersymmetry algebra to lowest order in
fermions). We have used most of the constraints to express some components of
the deformation tensors in terms of a few components of the embedding tensor
which we take to be independent and which are given in
Eq.~(\ref{eq:independent}). At that point we have started making contact with
the results of Ref.~\cite{Bergshoeff:2002nv}, since those independent
components are precisely the 8 possible deformations identified there. All of
them have a higher-dimensional origin discussed in detail in
Ref.~\cite{Bergshoeff:2002nv}. The field strengths, gauge transformations and
supersymmetry transformations of the deformed theory, written in terms of the
independent deformation tensors, are collected in Appendix~\ref{d9:sec-final}.

The 8 independent deformation tensors are still subject to quadratic
constraints, given in Eq.~(\ref{eq:irreduciblequadraticconstraints}), but
those constraints cannot be used to express analytically some of them in terms
of the rest, and, therefore, we must keep the 8 deformation parameters and
we must enforce these irreducible quadratic constraints.

In Section~\ref{sec-summary} we have used our knowledge of the global
symmetries (and corresponding Noether 1-forms), the independent deformation
tensors and the irreducible quadratic constraints of the theory, together with
the general arguments of Section~\ref{sec-magnetic} to determine the possible
7-, 8- and 9-forms of the theory (Table~\ref{tab:789formweights}), which are
dual to the Noether currents, independent deformation tensors and irreducible
quadratic constraints. We have compared this spectrum of higher-rank forms
with the results of Refs.~\cite{Bergshoeff:2010xc,Bergshoeff:2011zk}, based on
$E_{11}$ level decomposition. We have found that, in the sector unrelated to
the trombone symmetry, which was excluded from that analysis, the
embedding-tensor formalism predicts one doublet of 9-forms less than the
$E_{11}$ approach. However, both predictions are not contradictory: the extra
doublet of 9-forms may not survive the deformations on which the
embedding-tensor formalism is built: new 9-form St\"uckelberg shifts
proportional to the deformation parameters may occur that can be used to
eliminate it so only one combination of the two 9-form doublets survives. This
mechanism is present in the $N=2$ $d=4,5,6$ theories \cite{Huebscher:2010ib},
although the physics behind it is a bit mysterious.

Such a powerful mechanism as the embedding tensor seems to be a suitable tool in the search of a complete catalog of gaugings for every supergravity theory in different dimensions. Depending on the aim of our research and how witty we use it, we can face different problems. The completion of this catalog of deformations is still a intriguing task that suggests to be addressed by using this technique, as recent results show \cite{Dall'Agata:2012bb}. Another problem that the embedding tensor simplifies is the search of vacua for these gauged supergravities.
The fact of having a scalar potential conveniently
 expressed in terms of $\vartheta$, together with techniques that translate our search from the moduli space to the flux background spaces \cite{Dibitetto:2011gm}, simplifies very much the exploration of vacua of a determined theory, as can be checked in \cite{Dibitetto:2012ia,Borghese:2012qm,Borghese:2012zs,Borghese:2013dja,Catino:2013ppa}.


The second part treats gauged supergravities and their origin from SS compactifications of higher-dimensional supergravities.
Once we have a tool that provides all the possible deformations of a given supergravity, we decided to use it to extend this classification to lower dimensional theories. We performed the orbit classification of maximal and half-maximal $D=9,8,7$ theories. The aim of this work is not only interesting by itself, but also results a reference to understand what orbits have a geometric origin, in the sense of arising from a SS compactification of a higher-dimensional theory. Since there is a mismatch between the existence of some gauged supergravities and the gaugings that arise from flux compactifications, several T duality constructions emerged to justify the information leak that occurs when a dimensional reduction procedure is done. Once again, we want to remark that the embedding tensor formalism is essential because it provides all the possible gaugings and guarantees the existence of no more than the ones found. In other case, we would work with a set of gaugings without being sure that of the existence of more gaugings and hence, the comparison with the flux compactification gaugings could not be performed in a systematic way.

In Chapter \ref{ch:duality_orbits} we have provided a litmus test to the notion of  non-geometry, by classifying the explicit orbits of consistent gaugings of different supergravity theories, and considering the possible higher-dimensional origins of these. The results turn out to be fundamentally different for the cases of U-duality orbits of maximal supergravities, and T-duality orbits of half-maximal theories.

In the former case we have managed to explicitly classify all U-duality orbits in dimensions $8 \leq D \leq 11$. This led to zero, one, four and ten discrete orbits in dimensions $D=11, 10,9$ and $8$, respectively, with different associated gauge groups. Remarkably, we have found that all of these orbits have a higher-dimensional origin via some geometric compactification, be it twisted reductions or compactifications on group manifolds or coset spaces. In our parlance, we have therefore found that all U-duality orbits are geometric. The structure of U-duality orbits is therefore dramatically different from the sketch of figure 1 in the introduction. Although a full classification of all orbits in lower-dimensional cases becomes increasingly cumbersome, we are not aware of any examples that are known to be non-geometric. It could therefore hold in full generality that all U-duality orbits are necessarily geometric.

This is certainly not the case for T-duality orbits of gaugings of half-maximal supergravities. In this case, we have provided the explicit classification in dimensions $7 \leq D \leq 10$ (where in $D=7$ we have only included three-form fluxes). The numbers of distinct families of orbits in this case are zero, one, three and eleven in dimensions $D=10,9,8$ and $7$, respectively, which includes both discrete and one-parameter orbits. A number of these orbits do not have a higher-dimensional origin in terms of a geometric compactification. Such cases are {\bf orbits 2} and {\bf 3} in $D=8$ and {\bf orbits 1, 2} and {\bf 3} in $D=7$ for $\alpha\neq 0$. Indeed, these are exactly the orbits that do not admit an uplift to the maximal theory. As proven in section~\ref{subsec:GDFT}, all such orbits necessarily violate the weak and/or strong constraints, and therefore need truly doubled backgrounds. Thus, the structure of T-duality orbits is very reminiscent of figure 1 in the introduction. Given the complications that already arise in these simpler higher-dimensional variants, one can anticipate that the situation will be similar in four-dimensional half-maximal supergravity.

Fortunately, the formalism of double field theory seems tailor-made to
generate additional T-duality orbits of half-maximal supergravity. Building on the recent generalization of the definition of
double field theory \cite{Grana:2012rr}, we have demonstrated that
all T-duality orbits, including the non-geometric ones in $D=7,8$,
can be generated by a twisted reduction of double field theory. We
have explicitly provided duality twists for all orbits. For locally-geometric orbits the twists only depend on the physical
coordinates $y$, while for the non-geometric orbits these
necessarily also include $\tilde y$. Again, based on our exhaustive
analysis in higher-dimensions, one could conjecture that also in
lower-dimensional theories, all T-duality orbits follow from this
generalized notion of double field theory.

At this point we would like to stress once more that a given orbit of gaugings can be generated from different twist orbits. Therefore, there is a degeneracy in the space of twist orbits giving rise to a particular orbit of gaugings. Interestingly, as it is the case of {\bf orbit 6} in $D=7$  for instance, one might find two different twist orbits reproducing the same orbit of gaugings, one  violating weak and strong constraints, the other one satisfying both. Our notion of a locally
geometric orbit of gaugings is related to the existence of at least one undoubled background giving rise to it. However, this ambiguity seems to be peculiar of gaugings containing $Q$ flux. These can, in principle, be independently obtained by
either adding a $\beta$ but no $\tilde{y}$ dependence (locally geometric choice, usually called T-fold), or by including non-trivial $\tilde{y}$
dependence but no $\beta$ (non-geometric choice) \cite{Aldazabal:2011nj}.

Another remarkable degeneracy occurs for the case of semi-simple
gaugings, corresponding to {\bf orbits 1 -- 3} in $D=7$. For the
special case of $\alpha = 0$, we have two possible ways of
generating such orbits from higher-dimensions: either a coset
reduction over a sphere or analytic continuations thereof, or a
duality twist involving non-geometric coordinate dependence.
Therefore $d$-dimensional coset reductions seem to be equivalent to
$2d$-dimensional twisted torus reductions (with the latter in fact
being more general, as it leads to all values of $\alpha$).
Considering the complications that generally arise in proving the
consistency of coset reductions, this is a remarkable reformulation
that would be interesting to understand in more detail. Furthermore,
when extending the notion of double field theory to type II and
M-theory, this relation could also shed new light on the consistency
of the notoriously difficult four-, five- and seven-sphere
reductions of these theories.

Our results mainly focus on Scherk-Schwartz compactifications leading to gauged supergravities with vanishing $\xi_M$ fluxes. In addition, we have restricted to the NSNS sector and ignored $\alpha'$-effects.
Also, we stress once again that relaxing the strong and weak constraints
is crucial in part of our analysis. If we kept the weak constraint, typically the Jacobi identities would lead to backgrounds satisfying also the strong constraint \cite{Grana:2012rr}.
However, from a purely (double) field theoretical analysis the weak constraint is not necessary. A sigma model
analysis beyond tori would help us to clarify the relation between DFT without
the weak and strong constraints and string field theory on more
general backgrounds. We hope to come back to this point in the
future.

At this point, we wonder whether we could generalize this study to lower dimensions. Unfortunately, this is a considerable more complicated goal, due to how the global symmetry groups quickly grow. This means that the classification of the orbits is extraordinarily difficult. However, some questions based on some insights of our results could be set out. What is the relation between geometric orbits and maximal supergravities? That is, is there any underlying reason why the maximal theories analyzed only host geometric orbits? On the other hand, we wonder whether all the gaugings of half-maximal theories have a description in terms of DFT. What about 1/4-BPS states? Is DFT powerful enough to reproduce those solutions? Do these states violate even the relaxed version of the strong constraint? What about the supersymmetric completion of DFT? This is an issue that has already been addressed \cite{Berman:2013cli}. Finally, a sizzling problem is the generalization of DFT towards the M theory goal. Some recent constructions have recently been proposed \cite{Aldazabal:2013mya,Berman:2013uda,Geissbuhler:2013uka}.

The third part of the manuscript treats the multicenter black hole solutions in $N=2$ theories. Despite   different solutions
 have been  working  out since long time ago,
it is not trivial to find a set of parameters that satisfy
the physical constraints of these solutions.

In Chapter \ref{ch:bh}, we have presented a systematic study of general, stationary,
multicenter black hole solutions in ungauged four dimensional Einstein-Maxwell  $N=2$ supergravity theories minimally
coupled to  scalars, \emph{i.e.}
theories with quadratic prepotentials.
An important part of our analysis  has been based on the
  matrices  $\ssigma_F,\ssigma_N$ and their symplectic adjoints.
These  matrices are isometries of the
symplectic bilinear form. Their adjoints with respect to
the symplectic product
$\ssigmaad_N,\ssigmaad_F$, which fulfills the
property $\ssigma^2=-\mathds{1}$,  are  shown to lay inside the
Lie algebra of the isometry group, they are such
that $\ssigma+\ssigmaad=0$. They are  ``unitary'',
$\ssigma \ssigmaad=\mathds{1}$, with respect to the symplectic product.
Inner  products, $g,h$, are defined.
The three defined structures $(g, \omega, \ssigma)$
 form a compatible triple, each structure can be specified
by the two others.
The symplectic $2n_v+2$ dimensional space $W$ is decomposed
into eigenspaces of the matrix $\ssigma$. Projection
operators over these subspaces are considered.

Using the properties of these matrices, it is shown
in particular that symplectic vectors (for which a
\emph{stabilization} equation relating their imaginary
and real parts, $\re{X}=\ssigma\im{X}$ is valid) are
inside the subspace $W^-$, one  eigenspace of  the
matrix $\ssigma$.

We derive using pure algebraic properties,
some alternative expressions for the attractor
equations, \eqref{eq48} or \eqref{eq311zz}. In
this form, the attractor equations simply
equal (a multiple of) the vector $V$,
 which  lies on the subspace $W^-$,  with the part
of the charge  vector which lies on such a subspace.
We show some
properties of the central charge modulus which
can be expressed as a norm of a charge vector induced
by the inner product $g$.

Similarly, the values of the scalars at the fixed points
and at infinity
are given by  explicit expressions, \eqref{eq416} and \eqref{eq3434}, respectively.
By these formulas, the 
values of the scalar fields at the fixed points and at infinity are given  in terms of the projection of the
charges into the ei\-gen\-spa\-ces of the
matrix $\ssigma$.

Supposing a generic multicenter  ansatz, \eqref{eq216},
(which depends on  the center charges $q_a$ and the value at
infinity $\I_\infty$) and a new form of the attractor equations,
we have  derived, or rederived in a simple  way, different
relations.
The   scalar field solutions are explicitly given by
\eqref{eq514}-\eqref{eq514s}. In particular, we study some
properties of configurations for which
$z_\infty^\alpha=z_{f}^\alpha (Q)$. For these configurations,
the scalar charges vanish, \emph{cf.} \eqref{eq47}, and the
 vector $\I_\infty$ is  of the form
 \begin{align}
 \I_\infty
 &=
 \frac{Q}{\sqrt{ \sp{\ssigma Q}{Q}}}
\, .
 \end{align}
In fact,  the vanishing of the
scalar charges  is shown to be equivalent to the vanishing
of the quantities $(z_\infty^\alpha - z_{f}^\alpha (Q))$.
This is in close
analogy with the single center case, in which the vanishing of the
scalar charges is a necessary and sufficient condition for
the double extremality of the black hole \cite{Gibbons:1996af}.

The study of the near horizon and infinity geometry of
the black hole lead us to the consideration of  the area-like
quantities
$A_{ab}=\sp{\ssigma q_a}{q_b}$ and $\atot=\sum_{ab} A_{ab}$,
in addition to the horizon areas
$\ah_{,a}=\sp{\ssigma q_a}{q_a}$.
The metric element is written as (\emph{cf.} \eqref{eq415})
\begin{align}
-g_{rr}
&=
\sp{\R}{\I}
\nonumber\\
&=
1+2\sum_b\frac{M_b}{|\vecr-\vecr_b|}
+\sum_{a,b}\frac{  A_{ab}}{|\vecr-\vecr_a||\vecr-\vecr_b|}
\, ,
\end{align}
which is  positive  and  finite for any finite $|\vecr|$
if the matrix $(A_{ab})$ is  definite positive.
This is guaranteed by the fact that this matrix is
the Gram matrix of a set of (linearly independent)
vectors $q_a$  with the \emph{inner product} $g$.

We  proceed to the minimization of
$M_{ADM}(z_\infty^\alpha)$
($M_{ADM}^2= \sp{\ssigma \I_\infty}{Q}$),
\begin{align}
\left.\frac{\partial M_{ADM}}{\partial z_\infty^\alpha}\right|_{z_{min}} &= 0
\, ,
\end{align}
with respect to the scalars at infinity for a given
charge configuration. We can show that, for a
given charge configuration, we have
\begin{align}
| Z_\infty|^2
&=M_{ADM}^2\le \sp{\ssigma Q}{Q} =\atot
\, ,
\end{align}
where the equality appears at the minimum
$\left(z_\infty^\alpha\right)_{min}=z_{f}(Q)$.
The proof of this relation
is a simple application of the
 Cauchy-Schwartz inequality to
the inner product $\sp{\ssigma X}{Y}$ and the
asymptotic flatness condition.

The near horizon geometry is
completely determined in terms of the individual
horizon areas $\ah_{,a}=\sp{\ssigma q_a}{q_a}$.
The total horizon area $\ah$ is the sum
of the areas of its disconnected parts
\begin{align}
\ah
&=
\sum_a \ah_{,a}
=
\sum_a \sp{\ssigma q_a}{q_a}
\nonumber\\
&
=
2 \sum_a | Z_{f,a}|^2
\, .
\end{align}

We  use expansions of different quantities
in terms  of symplectic charge and extra vectors
to get a series of different results. We get, for example
different   bounds of the  quantities $\ah,\atot$.
For the case of two centers  with
charges $q_{1,2}$ the relation between both quantities is
given by
$\atot-\ah= 2 \sp{\ssigma q_1}{q_2}$, \emph{cf.} \eqref{eq430}.
Using some simple general arguments, we arrive to the bound
\begin{eqnarray}
\ah \le \atot &\le& 2 \ah
\, .
\end{eqnarray}


Finally, we have studied  diverse properties and given
some explicit expression of the quantity $\I_\infty$ by
expanding this vector in a certain symplectic basis of the
form \eqref{eq51},
\begin{eqnarray}
\I_\infty
&=&\alpha q+\beta\ssigma q+\gamma s +\epsilon\ssigma s
\, .
\end{eqnarray}
We arrive to the expression for the central charge
\begin{align}
| Z_\infty |^2
&=
M_{ADM}^2+N^2
=
\sp{\ssigma q}{q}(1-\Delta^2)
\, ,
\end{align}
where $\Delta^2=\gamma^2+\epsilon^2$.
The  condition
$| Z_\infty |=M_{ADM}=\sp{\ssigma q}{q}$ is
 fulfilled
if $N=0$ and $\Delta=0$.
The vanishing of
parameter $\Delta$ is equivalent to
demanding  $\I_\infty$ to be fully contained in the
subspace $\text{Span}(q,\ssigma q)$.
We finally arrive to an explicit expression for the
solution ansatz $\I$, which for this case results
\begin{eqnarray}
\I  &=& \frac{q}{M_{ADM}}\left (1+ \frac{M_{ADM}}{r}\right)
\, .
\end{eqnarray}
As a consequence of having $\I_\infty=q/M_{ADM}$,
the scalar fields $z^\alpha$ are constant everywhere
and equal to their values at the fixed point.
In this expression
the `unphysical' vector $\I$ is written in terms of 
the physical quantities $q$ and $M_{ADM}$, which can be input 
by hand from the beginning.

The projection of any symplectic vector that appears in the
theory (for example, a subset of the charge vectors
themselves or vectors characterizing the black hole ansatz
at infinity) in terms of these new bases might be of general
interest.
The use of this projection, as it has been shown here, allows the understanding of questions as the entropy effects in the fragmentation of a single center black hole into a multicenter one. It also simplifies the study of the extremality of the solutions in terms, for example, of simple dimensional considerations of each of the charge-longitudinal and transversal subspaces.


In this study, we have focused on minimal
coupling theories with quadratic prepotentials.
It is of interest to study  to which extent, and which
modifications are needed, to apply the main
techniques, properties and expressions presented here
 to the study of extremal and non extremal solutions in
theories with general prepotentials (where the
matrix $\ssigma$ is not constant)
or even theories without them.

\begin{appendices}

\chapter{Nuts and bolts: T-duality}
\label{sec:Tduality}

In the framework of supergravities considered as low energy 
effective field theories of string theories, 
the global symmetries of the SUGRAs are seen to 
correspond to \emph{dualities} of the string theories \cite{Hull:1994ys}.

Some of these string dualities are essentially perturbative and 
the worldsheet approach is valid to be studied. 
For instance, \emph{T-duality} \cite{Giveon:1994fu}, 
that relates string theories compactified on circles of radius $R$ and 
dual radius $R'=1/R$, is an exact symmetry at all orders in string 
perturbation theory \cite{Alvarez:1989ad}. 
However, the so-called \emph{S-duality}, is non-perturbative in the 
string coupling constant  and cannot be studied using the 
standard worldsheet approach. 
Finally, \emph{U-duality} is another duality that includes S- and T- 
duality and is considered directly related to the existence of the so-called M theory.



We will  show some basic ideas of T-duality
in the next 
paragraphs.


\subsubsection{The bosonic string}
We will restrict to the string common sector. We will follow \cite{Bergshoeff:1994cb,Ortin:2004ms}. Since T-duality relates different theories compactified on a circle, we will choose the effective action \eqref{eq:effective_ST} as the one on which to perform the dimensional reduction. We will get a $D=\hat D-1$ dimensional theory that will enjoy this duality. Let us assume the following standard KK reduction ansatz,
\begin{align}
\label{eq:bosonic_ansatz}
\hat e_{\hat\mu}{}^{\hat a}
&=
\left(\begin{array}{cc}
e_\mu{}^a & k A_\mu 
\\
0 & k
\end{array}\right)
\, ,
&
\phi
&=
\hat \phi-\frac{1}{2}\ln k
\, ,
\nonumber
\\
\hat B_{\mu\nu}
&=
B_{\mu\nu}
-A_{[\mu}B_{\nu]}
\, ,
&
\hat B_{\mu z} 
&=
B_\mu 
\, ,
\end{align}
where $\hat\mu=\{\mu,z\}$, \emph{i.e.} hatted indices and fields are defined on $\hat D$ dimensions and the unhatted ones correspond to $D$ dimensions. We will refer as $z$ the compactified coordinate.
After integrating over the compact coordinate, the reduced effective action is
\begin{align}
S
&\sim
\int d^Dx\sqrt{|g|} e^{-2\phi}\left(
	R
	-4(\partial\phi)^2
	+\frac{1}{2\cdot 3!} H^2
	+(\partial\log k)^2
	-\frac{1}{4}k^2 F^2(A)
	-\frac{1}{4}k^{-2}F^2(B)
	\right]
\, ,
\end{align}
where $F(A)$ and $F(B)$ are the field strengths of the vector fields $A_\mu$ and $B_\mu$, respectively. We can check the invariance of this action under the transformation rules
\begin{align}
A_\mu 
&\rightarrow
B_\mu 
\, ,
&
B_\mu 
&\rightarrow
A_\mu 
\, ,
&
k
&\rightarrow
k^{-1}
\, ,
\label{eq:T_dual_bosonic}
\end{align}
so that the KK scalar gets inverted and the KK vector and the winding vector are interchanged. Two interpretations can be done: first, we compactify a string background, T dualize it, and decompactify it into a different background. Second, we have two different compactifications of a given background; these compactifications give the same $D$-dimensional background and thus, are dual.

The way in that these two backgrounds are related is described by an isometry. These field relations are known as \emph{Buscher's rules} \cite{Buscher:1985kb,Buscher:1987sk,Buscher:1987qj}.
\begin{align}
\hat e'^a{}_z
&=
\mp \frac{\hat e^a{}_z}{\hat g_{zz}}
\, ,
&
\hat e'^a{}_\mu 
&=
\hat e^a{}_\mu 
-\frac{\hat g_{\mu z}\pm\hat B_{\mu z}}{\hat g_{zz}}
\, ,
\nonumber
\\
\hat B'_{\mu z}
&=
\frac{\hat g_{\mu z}}{\hat g_{zz}}
\, ,
&
\hat B'_{\mu\nu}
&=
\hat B_{\mu\nu}
+2\frac{\hat g_{[\mu|z|}\hat B_{\nu] z}}{\hat g_{zz}}
\, ,
\\
\hat \phi'
&=
\hat \phi
-\frac{1}{2}\ln|\hat g_{zz}|
\, .
\nonumber
\end{align}

Now, at the string level, 
let us study T-duality applied to the $\sigma$-model of the bosonic string introduced in \eqref{eq:gen_ST_action}, without considering the dilaton term, since it does not play any relevant role in this classic approach. Let us assume \eqref{eq:gen_ST_action} with hatted fields running over hatted indices. Then, decomposing the $\hat D$-dimensional fields into $D$-dimensional fields using \eqref{eq:bosonic_ansatz}, we have
\begin{align}
S
&=
-\frac{T}{2}\int d^2\sigma \sqrt{|\gamma|}\left[\gamma^{ij}g_{ij}-k^2F^2\right]
+\frac{T}{2}\int d^2\sigma \sqrt{|\gamma|}\epsilon^{ij}\left[B_{ij}+A_iB_j-2F_iB_j\right]
\, ,
\end{align}
where $g_{ij}$, $B_{ij}$, $A_i$, $B_i$ are the pullbacks of the $D$-dimensional metric, KR 2-form, KK and winding vectors respectively. $F_i$ is the field strength of the $Z$ coordinate,
\begin{align}
F_i
&=
\partial_i Z
+A_i
\, ,
\end{align}
which reflects the following shift invariance:
\begin{align}
\delta_\Lambda Z 
&=
-\Lambda(x)
\, ,
\nonumber
\\
\delta_\Lambda A_\mu
&=
\partial_\mu \Lambda
\, .
\end{align}
This invariance implies the following conserved current
\begin{align}
P_z{}^i
&=
T(k^2F^i-\star B^i)
\, ,
\end{align}
whose associated magnetic-like conserved current is
\begin{align}
W_z{}^i
&=
T\star F^i-\star A^i
\, .
\end{align}
Their associated charges are the momentum of the string in the compact dimension and the winding number, respectively. Then, if we perform a Poincar\'e duality transformation on the $Z$ coordinate, $Z\rightarrow Z'$, by using the Bianchi identity of $F_i$ and its equation of motion, we have 
\begin{align}
 S'
&=
-\frac{T}{2}\int d^2\sigma \sqrt{|\gamma|}\left[\gamma^{ij}g_{ij}-k^{-2}F'^2\right]
+\frac{T}{2}\int d^2\sigma \sqrt{|\gamma|}\epsilon^{ij}\left[B_{ij}+B_iA_j-2F'_iA_j\right]
\, ,
\end{align}
where 
\begin{align}
F'_i
&=
\partial_i Z'+B_i
\, .
\end{align}
This action coincides with the original one when we make the field replacements \eqref{eq:T_dual_bosonic}. We find that its conserved currents, $P_{Z'}^i$ and $W_{Z'}^i$, are closely related to those of the original theory,
\begin{align}
P_{Z'}^i
&=
W_Z{}^i
\, ,
&
W_{Z'}^i
&=
P_Z{}^i
\, .
\end{align}

Thus, we summarize that T-duality inverts the compactification radius and interchanges momentum modes with winding modes, leaving invariant the mass spectrum and performing a parity transformation on the right-moving modes.

For type II superstrings, this parity transformation changes the chirality of the spinors and the overall result is that the $N=(1,1)$ type IIA  theory can be mapped into the $N=(2,0)$ type IIB version. This relation holds for any value of the radius, in particular it relates the limits $R\rightarrow0$ and $R\rightarrow \infty$. For the case of $N=2A$ and $N=2B$ supergravity theories, 
there is a discrete symmetry relating the two supergravity theories when both of them are reduced to 9 dimensions \cite{Bergshoeff:1995cg}. A generalization of the Buscher's rules can be established \cite{Bergshoeff:1995as,Meessen:1998qm} when one performs dimensional reductions from $N=2A$ and $N=2B$ to $D=9$ and identifies the same fields from the two different reduction schemes \cite{Ortin:2004ms},
\begin{align}
\hat J_{\mu\nu}
&=
\hat g_{\mu\nu}
-\frac{\hat g_{\mu z} \hat g_{\nu z}-\hat B_{\mu z} \hat B_{\nu z}}{\hat g_{zz}}
\, ,
&
\hat J_{\mu y}
&=
\frac{\hat B_{\mu z}}{\hat g_{zz}}
\, ,
\nonumber
\\
\hat B_{\mu\nu}
&=
\hat B_{\mu\nu}
+\frac{\hat g_{\mu z} \hat B_{\nu z}-\hat B_{\mu z} \hat g_{\nu z}}{\hat g_{zz}}
\, ,
&
\hat B_{\mu y}
&=
\frac{\hat g_{\mu z}}{\hat g_{zz}}
\, ,
\nonumber
\\
\hat \varphi
&=
\hat\phi
-\frac{1}{2}\ln |\hat g_{zz}|
\, ,
&
\hat J_{yy}
&=
\frac{1}{\hat g_{zz}}
\, ,
\label{eq:Buscher_2AB}
\end{align}
\vspace{-7mm}
\begin{align*}
\hat C^{(2n)}_{\mu_1\cdots\mu_{2n}}
&=
\hat C^{(2n+1))}_{\mu_1\cdots\mu_{2n}z}
+2n\hat B_{[\mu_1|z|}\hat C^{(2n-1)}_{\mu_2\cdots\mu_{2n}]}
-2n(2n-1)\frac{\hat B_{[\mu_1|z|}\hat g_{\mu_2|z|}\hat C^{(2n-1)}_{\mu_3\cdots\mu_{2n}]z}}{g_{zz}}
\, ,
\\
\hat C^{(2n)}_{\mu_1\cdots\mu_{2n-1}y}
&=
-\hat C^{(2n-1))}_{\mu_1\cdots\mu_{2n-1}}
+(2n-1)\frac{\hat g_{[\mu_1|z|}\hat C^{(2n-1)}_{\mu_2\cdots\mu_{2n-1}]z}}{g_{zz}}
\, .
\end{align*}

On the other hand, T-duality effects on the heterotic superstrings result in the transformation laws of the heterotic whose gauge group is $E_8\times E_8$ into the heterotic theory with $SO(32)$ as a gauge group, and vice versa \cite{Ginsparg:1986bx}.

T-duality in type I string theory is even more subtle. We can obtain the effective action of type I by considering type IIB and truncating it using one of its $\mathbb{Z}_2$ symmetries plus the inclusion of an O9-plane and 32 D9-branes.\footnote{Reference \cite{Ortin:2004ms} pedagogically shows how to do it.} The T-duality between type IIB and type IIA theories implies the existence of the so-called \emph{type I'} \cite{Polchinski:1996na}, which can be interpreted as a rotation of the space where we compactify. This implies the interchange of Neumann and Dirichlet boundary conditions for certain coordinates.

The examples of T-duality that we have discussed are only the tip of a mathematical iceberg: there exist additional dualities known as \emph{mirror symmetries}, in which different 10-dimensional string theories compactified on Calabi-Yau manifolds are related to each other \cite{Candelas:1993dm}.

\chapter{Gaugings in $N=2$ $D=9$ supergravity}
\label{app:gaugings_d9}

\section{Conventions}
\label{d9:app-conventions}

We follow the conventions of Ref.~\cite{Bergshoeff:2002nv}.  In particular, we
use mostly plus signature $(-,+,\cdots ,+)$ and the gamma matrices satisfy

\begin{equation}
\gamma^{*}_{a}= -\gamma_{a}\, ,
\hspace{1cm}
\gamma_{a}=\eta_{aa}\gamma_{a}^{\dagger}\, .
\end{equation}

The Dirac conjugate of a spinor $\epsilon$ is defined by

\begin{equation}
\bar{\epsilon} \equiv \epsilon^{\dagger}\gamma_{0}\, .  
\end{equation}

Then, we have

\begin{equation}
 \label{eq:anbn}
\begin{array}{rcl}
(\bar{\epsilon}\gamma^{(n)}\lambda)^{*}
 & = & 
a_{n}\bar{\epsilon}^{*}\gamma^{(n)}\lambda^{*}\, ,
\\
& & \\
(\bar{\epsilon}\gamma^{(n)}\lambda)^{*}
 & = & 
b_{n}\bar{\lambda}\gamma^{(n)}\epsilon \, ,
\\
\end{array}
\end{equation}

\noindent
where the signs $a_{n}$ and $b_{n}$ are  given in Table~\ref{tab:anbn}
  
\begin{table}
    \centering
    \begin{tabular}{ccccccccccc}
\hline
$n$     & 0   & 1   & 2   & 3   & 4   & 5  & 6   & 7   & 8   & 9   \\
\hline\hline
$a_{n}$ & $-$ & $+$ & $-$ & $+$ & $-$ & $+$ & $-$ & $+$ & $-$ & $+$ \\        
$b_{n}$ & $+$ & $-$ & $-$ & $+$ & $+$ & $-$ & $-$ & $+$ & $+$ & $-$ \\
\hline
   \end{tabular}
   \caption{Values of the coefficients $a_{n}$ and $b_{n}$ defined in Eqs.~(\ref{eq:anbn}).}
    \label{tab:anbn}
\end{table}


\subsection{Spinor bilinears}
\label{d9:sec-bilinears}

We define the following real bilinears of the supersymmetry parameters
$\epsilon_{1}$ and $\epsilon_{2}$:

\begin{eqnarray}
\bar{\epsilon}_{2}\epsilon_{1}
& \equiv &
a+ib\, ,
\\
& & \nonumber \\   
\bar{\epsilon}_{2}\epsilon^{*}_{1}
& \equiv &
c+id\, ,
\\
& & \nonumber \\   
\bar{\epsilon}_{2}\gamma_{\mu_{1}\cdots \mu_{n}}\epsilon_{1}
& \equiv &
\xi_{\mu_{1}\cdots \mu_{n}}+i\zeta_{\mu_{1}\cdots \mu_{n}}\, ,
\\
& & \nonumber \\   
\bar{\epsilon}_{2}\gamma_{\mu_{1}\cdots \mu_{n}}\epsilon_{1}^{*}
& \equiv &
\sigma_{\mu_{1}\cdots \mu_{n}}+i\rho_{\mu_{1}\cdots \mu_{n}}\, ,
\end{eqnarray}


\section{Relation with other conventions}
\label{d9:sec-relationwithotherconventions}

The electric fields used in this paper are related to those used in
Ref.~\cite{Meessen:1998qm} (which uses a mostly minus signature) as
follows:

\begin{eqnarray}
K 
& = & 
e^{\frac{\sqrt{7}}{3}\varphi}\, ,
\\
& & \nonumber \\
\lambda \equiv C^{(0)} +ie^{-\varphi} 
& = & 
\tau \equiv \chi +ie^{-\phi}\, ,
\\
& & \nonumber \\ 
A_{(1)} 
& = &
A^{0}\, ,
\\
& & \nonumber \\
\mathbf{A}_{(1)} 
& = &
A^{\mathbf{i}}\, ,
\\
& & \nonumber \\
\mathbf{A}_{(2)} 
& = &
B^{i} +\tfrac{1}{2}A^{0\mathbf{i}}\, ,
\\
& & \nonumber \\
A_{(3)} 
& = &
-C +\tfrac{1}{2}\varepsilon_{\mathbf{i}j}A^{\mathbf{i}}\wedge B^{j} 
-\tfrac{1}{12} \varepsilon_{\mathbf{ij}} A^{0\mathbf{ij}}\, ,
\\
& & \nonumber \\
A_{(4)} 
& = &
-\tilde{C} +C\wedge A^{0} -\tfrac{1}{4}\varepsilon_{i\mathbf{j}}B^{i}\wedge
A^{0\mathbf{j}}\, .
\end{eqnarray}

\noindent
The field strengths are related by

\begin{eqnarray}
F_{(2)} 
& = &
F^{0}\, ,
\\
& & \nonumber \\
\mathbf{F}_{(2)} 
& = &
F^{\mathbf{i}}\, ,
\\
& & \nonumber \\
\mathbf{F}_{(3)} 
& = &
H^{i}\, ,
\\
& & \nonumber \\
F_{(4)} 
& = &
-G\, ,
\\
& & \nonumber \\
F_{(5)} 
& = &
-\tilde{G}\, .
\end{eqnarray}

The relation with the fields used in Ref.~\cite{Bergshoeff:2002nv}
(which also uses mostly plus signature) is given by (our fields are in
the r.h.s.~of these equations)

\begin{eqnarray}
B^{i}
& = &
-(B^{i} +\tfrac{1}{2}A^{0\mathbf{i}})\, ,   
\\
& & \nonumber \\
C
& = & 
-(C -\tfrac{1}{6}\varepsilon_{\mathbf{ij}}A^{0\mathbf{ij}})\, ,
\end{eqnarray}

\noindent
while the field strengths are related by 

\begin{eqnarray}
H^{i} & = & -H^{i}\, ,\\
& & \nonumber \\
G & = & -G\, .
\end{eqnarray}

\noindent
The rest of the fields are identical.

\section{Noether currents}
\label{d9:sec-noether}

The Noether 1-form currents of the undeformed theory $j_{A}$ are given by

\begin{align}
\star j_{m}
& =
\star d\mathcal{M}_{ij}\left(\mathcal{M}^{-1}\right)_{jk}T_{mi}{}^k
+e^{\frac{4}{\sqrt7}\varphi}(
\mathcal{M}^{-1}_{\mathbf{ij}})T_{m\mathbf{k}}{}^\mathbf{i} 
A^\mathbf{k}\wedge \star F^\mathbf{j}
\nonumber \\
& \nonumber \\
&
+T_{mk}{}^i\left[ 
	e^{-\frac{1}{\sqrt7}\varphi}\mathcal{M}^{-1}_{ij}\left(
		B^k-\tfrac{1}{2}A^{0k}
	\right)\wedge\star H^j
+\tfrac{1}{2}\varepsilon_{ij}\left(
	-2e^{\frac{2}{\sqrt7}\varphi}A^\mathbf{j}\wedge B^k\wedge\star G
\right.\right.
\nonumber \\
& \nonumber \\
&
\left.\left.
\phantom{e^{\frac{2}{\sqrt7}\varphi}}
	+\left(
		B^j-A^{0j}
	\right)\wedge B^k\wedge G
	+\varepsilon_{ln}A^l\wedge B^{jk}\wedge\left(
		H^n-\tfrac{1}{2}A^n\wedge F^0
	\right)
\right.\right.
\nonumber \\
& \nonumber \\
&
\left.\left.
\phantom{e^{\frac{2}{\sqrt7}\varphi}}
	+\tfrac{1}{4}\varepsilon_{ln}A^{0ln}\wedge B^k\wedge H^j
\right)\right]\, ,
\end{align}

\begin{align}
\star j_4
&=
\tfrac{6}{\sqrt7}\star d\varphi
+3\left[
	e^{\frac{4}{\sqrt7}\varphi}A^0\wedge\star F^0+e^{-\frac{1}{\sqrt7}\varphi}\mathcal{M}^{-1}_{ij}\left(
		B^i+\tfrac{1}{2}A^{0i}
	\right)\wedge\star H^j
	+e^{\frac{2}{\sqrt7}\varphi}\left(
		C-\tfrac{1}{6}\varepsilon_\mathbf{ij}A^{0\mathbf{ij}}
	\right)\wedge\star G
\right.
\nonumber \\
& \nonumber \\
&
\left.
\phantom{e^{\frac{2}{\sqrt7}\varphi}}
	+A^0\wedge\left(
		C+\varepsilon_{\mathbf{i}j}A^{\mathbf{i}}\wedge B^j
	\right)\wedge G
\right]
+\tfrac{3}{2}\varepsilon_{ij}\left[
	\left(
		-C
		+\varepsilon_{kl}A^k\wedge B^l
		-\tfrac{7}{12}\varepsilon_{kl}A^{0kl}
	\right)\wedge B^i\wedge H^j
\right.
\nonumber \\
& \nonumber \\
&
\left.
	-\tfrac{3}{2}A^{0i}\wedge C\wedge H^j
	+\left(
		A^i\wedge B^j
		-\tfrac{1}{2} A^{0ij}
	\right)\wedge F^0\wedge C
\right]\, ,
\end{align}

\begin{align}
\star j_5
& =
\tfrac{\sqrt7}{4}\star d\varphi
-\tfrac{3}{8}\star\frac{\tau d\bar\tau+\textrm{c.c.}}{(\Im\textrm{m}\tau)^2}
+e^{\frac{4}{\sqrt7}\varphi}T_{50}{}^0 A^0\wedge\star F^0
+e^{\frac{3}{\sqrt7}\varphi}T_{5\mathbf{k}}{}^\mathbf{i}\mathcal{M}^{-1}_{ij}A^\mathbf{k}\wedge\star F^\mathbf{j}
\nonumber \\
& \nonumber \\
&
+e^{-\frac{1}{\sqrt7}\varphi}\mathcal{M}^{-1}_{ij}\left[
		T_{5k}{}^i\left(
			B^k-\tfrac{1}{2}A^{0k}
		\right)
+\tfrac{1}{4}A^{0i}
\right]\wedge\star H^j
\nonumber \\
& \nonumber \\
&
+e^{\frac{2}{\sqrt7}\varphi}\left(
	T_5  C
	-\tfrac{1}{12}\varepsilon_{ij}A^{0ij}
	-T_{5k}{}^i\varepsilon_{ij}\left(
		A^k\wedge B^j
		-\tfrac{1}{6}A^{0kj}
	\right)
\right)\wedge\star G
\nonumber \\
& \nonumber \\
&
+\tfrac{1}{4}\varepsilon_{ij}\left[
	T_{5k}{}^i\left(
		-2B^{jk}
		+3A^{0j}\wedge B^k
		-5A^{0k}\wedge B^j
	\right)
	-\tfrac{1}{2}A^{0i}\wedge B^j
	\right]\wedge G
\nonumber \\
& \nonumber \\
&
+\tfrac{1}{4}\varepsilon_{ij}\left[
	T_{5k}{}^i\left(
		+2\varepsilon_{ln}A^l\wedge B^{nk}
		-\varepsilon_{ln}A^{0ln}\wedge B^k
	\right)
	-T_5\left(6A^{0i}+B^i\right)\wedge C
	-\tfrac{1}{12}\varepsilon_{kl}A^{0kl}\wedge B^i
\right]\wedge H^j
\nonumber \\
& \nonumber \\
&
+\varepsilon_{ij}\varepsilon_{ln}T_{5k}{}^i\left[
	\tfrac{5}{6}A^{0jk}\wedge B^l
	-A^{0lj}\wedge B^k
	+\tfrac{1}{2}A^k\wedge B^{jl}
\right]\wedge H^n
\nonumber \\
& \nonumber \\
&
+T_5\left[
	A^0\wedge C\wedge G
	+\tfrac{1}{2}\varepsilon_{ij}
		\left(B^j+\tfrac{1}{2}A^{0j}\right)\wedge A^i\wedge F^0\wedge C		
\right]
\end{align}


\section{Final results}
\label{d9:sec-final}

In this Appendix we give the final form of the deformed covariant field
strengths, covariant derivatives, gauge and supersymmetry transformations in
terms of the independent deformation parameters given in
Eq.~\ref{eq:independent}. We must bear in mind that they are assumed to
satisfy the irreducible quadratic constraints given in
Eq.~(\ref{eq:irreduciblequadraticconstraints}) and only then the field
strengths etc.~have the right transformation properties.

The covariant derivatives of the scalar fields are given by

\begin{align}
\mathfrak{D} \varphi
& =
-\tfrac{137}{24\sqrt{7}}\vartheta_{0}{}^{5} A^{0}
+\left(
-\tfrac{\sqrt{7}}{4}\vartheta_{\mathbf{i}}{}^{4}
+\tfrac{6}{\sqrt{7}}\vartheta_{\mathbf{i}}{}^{5}
\right)
A^{\mathbf{i}}\, ,
\\
& \nonumber \\
\mathfrak{D} \tau
& =
\vartheta_{0}{}^{m} k_{m}{}^{\tau} A^{\mathbf{1}}
-\tfrac{3}{4}\vartheta_{0}{}^{5}\tau A^{0}
+\tfrac{3}{4}\left(
\vartheta_{\mathbf{1}}{}^{5}\tau
+\vartheta_{\mathbf{2}}{}^{5}
\right)
\left(A^{\mathbf{1}}-\tau A^{\mathbf{2}}
\right)\, ,
\end{align}

\noindent
and their gauge transformations are explictly given by 

\begin{align}
\delta_{\Lambda} \varphi
& =
-\tfrac{137}{24\sqrt{7}}\vartheta_{0}{}^{5}\Lambda^{0}
+\left(
-\tfrac{\sqrt{7}}{4}\vartheta_{\mathbf{i}}{}^{4}
+\tfrac{6}{\sqrt{7}}\vartheta_{\mathbf{i}}{}^{5}
\right)
\Lambda^{\mathbf{i}}\, ,
\\
& \nonumber \\
\delta_{\Lambda} \tau
& =
\vartheta_{0}{}^{m} k_{m}{}^{\tau} \Lambda^{0}
-\tfrac{3}{4}\vartheta_{0}{}^{5}\tau\Lambda^{0}
+\tfrac{3}{4}
\left(
\vartheta_{\mathbf{1}}{}^{5}\tau
+\vartheta_{\mathbf{2}}{}^{5}
\right)
\left(
\Lambda^{\mathbf{1}}
-\tau\Lambda^{\mathbf{2}}
\right)\, .
\end{align}

The deformed $p$-form field strengths are given by 

\begin{eqnarray}
F^{0}
& = &
dA^{0}
-\tfrac{1}{2}\left(
3\vartheta_{\mathbf{i}}{}^{4}
+\tfrac{1}{2}\vartheta_{\mathbf{i}}{}^{5}
\right)A^{0\mathbf{i}}
+\left(
3\vartheta_{\mathbf{i}}{}^{4}
+\tfrac{1}{2}\vartheta_{\mathbf{i}}{}^{5}
\right)B^{\mathbf{i}}\, ,
\\
& & \nonumber \\
F^{\mathbf{i}}
& = &
dA^{\mathbf{i}}
+\tfrac{1}{2}\left(
\vartheta_{0}{}^{m}(T_{m}^{(3)})_{\mathbf{j}}{}^{\mathbf{i}}A^{0\mathbf{j}}
-\tfrac{3}{4}\delta_{\mathbf{1}}{}^\mathbf{i}\vartheta_{0}{}^{5} A^{0\mathbf{1}}
+\tfrac{3}{2}\varepsilon^{\mathbf{ij}}\vartheta_{\mathbf{j}}{}^{5} A^{\mathbf{12}}
\right)
\nonumber \\
& & \nonumber \\
& &
+\vartheta_{0}{}^{m}(T_{m}^{(3)})_{\mathbf{j}}{}^{\mathbf{i}} B^{j}
-\tfrac{3}{4}\delta_{1}{}^{\mathbf{i}}\vartheta_{0}{}^{5}B^{1}\, ,
\end{eqnarray}

\begin{eqnarray}
H^{i}
& = & 
\mathfrak{D}B^{i}
+\tfrac{1}{2}
\left(
A^{0}\wedge dA^{\mathbf{i}}+A^{\mathbf{i}}\wedge dA^{0}
\right)
+\tfrac{1}{6}
\varepsilon^{i\mathbf{j}}
\left(
3\vartheta_\mathbf{j}{}^{4}
+\tfrac{1}{2}\vartheta_{\mathbf{j}}{}^{5}
\right)A^{0\mathbf{12}}
\nonumber \\
& & \nonumber \\
& & 
+\varepsilon^{i\mathbf{j}} \left(
3\vartheta_{\mathbf{j}}{}^{4}
-\tfrac{1}{4}\vartheta_{\mathbf{j}}{}^{5}
\right)C\, ,
\\
& & \nonumber \\
G
& = &
\mathfrak{D}C
-\varepsilon_{\mathbf{i}j} \left[ F^{\mathbf{i}}\wedge B^{j} 
-\tfrac{1}{2} \delta^{j}{}_{\mathbf{j}}
\left(A^{\mathbf{i}}\wedge dA^{\mathbf{j} }
-\tfrac{1}{3}d(A^{0\mathbf{ij}}) \right) \right]
\nonumber \\
& & \nonumber \\
& &
+\tfrac{1}{2}\left(
\varepsilon_{ij}\vartheta_0{}^m(T_m^{(2)})_k{}^i B^{jk}
-\tfrac{3}{4}\vartheta_0{}^5 B^{12}
\right)
+Z\tilde C\, ,
\end{eqnarray}

\noindent
where the covariant derivatives acting on the different fields are given by 

\begin{eqnarray}
\mathfrak{D}B^{i}
& = &
dB^{i}
+\vartheta_{0}{}^{m}(T_{m}^{(2)})_{j}{}^{i} A^{0}\wedge B^{j}
-\tfrac{3}{4}\delta_{1}{}^{i}\vartheta_{0}{}^{5} A^{0}\wedge B^{1}
\nonumber \\
& & \nonumber \\
& &
+\left(
3\vartheta_{\mathbf{k}}{}^{4}-\tfrac{1}{4}\vartheta_{\mathbf{k}}{}^{5}
\right)
A^{\mathbf{k}}\wedge B^{i}
+\tfrac{3}{4}\delta_{\mathbf{j}}{}^{i}\vartheta_{k}{}^{5} A^{\mathbf{j}} \wedge B^{k}\, ,
\\
& & \nonumber \\
\mathfrak{D}C
& = & 
dC
-\tfrac{3}{4}\vartheta_{0}{}^{5} A^{0}\wedge C
+\left(
3\vartheta_{\mathbf{i}}{}^{4}
-\tfrac{1}{4}\vartheta_{\mathbf{i}}{}^{5}
\right)A^{\mathbf{i}}\wedge C\, .
\end{eqnarray}

The field strengths transform covariantly under the gauge transformations

\begin{eqnarray}
\delta_{\Lambda} A^{0}
& = & 
-\mathfrak{D}\Lambda^{0}
+\left(
3\vartheta_{\mathbf{i}}{}^{4}
+\tfrac{1}{2}\vartheta_{\mathbf{i}}{}^{5}
\right)\Lambda^{i}\, ,
\\
& & \nonumber \\
\delta_{\Lambda} A^{\mathbf{i}}
& = & 
-\mathfrak{D}\Lambda^{\mathbf{i}}
+\vartheta_{0}{}^{m}(T_{m}^{(3)})_{\mathbf{j}}{}^{\mathbf{i}}\Lambda^{j}
-\tfrac{3}{4}\delta_{1}{}^{\mathbf{i}}\vartheta_{0}{}^{5}\Lambda^{1}\, ,
\\
& & \nonumber \\
\delta_{\Lambda} B^{i}
& = & 
-\mathfrak{D}\Lambda^{i}
+F^{0}\wedge \Lambda^{\mathbf{i}}
+F^{\mathbf{i}} \Lambda^{0}
+\tfrac{1}{2}\left(
A^{0}\wedge \delta_{\Lambda} A^{\mathbf{i}}
+A^{\mathbf{i}}\wedge \delta_{\Lambda} A^{0}
\right)
\nonumber \\
& & \nonumber \\
& & 
+\varepsilon^{i\mathbf{j}}\left(
3\vartheta_{\mathbf{j}}{}^{4}
-\tfrac{1}{4}\vartheta_{\mathbf{j}}{}^{5}
\right)\Lambda\, ,
\\
& & \nonumber \\
\delta_{\Lambda} \left( C -\tfrac{1}{6}\varepsilon_{\mathbf{ij}}A^{0\mathbf{ij}}\right)
& = &
-\mathfrak{D}\Lambda
-\varepsilon_{\mathbf{i}j}\left(
 \Lambda^{\mathbf{i}}H^{j}
+F^\mathbf{i}\wedge \Lambda^{j}
-\delta_{\Lambda} A^{\mathbf{i}}\wedge B^{j}
\right)
\nonumber \\
& & \nonumber \\
& &
-\tfrac{1}{2}\varepsilon_{\mathbf{ij}}A^{0\mathbf{i}}\delta_{\Lambda}A^{\mathbf{j}}
+Z \tilde{\Lambda}\, ,
\end{eqnarray}

\noindent
where the covariant derivatives of the different gauge parameters are given by 

\begin{eqnarray}
\mathfrak{D}\Lambda^{0}
& = & 
d\Lambda^{0}
+\left(
3\vartheta_{\mathbf{i}}{}^{4}
+\tfrac{1}{2}\vartheta_{\mathbf{i}}{}^{5}
\right)A^{\mathbf{i}}\Lambda^{0}\, ,
\\
& & \nonumber \\
\mathfrak{D}\Lambda^{\mathbf{i}}
& = &
d\Lambda^{\mathbf{i}}
+\vartheta_{0}{}^{m}(T_{m}^{(3)})_{\mathbf{j}}{}^{\mathbf{i}}A^{0}\Lambda^{\mathbf{j}}
-\tfrac{3}{4}\delta_{1}{}^\mathbf{i}\vartheta_{0}{}^{5} A^{0}\Lambda^{\mathbf{1}}
+\tfrac{3}{4}\varepsilon^{\mathbf{ij}}\varepsilon_{\mathbf{kl}}\vartheta_{\mathbf{j}}{}^{5}
A^{\mathbf{k}}\Lambda^{\mathbf{l}}\, ,
\\
& & \nonumber \\
\mathfrak{D}\Lambda^{i}
& = &
d\Lambda^{i}
+\vartheta_{0}{}^{m}(T_{m}^{(2)})_{j}{}^{i} A^{0} \wedge \Lambda^{j}
+\left(
3\vartheta_{\mathbf{k}}{}^{4}
-\tfrac{1}{4}\vartheta_{\mathbf{k}}{}^{5}
\right)
A^{\mathbf{k}}\wedge \Lambda^{i}
\nonumber \\
& & \nonumber \\
& &
+\tfrac{3}{4}\delta_{\mathbf{j}}{}^{i}\vartheta_{\mathbf{k}}{}^{5}
A^{\mathbf{j}}\wedge \Lambda^{k}\, ,
\\
& & \nonumber \\
\mathfrak{D}\Lambda
& = &
d\Lambda
-\tfrac{3}{4}\vartheta_{0}{}^{5} A^{0}\wedge\Lambda
+\left(
3\vartheta_{\mathbf{i}}{}^{4}
-\tfrac{1}{4}\vartheta_{\mathbf{i}}{}^{5}
\right) A^{\mathbf{i}}\wedge\Lambda\, .
\end{eqnarray}

The supersymmetry transformation rules of the fermion fields are given by 

\begin{eqnarray}
\delta_{\epsilon}\psi_{\mu}
& = & 
\mathfrak{D}_{\mu}\epsilon
+f\gamma_{\mu} \epsilon
+k\gamma_{\mu} \epsilon^{*}
+\tfrac{i}{8\cdot 2!}e^{-\frac{2}{\sqrt{7}}\varphi}
\left(\tfrac{5}{7}\gamma_{\mu}\gamma^{(2)} 
-\gamma^{(2)}\gamma_{\mu} \right)F^{0}\epsilon  
\nonumber \\
& & \nonumber \\
& & 
-\tfrac{1}{8\cdot 2!}e^{\frac{3}{2\sqrt{7}}\varphi+\frac{1}{2}\phi}
\left(\tfrac{5}{7}\gamma_{\mu}\gamma^{(2)} 
-\gamma^{(2)}\gamma_{\mu} \right)(F^{1}-\tau F^{2})\epsilon^{*}  
\nonumber \\
& & \nonumber \\
& & 
-\tfrac{i}{8\cdot 3!}e^{-\frac{1}{2\sqrt{7}}\varphi}
\left(\tfrac{3}{7}\gamma_{\mu}\gamma^{(3)} 
+\gamma^{(3)}\gamma_{\mu} \right)(H^{1}-\tau H^{2})\epsilon^{*}  
\nonumber \\
& & \nonumber \\
& & 
-\tfrac{1}{8\cdot 4!}e^{\frac{1}{\sqrt{7}}\varphi}
\left(\tfrac{1}{7}\gamma_{\mu}\gamma^{(4)} 
-\gamma^{(4)}\gamma_{\mu} \right)G \epsilon\, ,
\end{eqnarray}

\begin{eqnarray}
\delta_{\epsilon}\tilde{\lambda}
& = & 
i\not\!\!\mathfrak{D} \varphi\epsilon^{*} +\tilde{g}\epsilon  +\tilde{h}\epsilon^{*}
-\tfrac{1}{\sqrt{7}}e^{-\frac{2}{\sqrt{7}}\varphi}\not\! F^{0}\epsilon^{*}
-\tfrac{3i}{2\cdot 2!\sqrt{7}}e^{\frac{3}{2\sqrt{7}}\varphi +\frac{1}{2}\phi}
(\not\! F^{1}-\tau^{*}\not\! F^{2})\epsilon
\nonumber \\
& & \nonumber \\
& & 
-\tfrac{1}{2\cdot 3!\sqrt{7}}e^{-\frac{1}{2\sqrt{7}}\varphi +\frac{1}{2}\phi}
(\not\!\! H^{1}-\tau^{*}\not\!\! H^{2})\epsilon
-\tfrac{i}{4!\sqrt{7}}e^{\frac{1}{\sqrt{7}}\varphi}\not\! G\epsilon^{*}\, ,
\end{eqnarray}

\begin{eqnarray}
\delta_{\epsilon}\lambda
& = & 
-e^{\phi}\not\!\!\mathfrak{D} \tau\epsilon^{*} +g\epsilon  +h\epsilon^{*}
-\tfrac{i}{2\cdot 2!}e^{\frac{3}{2\sqrt{7}}\varphi +\frac{1}{2}\phi}
(\not\! F^{1}-\tau\not\! F^{2})\epsilon
\nonumber \\
& & \nonumber \\
& & 
+\tfrac{1}{2\cdot 3!}e^{-\frac{1}{2\sqrt{7}}\varphi +\frac{1}{2}\phi}
(\not\!\! H^{1}-\tau\not\!\! H^{2})\epsilon\, ,
\end{eqnarray}

\noindent
where

\begin{eqnarray}
\mathfrak{D}_{\mu}\epsilon  
 & = & 
\left\{ 
\nabla_{\mu}
+\tfrac{i}{2}
\left[
\tfrac{1}{2}e^{\phi}
\mathfrak{D}^{5}_{\mu}\chi
+A^{I}{}_{\mu}\vartheta_{I}{}^{m}\mathcal{P}_{m}
\right]
+\tfrac{9}{14}\gamma_{\mu}\not\!\!A^{I}\vartheta_{I}{}^{4}
\right\}\epsilon\, ,
\\
& & \nonumber \\
\mathfrak{D}^{5}_{\mu}\chi
& = & 
\partial_{\mu}\chi
-\tfrac{3}{4}A^{I}{}_{\mu}\vartheta_{I}{}^{5} \chi\, ,
\end{eqnarray}

\noindent
and where the fermion shifts are given by 

\begin{eqnarray}
f
& = &
\tfrac{1}{14}e^{\frac{2}{\sqrt{7}}\varphi}
\left(
\vartheta_{0}{}^{m}\mathcal{P}_{m}
+\tfrac{3i}{2}\vartheta_{0}{}^{5}
\right)\, ,
\\
& & \nonumber \\
k
& = &
-\tfrac{9i}{14}e^{-\frac{3\varphi}{2\sqrt{7}}+\frac{\phi}{2}}
\left( \vartheta_{\mathbf{1}}{}^{4}\tau +\vartheta_{\mathbf{2}}{}^{4}
\right)\, ,
\\
& & \nonumber \\
\tilde{g}
& = &
e^{-\frac{3\varphi}{2\sqrt{7}}+\frac{\phi}{2}}
\left[
\tfrac{6}{\sqrt{7}}
\left( 
\vartheta_{\mathbf{1}}{}^{4}\tau^{*} +\vartheta_{\mathbf{2}}{}^{4}
\right)
+\tfrac{\sqrt{7}}{4}
\left(
\vartheta_{\mathbf{1}}{}^{5}\tau^{*} +\vartheta_{\mathbf{2}}{}^{5}
\right)
\right]\, ,
\\
& & \nonumber \\
\tilde{h}
& = &
\tfrac{4}{\sqrt{7}}e^{\frac{2}{\sqrt{7}}\varphi}
\left(
\tfrac{3}{16}\vartheta_{0}{}^{5}+\vartheta_{0}{}^{m}\mathcal{P}_{m}
\right)\, ,
\\
& & \nonumber \\
g
& = &
\tfrac{3}{4}e^{-\frac{3\varphi}{2\sqrt{7}}+\frac{\phi}{2}}
\left(
\vartheta_{\mathbf{1}}{}^{5} \tau +\vartheta_{\mathbf{2}}{}^{5}
\right)\, ,
\\
& & \nonumber \\
h
& = &
i e^{\frac{2\varphi}{\sqrt{7}}+\phi}
\left(
\vartheta_{0}{}^{m} k_{m}{}^{\tau}
-\tfrac{3}{4}\vartheta_{0}{}^{5} \tau
\right)\, .
\end{eqnarray}

The supersymmetry transformations of the bosonic fields are

\begin{eqnarray}
\delta_{\epsilon}\varphi 
& = & 
-\tfrac{i}{4}\bar{\epsilon}\tilde{\lambda}^{*}+\mathrm{h.c.}\, ,
\\
& & \nonumber \\  
\delta_{\epsilon}\tau 
& = & 
-\tfrac{1}{2}e^{-\phi}\bar{\epsilon}^{*}\lambda\, ,
\end{eqnarray}

\begin{eqnarray}
\delta_{\epsilon}A^{0}{}_{\mu}
& = & 
\tfrac{i}{2}e^{\frac{2}{\sqrt{7}}\varphi}\bar{\epsilon}
\left(\psi_{\mu} -\tfrac{i}{\sqrt{7}}\gamma_{\mu}\tilde{\lambda}^{*}\right) 
+\mathrm{h.c.}
\, ,
\\ 
& & \nonumber \\ 
\delta_{\epsilon}A^{\mathbf{1}}{}_{\mu}
& = & 
\tfrac{i}{2}\tau^{*}e^{-\frac{3}{2\sqrt{7}}\varphi+\frac{1}{2}\phi}
\left(
\bar{\epsilon}^{*}\psi_{\mu} 
-\tfrac{i}{4}\bar{\epsilon}\gamma_{\mu}\lambda
+\tfrac{3i}{4\sqrt{7}}\bar{\epsilon}^{*}\gamma_{\mu}\tilde{\lambda}^{*}
\right)
+\mathrm{h.c.}
\, ,
\\ 
& & \nonumber \\ 
\delta_{\epsilon}A^{\mathbf{2}}{}_{\mu}
& = & 
\tfrac{i}{2}e^{-\frac{3}{2\sqrt{7}}\varphi+\frac{1}{2}\phi}
\left(
\bar{\epsilon}^{*}\psi_{\mu} 
-\tfrac{i}{4}\bar{\epsilon}\gamma_{\mu}\lambda
+\tfrac{3i}{4\sqrt{7}}\bar{\epsilon}^{*}\gamma_{\mu}\tilde{\lambda}^{*}
\right)
+\mathrm{h.c.}
\end{eqnarray}

\begin{eqnarray}
\delta_{\epsilon}B^{1}
& = &
\tau^{*} e^{\frac{1}{2\sqrt{7}}\varphi +\frac{1}{2}\phi} 
\left[ 
\bar{\epsilon}^{*}\gamma_{[\mu}\psi_{\nu]} 
-\tfrac{i}{8}\bar{\epsilon}\gamma_{\mu\nu}\lambda 
-\tfrac{i}{8\sqrt{7}}\bar{\epsilon}^{*}\gamma_{\mu\nu}\tilde{\lambda}^{*}
\right]   
+\mathrm{h.c.}
\nonumber \\
& & \nonumber \\
& & 
-\delta^{1}{}_{\mathbf{i}}\left(A^{0}{}_{[\mu|}\delta_{\epsilon}A^{\mathbf{i}}{}_{|\nu]}
+A^{\mathbf{i}}{}_{[\mu|}\delta_{\epsilon}A^{0}{}_{|\nu]}\right)\, ,
\\
& & \nonumber \\  
\delta_{\epsilon}B^{2}
& = &
e^{\frac{1}{2\sqrt{7}}\varphi +\frac{1}{2}\phi} 
\left[ 
\bar{\epsilon}^{*}\gamma_{[\mu}\psi_{\nu]} 
-\tfrac{i}{8}\bar{\epsilon}\gamma_{\mu\nu}\lambda 
-\tfrac{i}{8\sqrt{7}}\bar{\epsilon}^{*}\gamma_{\mu\nu}\tilde{\lambda}^{*}
\right]   
+\mathrm{h.c.}
\nonumber \\
& & \nonumber \\
& & 
-\delta^{2}{}_{\mathbf{i}}
\left(A^{0}{}_{[\mu|}\delta_{\epsilon}A^{\mathbf{i}}{}_{|\nu]}
+A^{\mathbf{i}}{}_{[\mu|}\delta_{\epsilon}A^{0}{}_{|\nu]}\right)\, ,
\end{eqnarray}

\begin{eqnarray}
\delta_{\epsilon}C_{\mu\nu\rho} 
& = & 
-\tfrac{3}{2} e^{-\frac{1}{\sqrt{7}}\varphi} \bar{\epsilon} \gamma_{[\mu\nu}
\left(\psi_{\rho]} +\tfrac{i}{6\sqrt{7}}\tilde{\lambda}^{*}
\right)   
+\mathrm{h.c.}
\nonumber \\
& & \nonumber \\  
& & 
+ 3\delta_{\epsilon}A^{I}{}_{[\mu|}
\left(
g_{Ii}B^{i}{}_{|\nu\rho]}
+\tfrac{2}{3} h_{IJ}{}^{i} g_{Ki} A^{JK}{}_{|\nu\rho]}
\right)\, .
\end{eqnarray}
\chapter{Duality orbits of non-geometric fluxes}
\label{app:duality_orbits}

\section{Different solvable and nilpotent gaugings}\label{appendix_A}

In section~\ref{subsec:Half_Max7} we have studied the T-duality
orbits of gaugings in half-maximal $D=7$ supergravity and for each
of them, we identified the gauge algebra and presented the results in
table~\ref{orbits_halfmax7}. Since there is no exhaustive
classification of non-semisimple algebras of dimension 6, we would
like to explicitly give the form of the algebras appearing in
table~\ref{orbits_halfmax7}.

\subsection*{Solvable algebras}

\subsubsection*{The CSO($2,0,2$) and CSO($1,1,2$) algebras}

The details about these algebras can be found in
ref.~\cite{deRoo:2006ms}; we summarise here some relevant facts.

The six generators are labelled as
$\{t_{0},\,t_{i},\,s_{i},\,z\}_{i=1,2}$, where $t_{0}$ generates
SO($2$) (SO($1,1$)), under which $\{t_{i}\}$ and $\{s_{i}\}$
transform as doublets
\be
\begin{array}{cccc}
\left[t_{0},\,t_{i}\right]\,=\,{\epsilon_{i}}^{j}\,t_{j} & , &
\left[t_{0},\,s_{i}\right]\,=\,{\epsilon_{i}}^{j}\,s_{j} & ,
\end{array}
\ee
where the Levi-Civita symbol ${\epsilon_{i}}^{j}$ has one index
lowered with the metric $\eta_{ij}\,=\,$diag$(\pm 1,1)$ depending on
the two different signatures. $z$ is a central charge appearing in
the following commutators
\be \left[t_{i},\,s_{j}\right]\,=\,\delta_{ij}\,z\ . \ee
The Cartan-Killing metric is diag($\mp 1, \underbrace{0, \cdots,
0}_{\textrm{6 times}}$), where the $\mp$ is again related to the two
different signatures.

\subsubsection*{The $\mathfrak{f}_{1}$ and $\mathfrak{f}_{2}$ algebras}

These are of the form Solv$_{4}\,\times\,$U$(1)^{2}$. The 4
generators of Solv$_{4}$ are labeled by
$\{t_{0},\,t_{i},\,z\}_{i=1,2}$, where $t_{0}$ generates SO($2$)
(SO($1,1$)), under which $\{t_{i}\}$ transform as a doublet
\be \left[t_{0},\,t_{i}\right]\,=\,{\epsilon_{i}}^{j}\,t_{j}\ , \ee
\be \left[t_{i},\,t_{j}\right]\,=\,\epsilon_{ij}\,z\ . \ee
The Cartan-Killing metric is diag($\mp 1, \underbrace{0, \cdots,
0}_{\textrm{6 times}}$).

\subsubsection*{The $\mathfrak{h}_{1}$ and $\mathfrak{h}_{2}$ algebras}

The 6 generators are $\{t_{0},\,t_{i},\,s_{i},\,z\}_{i=1,2}$ and
they satisfy the following commutation relations
\be
\begin{array}{lclc}
\left[t_{0},\,t_{i}\right]\,=\,{\epsilon_{i}}^{j}\,t_{j} & , &
\left[t_{0},\,s_{i}\right]\,=\,{\epsilon_{i}}^{j}\,s_{j}\,+\,t_{i} &
, \\[2mm]
\left[t_{i},\,s_{j}\right]\,=\,\delta_{ij}\,z & , &
\left[s_{i},\,s_{j}\right]\,=\,\epsilon_{ij}\,z & .
\end{array}
\ee
The Cartan-Killing metric is diag($\mp 1, \underbrace{0, \cdots,
0}_{\textrm{6 times}}$).

\subsubsection*{The $\mathfrak{g}_{0}$ algebra}

The 6 generators are $\{t_{0},\,t_{I},\,z\}_{I=1,\cdots,4}$, where
$t_{0}$ transforms cyclically the $\{t_{I}\}$ amongst themselves
such that
\be
\left[\left[\left[\left[t_{I},\,t_{0}\right],\,t_{0}\right],\,t_{0}\right],\,t_{0}\right]\,=\,t_{I}\
, \ee
and
\be \left[t_{1},\,t_{3}\right]\,=\,\left[t_{2},\,t_{4}\right]\,=\,z\
.\ee
Note that this algebra is solvable and not nilpotent even though its
Cartan-Killing metric is \emph{completely zero}.

\subsection*{Nilpotent algebras}

\subsubsection*{The CSO($1,0,3$) algebra}

The details about this algebra can be again found in
ref.~\cite{deRoo:2006ms}; briefly summarizing, the 6 generators are
given by  $\{t_{m},\,z^{m}\}_{m=1,2,3}$ and they satisfy the
following commutation relations
\be \left[t_{m},\,t_{n}\right]\,=\,\epsilon_{mnp}\,z^{p}\ ,\ee
with all the other brackets being vanishing. The order of nilpotency
of this algebra is 2.

\subsubsection*{The $\mathfrak{l}$ algebra}

The 6 generators $\{t_{1},\cdots,\,t_{6}\}$ satisfy the following
commutation relations
\be
\begin{array}{lclclc}
 \left[t_{1},\,t_{2}\right]\,=\,t_{4} & , &
 \left[t_{1},\,t_{4}\right]\,=\,t_{5} & , &
 \left[t_{2},\,t_{4}\right]\,=\,t_{6} & .
\end{array}
\ee
The corresponding central series reads
\be
\begin{array}{lclclcl}
 \left\{t_{1},\,t_{2},\,t_{3},\,t_{4},\,t_{5},\,t_{6}\right\} &\supset&
 \left\{t_{4},\,t_{5},\,t_{6}\right\} &\supset&
 \left\{t_{5},\,t_{6}\right\} &\supset& \left\{0\right\}\ ,
\end{array}
\ee
from which we can immediately conclude that its nilpotency order is
3.

\section{SO($2,2$) and SO($3,3$) 't Hooft symbols}\label{appendix_B}

In section~\ref{sec:DFT} we discuss the origin of a given flux
configuration from DFT backgrounds specified by twist matrices $U$.
The deformations of half-maximal supergravity in $D=10-d$ which can
be interpreted as the gauging of a subgroup of the T-duality group
O($d,d$) can be described by a 3-form of O($d,d$) $f_{ABC}$ which
represents a certain (non-)geometric flux configuration.

In $D=8$ and $D=7$, the T-duality group happens to be isomorphic to
SL($2)\,\times\,$SL($2$) and SL($4$) respectively. As a consequence,
in order to explicitly relate flux configurations and embedding
tensor orbits, we need to construct the mapping between T-duality
irrep's and irrep's of SL($2)\,\times\,$SL($2$) and SL($4$)
respectively.

\subsection*{From the $(\textbf{2},\textbf{2})$ of SL($2)\,\times\,$SL($2$) to the $\textbf{4}$ of SO($2,2$)}

The 't Hooft symbols $\left(G_A\right)^{\alpha i}$ are invariant
tensors which map the fundamental representation of SO($2,2$) (here
denoted by $A$), into the $(\textbf{2},\textbf{2})$ of
SL($2)\,\times\,$SL($2$)
\be v^{\alpha i}\,=\,\left(G_A\right)^{\alpha i}\,v^{A}\,,
\label{vec2vecvec}\ee
where $v^{A}$ denotes a vector of SO($2,2$) and the indices $\alpha$
and $i$ are raised and lowered by means of $\epsilon_{\alpha\beta}$
and $\epsilon_{ij}$ respectively. $\left(G_A\right)^{\alpha i}$ and
$\left(G_A\right)_{\alpha i}$ satisfy the following identities
\bea
&&\left(G_A\right)_{\alpha i}\,\left(G_B\right)^{\alpha i}\,=\,\eta_{AB}\ ,\\
&&\left(G_A\right)^{\alpha i}\,\left(G^{A}\right)^{\beta
j}\,=\,\epsilon^{\alpha \beta}\,\epsilon^{ij}\ , \eea
where $\eta_{AB}$ is the SO($2,2$) metric.

After choosing light-cone coordinates for SO($2,2$), our choice for
the tensors $\left(G_A\right)^{\alpha i}$ is the following
\begin{align}
\left(G_1\right)^{\alpha i} &= \left(
  \begin{array}{cc}
  0 & 0 \\
  0 & 1
  \end{array}
  \right)\,\,,
& \left(G_2\right)^{\alpha i} &= \left(
  \begin{array}{cc}
  0 & 1 \\
  0 & 0
  \end{array}
  \right)\,\,,
\\
\left(G_{\bar{1}}\right)^{\alpha i} &= \left(
  \begin{array}{cc}
  1 & 0 \\
  0 & 0
  \end{array}
  \right)\,\,,
& \left(G_{\bar{2}}\right)^{\alpha i} &= \left(
  \begin{array}{cc}
  0 & 0 \\
  -1 & 0
  \end{array}
  \right)\,\,.
\end{align}
By making use of the mapping \eqref{vec2vecvec}, we can rewrite the
structure constants $(X_{\alpha i})_{\beta j}{}^{\gamma k}$ as a
3-form of SO($2,2$) as follows:
\begin{align}
 f_{ABC}
&= (X_{\alpha i})_{\beta j}{}^{\gamma k} (G_{A})^{\alpha i}
(G_{B})^{\beta j} (G_{C})_{\gamma k}\,\,. \label{X2f_D=8}
\end{align}

\subsection*{From the $\textbf{6}$ of SL($4$) to the $\textbf{6}$ of SO($3,3$)}

The 't Hooft symbols $\left(G_A\right)^{mn}$ are invariant tensors
which map the fundamental representation of SO($3,3$), \emph{i.e.}
the \textbf{6} into the anti-symmetric two-form of SL($4$)
\be v^{mn}\,=\,\left(G_A\right)^{mn}\,v^{A}\,,
\label{vec2two-form}\ee
where $v^{A}$ denotes a vector of SO($3,3$). The two-form irrep of
SL($4$) is real due to the role of the Levi-Civita tensor relating
$v^{mn}$ to $v_{mn}$
\be v_{mn}\,=\,\frac{1}{2}\,\epsilon_{mnpq}\,v^{pq}\,. \ee
The 't Hooft symbols with lower SL($4$) indices
$\left(G_A\right)_{mn}$ carry out the inverse mapping of the one
given in \eqref{vec2two-form}. The tensors $\left(G_A\right)^{mn}$
and
$\left(G_A\right)_{mn}=\frac{1}{2}\,\epsilon_{mnpq}\,\left(G_A\right)^{pq}$
satisfy the following identities
\bea
&&\left(G_A\right)_{mn}\,\left(G_B\right)^{mn}\,=\,2\,\eta_{AB}\,,\\
&&\left(G_{A}\right)_{mp}\,\left(G_{B}\right)^{pn}\,+\left(G_{B}\right)_{mp}\,\left(G_{A}\right)^{pn}\,=\,-\delta_m^{n}\,\eta_{AB}\,,\\
&&\left(G_A\right)_{mp}\,\left(G_B\right)^{pq}\,\left(G_C\right)_{qr}\,\left(G_D\right)^{rs}\,\left(G_E\right)_{st}\,\left(G_F\right)^{tn}\,=\,\delta_m^{n}\,\epsilon_{ABCDEF}\,,
\eea
where $\eta_{AB}$ and $\epsilon_{ABCDEF}$ are the SO($3,3$) metric
and Levi-Civita tensor respectively.

After choosing light-cone coordinates for SO($3,3$) vectors, our
choice of the 't Hooft symbols is
\begin{align}
\left(G_1\right)^{mn}
&=
\left(
  \begin{array}{cccc}
  0 & -1 & 0 & 0\\
  1 & 0 & 0 & 0\\
  0 & 0 & 0 & 0\\
  0 & 0 & 0 & 0
  \end{array}
  \right)\,\,,
&
\left(G_2\right)^{mn}
&=
\left(
  \begin{array}{cccc}
  0 & 0 & -1 & 0\\
  0 & 0 & 0 & 0\\
  1 & 0 & 0 & 0\\
  0 & 0 & 0 & 0
  \end{array}
  \right)\,\,,
\\
\left(G_3\right)^{mn}
&=
\left(
  \begin{array}{cccc}
  0 & 0 & 0 & -1\\
  0 & 0 & 0 & 0\\
  0 & 0 & 0 & 0\\
  1 & 0 & 0 & 0
  \end{array}
  \right)\,\,,
&
\left(G_{\bar1}\right)^{mn}
&=
\left(
  \begin{array}{cccc}
  0 & 0 & 0 & 0\\
  0 & 0 & 0 & 0\\
  0 & 0 & 0 & -1\\
  0 & 0 & 1 & 0
  \end{array}
  \right)\,\,,
\\
\left(G_{\bar2}\right)^{mn}
&=
\left(
  \begin{array}{cccc}
  0 & 0 & 0 & 0\\
  0 & 0 & 0 & -1\\
  0 & 0 & 0 & 0\\
  0 & 1 & 0 & 0
  \end{array}
  \right)\,\,,
&
\left(G_{\bar3}\right)^{mn}
&=
\left(
  \begin{array}{cccc}
  0 & 0 & 0 & 0\\
  0 & 0 & 1 & 0\\
  0 & -1 & 0 & 0\\
  0 & 0 & 0 & 0
  \end{array}
  \right)\,\,.
\end{align}

Thus, we can rewrite the structure constants in the \textbf{6},
$(X_{mn})_{pq}{}^{rs}$, arising from (\ref{gen_half-max}) as a
3-form of SO($3,3$) as follows:
\begin{align}
 f_{ABC}
&= (X_{mn})_{pq}{}^{rs} (G_{A})^{mn} (G_{B})^{pq} (G_{C})_{rs}\,\,.
\label{X2f_D=7}
\end{align}
\end{appendices}

\chapter*{Publications\markboth{Publications}{Publications}}
\addcontentsline{toc}{chapter}{Publications}
\label{publications}

List of publications arising from the research done during the PhD period.

\begin{enumerate}

\item
J.~Fernandez-Melgarejo and E.~Torrente-Lujan, ``{N=2 SUGRA BPS Multi-center
  solutions, quadratic prepotentials and Freudenthal transformations},'' to appear in JHEP.
\href{http://arxiv.org/abs/1310.4182}{{\tt arXiv:1310.4182 [hep-th]}}.

\item
J.~Fernandez-Melgarejo, T.~Ortin, and E.~Torrente-Lujan, ``{Maximal Nine
  Dimensional Supergravity, General gaugings and the Embedding Tensor},''
  \href{http://dx.doi.org/10.1002/prop.201200039}{{\em Fortsch.Phys.} {\bf 60}
  (2012)  1012--1018},
\href{http://arxiv.org/abs/1209.3774}{{\tt arXiv:1209.3774 [hep-th]}}.

\item
G.~Dibitetto, J.~Fernandez-Melgarejo, D.~Marques, and D.~Roest, ``{Duality
  orbits of non-geometric fluxes},''
  \href{http://dx.doi.org/10.1002/prop.201200078}{{\em Fortsch.Phys.} {\bf 60}
  (2012)  1123--1149},
\href{http://arxiv.org/abs/1203.6562}{{\tt arXiv:1203.6562 [hep-th]}}.

\item
E.~A. Bergshoeff, J.~Fernandez-Melgarejo, J.~Rosseel, and P.~K. Townsend, ``{On
  'New Massive' 4D Gravity},''
  \href{http://dx.doi.org/10.1007/JHEP04(2012)070}{{\em JHEP} {\bf 1204} (2012)
   070},
\href{http://arxiv.org/abs/1202.1501}{{\tt arXiv:1202.1501 [hep-th]}}.

\item
L.~Granda, E.~Torrente-Lujan, and J.~Fernandez-Melgarejo, ``{Non-minimal
  kinetic coupling and Chaplygin gas cosmology},''
  \href{http://dx.doi.org/10.1140/epjc/s10052-011-1704-4}{{\em Eur.Phys.J.}
  {\bf C71} (2011)  1704},
\href{http://arxiv.org/abs/1106.5482}{{\tt arXiv:1106.5482 [hep-th]}}.

\item
J.~Fernandez-Melgarejo, T.~Ortin, and E.~Torrente-Lujan, ``{The general
  gaugings of maximal d=9 supergravity},''
  \href{http://dx.doi.org/10.1007/JHEP10(2011)068}{{\em JHEP} {\bf 1110} (2011)
   068},
\href{http://arxiv.org/abs/1106.1760}{{\tt arXiv:1106.1760 [hep-th]}}.

\item
M.~Picariello, B.~Chauhan, C.~Das, Fernandez-Melgarejo, D.~Montanino, {\em et
  al.}, ``{Neutrino Dipole Moments and Solar Experiments},''
\href{http://arxiv.org/abs/0907.0637}{{\tt arXiv:0907.0637 [hep-ph]}}.

\end{enumerate}

\chapter*{Acknowledgements\markboth{Acknowledgements}{Acknowledgements}}
\addcontentsline{toc}{chapter}{Acknowledgements}

I would very much like to thank my supervisor Emilio Torrente for his guidance, enthousiasm, patience and uncommon willingness.

I am indebted to Tom\'as Ort\'{\i}n for his friendly advice and help in various respects. 

My deepest thanks to Pepe Bernab\'eu, for his moral support at the beginning of this period.

I would like to thank my collaborators Eric Bergshoeff, David Berman and Diederik Roest for pleasant and fruitful collaborations. 

In addition, I would like to thank the members of my Ph.D. committee, Eric Bergshoeff, David Berman, Julio Guerrero, Yolanda Lozano and Tom\'as Ort\'{\i}n, for their efforts.

I am grateful to the Physics Department of the University of Murcia for the
opportunity to perform this research during the last four years and to the CTN Groningen, the Physics Department of Queen Mary University of London, IFT-UAM/CSIC, the Theoretical Physics Department of the University of Valencia, IFIC and CERN for their warm hospitality while part of this thesis was done.

This work has been supported by the Spanish Ministry
of Education FPU grant AP2008-00919.

\addcontentsline{toc}{chapter}{References}
\markboth{\MakeUppercase{Abstract}}{}

\bibliography{references}		
\bibliographystyle{utphys}	


\end{document}